\documentclass[journal=chreay,manuscript=review]{achemso}

\usepackage[version=3]{mhchem} 

\def\lapprox{{_<\atop{^\sim}}}

\author{Ewine F.\ van Dishoeck} 
\altaffiliation{Max Planck Institute
  for Extraterrestrial Physics, Garching, Germany}
\email{ewine@strw.leidenuniv.nl} 
\affiliation[Leiden University]
{Leiden Observatory, Leiden University, P.O. Box 9513, 2300 RA The
  Netherlands} 
\author{Eric Herbst} \email{eh2ef@virginia.edu}
\affiliation[University of Virginia] {Department of Chemistry,
  University of Virginia, PO Box 400319, Charlottesville, VA 22904, USA}
\author{David A. Neufeld} \email{neufeld@pha.jhu.edu}
\affiliation[Johns Hopkins University] {Department of Physics and Astronomy,
  Johns Hopkins University, 3400 North Charles Street, Baltimore, MD 21218, USA}

\title[]
{Interstellar water chemistry: from laboratory to observations}

\begin{document}


\newpage

\setcounter{tocdepth}{3}
\tableofcontents



\newpage

\section{Introduction}

Water is a fundamental molecule in the universe. On Earth, water
exists in all three phases: gas, liquid and solid (ice). Liquid water
facilitates many chemical reactions as a solvent. Much of the geology
on Earth is shaped by water. Life on Earth, and similar life elsewhere
on Earth-like planets in space, would not be possible without water.

Water consists of hydrogen and oxygen atoms, which are the first and
third most abundant elements in the universe. According to the
standard cosmology \citep{Steigman07,WMAP13,Planck13}, hydrogen was
formed immediately after the Big Bang, some 13.8 billion years ago
(13.8 Gyr). After about 1 microsecond, protons and neutrons emerged
from the cooling quark-gluon plasma, and after 1 second the baryonic
content was frozen. Primordial nucleosynthesis produced D, He (the
second most abundant element) and Li nuclei after about 3 minutes, and
they recombined to neutral atoms after about $10^5$ yr (0.1
Myr). Heavier atoms including oxygen were produced much later in the
evolution of the universe (at least a few $10^8$ yr), by nuclear
fusion of H and He in the central parts of massive stars. When these
stars have exhausted their supply of nuclear fuel, they develop a wind
or explode as a supernova, and enrich the space between the stars with
heavy elements.

\begin{figure}
  \includegraphics[angle=0,width=0.7\textwidth]{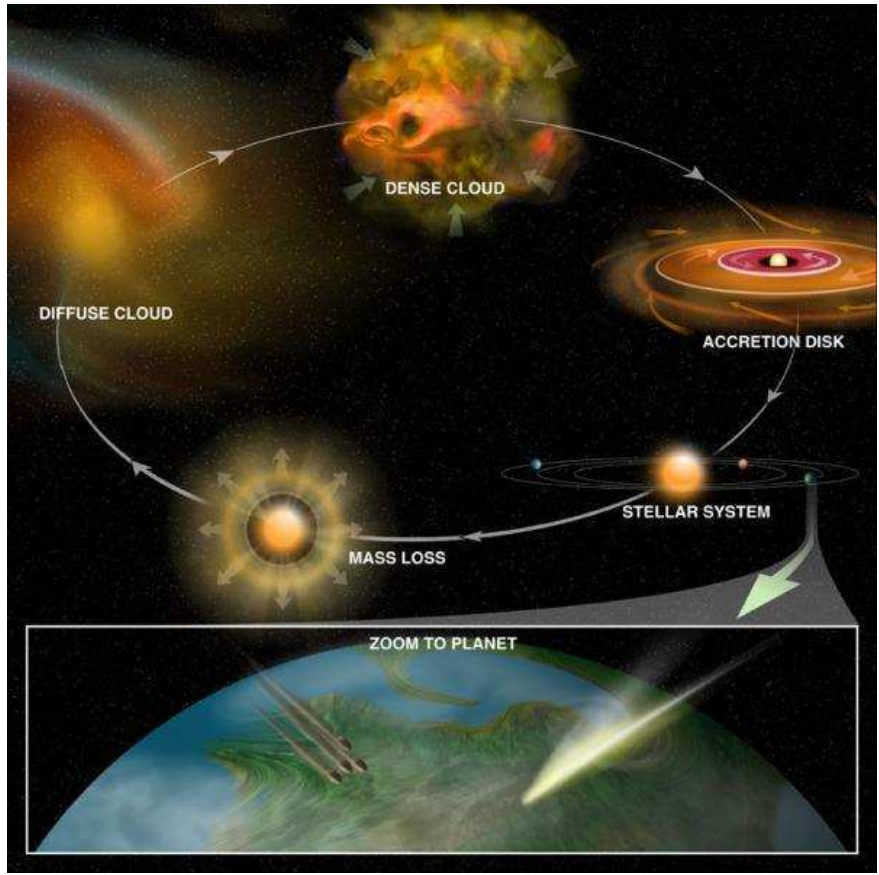}
  \caption{Schematic of different types of interstellar clouds and the
    cycle of the formation and death of stars. Reprinted with
    permission from Bill Saxton, NRAO/AUI/NSF.}
  \label{fig:cycle}
\end{figure}

Even after several generations of stars, the abundance of hydrogen
atoms is still some 2000 times higher than that of oxygen atoms in our
solar neighborhood.  Molecules such as water form in localized regions
of cold and dense gas and dust between the stars (see
Fig.~\ref{fig:cycle}). Typical densities in such interstellar clouds
are 10$^4$ hydrogen molecules per cm$^3$ and temperatures can be as
low as 8 K (pressure $\sim 10^{-14}$ mbar). The clouds also contain
small 0.001--0.1 $\mu$m sized solid dust particles or `grains',
consisting of amorphous silicates and carbonaceous material. Grains
are important because they absorb and scatter the ultraviolet (UV)
radiation produced by stars and thereby protect molecules from
dissociating photons. Reactions on the surfaces of the dust particles
promote the formation of molecules, especially of H$_2$ and other
hydrogen-rich species like H$_2$O.

Interstellar space is a gigantic ultra-high vacuum laboratory with
densities low enough that molecules can form only through kinetic
two-body processes in the gas phase.  H$_2$O is one of the simplest molecules
consisting of hydrogen and oxygen atoms. Hence one would expect the
chemistry of water in space to be straight-forward and to have been
fully understood and characterized decades ago. While the basic
framework of interstellar water chemistry was indeed established in
the 1970's and recognized to involve both gas-phase
\citep{Solomon72,Herbst73} and solid-state chemistry
\citep{VandeHulst96,Allen77,Tielens82}, many of the key chemical
processes have been measured in the laboratory only in the last
decade. Tests of these chemical processes are possible only now thanks
to a number of powerful telescopes culminating with the {\it Herschel
  Space Observatory}, which have high enough sensitivity, spatial or
spectral resolution to detect water in various environments
\citep{Cernicharo05,Melnick09,Hollenbach09,Bergin12}.

\begin{figure}
\includegraphics[angle=0,width=0.6\textwidth]{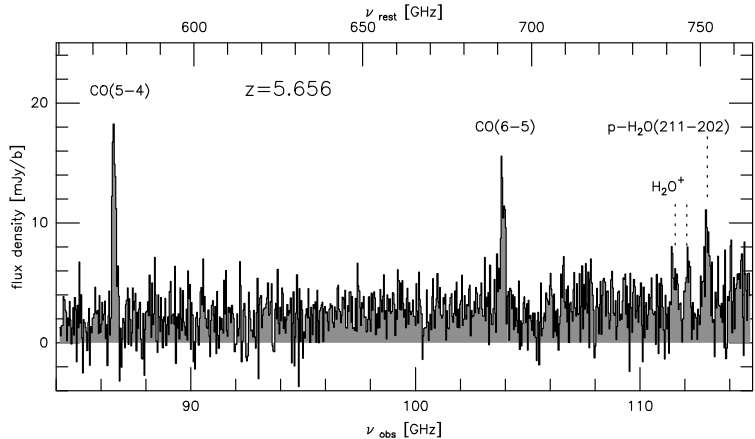}
\caption{ ALMA spectrum of the distant galaxy SPT0346-52 showing
  detections of pure rotational lines of neutral and ionized water,
  together with CO. The galaxy has a redshift $z$=5.66 (transition
  shifted to $\nu_{\rm obs}=\nu_{\bf rest}/(1+z)$). This redshift
  corresponds to an age of just 1.0 Gyr since the Big Bang. Reprinted
  with permission from Reference 16. Copyright 2013 American Astronomical
  Society.}
  \label{fig:waterhighz}
\end{figure}

\subsection{Water observations}

Interstellar water vapor was discovered in 1969 in the Orion nebula by
the group of Charles Townes \citep{Cheung69}. This detection was
somewhat accidental since it was found that water can emit anomalously
strong radiation at 22 GHz (1.4 cm) via the maser process (see
\S~\ref{sect:masers}). This self-amplifying process produces very
bright and sharp lines which can be observed even in external
galaxies. In fact, the best evidence for the existence of a black hole
outside our Milky Way is based on tracking accurately the motions of
water masers in the galaxy NGC 4258 \citep{Miyoshi95}.

\begin{figure}
  \includegraphics[angle=0,width=0.7\textwidth]{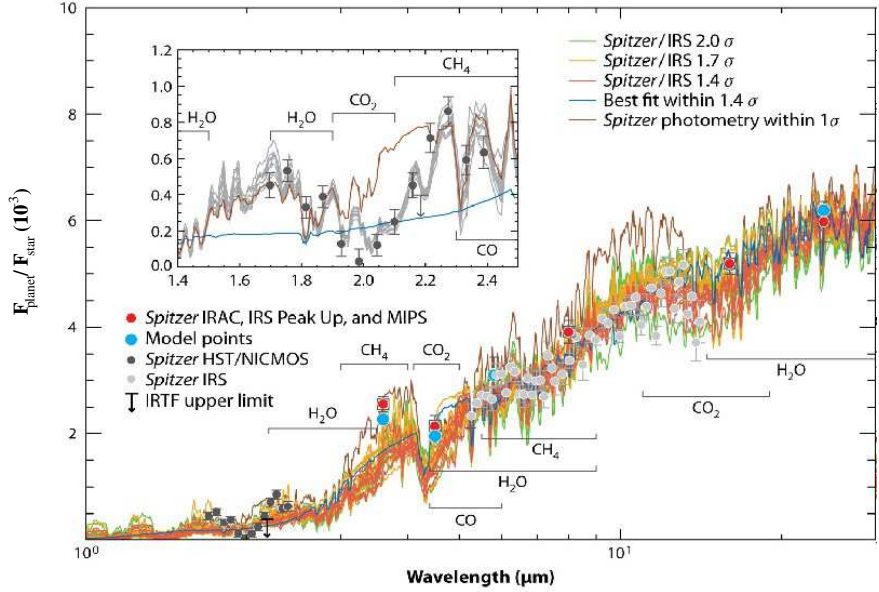}
  \caption{ Spectrum of the exoplanet HD 189733b observed by various
    instruments (black and blue dots and squares with error bars),
    together with various model spectra suggesting the presence of
    water in absorption. Adapted with
    permission from Reference 17, copyright Annual Reviews and
    from Reference 18, copyright 2009 American Astronomical Society.}
  \label{fig:waterplanet}
\end{figure}

While this unexpected discovery and application of water is
interesting in itself, it tells us little about the actual chemistry
and abundance of water in space. To probe the bulk of gaseous water,
observations of lines that are thermally excited and do not exhibit
population inversion are needed. Observations of the majority of these
pure rotational lines are blocked from the ground by the water that is
present in the Earth's atmosphere. Hence, space missions, starting
with the {\it Infrared Space Observatory} (ISO) and followed by the
{\it Submillimeter Wave Astronomy Satellite} (SWAS), the Swedish-led
satellite Odin and finally {\it Herschel}, have been crucial for our
understanding of the water chemistry (see Table~\ref{tab:telescopes}).
Non-masing water emission has now been detected in many environments
ranging from diffuse clouds to dense planet-forming disks around young
stars in the Milky Way and from nearby galaxies out to the highest
redshifts. Water is also found in cometary comae, the atmospheres of
planets in our own solar system and even those of extra-solar planets,
or exo-planets\cite{Birkby13} (see Figures~\ref{fig:waterhighz} from
\citealt{Weiss13} and \ref{fig:waterplanet} adapted from
\citealt{Seager10} and \citealt{Madhusudhan09} for two extreme
examples of water in the distant and nearby universe, respectively).

At the low densities and high vacuum conditions in space, water exists
either as a gas or in the solid state as ice. Liquid water can only
occur under relatively higher pressures on large solid bodies such as
asteroids or planets; its triple point is at 273 K and 6.1 mbar. Water
gas freezes out as ice around 100 K under interstellar conditions
\citep{Fraser01}. Water ice was detected in 1973 in the infrared
spectra of protostars forming deep inside molecular clouds
\citep{Gillett73}, 
and is now found in dense interstellar clouds throughout our own
and external galaxies \citep{Whittet03}.

This review provides first an overview of the techniques to observe
water in space (\S 1.2), of the types of clouds in which water is
observed (\S 1.3--1.4) and of water spectroscopy and excitation (\S
2). The bulk of the review discusses the various chemical processes
that lead to the formation of water, both in the gas and on the
surfaces of grains (\S 3). Many of these processes also apply to other
molecules, but the focus here is on water. Finally, the different
chemical routes to water are tested against a wide range of
observations, from tenuous molecular clouds in which UV radiation
penetrates to high density regions in which stars and planets are
currently forming (\S 4).  The discussion is limited to H$_2^{16}$O
and its $^{18}$O and $^{17}$O isotopologues. The chemistry of HDO is
similar to that of H$_2$O in many aspects, but there are also
important differences due to fractionation reactions which can
selectively enhance HDO.  Although interesting by themselves,
discussion of these processes is outside the scope of this review. HDO
is summarized in recent reviews by Caselli \& Ceccarelli
\citep{Caselli12aar} and by Ceccarelli et al.\ \citep{Ceccarelli14}.

\begin{figure}
\includegraphics[angle=0,width=0.7\textwidth]{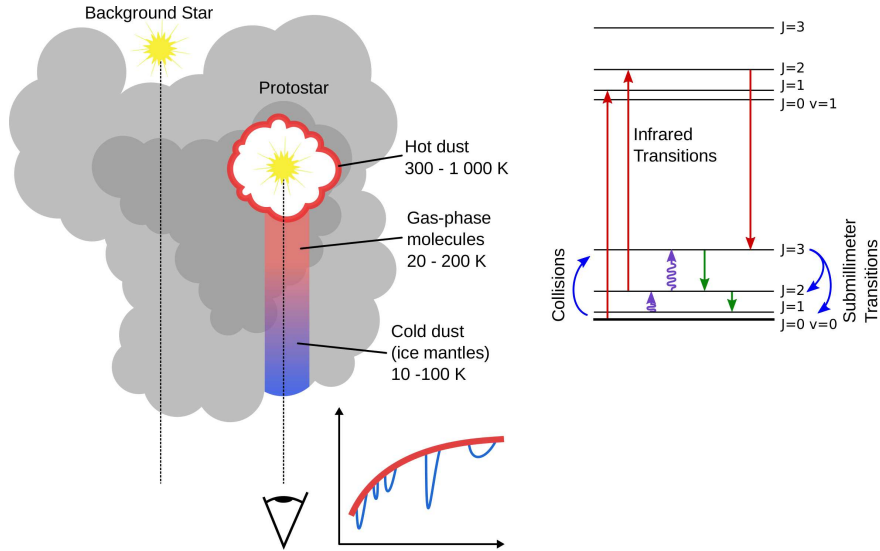}
\caption{ Left: Schematic energy level diagram showing typical pure
  rotational submillimeter and vibration-rotation near-infrared lines,
  together with various radiative and collisional processes that
  excite the molecules. Right: Cartoon illustrating infrared
  absorption line observations of ice or gas toward an embedded
  protostar or a background star. The hot (300--1000 K) dust near the
  star provides the continuum against which cold ice and gas can be
  seen in absorption. Absorption can also occur against cool
  background stars.}
  \label{fig:absem}
\end{figure}

\subsection{Astronomical techniques}
\label{sect:telescopes}

Water can be observed through various transitions over a range of
wavelengths (see \S 2), which require different telescopes,
instruments and detectors as summarized in
Table~\ref{tab:telescopes}. Throughout this paper, we will use the
astronomical convention to indicate submillimeter lines through their
frequencies in GHz and far-infrared lines through their wavelengths in
$\mu$m. Laboratory spectroscopists usually adopt either GHz or
wavenumbers, cm$^{-1}$. For reference, 200/100/50 $\mu$m correspond to
50/100/200 cm$^{-1}$ and 1500/3000/6000 GHz, respectively.

Astronomers express the spatial resolution that can be achieved by
their telescopes in units of seconds of arc, or arcseconds. One arcsec
is 1/3600 of a degree or 1/206265 of a radian ($\pi$/180/3600).  The
nearest interstellar clouds are at a distance of $\sim$ 150 parsec,
with parsec a distance unit employed by astronomers. A parsec is
defined as the distance from the Sun under which the Sun-Earth system
is seen with an angle of 1 arcsec. In practice, 1 pc = $3.086\times
10^{18}$ cm or 3.261 lightyear. Another distance unit often used is
the Astronomical Unit (AU, not to be confused with atomic unit or
arbitrary unit), which is defined as the average Sun-Earth distance, 1
AU= $1.496\times 10^{13}$ cm.  At a distance of 100 pc, 1 arcsec
($''$) corresponds to 100 AU diameter. Thus, a typical beam (spatial
resolution element) of {\it Herschel} of 20$''$ on the sky encompasses
a 3000 AU (0.015 pc) diameter region if one of the closest
star-forming regions at 150 pc distance is observed. For comparison,
the mean orbital radius of Neptune in our own Solar system is at
$\sim$30 AU, so to study a planet-forming region in the nearest clouds
requires a spatial resolution of 0.2$''$. Such high spatial
resolution is only just being achieved by the Atacama Large
Millimeter/submillimeter Array (ALMA). In the following, the main
telescopes for observing water are briefly discussed.

\subsubsection{Description of telescopes}

The {\it Infrared Space Observatory} (ISO) was a cooled 0.6m infrared
telescope in space covering the 2.5--240 $\mu$m range and operative
from 1995--1998. It had two spectrometers on board: the
Short-Wavelength-Spectrometer (SWS, 2.5--45 $\mu$m)\citep{deGraauw96}
and the Long-Wavelength-Spectrometer (LWS, 45--197
$\mu$m)\citep{Clegg96}. Each instrument had a low-resolution grating
and a high-resolution Fabry-Perot mode, which provided modest
resolving powers $R=\lambda/\Delta \lambda$ ranging from $\sim 200$ up
to $\sim 10^4$. The mid- to far-infrared wavelength range is
particularly rich in rotational transitions of H$_2$O and related
hydrides, as well as vibrational bands of gas and ice. ISO
indeed provided a first view of the wealth of H$_2$O features and its
gas-grain chemistry \citep{vanDishoeck04,Cernicharo05}.

\begin{table}
 \caption{Important telescopes for water observations}
  \begin{tabular}{llcc}
    \hline
    Telescope & Wavelength / & Spectral resol-& Spatial  \\
              & Frequency    & ving power $R$ & resolution$^a$   \\
    \hline
    \smallskip
    SWAS & 557 GHz & $10^6$ & $3.2'\times 4.0'$   \\
    Odin & 557 GHz & $10^6$ & $2'$  \\
    Ground cm & 22 GHz & $10^6$ & 0.001--few$''$ \\
    Ground mm & Many, e.g., 183, 380 GHz & $10^7$ & 0.2--20$''$ \\
              & 203$^b$, 391$^b$, 692$^b$ & $10^7$ & 0.2--20$''$ \\
    {\it Herschel}-HIFI & 480-1250 / & $10^7$ & 44$''$--17$''$  \\
                  & 1410-1910 GHz & $10^7$ & 15$''$--9$''$ \\
    {\it Herschel}-PACS & 55-210 $\mu$m & $(1-5)\times  10^3$ & 9.4$''$ \\
    {\it Herschel}-SPIRE & 200-670 $\mu$m & $\sim 10^3$ & 17--42$''$ \\
    {\it Spitzer} & 10--38 $\mu$m & 600 & $10''$ \\
    ISO-SWS & 2.5--45 $\mu$m & 2000, 20000 
         & $14''\times 20''$ to $17\times 40''$\\
    ISO-LWS & 45--197 $\mu$m & 200, 10000 & $\sim 80''$ \\
    Ground 4-10 m optical/ & 2.8--3.3 $\mu$m& $\leq 10^5$ & $\leq 1''$ \\
    \ \ \ \ \ \ \ \ \ \ \ \ \ infrared & 11--14 $\mu$m& $10^4$ & $ 1''$ \\
    \hline
  $^a$ 1 arcmin = 1$'$ = 60$''$.
  $^b$ H$_2^{18}$O.  
  \end{tabular}
  \label{tab:telescopes}
\end{table}

SWAS\citep{Melnick00}, launched in 1998, and Odin\citep{Nordh03},
launched in 2001, were pioneering submillimeter satellites with
antennas of $\sim$0.6 and 1.1m diameter, respectively, operating
primarily in the 1998--2007 period. Both missions were specifically
designed to observe the lowest pure rotational $1_{10}-1_{01}$
transition of H$_2$O at 557 GHz. The receivers were based on the
heterodyne technique, in which the high frequency signal from the sky
is down converted by mixing with a local oscillator signal tuned to a
nearby frequency. The difference signal has a much lower frequency
which can be amplified with standard radio techniques. Heterodyne
receivers naturally have very high spectral resolving power
$R>10^6$. Since the velocity resolution $\Delta$v=$c \Delta
\lambda/\lambda$ with $c$ the speed of light, this value of $R$
corresponds to $<0.3$ km s$^{-1}$ in velocity units. Typical observed
widths of water lines range from a few km s$^{-1}$ up to 100 km
s$^{-1}$, so these spectra fully resolve the kinematics of the gas.

The {\it Herschel Space Observatory}\citep{Pilbratt10} was a 3.5m
diameter telescope, operative from mid-2009 to mid-2013.  It had three
instruments which provided imaging and spectroscopic capabilities in
the 55-672 $\mu$m wavelength range.  As for the earlier missions,
observing water in space was one of the main drivers for the design of
{\it Herschel}. Because the instruments were housed in a superfluid
helium cryostat and the telescope was bigger, the sensitivity of {\it
  Herschel} was much higher than that of SWAS and Odin.  Moreover,
because the sharpness of the images scales inversely with the diameter
of the telescope, the spatial resolution provided by {\it Herschel}
was up to an order of magnitude higher than that of the earlier
missions.

Of the three {\it Herschel} instruments, the results from the
Heterodyne Instrument for the Far Infrared (HIFI) will be highlighted
most in this review \citep{deGraauw10}. HIFI was a very high
resolution ($R$ up to $10^7$) heterodyne spectrometer covering the
490--1250 GHz (600--240 $\mu$m; 16--42 cm$^{-1}$) and 1410--1910 GHz
(210--157 $\mu$m; 47--64 cm$^{-1}$) bands. HIFI observed a single
pixel on the sky at a time.  The Photodetector Array Camera and
Spectrometer (PACS) consisted of a camera and a medium resolution
imaging spectrometer for wavelengths in the range 55--210 $\mu$m (180--47
cm$^{-1}$) \citep{Poglitsch10}. The PACS spectrometer obtained spectra
simultaneously over a limited wavelength range at each pixel of a
$5\times 5$ array. The detector consisted of a $25\times 16$ pixel
Ge:Ga photoconductor array. The Spectral and Photometric Imaging
REceiver (SPIRE) was a camera and a low resolution Fourier Transform
Spectrometer for wavelengths in the range 194-672 $\mu$m (51--15
cm$^{-1}$) using bolometers to detect the radiation.

Most of the water line radiation from space does not reach telescopes
on the ground, however, because it is absorbed by the abundant water
in the Earth's atmosphere.  However, high-lying maser lines of
H$_2^{16}$O such as the $3_{13}-2_{20}$ 183 GHz and $4_{14}-3_{21}$
380 GHz lines, as well as the well known $6_{16}-5_{23}$ line at 22
GHz (1 cm) through which interstellar water was originally discovered,
can be observed from Earth \citep{Cernicharo90}.

Another option is to observe water isotopologs. About 1 in 500 water
molecules in space contain the heavier $^{18}$O isotope.  Some
signatures from H$_2^{18}$O, such as the $3_{13}-2_{20}$ line at 203
GHz, the $4_{14}-3_{21}$ line at 391 GHz and the $5_{32}-4_{41}$ line
at 692 GHz, are able to penetrate the Earth's atmosphere even if
non-masing, and these lines have been detected with ground-based
telescopes such as the 10m Caltech Submillimeter Observatory (CSO),
the 15m James Clerk Maxwell Telescope (JCMT), and the IRAM 30m
telescope. Interferometers are even more powerful: these consist of
arrays of antennas of which the signals are combined to produce a
single, sharper image. Examples include the IRAM Plateau the Bure
Interferometer (PdBI) (6$\times 15$m dishes), the SubMillimeter Array
(SMA) (8$\times 6$m) and most importantly the recently inaugurated
ALMA (54$\times$12m + 12$\times$7 m). Since these ground-based
telescopes and arrays are much bigger and can see up to a 1000 times
sharper than any of the current or past satellites, they allow
astronomers to zoom in on the astronomical sources and map the
location of water isotopologues in detail
\citep{Jorgensen10}. However, ground-based facilities can only observe
thermal lines from warm water ($>$100 K), not from cold water.

The {\it Spitzer Space Telescope} was launched in 2003 and performed
mid-infrared spectroscopic observations from 5--38 $\mu$m until the
mission ran out of coolant in mid-2009. The IRS instrument had modest
resolving power $R=600$ from 10--38 $\mu$m, but only $R\sim 50$ from
5--10 $\mu$m \citep{Houck04}. However, its sensitivity surpassed that
of ISO by nearly three orders of magnitude due to the new generation
of large format Si:As array detectors. {\it Spitzer} was therefore
able to open up a completely new view of warm water in planet-forming
regions by observations of highly excited pure rotational lines. It
was also very well suited to detect solid state bands of water ice in
sources similar to those that led to our own solar system.

A variety of large (4--10m) ground-based optical telescopes have near-
and mid-infrared spectrometers that can observe the vibration-rotation
lines of hot water near 3 $\mu$m as well as highly-excited pure
rotational lines near 10 $\mu$m. Examples of instruments are
Keck-NIRSPEC ($R$=25000), VLT-CRIRES ($R=10^5$), VLT-VISIR ($R=10^4$)
and Gemini-Michelle and -TEXES. Most of them are echelle grating
spectrometers using large format infrared detector arrays in which one
array dimension is used for the spectral range and the other array
dimension for the spatial domain. Thus, these `long-slit'
spectrometers also provide information on the spatial distribution of
the emission on subarcsec scale. In the future, the EXES mid-infrared
instrument ($R=10^5$) on the Stratospheric Observatory for Infrared
Astronomy (SOFIA) \citep{Sofia12}, a 2.7m telescope onboard a B747
airplane, will be a powerful facility as well for water observations
since it flies at 13.7 km above much of the Earth's atmosphere.

At near- and mid-infrared wavelengths, cold ices and gases can be seen
in absorption against a bright continuum. This continuum can be
provided by hot dust close to either a young or old star but it can also be
due to a background star (see Fig.~\ref{fig:absem}).  Absorption
observations probe only a pencil-beam line-of-sight toward the source.
In contrast, submillimeter and far-infrared lines of warm and cold
gaseous water are generally in emission, and can thus be
mapped. Cold water gas can also absorb against line emission
from warmer water at larger distances.

The bound-bound electronic transitions of H$_2$O at UV wavelengths can
in principle be observed using high spectral resolution instruments
(up to $10^5$) on the {\it Hubble Space Telescope} (HST) and the {\it Far
  Ultraviolet Space Explorer} (FUSE), in absorption along the line of
sight to a bright star. However, there are no detections of
interstellar water at UV wavelengths to date \citep{Spaans98}.

\subsection{Types of interstellar and circumstellar clouds}
\label{sect:clouds}

The interstellar medium has a complex and inhomogeneous structure, the
denser concentrations of which are called clouds (see book by
Tielens\citep{Tielens05} and \citealt{Tielens13} for broad
overviews). A summary of typical densities, temperatures and
extinctions (see \S~\ref{sect:radiation}) of various cloud types
discussed here is presented in Table~\ref{tab:clouds}.  The regions
range from low density diffuse clouds in which UV radiation can
readily penetrate to cold dense cores on the verge of collapse to form
a star to the envelopes around dying stars
(Figure~\ref{fig:cycle}). Diffuse and translucent clouds are low
density examples of PhotoDissociation or Photon-Dominated Regions
(PDRs) \citep{Tielens85}; traditional PDR examples are dense clouds
illuminated by a nearby bright star such as seen in Orion. The
terminology indicates that the physics and chemistry are controlled by
far-ultraviolet (FUV, 912--2000 \AA) radiation. In case the nearby
source emits X-rays, the region is called an XDR.

The principal ingredient of clouds is hydrogen, followed by helium
(about 0.10 by number). Oxygen, carbon and nitrogen are present with
abundances of $5\times 10^{-4}$, $3\times 10^{-4}$ and $1\times
10^{-4}$, respectively. The small silicate and carbonaceous dust
grains make up $\sim$1\% of a cloud by mass but only $\sim 10^{-12}$
by number. Hydrogen is primarily in molecular form in the regions
listed in Table~\ref{tab:clouds}, but is difficult to observe
directly. Astronomers therefore usually observe the strong lines from
the abundant CO molecule and its isotopologues as a tracer of H$_2$,
even though this naturally introduces uncertainties.

\begin{figure}
\includegraphics[angle=0,width=0.4\textwidth]{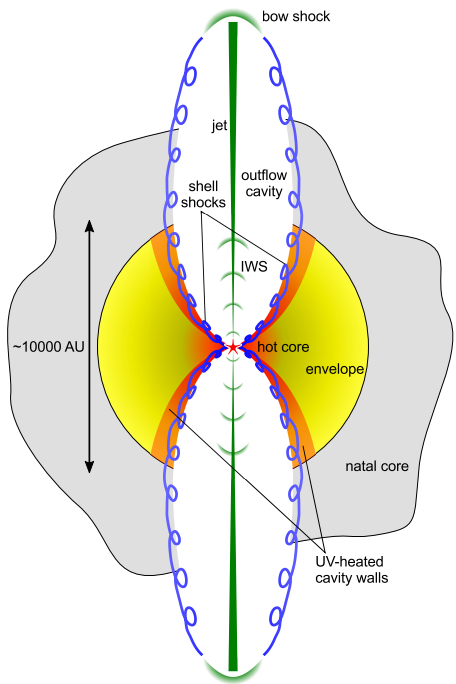}
\caption{Cartoon of a protostellar envelope with the different
  physical components and their nomenclature indicated. IWS stands for
  internal working surfaces. The indicated scale is appropriate for a
  low-mass protostar. On this scale, the $\sim$100 AU radius disk
  surrounding the protostar is not visible. Reprinted with permission from
  Reference 40. Copyright 2011 The University of Chicago
  Press.}
  \label{fig:protostar}
\end{figure}

Much of this review discusses water associated with star-forming
regions in which a cloud is collapsing under its own gravity to form
one or more protostars. In this stage, the young star is deeply
embedded in its natal cloud and is still accreting material from the
envelope and thus growing in mass. The luminosity from the young star
heats the envelope and sets up a radial gradient in temperature, with
values in the inner envelope (sometimes also called `hot core') well
above 100 K and temperatures in the colder outer regions down to 10
K. Because the cloud always has some angular momentum, material cannot
continue to fall in radially but part of it ends up in a rotating disk
around the young star (Fig.~\ref{fig:cycle}).  The protostar also
develops jets and winds which can escape in a direction perpendicular
to the disk. When these high-velocity jets and winds interact with the
quiescent surrounding envelope and cloud, they create shocks and
material can be entrained in so-called bi-polar outflows.
Figure~\ref{fig:protostar} \citep{vanDishoeck11} illustrates the
different protostellar components. Although this cartoon refers to an
isolated low-mass protostar (mass $<$2 M$_{\odot}$, luminosity $<10^2$
L$_{\odot}$) which will develop into a star like our Sun, the same
physical components are also found in high-mass protostellar
environments forming stars such as those seen in Orion (mass $<$8
M$_{\odot}$, luminosity $>10^4$ L$_{\odot}$).

\begin{table}
  \caption{Types of interstellar and circumstellar clouds and their physical 
   characteristics}
  \begin{tabular}{lcccl}
    \hline
   \smallskip
    Name & Typical & Typical & Typical & Examples \\
         & Density (cm$^{-3}$)& Temperature (K) & $A_V$ (mag)$^a$  \\
    \hline
   \smallskip
    Diffuse cloud& $10^2$ & 30--100 & $\leq$1 & $\zeta$ Oph  \\
    Translucent cloud& $10^3$ & 15--50 & 1--5 & HD 154368  \\
    Dense PDR          & $10^4-10^5$ & 50--500 & $<$10 & Orion Bar \\
    Cold dense cloud& $10^4-10^5$ & 10--20 & $>$10 & Taurus cloud  \\
    Prestellar core & $\geq 10^5$ & 8--15 & 10--100 & L1544  \\
    Prostellar envelope \\
    \ \ \ \ Cold outer  & $10^4-10^7$ & 8--100 & 10--100 & NGC 1333 IRAS4A\\ 
    \ \ \ \ Warm inner/ & $10^7-10^9$ & $\geq$100 & 100--1000 & W3 IRS5 \\
    \ \ \ \ Hot core   && && Orion hot core\\
    Shock & $10^4-10^5$ & 200--2000 & $\leq$few & L1157 B1\\
    Protoplanetary disk & & \\
    \ \ \ \ Outer & $10^6-10^{10}$ & 10--500 & 1--100 & TW Hya, HD 100546 \\
    \ \ \ \ Inner & $10^9-10^{15}$ & 100--3000 & 1--1000 & AS 205 \\
    AGB envelope & & \\
    \ \ \ \ Outer & $<10^{8}$ & 10--100 & 1--50 & IRC+10216 (C-rich)\\
    \ \ \ \ Inner & $10^{10}-10^{13}$ & 100--2000 & 50--1000 &  VW CMa (O-rich)\\
    \hline
  \end{tabular}
 $^a$ See \S~\ref{sect:radiation} for definition  
\label{tab:clouds}
\end{table}

With time ($\sim$1 Myr), the opening angle of the wind increases and
the envelope material is gradually dispersed, revealing a young
pre-main sequence star surrounded by a so-called protoplanetary disk.
The term `pre-main sequence' indicates that the young star is not yet
in its stable hydrogen-burning stage, although nuclear fusion of
deuterium nuclei does occur. The young star emits UV radiation from
the stellar corona as well as from shocked accreting material that is
being funneled from the disk onto the star. Once accretion stops and
the disk becomes less turbulent, the sub~$\mu$m-sized grains coagulate
to larger and larger particles and settle to the midplane (few
Myr). There they can form kilometer-sized planetesimals that interact
gravitationally to form (proto)planets and eventually a full planetary
system that may or may not resemble our planetary system (up to 100
Myr). Within our solar system, comets and asteroids are left-over
planetesimals that did not make it into a planet and were scattered
and preserved in the outer regions of the disk or `nebula' out of
which our solar system formed. The star itself reaches the main
sequence after a few $\times 10^7$ yr.

At the end of their life cycle, low-mass stars have exhausted their
nuclear fuel and they are no longer stable but expand and lose part of
their mass (Fig.~\ref{fig:cycle}). These evolved Asymptotic Giant
Branch (AGB) stars are therefore surrounded by circumstellar
envelopes, i.e., extended dense shells of molecular material which can
also contain water.  The temperatures are high, 2000--3000 K near the
stellar photosphere, but drop with radius down to 10 K at the outer
edge.  Depending on the type of star, the envelopes can be either
carbon-rich (with an elemental abundance of carbon exceeding that of
oxygen) or oxygen-rich.

High mass young stars evolve much faster than their low-mass
counterparts and are still embedded in their natal clouds when they
reach the hydrogen-burning main sequence phase, after $\sim 10^5$ yr
\citep{Mottram11}. Thus, they do not have an optically visible
pre-main sequence stage and the existence of disks around high-mass
young stars is still heavily debated.  High mass stars emit copious UV
photons that can ionize atomic hydrogen leading to a fully ionized H
II region surrounding the young star once it has reached the main
sequence.  At the end of their lives they explode as supernovae with
their cores ending up as black holes or neutron stars.

\subsection{Types of radiation fields and ionization fraction}
\label{sect:radiation}

\begin{figure}
\includegraphics[angle=-90,width=0.65\textwidth]{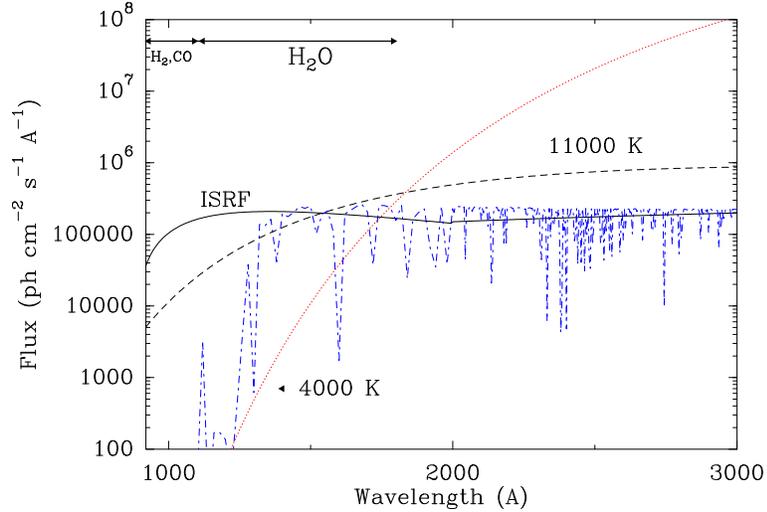}
  \caption{Comparison of the general interstellar radiation field of
    Draine (1978) (extended for $\lambda > 2000$ \AA\ using
    \cite{vanDishoeck82}) with various stellar radiation fields
    scaled to have the same integrated intensity from 912--2000 \AA.
    The scaled stellar atmosphere model radiation field of a B9 star
    \cite{Hauschildt99} ($T_*\approx 11000$~K) is included as well
    (dash-dotted).  The wavelength range where the photodissociation
    of H$_2$O occurs is indicated. Based on
    Reference 45.}
  \label{fig:radfield}
\end{figure}

UV radiation plays an important role in interstellar chemistry, both
in destroying chemical bonds and in liberating molecules from the ice.
The UV radiation impinging on an interstellar cloud is often
approximated as the average radiation produced by all stars from all
directions in its neighborhood. The average interstellar radiation
field (ISRF) has been characterized by \citet{Habing68} and
\citet{Draine78} to have an intensity of about $I_0=10^8$ photons
cm$^{-2}$ s$^{-1}$, with a relatively flat spectrum between 912 and
2000 \AA\ (Fig.~\ref{fig:radfield}) \citep{Habing68}. The threshold of
912 \AA\ corresponds to the ionization potential of atomic H at 13.6
eV; because of its high abundance, virtually all photons with energies
above 13.6 eV are absorbed by H and do not affect the chemistry.

If a cloud is located close to a bright star, the radiation from the
star itself can dominate over that of the general ISRF. For hot O and
B-type stars with effective photospheric temperatures of
$T_*$=20000--50000~K, the shape of the radiation field is not very
different from that of the ISRF. However, cooler A, F, G, K and M-type
stars with $T_*$=3000--10000 K will have many fewer FUV photons to
dissociate molecules. In some sources (e.g., young stars, shocks),
Lyman $\alpha$ radiation at 1216 \AA \ dominates the radiation
field. Water has a substantial photodissociation cross section at this
wavelength, whereas other molecules like CO, N$_2$ and CN do not (see
Tables and discussion in \citealt{Bergin03,vanDishoeck06photo}).

Dust particles attenuate the UV radiation with depth into a cloud
thereby shielding molecules from the harshest radiation.  The dust
optical depth at wavelength $\lambda$ is given by\citep{Tielens05}
$\tau_d (\lambda)=n_d C_{\rm ext} (\lambda) L$ where $n_d$ is the dust
density in cm$^{-3}$, $C_{\rm ext}$ the extinction cross section in
cm$^2$ and $L$ the pathlength in cm. The extinction is the sum of
absorption and scattering processes. Astronomers usually measure the
extinction at visual wavelengths, $A_V$, which is defined as 1.086
times $\tau_d$ at 5500 \AA. The intensity decreases as $I_{5500}=I_0
10^{-0.4 A_V}$ with depth into a cloud. The steepness of this decline
increases toward UV wavelengths and depends on the grain properties
such as size, composition, shape and scattering characteristics
\citep{Roberge91}. The observed extinction curve from infrared to UV
wavelengths implies that there must be a large range of grain
sizes. Most of the dust mass is in $\sim$0.1 $\mu$m grains, but most
of the surface area is in much smaller grains, down to 0.001 $\mu$m or
less. These smaller grains dominate the absorption and scattering of
UV radiation.

For a typical interstellar grain size distribution, UV radiation
ceases to be important at $A_V\approx 5$ mag, when the intensity at
visible wavelengths has declined by a factor of 100 and that at UV
wavelengths by a factor of at least $10^4$. Because the extinction
increases monotonically with path length $L$, $A_V$ (in units of
magnitudes) is often used as a measure of depth into a cloud. Another
quantity often used by astronomers is that of column density in
cm$^{-2}$, i.e., the number density $n$ in cm$^{-3}$ of a species
integrated along a path, $N=\int n \ {\rm d}L$. The relation between
extinction and the column density of hydrogen nuclei is found
empirically to be \citep{Bohlin78,Rachford09} $N_{\rm
  H}=N$(H)+2$N$(H$_2$)=$1.8\times 10^{21}$ $A_V$ cm$^{-2}$, based on
observations of diffuse clouds where both $N$(H), $N$(H$_2$) and $A_V$
are neasured directly.

Cosmic ray particles, i.e., highly energetic atomic nuclei with $>$MeV
energies, penetrate even the densest clouds and provide the required
level of ionization to kick-start the chemistry. The resulting ions
can react rapidly with neutral molecules down to very low temperatures
as long as the reactions are exothermic and have no activation barrier
(see \S 3.1).  The cosmic rays also maintain a low level of UV
radiation by interacting with hydrogen \citep{Prasad83}. The
ionization of H and H$_2$ produces energetic secondary electrons which
can bring H$_2$ into excited electronic states. These states
subsequently decay through spontaneous emission, mostly in the H$_2$
$B-X$ Lyman and $C-X$ Werner bands, producing a UV spectrum consisting
of discrete lines and a weak continuum in the 900--1700 \AA\ range
\citep{Gredel89}. The flux of internally generated UV photons is
typically $10^4$ photons cm$^{-2}$ s$^{-1}$ but depends on the energy
distribution of the cosmic rays (see Fig.~4 of \citealt{Shen04}).

Some cosmic sources (e.g., hot matter near young stars or black holes)
also produce X-rays which can impact the chemistry. However, in
practice their chemical effects are similar to those of cosmic rays
\citep{Bruderer09a}.

Interstellar clouds are largely neutral. In diffuse clouds, the UV
radiation from the ISRF can ionize atomic carbon because its first
ionization potential is less than 13.6 eV, in contrast with oxygen and
nitrogen. Since the abundance of gas-phase carbon with respect to
hydrogen is about $10^{-4}$, this also sets the maximum electron
fraction in the cloud (about 2/3 of the carbon budget is locked up in
carbonaceous grains). With depth into the cloud, the UV radiation
decreases and carbon is converted from atomic into molecular
form. Around $A_V$=5 mag, cosmic rays take over as the main ionizing
agent at a rate denoted by $\zeta$ in s$^{-1}$.  The resulting
ionization fraction depends on the detailed chemistry and grain
physics but is typically $10^{-8}-10^{-7}$ and scales as
$(n/\zeta)^{-1/2}$.

\section{Water spectroscopy and excitation}
\label{sect:spectroscopy}

Water has dipole-allowed pure rotational, vibrational and electronic
transitions which occur at far-infrared/submillimeter, infrared and
ultraviolet wavelengths, respectively. Water has three vibrational
modes: the $\nu_1$ symmetric stretch centered at 2.7 $\mu$m, the
$\nu_2$ bending mode at 6.2 $\mu$m and the $\nu_3$ asymmetric stretch
at 2.65 $\mu$m. Astronomical examples of each of the types of
transitions are presented.

\subsection{Pure rotational transitions}

\begin{figure}
\includegraphics[angle=0,width=0.75\textwidth]{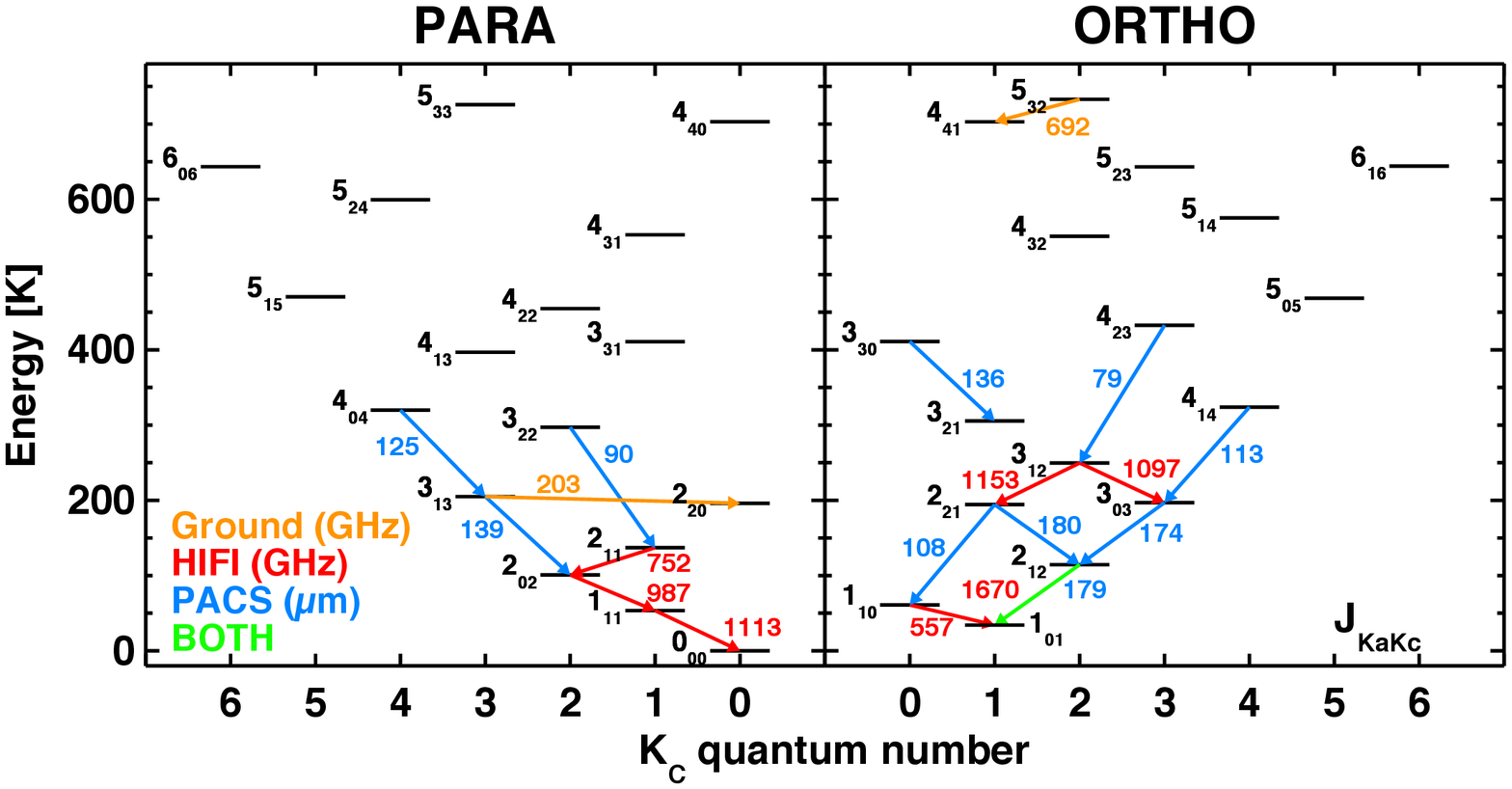}
\caption{Energy levels of ortho- and para-H$_2$O, with some important
  transitions observed with {\it Herschel}-HIFI (in GHz) and PACS (in
  $\mu$m) indicated. Adapted with permission from
  Reference 40. Copyright 2011 The University of Chicago
  Press.}
\label{fig:energy}
\end{figure}

H$_2$O is an asymmetric rotor with an irregular set of energy levels
characterized by quantum numbers $J_{K_A K_C}$, where $J$ indicates
the total rotational quantum number of the molecule and $K$ the
projections of $J$ on the principal axes of inertia. For asymmetric
rotors like H$_2$O, $K_A$ and $K_C$ indicate the $K$ values at the
prolate and oblate symmetry limits to which the level
correlates. Because of the nuclear spin statistics of the two hydrogen
atoms, the H$_2$O energy levels are grouped into ortho ($K_A+K_C=$odd,
parallel nuclear spins of the H nuclei) and para ($K_A+K_C=$even;
antiparallel nuclear spins of H nuclei) ladders
(Fig.~\ref{fig:energy}). Radiative transitions between these two
ladders are forbidden to high order, so that only chemical reactions
can effectively transform one form of water into the other by
exchanging an H atom.  The energy levels within each ladder are
populated by a combination of collisional and radiative excitation and
de-excitation processes (see \S~\ref{sect:excitation}). The radiative
rates are governed by the electric dipole moment of water,
$\mu_D$=1.85 Debye (6.17$\times 10^{-30}$ C-m).

There are two main publicly available data bases that summarize the
transition frequencies, transition strengths or Einstein $A$
coefficients, and statistical weights for astronomically relevant
molecules like water: the Jet Propulsion Laboratory catalog (JPL)
\citep{Pickett98} {\tt spec.jpl.nasa.gov}; and the Cologne
Database for Molecular Spectroscopy \citep{Muller01,Muller05} {\tt
  www.astro.uni-koeln.de/cdms/catalog}.

The molecular data for H$_2$O in the vibrational ground state are well
known up to very high $J$-values from laboratory work starting more than 30
years ago \citep{Delucia72,Polyansky97,Toth99}.  For pure
rotational transitions within $\nu_2$, 2$\nu_2$, $\nu_1$ and
$\nu_3$ vibrationally excited states, new measurements \citep{Yu12} and
intensities \citep{Coudert08} have recently become available.

\begin{figure}
\includegraphics[angle=0,width=0.8\textwidth]{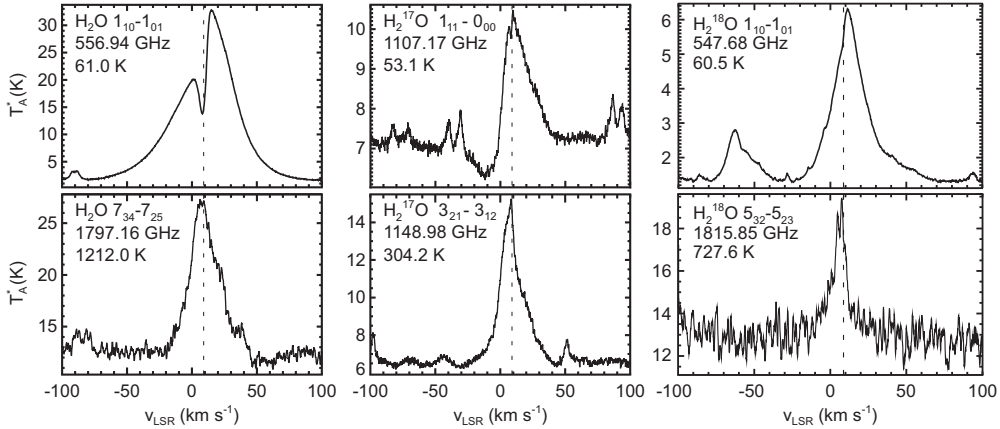}
\caption{Examples of pure rotational lines of H$_2$O and its
  isotopologues observed with {\it Herschel}-HIFI toward the Orion-KL
  molecular cloud. The quantum numbers, frequencies and upper energy
  level of the transitions are indicated. The velocity scale is in the
  so-called Local Standard of Rest (LSR) frame; the dashed line
  indicates the 9 km s$^{-1}$ velocity of the cloud. In this and other
  submillimeter heterodyne spectra, the units of intensity are in
  (antenna or main-beam) temperature $T_A$ or $T_{mb}$ in Kelvin, as
  commonly used by radio astronomers. The Rayleigh-Jeans equation
  relates this temperature to intensity $I$ through
  $T_{mb}$=$c^2/(2k_B\nu^2) I_\nu$ where $c$ is the speed of light,
  $k_B$ is the Boltzmann constant and $\nu$ is frequency. Adapted
  with permission from Reference 67. Copyright 2010 
  American Astronomical Society.}
\label{fig:melnick}
\end{figure}

The spectroscopy of the important isotopologues H$_2^{18}$O and
H$_2^{17}$O is less well covered in the laboratory and new
measurements of transitions in vibrationally excited states are
warranted. The current line lists derive primarily from older work
\citep{Delucia72,Delucia75,Steenbeck71,Johns85,Guelachvili83}.

\begin{figure}
  \includegraphics[width=0.7\textwidth]{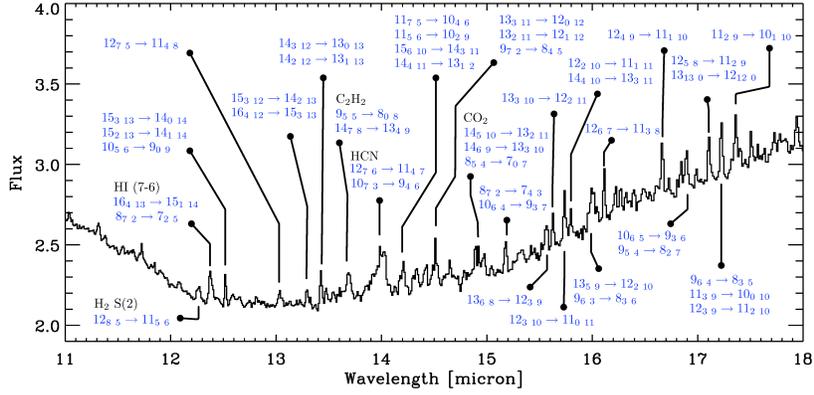}
  \caption{{\it Spitzer} spectrum of the protoplanetary disk around
    the young star RNO 90 illustrating the detection of many highly
    excited pure rotational water lines in space. Some representative
    identifications are indicated; many features are blends at the low
    spectral resolving power $R=600$ of the {\it
      Spitzer}-IRS. Transitions refer to $J_{K_aK_c}$ in the ground
    vibrational state of H$_2^{16}$O unless otherwise indicated
    Reprinted with permission from Reference 74. Copyright 2010
    American Astronomical Society.}
  \label{fig:spitzer}
\end{figure}

Many of the pure rotational lines of water and its isotopologues have
been detected at submillimeter and far-infrared wavelengths toward
bright sources such as the Orion molecular cloud
\citep{Melnick10,Herczeg12,Coutens12,Neill13} (see
Fig.~\ref{fig:melnick} for example).  Very highly excited pure
rotational lines up to $J$=18 ($E_u/k\approx 5000$~K) are found at
mid-infrared wavelengths in low-resolution {\it Spitzer Space
  Telescope} data at 10--38 $\mu$m
\citep{Watson07,Carr08,Salyk08,Pontoppidan10}.  An example spectrum of
an infrared spectrum of a protoplanetary disk is presented in
Fig.~\ref{fig:spitzer}. In contrast with the submillimeter spectra,
the mid-infrared lines are generally not resolved so that most
features are blends of lines.

\subsection{Vibrational transitions: gas and ice}

\begin{figure}
\includegraphics[angle=-90,width=0.5\textwidth]{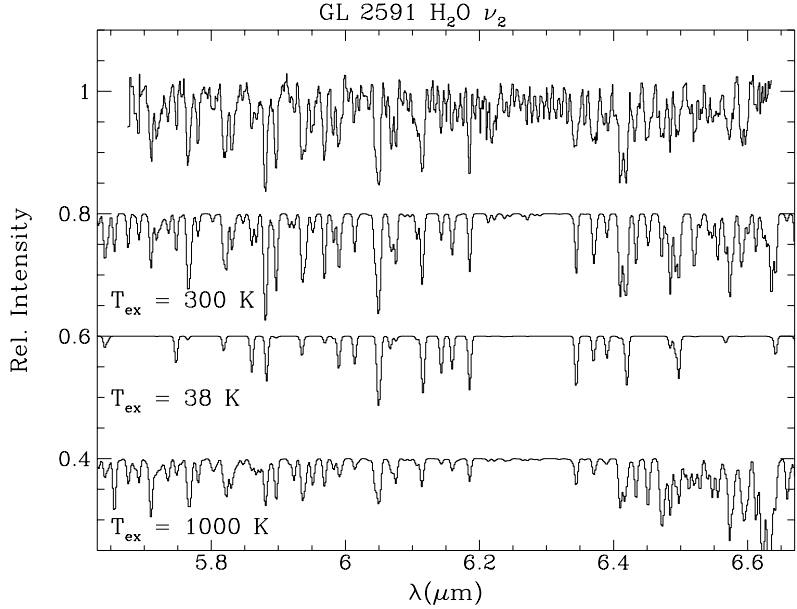}
\caption{Vibration-rotation lines of H$_2$O in the $\nu_2$ band
  observed with ISO-SWS in absorption toward the high-mass protostar
  AFGL 2591. The normalized spectrum is compared with simulated
  spectra for various excitation temperatures, with 300 K providing
  the best fit. The model spectra are offset vertically for
  clarity. Even at this low spectral resolving power of $R\approx
  2000$, the data can distinguish between different models. Reprinted
  with permission from Reference 83.  Copyright 1996 European Southern
  Observatory. }
\label{fig:gl2591}
\end{figure}

The vibration-rotation transitions of water at infrared wavelengths
have been studied for many decades in the laboratory
\citep{Flaud75,Camy76} and all the relevant molecular data are
summarized in the HITRAN database \citep{Hitran09} at {\tt
  www.cfa.harvard.edu/hitran}.  Most recently, line lists appropriate
for temperatures up to several thousand K and including higher
vibrational transitions have been published for water and its
isotopologues
\citep{Barber06,Shirin06,Toth99,Voronin10,Brunken07,Tennyson12} and
are posted at {\tt www.exomol.com/molecules/H2O.html}. The higher
temperature data are particularly important for exoplanets and cool
stellar atmospheres. An example of an observed vibration-rotation
spectrum at low spectral resolution toward a high-mass protostar is
presented in Fig.~\ref{fig:gl2591} \cite{Helmich96}.

\begin{figure}
  \includegraphics[angle=-90,width=0.5\textwidth]{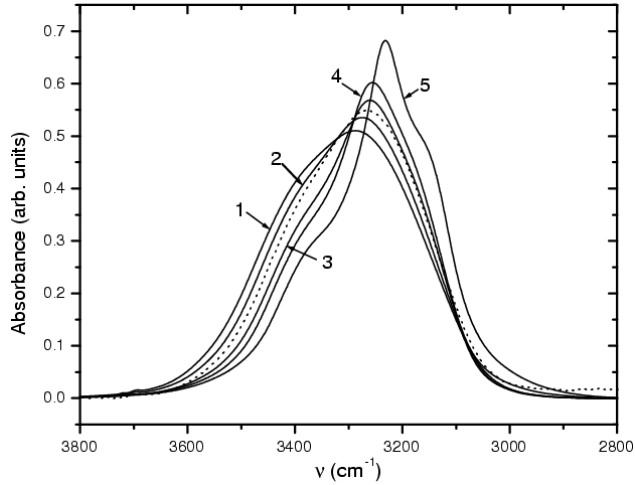}
  \caption{The OH stretching mode at 3.3 $\mu$m of a sample of pure
    water ice as deposited on a quartz substrate at 12 K (dashed
    line), compared with the spectra of water ice deposited on a CsI
    substrate (full lines): 1) after deposition at 12 K; 2) after
    warm-up to 50 K; 3) to 80 K; 4) to 120 K, 5) to 160 K. Note the
    appearance of a sharp peak due to crystallization at 160~K. Under
    interstellar conditions at much lower pressures and slower warm-up
    rates, the phase transitions shift to lower temperatures (see
    \S~\ref{sect:desorption}). Reprinted with permission from Reference 92.
    Copyright 2002 European Southern Observatory.}
  \label{fig:schutte}
\end{figure}

There is also a rich literature on laboratory spectroscopy of water
ice, both for interstellar and solar system applications \citep[see
reviews in][]{DHendecourt99,Gudipati13}.  In contrast with low
pressure gas-phase spectra, the solid-state water spectra have no
rotational substructure and are very broad, with profiles that depend
on the morphology, temperature, thermal history and environment of the
water molecules
\citep[e.g.,][]{Tielens83,Smith89,Hudgins93,Jenniskens95,Stevenson99,Kimmel01}. For
example, the spectrum of crystalline water ice has a sharp feature
around 3200 cm$^{-1}$ (3.1 $\mu$m) that is lacking in amorphous water
ice (Fig.~\ref{fig:schutte})\citep{Schutte02}.  Most water ice in the
universe is actually thought to be in a high-density amorphous ice
form which does not occur naturally on Earth
\citep{Jenniskens94,Jenniskens95}.  Porous ices have dangling OH bonds
that absorb around 3700 cm$^{-1}$ (2.70 $\mu$m) \citep{Ehrenfreund96},
but are not seen in space.  In interstellar ices, water is mixed with
other species such as CO and CO$_2$, which can block the dangling OH
bands and affect both the line profiles and intensities, as
illustrated by laboratory studies for the 6 $\mu$m bending mode
\citep{Bouwman07}.  The far-infrared librational modes of water ice at
45 and 63 $\mu$m have been measured as well
\citep{Moore92,Moore94}. Laboratory spectra for fitting astronomical
data can be downloaded from various websites such as the NASA-Ames ice
database at {\tt www.astrochem.org/db.php} and the Leiden ice database
at {\tt www.strw.leidenuniv.nl/$\sim$lab}.

\begin{figure}
\begin{minipage}{0.475\textwidth}
\includegraphics[angle=90,width=\textwidth]{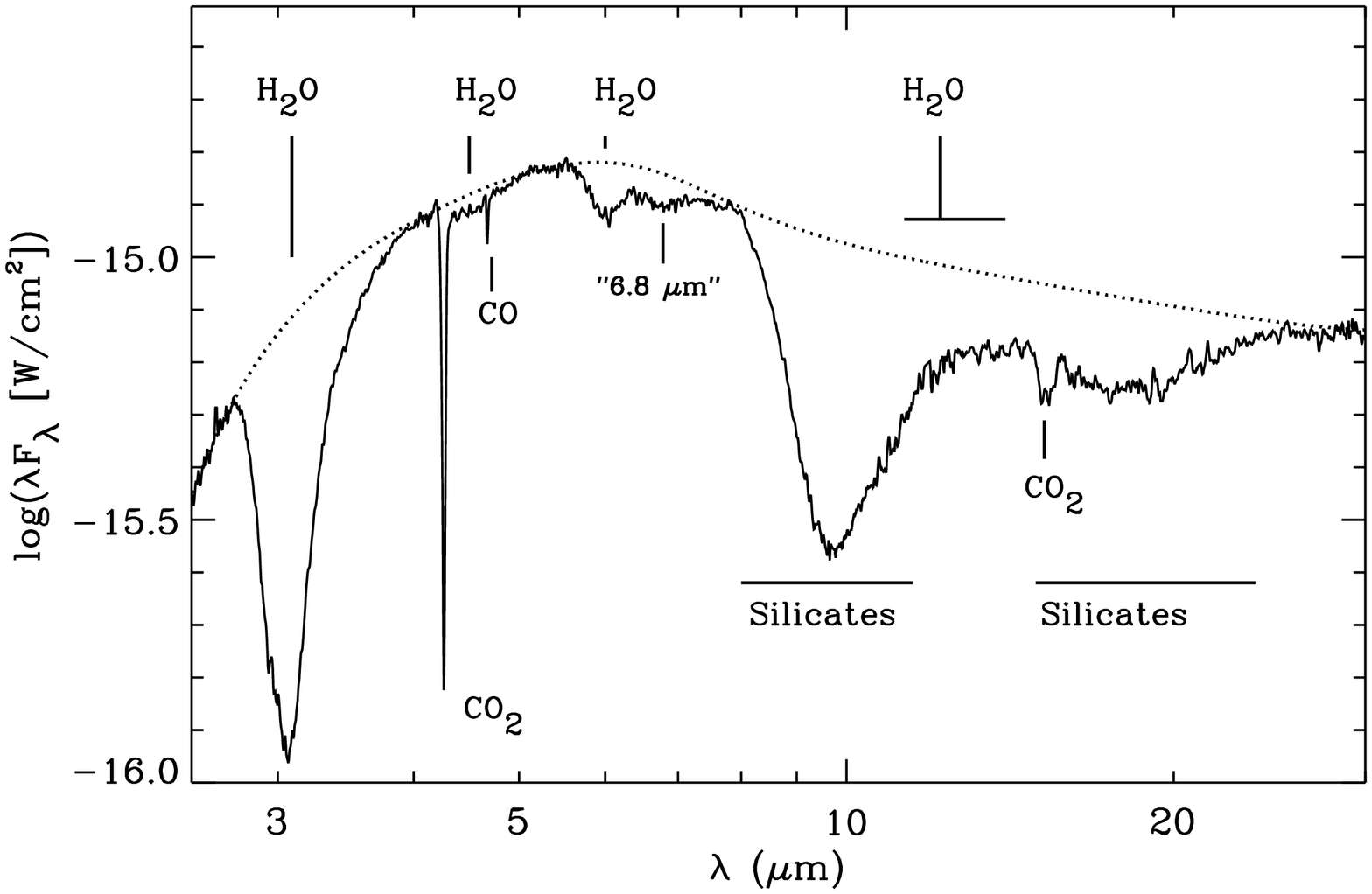}
\end{minipage}\hfill%
\begin{minipage}{0.52\textwidth}
\includegraphics[angle=90,width=\textwidth]{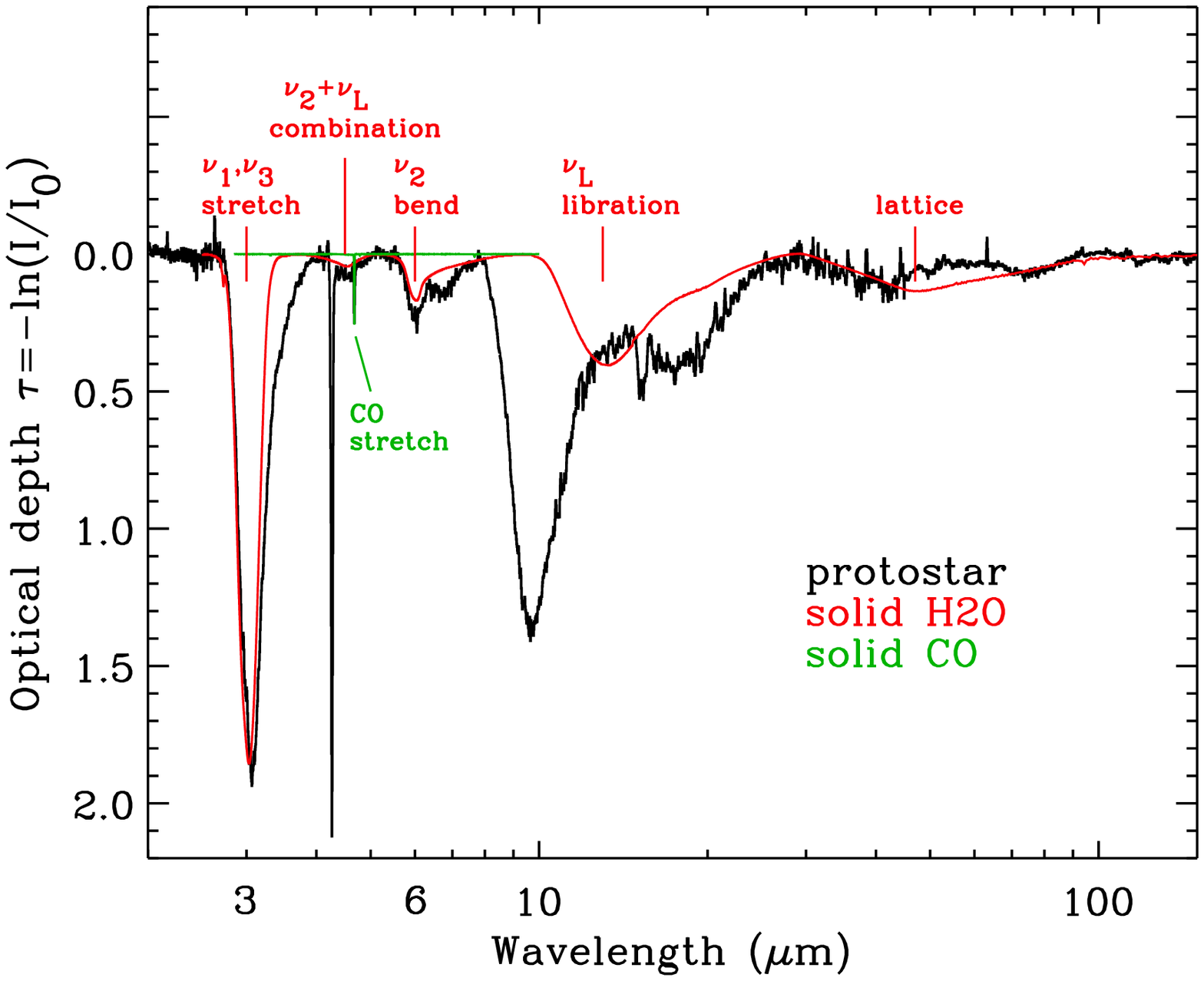}
\end{minipage}\hfill%
\caption{Left: Detection of water ice at 3, 4.5, 6 and 11 $\mu$m along
  the line of sight toward the protostar Elias 29 obtained with the
  ISO-SWS. The dotted line indicates the source continuum which acts
  as the `lamp' against which the water ice bands are seen in
  absorption. Right: Continuum-divided spectrum on an optical depth
  scale overlaid with a laboratory spectrum of H$_2$O ice at $T$=10 K
  from \citet{Hudgins93}. Reprinted with permission from Reference
  100. Copyright 2008 Europen Southern Observatory.}
  \label{fig:waterice}
\end{figure}

Water ice has been observed both from the ground at 3 $\mu$m and in
space up to long wavelengths with a wide variety of instruments
\citep{Boogert00,Gibb04,Boogert08} (Fig.~\ref{fig:waterice}).  In most cases,
the absorption is against the hot dust surrounding a protostar
embedded within the cloud, but there is an increasing data set on
water ice toward background sources \citep{Whittet07,Boogert11}
(Fig.~\ref{fig:absem}). The latter situation allows pristine water ice
to be probed, unaffected by heating or radiation from the
protostar. Water ice has also been seen in the spectra of evolved
stars \citep{Soifer81,Sylvester99} and toward many sources in external
galaxies \citep{Spoon04,Sajina09,Shimonishi10}.  The observed ice
spectra are generally consistent with compact amorphous water ice,
although the actual interstellar ice porosity is poorly constrained
\citep{Bossa12}. Only few sources show the signature of crystalline
water ice \citep[e.g.,][]{Aitken88,Schegerer10}.

\subsection{Electronic transitions}

The ground electronic state of water is non-linear with a bond angle
of 104.5$^o$ and O-H bond length of 0.95 \AA. The molecule has
C$_{2v}$ symmetry with a molecular orbital configuration
$1a_1^22a_1^23a_1^21b_1^21b_2^2$ making up the $\tilde X$ $^1A_1$
state. Its dissociation energy to H + OH is 5.1 eV. The first excited
electronic state to which dipole allowed transitions are possible is
$\tilde A$ $1 ^1B_1$ ($1b_1 \to 4a_1 (3s)$ excitation), which has a
vertical excitation energy of 7.5 eV. Strong absorptions therefore do
not start until $\sim$7 eV ($\sim$1800 \AA), well above the
dissociation energy. The $\tilde A$ state does not have bound energy
levels however: absorption leads directly to dissociation (see
\S~\ref{sect:photodissociation}). Similarly, the next dipole-allowed
excited state, the $\tilde B$ $^1A_1$ state ($3a_1\to 4a_1 (3s)$
excitation) at 9.8 eV, is largely dissociative. Bound electronic
states do not appear until higher energies and have largely Rydberg
character in which a bonding electron is promoted to a hydrogen-like
`diffuse' orbital such as the $4s$ orbital.

\begin{figure}
\includegraphics[angle=0,width=0.7\textwidth]{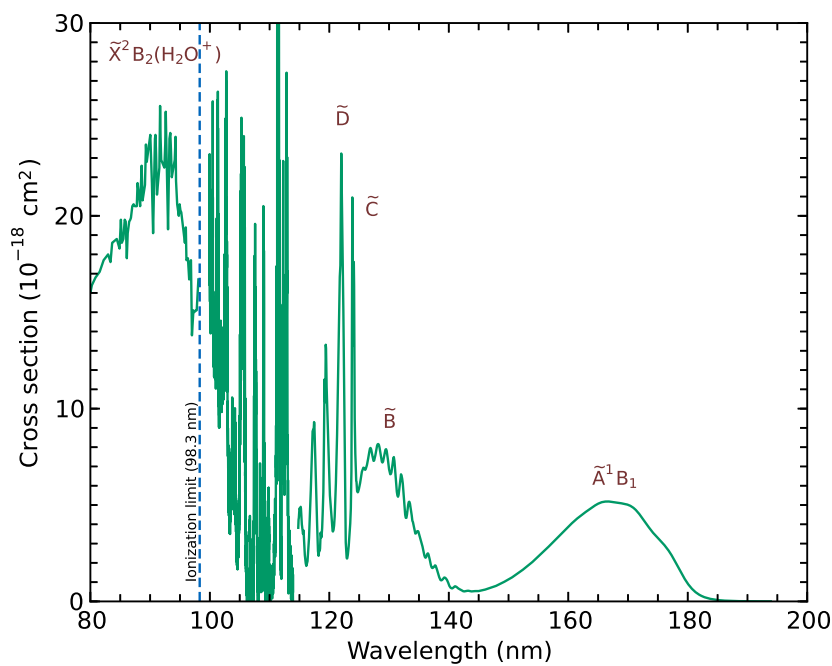}
\caption{Absorption spectrum of gaseous H$_2$O starting from the
  $\tilde X ^1A_1$ state.  Photodissociation occurs mainly through the
  $\tilde A$ and $\tilde B$ electronic states. Figure produced using data
  from the Reference 114.}
\label{fig:fillion}
\end{figure}

\begin{figure}
\includegraphics[angle=0,width=0.7\textwidth]{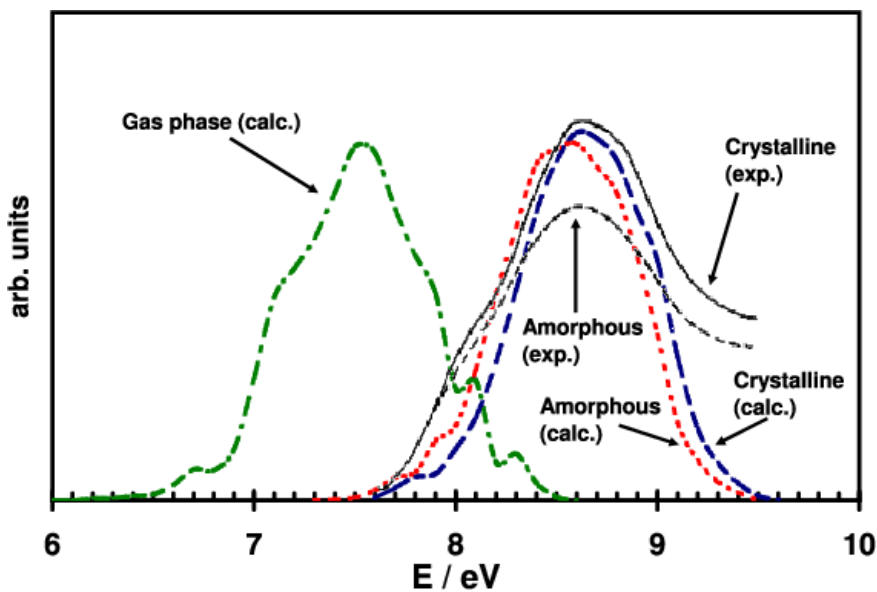}
\caption{ UV absorption spectrum of water ice compared with that of
  water vapor. Note the shift of the $\tilde A - \tilde X$ ice band to
  higher energies. Reprinted with permission from Reference
  115. Copyright 2008 European Southern Observatory.}
\label{fig:icevsgas}
\end{figure}

The water gas-phase absorption spectrum is well
known\citep{Yoshino96,Cheng99,Fillion01} and is available from the MPI
Mainz UV-VIS spectral atlas data base\cite{Mainz} at {\tt
 www.uv-vis-spectral-atlas-mainz.org}. Figure
\ref{fig:fillion} illustrates the H$_2$O ultraviolet absorption
spectrum, including the ionization limit at 12.61 eV (983 \AA). Since
most of the excited electronic states are dissociative, this figure
illustrates directly the photodissociation processes described further
in \S~\ref{sect:photodissociation}.  Figure \ref{fig:icevsgas}
compares the absorption spectrum of water ice with that of water
vapor; the ice band is shifted to shorter wavelengths due to the
slightly smaller dipole moment of water in the excited state leading
to less favorable interactions with the neighboring water molecules
\citep{Andersson06,Andersson08}. The EUV spectrum of H$_2$O excited by
electron impact has been measured by \citet{Ajello84}.

Searches for H$_2$O absorption into the $\tilde C$ $^1B_1$ state around 1240
\AA\ have been made for diffuse clouds \citep{Spaans98} but no lines
have been detected.  Similarly, no H$_2$O emission has yet been
detected at UV wavelengths.

\subsection{Rate coefficients for inelastic collisions}
\label{sect:collisions}

The most important collision partners of water in interstellar space
are ortho- and para-H$_2$, with electrons and H only significant in
specific regions and He contributing at a low level. The only method
to provide the thousands of state-to-state rate coefficients needed
for astronomical applications is through theory.  The process consists
of two steps. First, a multi-dimensional potential energy surface
describing the interactions between the colliders has to be computed
using state-of-the-art {\it ab initio} quantum chemical methods and
fitted to a convenient representation for use in the dynamics.
Second, the dynamics on this surface have to be calculated using
inelastic scattering calculations at a range of collision
energies. Full quantum calculations are used at low collision energies
whereas quasi-classical methods are often employed at the highest
energies.

Early studies used simplifications such as replacing H$_2$ with He
\citep{Green93}, including only a limited number of degrees of freedom
of the colliding system (i.e., keeping some angles or internuclear
distances frozen), or approximations in the scattering calculations
\citep{Phillips96}.  The most recent calculations have been performed
on a 5D potential surface and consider collisions of H$_2$O with both
para-H$_2$ and ortho-H$_2$ separately \citep{Dubernet09,Daniel11}.
This 5D potential is based on the full 9D potential surface computed
by \citet{Valiron08}.  The rate coefficients for pure rotational
transitions within $v=0$ have been computed with the quantum close
coupling method, even up to high energies when such calculations
become very computer intensive.  Interestingly, the cross sections for
collisions of ortho- and para-H$_2$O with ortho-H$_2$ are found to be
significantly larger than those with para-H$_2$ for low $J$, so that
explicit treatment of the ortho/para H$_2$ ratio in interstellar cloud
models is important.

The quantum close coupling calculations performed to date have
included the lowest 45 rotational states of both ortho- and
para-H$_2$O.  These include the states from which emission or
absorption has been observed in the vast majority of astronomical
observations.  However, a few studies of astrophysical absorption
\citep{Indriolo13water} and emission
\citep{Watson07,Meijerink09,Herczeg12} probe states of even higher
excitation (see Fig.~\ref{fig:spitzer}); here, extrapolation methods
must currently be employed to estimate the collisional rate
coefficients of relevance \citep{Faure08,Neufeld12}.

Direct comparison of absolute values of theoretical state-to-state
cross sections with experimental data is not possible for water, but
the accuracy of the 9D potential surface and resulting rate
coefficients has been confirmed implicitly through comparison with
other data sets\citep{Wiesenfeld10,Drouin12}, including differential
scattering experiments of H$_2$O with H$_2$ \citep{Yang10b} and the
presence of weakly bound states \citep{Avoird11a,Avoird11b,Avoird12}.
At higher temperatures ($>80$ K), the cross sections derived from
pressure broadening data\citep{Dick10} agree with theory within 30\%
or better for low $J$. Overall, the uncertainty in the most recent cross
sections obtained from full quantum calculations is estimated to range
from a few tens of \% to a factor of a few, depending on the
transition \citep{Daniel12}.

Several studies have investigated how the water model line intensities
from interstellar clouds depend on the choice and accuracy of the
collisional rate coefficients \citep{Crimier07,Crimier07b,Daniel12}.
For some transitions, such as the important $1_{10}-1_{01}$ 557 GHz
line, the uncertainty in the intensity scales linearly with that in
the cross section \citep{Grosjean03}.  For higher-lying lines, the
dependence is less direct due to optical depth effects and the
possibility that infrared pumping contributes (see
\S~\ref{sect:excitation}). The difference between model intensities
using the older quasi-classical trajectory results \citep{Faure07} and
the new quantum rate coefficients \citep{Dubernet09,Daniel11} is
typically less than 50\%, although some lines can show variations up
to a factor of 3, especially at lower temperatures and
densities. While substantial, these uncertainties generally do not
affect astronomical conclusions. Altogether, the collisional rate
coefficients have now reached such high accuracy that they are no
longer the limiting factor in the interpretation of the astronomical
water data. This conclusion is a testimony to the decade long effort
by molecular physicists and quantum chemists to determine them.

For vibration-rotation transitions of water, much more limited
information is available. A set of collisional rate coefficients
derived from vibrational relaxation data has been published
\citep{Faure08} but is accurate only to an order of magnitude. The
same study also considers H$_2$O-electron rate coefficients. Little is
known about H$_2$O-H collisional rate coefficients, which are important
in some astrophysical regions such as dissociative shocks
\citep{Hollenbach13}.

Collisional rate coefficients for H$_2$O and for other astrophysically
relevant species can be downloaded electronically from the BASECOL
database \citep{Dubernet13} at {\tt www.basecol.obspm.fr} and from the
LAMDA database \citep{Schoier05} at {\tt
  home.strw.leidenuniv.nl/$\sim$moldata/}.

\subsection{Water excitation and radiative transfer}
\label{sect:excitation}

The observed spectrum of interstellar water vapor, whether detected in
emission or absorption, depends upon the relative populations in the
various rovibrational states (the `level populations').  These, in
turn, are determined by a complex interplay of collisional and
radiative processes.  Under conditions of thermal equilibrium (TE),
the molecular motions are characterized by a Maxwell-Boltzmann
distribution with a single kinetic temperature, $T_{\rm kin}$, and
the water molecules are bathed in a blackbody radiation field with a
temperature $T_{\rm BB}$ equal to $T_{\rm kin}$. Then the fractional
population in any given rovibrational state, $i$, of energy $E_i$, is
determined purely by thermodynamic considerations, and is given by the
Boltzmann factor $f_i = \exp (-E_i/kT)/Q(T),$ where $g_i$ is the
degeneracy of the state and $Q(T)= \sum_i g_i \exp (-E_i/kT)$ is the
partition function.

In the interstellar medium, however, the radiation field is typically
far from equilibrium with the gas, and TE does not apply.  Were this
not the case, molecules would show no net emission or absorption of
radiation, and would therefore be undetectable.  Nevertheless, if the
gas density is sufficiently high, collisional processes can dominate
radiative processes and the relative level populations can attain a
state of `local thermodynamic equilibrium' (LTE), in which they are
described by a Boltzmann distribution at a single temperature, $T_{\rm
  kin}$.  

Radiative rates scale with the dipole-moment $\mu_D$ and transition
frequency $\nu$ as $\mu_D^2 \nu^3$.  Because water possesses a large
dipole moment and a large rotational constant, its spontaneous
radiative rates for dipole-allowed rotational transitions are
relatively large compared, for example, to those of the commonly
observed CO molecule ($\mu_D=1.85$ vs 0.1 Debye and rotational
constants 9--28 vs 1.9 cm$^{-1}$).  Thus, except in very dense regions
where stars are forming, the level populations typically show
departures from LTE and a detailed treatment of the molecular
excitation is needed.  In this case, the level populations quickly
reach a quasi-equilibrium in which each state is populated at the same
rate as it is depopulated. The resulting populations can still be
plotted in a Boltzmann diagram and fitted with a so-called excitation
temperature $T_{\rm ex}$ or $T_{\rm rot}$ (if only rotational levels
are involved), but $T_{\rm ex}$ has no physical meaning and can be
very different from the kinetic temperature $T_{\rm kin}$ of the
region.

The excitation of interstellar water vapor generally involves three
types of processes: (1) collisional excitation and de-excitation; (2)
absorption of -- and stimulated emission by -- the surrounding
radiation; and (3) spontaneous radiative decay (see
Fig.~\ref{fig:absem}).  In the first of these processes, inelastic
collisions with molecular hydrogen usually dominate; because the rate
coefficients for excitation by ortho- and para-H$_2$ can differ
substantially, the collisional excitation rates depend both upon the
H$_2$ density and the H$_2$ ortho-to-para ratio
(\S~\ref{sect:collisions}).  In environments where the electron
fractional abundance exceeds $\sim 10^{-4}$, electron impact
excitation can also be important, provided the electron density is
high enough for collisions to dominate over radiative excitation.

In the second of these processes (absorption and stimulated emission),
the relevant radiation field includes both external continuum
radiation and the line radiation emitted by nearby water molecules.  A
strong external radiation field, resulting from continuum emission
from warm dust, for example, can lead to pumping of pure rotational
transitions and vibrational pumping in a series of vibrational bands,
particularly the $\nu_2$ = 1 -- 0 band at 6.3 $\mu$m.  The effects of
line radiation from nearby water molecules complicates the problem
significantly, because the level populations of water molecules in one
location affect those in another.  Put another way, the equations of
statistical equilibrium that determine the level populations at each
point within an interstellar gas cloud must be solved simultaneously
with the equations of radiative transfer that determine the radiation
field at each point.

Several methods, of varying complexity, have been developed to solve
the excitation problem described above.  These include Monte-Carlo
simulations\citep{Bernes79}, iterative methods such as accelerated
Lambda iteration (ALI)\citep{Rybicki91}, hybrid methods, and escape
probability methods\citep{Hummer82} in which the spontaneous emission
of photons is assumed to be followed either by escape or local
reabsorption.  For media with a monotonic and large velocity
gradient\citep{Sobolev60} (LVG), escape probability methods are
preferred, as they are both accurate and computationally inexpensive;
for static media, however, there is a trade-off between computational
expense and accuracy.  Publicly-available computer programs that can
be used to model the excitation of - and emission from -- interstellar
water vapor and other molecules include RATRAN\citep{Hogerheijde00} (a
hybrid Monte-Carlo/ALI code), LIME\citep{Brinch10} (a code suited for
arbitrary 3D geometries) and Radex\citep{vanderTak07} (an escape
probability code).

In the absence of an external radiation field, the typical behavior of
the water rotational level populations is that departures from LTE are
most pronounced for the states of highest energy, as expected because
the spontaneous radiative decay rates tend to increase with energy. A
secondary effect is that the departures from LTE also increase with
quantum number $K_A$.  In this context, the set of states with lowest
$K_A$ (0 or 1) for given $J$ are sometimes called the `backbone'
states\citep{deJong73} (i.e., for ortho-water, the states $1_{01}$,
$2_{12}$, $3_{03}$, $4_{14}$, $5_{05}$, $6_{16}$....).  These states
tend to show the smallest departures from LTE because they possess the
fewest routes for spontaneous radiative decay.

\subsection{Masers}
\label{sect:masers}

\begin{figure}
\includegraphics[angle=90,width=0.8\textwidth]{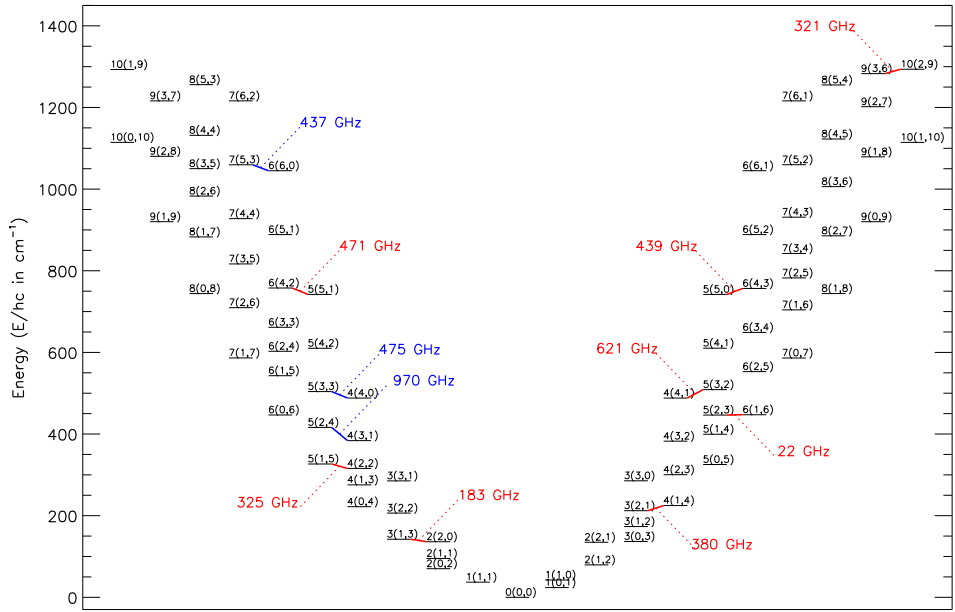}
\caption{Energy level diagram showing water maser transitions detected
  to date.  Red lines denote known interstellar water maser
  transitions, and blue lines denote transitions for which maser
  emission has been confirmed in evolved stars only. Reprinted with
  permission from Reference 152. Copyright 2013 American Astronomical
  Society.}
\label{fig:masers}
\end{figure}

One fascinating consequence of the secondary behavior described above
(i.e.\ the fact that the departures from LTE increase with $K_A$), is
that radiative transitions with $\Delta J = J_u - J_{\ell}=1$, $\Delta
K_A = +1$ between states of similar energy can exhibit a `population
inversion'.  Here, the population in the upper state, divided by its
degeneracy, exceeds that for the lower state; as a result, the
stimulated emission rate exceeds the absorption rate, leading to the
maser phenomenon in which the intensity of the radiation along a given
ray can increase exponentially.  Indeed, the first water transition
ever detected from the interstellar gas, the $6_{16}- 5_{23}$
transition near 22 GHz\citep{Cheung69}, exhibits strong maser action.
Interferometric observations\citep{Genzel86} of this transition reveal
large luminosities emerging from maser spots of size only $\sim
10^{13}$~cm; typical brightness temperatures (defined as the
temperature for which the Planck function yields an intensity equal to
that observed) often exceed $10^{10}$~K and in extreme
cases\citep{Garay89} can exceed $10^{15}$~K.  Over the 40 years since
the first detection of interstellar water maser emission in the 22 GHz
transition, several additional maser transitions have been found at
higher frequencies: these are listed in Table~\ref{tab:masers} and
marked on the energy level diagram shown in Figure~\ref{fig:masers}
\cite{Neufeld13maser}.  With the exception of the 621~GHz and 970~GHz
transitions, which were observed with the HIFI instrument on {\it
  Herschel}, and the 380~GHz transition, detected using the Kuiper
Airborne Observatory, all the transitions listed in
Table~\ref{tab:masers} can be observed from ground-based
observatories.

In the case of interstellar water masers, the transitions observed to
show maser action are precisely those for which population inversions
are predicted\citep{Neufeld91} in the absence of an external radiation
field.  Many of the interstellar water masers are likely produced by
collisional excitation in shocks \citep{Hollenbach13} associated with
forming stars.

Several additional maser transitions, observed in evolved stars but
not the interstellar medium, are most likely pumped by continuum
radiation.  These circumstellar maser transitions include several
transitions within the $\nu_2$ vibrational band: here, maser action is
apparently favored for $\Delta K_A = -1$.  A possible pumping scheme
has been discussed by Alcolea \& Menten\citep{Alcolea93}, but further
modeling is still needed to understand the excitation of
vibrationally-excited masers in circumstellar envelopes.

\begin{table*}[t]
\caption{Masing transitions of water vapor}
\begin{center}
\begin{tabular}{c c c l l }
\hline
\hline
Vibrational state & Transition & Frequency (GHz) & Source type$^a$ & Reference  \\
\hline
Ground 	  & $6_{16} -5_{23}$ & \phantom{0}22.235  	& ISM, CSE & Cheung et al.\citep{Cheung69} 	\\
Ground 	  & $3_{13} -2_{20}$ & 183.310 	& ISM, CSE     & Waters et al.\citep{Waters80} 	\\
Ground    & $10_{29}-9_{36}$\phantom{0} & 321.226 	& ISM, CSE & Menten et al.\citep{Menten90a} 	\\
Ground    & $5_{15} -4_{22}$ & 325.153 	& ISM, CSE & Menten et al.\citep{Menten90b} 	\\
Ground    & $4_{14} -3_{21}$ & 380.197 	& ISM, CSE & Phillips et al.\citep{Phillips80} 	\\
Ground    & $7_{53} -6_{60}$ & 437.347 	& ISM, CSE & Melnick et al.\citep{Melnick93} 	\\
Ground    & $6_{43} -5_{50}$ & 439.151 	& ISM, CSE & Melnick et al.\citep{Melnick93} 	\\
Ground    & $6_{42} -5_{51}$ & 470.889 	& CSE 	   & Melnick et al.\citep{Melnick93} 	\\
Ground    & $5_{33} -4_{40}$ & 474.689 	& CSE 	   & Menten et al.\citep{Menten08} 	\\
Ground    & $5_{32} -4_{41}$ & 620.701  & ISM, CSE & Neufeld et al.\citep{Neufeld13maser} 	\\
Ground    & $5_{24} -4_{31}$ & 970.315  & CSE 	   & Neufeld et al.\citep{Justtanont12} 	\\
\hline
$\nu_2=1$ & $4_{40}-5_{33}$  &  \phantom{0}96.261  & CSE	   & Menten \& Melnick\citep{Menten89} 	\\
$\nu_2=1$ & $5_{50}-6_{43}$  &  232.687	& CSE      & Menten et al.\citep{Menten06} 	\\
$\nu_2=1$ & $1_{10}-1_{01}$  &  658.006	& CSE      & Menten \& Young\citep{Menten06} 	\\
\hline
\end{tabular}
\end{center}
\label{tab:masers}
$^a$ ISM: interstellar medium; CSE: circumstellar envelopes of evolved stars
\end{table*}

In addition to the intrinsic interest of the maser phenomenon, the 22
GHz transition has proven to be a powerful astronomical tool.  For
example, within the Galaxy, the 22 GHz maser emission is used as a
beacon to locate regions where massive stars are forming
\citep{WalshA11}. Masers have also been detected in many external
galaxies, especially those with strong far infrared fluxes
\citep{Lo05,Surcis09}. Thanks to their extraordinary brightness, 22
GHz water masers can be observed with Very Long Baseline
Interferometry (VLBI), which can provide millarcsecond
resolution. Moreover, individual maser spots typically show very
narrow linewidths, allowing accurate line-of-sight velocities to be
inferred.  Thus, VLBI observations enable proper motion studies of
warm astrophysical gas and can provide geometric distance estimators,
yielding the distances and motions of star forming regions in our
Galaxy and of circumnuclear disks in active
galaxies\citep{Brunthaler06,Braatz10}. As noted in the introduction,
such observations have led to the best evidence yet obtained for the
existence of supermassive black holes in external
galaxies\citep{Miyoshi95}, as well as revised estimates for the size
of the Milky Way\citep{Reid09}.

\section{Water chemistry}

\begin{figure}
\includegraphics[angle=0,width=0.8\textwidth]{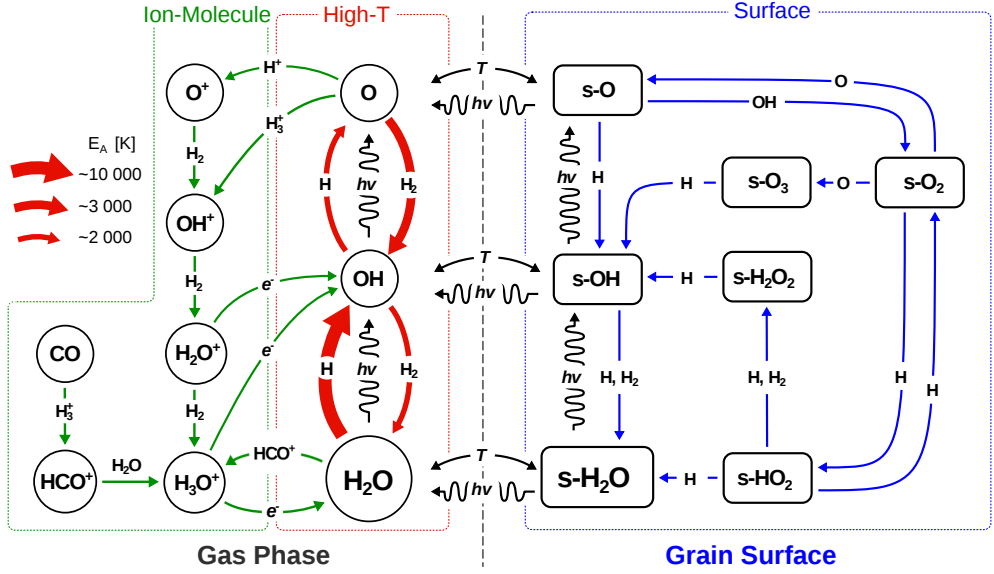}
\caption{Summary of the main gas-phase and solid-state chemical
  reactions leading to the formation and destruction of H$_2$O.  Three
  different chemical regimes can be distinguished: (i) ion-neutral
  chemistry, which dominates gas-phase chemistry at low $T$ (green);
  (ii) high-temperature neutral-neutral chemistry (red); and (iii)
  solid state chemistry (blue).  $s$-X denotes species X on the ice
  surfaces. Adapted with permission from Reference 40. Copyright 2011
  The University of Chicago Press.}
\label{fig:network}
\end{figure}

Figure~\ref{fig:network} presents an overview of the main reactions
involved in the water chemistry. Three distinct synthetic routes are
seen: (i) low-temperature ion-neutral chemistry ($\lapprox 100$ K);
(ii) high-temperature neutral-neutral chemistry; and (iii) surface
chemistry. In the following subsections, each of these types of
chemistries will be discussed in detail. Throughout this section,
$s$-X denotes a surface species.

\subsection{Low-temperature ion-neutral chemistry}
\label{sect:ion}

\subsubsection{General considerations}

Ion-neutral chemistry is the dominant gas-phase chemistry in cold
interstellar clouds, where the low temperatures limit the important
reactions to those that are rapid, exothermic, and without activation
energy barriers.  Figure~\ref{fig:network} shows the chain of
ion-molecule reactions leading to water. Such reactions between ions
and neutrals have strong long-range forces, leading to large rate
coefficients for product channels without
barriers\citep{Wakelam10ssr}.  For reactions in which the neutral has
no permanent dipole moment, the long-range force normally consists of
the attraction between the ionic charge and the dipole polarizability
of the neutral.  For such systems, a simple long-range capture model
known as the Langevin model leads to a temperature independent rate
coefficient $k_L$ (cm$^3$ s$^{-1}$) of \citep{Steinfeld99}

$$ k_L= 2\pi e \sqrt(\alpha/\mu) \approx 10^{-9} \eqno(1) $$ 
where $e$ is the electronic charge (esu), $\alpha$ is the
polarizability (cm$^3$), and $\mu$ (gr) is the reduced mass of the
reactants.  Typical values are $10^{-9}$ cm$^3$ s$^{-1}$. For neutral
species with a permanent dipole moment like water, a somewhat more
complex long-range capture approach shows that there is an
approximately inverse dependence on the square root of temperature
from roughly 10 K to 300 K, which then gradually turns into the
Langevin rate coefficient at higher temperatures.  In the
low-temperature limit, a theoretical approach, known as the trajectory
scaling method \citep{Wakelam10ssr,Su82,Maergoiz09,Woon09}, yields an
expression for the rate coefficient $k_{TS}$ of \citep{Steinfeld99}

$$ k_{TS}/k_L=0.4767x + 0.6200 \eqno(2)$$
in terms of a unitless parameter $x$, which is given by the equation

$$ x= \mu_D /(2\alpha kT)^{1/2}. \eqno(3) $$
Here $\mu_D$ (esu-cm) is the dipole moment of the neutral reactant.  In more
common units, the dipole moment for water is
6.17$\times 10^{30}$ C-m (1.85 Debye).  With a large dipole
moment and an ion of low mass; e.g., H$_3^+$, the predicted rate
coefficient for ion-neutral reactions at 10 K can exceed 10$^{-8}$
cm$^3$ s$^{-1}$.  When the temperature is high enough that $x < 2$,
another expression must be used for the rate coefficient
\citep{Maergoiz09}:

$$ k_{TS}/k_L=(x+0.5090)^2/10.526 + 0.9754. \eqno(4) $$
With this second expression, the ratio between the trajectory scaling
and Langevin rate coefficients eventually approaches unity as the
temperature is raised.  A large compendium of ion-neutral reactions
and rate coefficients measured at assorted temperatures (but rarely
under 100 K), shows that the long-range capture models are often quite
accurate although they do not tell us anything about the product
channels and their branching fractions \citep{Anicich03}.  In addition
to ion-neutral reactions, low-temperature chemical processes do
include selected neutral-neutral reactions involving at least one radical,
because  these processes possess no activation energy barrier
and even have a weak inverse temperature dependence in their rate
coefficient \citep{Smith04}.

While ion-neutral reactions produce an assortment of molecular ions,
both positively and, to a lesser extent, negatively charged, the final
step in the production of neutral molecules like H$_2$O normally
occurs via dissociative recombination reactions between positive
molecular ions and electrons.  The rate coefficients and products of
these highly exothermic and rapid reactions have been studied via both
bench-top experiments; viz., the flowing afterglow apparatus, and
large-scale; viz., storage ring methods
\citep{Mitchell06,Larsson08}. The rate coefficients are typically
10$^{-7}$ cm$^3$ s$^{-1}$ at room temperature and vary with the
inverse square root of temperature.  Product channels are often dominated by
significant dissociation, such as three-body channels (e.g.,
H$_3$O$^+$ + $e$ $\to$ OH + H + H) rather than breakage of just one
bond.

Ions are produced by a variety of processes, with cosmic ray
ionization being the most universal because the cosmic rays,
travelling near the speed of light with energies ranging upwards of 1
MeV to more than 1 GeV, are able to penetrate large column densities
of material, as discussed in \S 1.  The energy spectrum of cosmic rays
cannot be fully determined by measurements above the Earth, however,
because the solar wind and the Earth's magnetic field interfere with
the low-energy flux, which is the most important for ionization since
the cross section for ionization depends inversely on the
translational energy of the cosmic
rays\citep{Rimmer12,Padovani09,Padovani13}. 
Using estimates for the low-energy flux in unshielded interstellar
space and the penetration efficiency in diffuse and dense sources,
combined with chemical simulations, the first-order rate coefficient
$\zeta_{\rm H}$ for ionization of atomic hydrogen directly by cosmic
rays and secondarily by electrons produced by cosmic ray bombardment
is found to be as high as $10^{-15}$ s$^{-1}$ in diffuse clouds and at
the edge of denser sources\citep{McCall03,Indriolo12}.  The value of
$\zeta_{\rm H}$ is reduced to less than $10^{-16}$ s$^{-1}$ deeper
into the cloud by $A_V=10$ mag, and eventually drops to $\sim
10^{-17}$ s$^{-1}$ in the interior of dense
clouds\citep{Rimmer12,Padovani13b}.  Propagation effects associated
with Alfven waves (a magnetohydrodynamic phenomenon) can also play a
role in excluding cosmic rays from dense cloud interiors, even for
somewhat smaller shielding column densities.

\subsubsection{Low temperature gas-phase formation of H$_2$O}

The ion-neutral synthesis of gaseous water commences with the
formation of molecular hydrogen on the surfaces of dust particles (see
\S~\ref{sect:icechem}), after which H$_2$ is either ejected
immediately or sublimates within a short period, even at temperatures
as low as 10 K
\citep{Hollenbach71,Katz99,Chang05,Iqbal12,Cazaux02,Cazaux10err}. The
formation of H$_2$ occurs with high efficiency even in diffuse clouds.
Indeed, in some diffuse clouds with $A_V<1$ mag, approximately half of
the hydrogen has already been converted from atoms to molecules
\citep{vanDishoeck86}. Ionization of H$_2$ by cosmic ray protons and
secondary electrons occurs with a first-order rate coefficient
$\zeta_{\rm H_2}\approx 2 \zeta_{\rm H}$ and leads primarily to the
hydrogen ion, H$_2^+$, and electrons \citep{Rimmer12}. Other products
include\citep{Cordiner09} H, H$^+$, and even H$^-$.

Once H$_2^+$ is produced, it is rapidly converted into the triatomic
hydrogen ion by reaction with ubiquitous H$_2$:

$$ {\rm H_2^+ + H_2} \to {\rm H_3^+ + H} \eqno(5) $$ with a near
Langevin rate coefficient\citep{Anicich03} of $(1.7 - 2.1) \times
10^{-9}$ cm$^3$ s$^{-1}$. At an H$_2$ gas density of 10$^4$ cm$^{-3}$,
the time scale between reactive collisions with H$_2$ is then 14
hours, which is a short time in astronomical terms.  The H$_3^+$ ion
does not react with H$_2$, but is destroyed more slowly by reaction
with electrons (timescale about 50 yr) and with a variety of
abundant neutral atoms and molecules \citep{Wakelam12,Herbst08}.  Reaction
with atomic oxygen leads mainly to the transitory OH$^+$ ion at an
overall rate coefficient of $1.2 \times 10^{-9}$ cm$^3$ s$^{-1}$, with
a product branching fraction of 0.70, and to the water ion with a
branching fraction of 0.3 \citep{Anicich03}:

$$ {\rm O   +   H_3^+} \to  {\rm OH^+  +  H_2;   H_2O^+  +  H} \eqno(6) $$
The hydroxyl ion reacts rapidly with H$_2$ to form the water ion:

$$ {\rm OH^+  +   H_2}  \to {\rm  H_2O^+   +   H}  \eqno(7) $$
which then reacts with H$_2$ to form the saturated hydronium ion
(H$_3$O$^+$) + H.  Although the ions OH$^+$ and H$_2$O$^+$ are removed
rapidly by reaction with H$_2$, there are many sources in which these
ions can be detected as long as the clouds have a relatively high
H/H$_2$ fraction
\citep{Wyrowski10oh+,Gerin10,Benz10,Bruderer10,Gupta10,Wyrowski10,Menten11}
(see \S~\ref{sect:diffuse}).

The sequence of reactions leading to the hydronium ion can also start
with protons, which undergo a slightly endothermic charge transfer
reaction with oxygen atoms \citep{Steigman71}:

$$ {\rm H^+ + O} \to {\rm H + O^+} \eqno(8) $$ after which a reaction
with H$_2$ leads quickly to OH$^+$ and H.  The charge exchange route
is more efficient in diffuse clouds where at least 50\% of the
hydrogen is in the form of atoms and the temperature is high enough
(50 -- 100 K) that the weak endothermicity of 226 K can be
overcome. Measured values\citep{Anicich03} of the rate coefficient at
300 K range from $(6.0 - 6.8) \times 10^{-10}$ cm$^3$ s$^{-1}$. Computed
values are up to a factor of 2 lower at this temperature \citep{Spirko03}.

The hydronium ion is destroyed principally by dissociative
recombination with electrons.  Although the rate coefficient at low
temperatures is very large, the fractional electron abundance with
respect to hydrogen in cold dense clouds of $10^{-7} - 10^{-8}$ is so
low that the overall rate of reaction is not rapid.  Nevertheless, it
is this reaction that leads to water as well as to OH and possibly O:

$${\rm H_3O^+ + e} \to {\rm H_2O + H; OH + H_2; OH + 2H; O + H_2 + H}
  \eqno(9)$$

  A number of experiments have been undertaken on this
  system\citep{Jensen00,Buhr10,Neau00,Herd90}, and the currently
  accepted values for the overall rate coefficient as a function of
  temperature and the product branching fractions derive from storage
  ring experiments, the most recent in Heidelberg \citep{Buhr10}.  The
  branching fractions from this experiment (on D$_3$O$^+$) in the low
  temperature limit are 0.165 (D$_2$O + D), 0.125 (OD + D$_2$), 0.71
  (OD + D + D), and 0.0 (O + D$_2$ + D). These values are in excellent
  agreement with an earlier storage ring study in Stockholm
  \citep{Neau00}, but somewhat less so with an earlier result from a
  storage ring in Aarhus \citep{Jensen00} and a flowing after glow
  study \citep{Herd90}.  The Stockholm group found no isotopic effect,
  so that the Heidelberg results can be used for H$_3$O$^+$.  Note
  that water is not the major product channel in any of these
  experiments.

  The destruction of gaseous water is mainly by reactions with
  abundant ions such as H$_3^+$, HCO$^+$, C$^+$, H$^+$, and He$^+$.
  With its dipole moment of 1.85 Debye, the ion-neutral rate
  coefficients at low temperatures are considerably greater than the
  Langevin value.  The measured value \citep{Anicich03} for the
  reaction with H$_3^+$ at room temperature, 300 K, is $5 \times
  10^{-9}$ cm$^3$ s$^{-1}$, and the estimated value using the
  trajectory scaling model is also this value.  At a temperature of 10
  K, the predicted rate coefficient would be approximately a factor of
  4.6 greater.  The reactions of water with protonating ions; e.g.,
  H$_3^+$ and HCO$^+$, lead to H$_3$O$^+$, but dissociative
  recombination leads to species other than water on 83.5\% of
  collisions, so that the processes are not cyclic in the main and do
  lead to the destruction of water with a high efficiency.  The
  balance between formation and destruction of water in dense clouds
  leads to a prediction\citep{Smith04} of a small fractional abundance
  of water of $10^{-8}$ to a few$\times 10^{-7}$ in pure gas-phase
  models. This is consistent with observations of diffuse and
  translucent clouds but considerably greater than what is
  observed in cold dense clouds (see \S~\ref{sect:diffuse}).

\subsubsection{H$_2$O photodissociation}
\label{sect:photodissociation}

Destruction of neutrals such as water can occur not only by ions but
also via photo-induced processes. The photodissociation of H$_2$O in
the gas phase has been very well studied, both experimentally
\citep{Okabe78,vanHemert79,Yoshino96,Cheng99,Fillion01} and
theoretically \citep{Engel88,Engel92,Schinke93,vanHarrevelt01}. Many
of the excited electronic states of water are dissociative, i.e., the
potential is either fully repulsive leading to direct dissociation or
the potential is bound in one direction but dissociative in another
direction. Alternatively, a bound state can couple with a potential
surface of another symmetry leading to pre-dissociation. For the
interstellar radiation field, the photodissociation is dominated by
absorption into the $\tilde A$ state, which dissociates into H + OH in
the electronic ground state, $X ^2\Pi$ (see
Fig.~\ref{fig:fillion}). The OH molecules are produced with low
rotational excitation but with considerable vibrational excitation,
which increases with photon energy \citep{vanHarrevelt01}. At the
experimental energy of 7.87 eV (1576 \AA), close to the peak
absorption energy, the distribution\citep{Hwang99,Yang00} over
$v$=0, 1, 2, 3 and 4 levels of OH is 1.00 : 1.11 : 0.61 : 0.30 : 0.15.

Photodissociation through the $\tilde B$ state is important at higher
photon energies, including Lyman $\alpha$ at 1216 \AA. In this case,
the OH is produced not only in the ground state but also for a small
fraction in the excited electronic $A ^2\Sigma^+$ state. The OH $A$ state
will rapidly decay radiatively to the $X$ state. The OH is produced
vibrationally cold, with the $X ^2\Pi$ $v$=0, 1, 2, 3 and 4 populations
scaling as $\sim$0.60: 0.10 : 0.09 : 0.08 : 0.07 , with the fractions
only weakly dependent on photon energy
\citep{vanHarrevelt00,Fillion01}. In strong contrast with the $\tilde
A$ state, the photodissociation of water through the $\tilde B$ state
produces OH molecules with very high rotational excitation, with
levels up to $J$=70/2 populated for $v$=0. The precise $J$ values
depend sensitively on the energy above the OH(X) or OH(A)
threshold. Specifically, there is a strong preference for populating
the highest rotational product state for which the rotational barrier
energy is lower than the available photon energy, the so-called
`single $J$ phenomenon' observed experimentally
\citep{Harich00,vanHarrevelt01comp}. In \S~\ref{sect:comets}, specific
interstellar regions will be discussed where OH is produced
prominantly through this channel and where the `prompt emission' of
highly rotationally excited OH is observed.

A small fraction of absorptions into the higher electronic states of
H$_2$O can also lead to the O + H$_2$ or O + H + H products
\citep{Okabe78,Fillion01,vanHarrevelt01comp}. For the general
interstellar radiation field, the overall fraction is about 10\%
compared with the H + OH products.

There are two different treatments for translating the cross sections
into photodissociation and photoionization rates to be used in
astrochemical models.  The first is for radiation that originates
external to the source and can penetrate significantly, such as in
diffuse interstellar clouds and dense PDRs.  Here the rate (s$^{-1}$)
as a function of depth $A_V$ into the cloud is often fit to the
expression \citep{vanDishoeck88photo,Roberge91}

$$ k_{pd}= \alpha \exp{(-\gamma A_V)} \eqno(10)$$
where $A_V$ is the visual extinction, $\alpha$ is related to the
strength of the radiation field in the absence of extinction as well
as the absorption cross sections to the appropriate excited electronic
states (Fig.~\ref{fig:fillion}), and $\gamma$ is a unitless parameter
that takes into account the extinction at shorter wavelengths than the
visible, which are particularly important for photodestruction.  For
gaseous water in the unshielded average interstellar radiation
field\citep{Draine78}, the photodissociation parameter for the
production of OH and H is \citep{vanDishoeck06} $\alpha = 8.0 \times
10^{-10}$ s$^{-1}$ whereas the depth dependence is given by
$\gamma$=2.20.  See {\tt www.strw.leidenuniv.nl/$\sim$ewine/photo} for
more information. For clouds exposed to more intense radiation, this
rate is scaled by a factor $\chi$.

In dense regions shielded from stellar radiation, there is still a
remnant UV radiation field caused by secondary electrons produced by
cosmic ray bombardment, mainly of molecular hydrogen in the gas, as
discussed in \S~\ref{sect:radiation}.  The photodissociation rate
coefficient (s$^{-1}$) for this radiation source is fit to the
expression \citep{Gredel89}

$$ k_{pd} = \zeta_{\rm H2} A/[C^{\rm H}_{\rm ext} (1-\omega)] \eqno(11)$$ 
where the cosmic ray ionization rate $\zeta_{\rm H_2} $ (s$^{-1}$)
inside dense clouds is of order $10^{-17}-10^{-16}$ s$^{-1}$, $C^{\rm
  H}_{\rm ext}$ is the dust extinction cross section per H atom in
cm$^2$, $\omega$ is the dust albedo at UV wavelengths with a typical
value of 0.5, and $A$ contains the integral of the molecular
photodissociation cross sections with the cosmic-ray induced spectrum.
For water, the rate is calculated \citep{Gredel89} to be 970--980
$\zeta_{\rm H2}$ s$^{-1}$ for the production of OH and H.

The role of X-rays in the destruction of water and other neutrals can
also be important in regions near sources that produce copious amounts
of X-rays such as near the midplane of protoplanetary disks around cool T Tauri
stars \citep{Aikawa99} and the inner envelopes around low-mass
protostars \citep{Stauber06}.

\subsection{High-temperature gas-phase chemistry}
\label{sect:hightemp}

As gas-phase temperatures rise from the 10 K value of cold cores, the
relative importance of endothermic reactions and exothermic reactions
with activation energy increases.  The subject has been looked at anew
by \citet{Harada10}. Among the most important of such
reactions are those between neutral species and H$_2$, the most abundant
gas-phase molecule.  The first reactions of this type to `turn on'
involve radicals or atoms.  For the case of water production, the two
key reactions of this type are

$$ {\rm O   +   H_2} \to  {\rm    OH   +    H} \eqno(12) $$

and

$$ {\rm OH   +   H_2} \to  {\rm H_2O   +   H} \eqno(13) $$

Both of these reactions have been studied by many groups, and there
are theoretical and experimental studies in assorted temperature
ranges, as well as critical reviews of the
literature\citep{Baulch92,Wakelam12}. The rate coefficients determined
from the data are typically fit to a modified Arrhenius form with
three parameters:

$$ k(T)= \alpha (T/300 {\rm K})^\beta \exp(-\gamma/T) \eqno(14)$$

The 1992 review by Baulch et al.\citep{Baulch92} was a careful study
of the existing literature at the time; a rate coefficient $k$ (cm$^3$
s$^{-1}$) for reaction (12) of $3.44 \times
10^{-13}(T/300)^{2.67}\exp(-3160/T)$ was suggested for the temperature
range 300 K - 2500 K.  This reaction is endothermic by roughly 900 K
(1 kcal/mol = 503 K) but has a much larger barrier, more precisely a
value for $\gamma$ of 3160 K.  At a temperature of 300 K, the
recommended rate coefficient is $9.2 \times 10^{-18}$ cm$^3$ s$^{-1}$.
A more recent theoretical value of $7.1 \times 10^{-18}$ cm$^3$
s$^{-1}$ was computed \citep{Balakrishnan04} for a temperature of 300
K, in reasonable agreement with the recommended value.  If one
compares the rate of the reaction per O atom ($k_{12} n$(O)) with the
analogous rate for the initial reaction for the ion-molecule synthesis
(O + H$_3^+$), the O + H$_2$ rate becomes roughly equal to the
ion-molecule rate at 300 K with a typical fractional abundance of
$10^{-8}$ for H$_3^+$ with respect to H$_2$ in dense clouds.  At a
somewhat lower abundance of H$_3^+$, the temperature at which the
neutral-neutral reaction dominates\citep{Charnley97} is somewhat
lower, such as 230 K.  At temperatures higher than 300 K, the
neutral-neutral mechanism certainly dominates, if the OH product then
continues efficiently to water by reaction (13).  More recent
experiments for reaction (12) have mainly studied the high-temperature
limit ($>$ 1000 K).  For example, if the high temperature studies of
\citet{Javoy03} are extrapolated down to 1000 K, a rate coefficient
for O + H$_2$ of $\sim 9 \times 10^{-14}$ cm$^3$ s$^{-1}$ is obtained,
whereas the recommended value of Baulch et al. is $4 \times 10^{-13}$
cm$^3$ s$^{-1}$.  In summary, the 1992 recommended value still seems
to be reasonable over a wide range of temperatures relevant for
interstellar applications.

Regarding  the reaction of OH + H$_2$, which is highly exothermic
but possesses a considerable barrier, the most recent review
recommends a rate coefficient of $7.7 \times 10^{-12} \exp(-2100/T)$
cm$^3$ s$^{-1}$ over the temperature range 200-450 K
\citep{Atkinson04}, with a 300 K value of $6.7 \times 10^{-15}$.  The
1992 critical review of Baulch et al. recommends a value over the
temperature range 300-2500 K of $1.55 \times 10^{-12} (T/300)^{1.60}
\exp(-1660/T)$ cm$^3$ s$^{-1}$, leading to a rate coefficient of $6.1
\times 10^{-15}$ at room temperature.  Both of these rate coefficients
are smaller than the one from the expression adopted by Harada et al.
\citep{Harada10} ($k = 8.40 \times 10^{-13} \exp(-1040/T)$). The rate
coefficients must be compared with those for other competitive
reactions of OH with abundant species, such as OH + O, to determine
the efficiency of the neutral-neutral reaction route to water.  At
room temperature, the latter reaction has a rate
coefficient\citep{Atkinson04} of $3.5 \times 10^{-11}$ cm$^3$
s$^{-1}$. When multiplied by a standard O atom fractional abundance
with respect to H$_2$ of 10$^{-4}$, one gets a rate per OH of $3.5
\times 10^{-15}$ cm$^3$ s$^{-1}$, somewhat smaller than the rate per
OH of the OH + H$_2$ reaction.  Thus, it is likely that the
neutral-neutral reactions leading to water formation in the gas-phase
dominate at room temperature and above.

When the H$_{2}$ reactant is excited to its $v=1$ vibrational state,
as is the case in the outer layers of dense PDRs and disks (\S 4), the
rate coefficients of both reactions (12) and (13) are known to
increase by orders of magnitude at 300 K, since the vibrational energy
is sufficient to overcome the barrier if it is useable for that
purpose \citep{Balakrishnan04,Sultanov04,Han00}. Moreover much of the
excess energy becomes vibrational energy of the product.

Depending on the H/H$_2$ ratio in the gas, the back reactions of
equations (12) and (13) can of course also occur and drive water back
to oxygen. Under most astrophysical situations, however, the forward
reactions dominate because of the higher activation energies involved
(Fig.~\ref{fig:network}).

The neutral-neutral reaction network that forms water is particularly
important in shocks \citep{Draine83,Bergin98} (see
\S~\ref{sect:shocks}). Following the passage of shock waves, the
internal energy of the water molecules relaxes much more quickly than
the translational energy, so that the situation is a distinctly
non-thermal one ($T_{\rm rot}<< T_{\rm kin}$). In such situations, the
use of state-selective rate coefficients according to the actual water
excitation, rather than thermally averaged rate coefficients, would be
more appropriate although this is seldom done (see
\citealt{Balakrishnan04} for the effect on the O + H$_2$ reaction).

\subsection{Ice chemistry}
\label{sect:icechem}

\begin{figure}
\includegraphics[angle=0,width=0.6\textwidth]{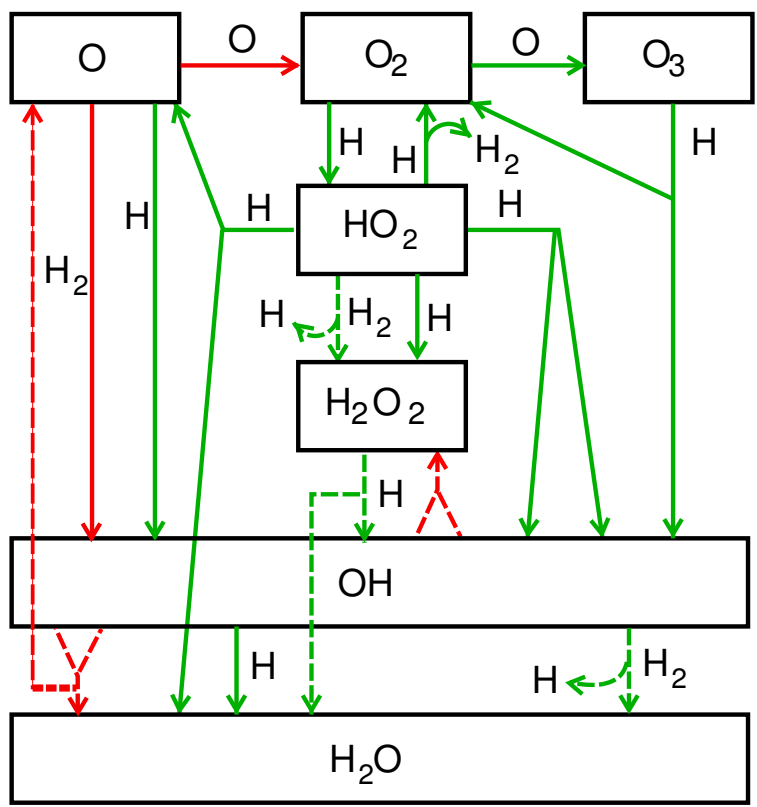}
\caption{Schematic representation of the surface chemistry reaction
  network leading to the formation of water ice, based on
  \citealt{Cuppen10} and \citealt{Lamberts13}. Green arrows refer to
  reactions that have been characterized by experiments or through
  simulations of experimental data. Red arrows refer to reactions that
  have not yet been studied experimentally or for which the
  experimental data are inconclusive.  Four cases are
  distinguished. (i) efficient, effectively barrierless ($<$500 K)
  reactions (green), (ii) reactions with a barrier above 500 K
  (dashed-green), (iii) reactions not studied experimentally (red),
  and (iv) reactions studied experimentally but for which results are
  inconclusive (red-dashed).}
\label{fig:icenetwork}
\end{figure}

\subsubsection{General considerations}

For most sources other than those at high temperatures, it is
important to include the chemistry that occurs on dust grains,
especially for water. Surface chemistry is coupled to the gas-phase
chemistry via accretion and desorption processes, the latter both
thermal and non-thermal.  Thermal desorption is another term for sublimation,  
while non-thermal processes include photodesorption,
which has recently been studied in the laboratory and through
classical molecular dynamics simulations (see below).

Technically, the term sublimation refers to solid to gas conversion
under complete equilibrium conditions, where both desorption and
accretion occur at equal rates. Thermal desorption refers to
desorption from a surface that occurs into a vacuum. Since desorption
always occurs in competition with accretion under astrophysical
conditions, we use primarily the term sublimation to describe the
sold-gas transition. The term thermal desorption is normally used to
describe temperature-programmed desorption experiments in the
laboratory.

In addition to the formation of H$_2$, low-temperature grains provide
the surfaces for reactions that form water ice and a number of other
ices, with methanol the most complex yet observed
\citep{Gibb04,Oberg11}.  Since water is the dominant observed ice in
dense interstellar clouds and protostellar regions, the chemistry of
its formation has been studied extensively.  Direct accretion from the
gas in dense regions can only explain the abundance of CO ice among
the dominant interstellar ices.  So, ice chemistry is needed to
convert abundant accreting species such as oxygen atoms into water.
As depicted in Figures~\ref{fig:network} and \ref{fig:icenetwork},
three well-studied major routes exist for the low-temperature
formation of water ice\citep{Tielens82}, the simplest of which is the
hydrogenation of atomic oxygen, with more complex routes starting from
$s$-O$_2$ and $s$-O$_3$.  Both of these molecular species are formed
via surface reactions, in which $s$-O atoms associate.  Hydrogenation
occurs via reaction of the oxygen species with atomic hydrogen.  In
addition to this hydrogenation route, a new study indicates that water
ice can also be formed by the reaction of $s$-OH and $s$-H$_2$ on
grains via a tunneling process \citep{Oba12} even though it has a
barrier of 2100 K.

At low temperatures, the dominant mechanism for reactions on surfaces
is thought to be the so-called Langmuir-Hinshelwood (LH) process, in
which surface species, bound only by weak `physisorption'
forces, diffuse among so-called lattice sites of minimum potential
energy either by a classical random walk or by tunneling under
barriers between sites.  Unlike the gas-phase, the dominant reactions
on surfaces and ices, as they build up, are association reactions (A +
B $\to$ AB* $\to$ AB) in which rapid energy transfer from the complex (AB*)
to the grain allows the complex to stabilize. For cold grains, only
the most weakly bound species, such as atomic hydrogen and oxygen, can
diffuse, whereas at temperatures above 20--30 K, the more volatile heavier
species begin to diffuse and desorb thermally.  The LH mechanism can
be modeled by rate equations similar to those for the gas-phase
chemistry, in which the rate coefficients are related to the rate of
diffusion of both reactant partners.  

It is important to distinguish the chemistry that occurs on bare
surfaces (silicates and amorphous carbon) from the chemistry that
occurs on ice layers that build up in cold dense sources. If most of
the available oxygen is converted to water ice, the number of water
ice monolayers on each grain becomes of order hundred.  For such thick
layers, the underlying grain surface does not matter anymore for the
chemistry. Note also that ice chemistry can occur not only on the
surface but also in layers under the topmost one, where both bulk
diffusion and chemistry in pores can be competitive.  For these lower
layers, a simple LH mechanism such as described below may not be
relevant, and other possibilities have been suggested, such as the
swapping in position of nearby species \citep{Fayolle11co2}.

To discuss the water chemistry that occurs on the surfaces of ice
mantles of interstellar grains, we first need to consider rate
coefficients for accretion onto the grains and desorption from them as
well as those for assorted chemical processes.  In the following
discussion, we use upper case $K$ for the rate coefficients involving
grains to distinguish them from gas-phase rate coefficients.  Unless
stated to the contrary, we will use concentration units of number of
species per dust particle for adsorbed species.  More normal units for
laboratory processes are monolayers, and areal concentrations.  With
these conventions, the first-order accretion rate coefficient for a
given gas phase species, $K_{\rm acc}$ (s$^{-1}$), is given by 

$$K_{\rm acc}= \xi \sigma_d {\rm v}  \eqno(15) $$ 
where $\xi$ is the efficiency of sticking, $\sigma_d$ is the grain
cross section, and v is the thermal velocity of the accreting species
A.  The accretion rate (s$^{-1}$) for the number of species A per
grain is then the product of this rate coefficient and the gas-phase
concentration of A, $n$(A).  Other units for the accretion rate of
species A are easily obtained.  If it is desired to express the
accretion rate in terms of areal concentration per second (cm$^{-2}$
s$^{-1}$), then the rate coefficient must be multiplied by $n$(A) and
also divided by the surface area of the grain.  To convert this
expression to monolayers per second, one must divide the expression
for the rate in areal concentration per second by the lattice site
density, $N$, which lies in the vicinity of 10$^{15}$ cm$^{-2}$
depending upon the surface \citep{Katz99}.  Finally, if it is desired
to determine the volume rate of accretion, for use in gas-grain chemical
simulations of interstellar sources, one must multiply the rate coefficient by both the
concentration of species A and the grain concentration.  The sticking
efficiency $\xi$ is normally estimated from the theoretical expression
of Hollenbach \& Salpeter \citep{Hollenbach71} and has been studied
explicitly for H on water ice by molecular dynamics simulations
\citep{Buch91,Alhalabi07}. It varies strongly with temperature but is
close to unity at 10 K.

The thermal desorption/sublimation rate coefficient $K_{\rm subl}$
(s$^{-1}$) is given by the first-order Wigner-Polanyi equation
\citep{Wakelam10ssr}

$$ K_{\rm subl}= \nu \exp(-E_D/k_BT) \eqno(16) $$
In this expression, $\nu$ is the so-called trial frequency, which
corresponds roughly to the frequency of vibration of a physisorbed
molecule such as water trapped in a lattice site ($\sim 10^{12}$
s$^{-1}$), $k_B$ is the Boltzmann constant, while $E_D$ is the
desorption energy, or energy needed to remove a water molecule from
the ice into the gas phase.  The desorption energy of course depends
on the ice or bare surface from which desorption occurs.  For a water
molecule on an ice dominated by water, a commonly used value based on
laboratory studies\citep{Fraser01} is $E_D/k_B \approx 5600-5700$ K,
where this relatively high value for a physisorbed species is caused
at least in part by hydrogen bonding.

Non-thermal desorption mechanisms include photodesorption, desorption
following chemical reaction, and cosmic ray-induced desorption. The
photodesorption rate coefficient is normally reported in terms of an
efficiency per incident photon ($\sim 10^{-3}$) multiplied by the flux
(cm$^{-2}$ s$^{-1}$) of photons in the far UV portion of the spectrum
striking a grain.  To achieve units of molecules s$^{-1}$, the product
must be divided by the site density of the surface ($\sim 10^{15}$
cm$^{-2}$).  The overall rate of photodesorption per grain would then
be obtained by multiplying the rate coefficient by the number of
molecules on the surface or within a few monolayers of the surface.
Laboratory experiments with monochromatic light sources indicate that
photodesorption can either be a discrete or continuous process
depending upon the excited electronic state of the adsorbate initially
excited. Water ice provides an example of a continuous process
\citep{Oberg09h2o} (see Fig.~\ref{fig:icevsgas}), whereas CO ice provides
an example of a mainly discrete one \citep{Fayolle11}.

Other mechanisms for non-thermal desorption have been less explored.
Nevertheless, the results of a very recent laboratory study of the
surface reaction between $s$-D and $s$-OD, which produces deuterated
water on grains, indicates that more than 90\% of the product is
released into the gas phase \citep{Dulieu13}, although this fraction
depends strongly on the underlying surface. This high percentage does
not hold universally for other reactions, however, and the details of
so-called reactive or chemical desorption are not well understood.

It is normally assumed that the photodissociation rate for ice species
is the same as for the gas unless there is strong evidence to the
contrary, as does occur for water ice \citep{Andersson06}
(Fig.~\ref{fig:icevsgas}).  Also, unlike photodesorption, which is
mainly a surface to near-surface process
\citep{Andersson08,Oberg09h2o}, photodissociation can occur throughout
the entire ice mantle up to $\sim$100 monolayers. For water ice, a
small photoionization channel leading to H$_2$O$^+$ in the gas is not
included in ice models, but is added to the $s$-OH + $s$-H production
channel because it is assumed that the newly formed ion will undergo
dissociative recombination on the negatively charged grain surface.
Photodissociation can be followed by reactive desorption if the
products are reactive radicals.  For example, the photodissociation
products $s$-OH + $s$-H produced in the top ice layers can recombine
to form water once again, with a fraction being ejected from the
surface (see \S~\ref{sect:desorption}).

The LH diffusive rate coefficient is based on the site-to-site hopping
rate $K_{\rm hop}$ to a nearest neighbor site, which is similar to the
sublimation rate except that the diffusive barrier $E_b$ takes the
place of the desorption energy: viz.,

$$ K_{\rm hop}= \nu \exp(-E_b/k_BT) \eqno(17)    $$
where $\nu$ is once again the trial frequency.  The rate coefficient
is first-order and is used as written with concentration units of
molecules per grain.  In some treatments, tunneling under the
diffusion barrier is included for atomic hydrogen \citep{Iqbal12}. The diffusion
rate equivalent to a journey over an entire grain, $K_{\rm diff}$, is just the
hopping + tunneling rate divided by the number of lattice sites on the
grain $N$.  The LH rate coefficient for the reaction between, e.g., $s$-OH
and $s$-H is obtained by
calculating the rate at which the two diffusing reactants find
themselves in the same lattice site \citep{Herbst08}. For reactions without chemical
activation energy, the reaction is then assumed to occur
instantaneously, whereas for reactions with chemical activation
energy, a factor $\kappa <1 $ is included to take into account the more rapid of
the two processes of tunneling under the activation barrier
(transition state) or hopping over it.

The overall expression for $K_{\rm LH}$ for the surface reaction between
      $s$-OH and $s$-H is then given by 

$$ K_{\rm LH}=\nu \kappa (K_{\rm diff}(s{\rm -H}) + K_{\rm diff}(s{\rm -OH})), \eqno(18)$$
if the units for the concentration of $s$-OH and $s$-H are numbers of
species per grain.  Other possible units, which will involve an
additional factor in the expression for the LH rate coefficient,
include monolayers per grain, and actual volume concentrations, which
requires knowledge of the volume concentration of grains.  For the
radicals OH and H, $\kappa=1$, while for the reaction $s$-OH +
$s$-H$_2$ $\to$ $s$-H$_2$O + $s$-H, another pathway to form water,
$\kappa < 1$ because of chemical activation energy \citep{Oba12}.  It
should be noted that the role of activation energy for surface
reactions differs from that in the analogous gas phase process, both
because the actual values need not be the same, and because for the
surface process, there is a competition involving the two types of
barriers --- diffusive and chemical.  Thus, if the diffusion barrier
is higher than the chemical barrier, the two reactants will have many
chances to hop over or tunnel under the chemical barrier before they
diffuse away.  An analogous process in a gas-phase reaction involves a
long-lived complex, which is trapped statistically.
  
It is important to realize that the above formulation ignores the
option of the direct (Eley-Rideal) process in which the reactant lands
on top of the surface molecule, or the `hot atom' processs in which
the reactant reaches the surface with extra translational energy. More
general limitations of the above formulation are discussed in
\S~\ref{sect:advanced}.

Although relevant laboratory experiments on the build-up of ices such
as water on cold surfaces have been undertaken increasingly in the
last decade, it is still the case that very few quantitative studies
of actual rate coefficients are available \citep{Wakelam10ssr}. The
experiments typically deposit several layers of one of the reactants
on a cold substrate (around 10--30 K) and then bomdard the ice with
the second reactant, usually atomic H or O. The reaction is then
monitored {\it in situ} through infrared spectroscopy of the products
in the ice. For example, in experiments of O$_2$ ice bombarded by
H-atoms\citep{Ioppolo08}, the 3 $\mu$m features of H$_2$O$_2$ and
H$_2$O appear after minutes (for a typical H-atom flux of $\sim
10^{13}$ atoms cm$^{-2}$ s$^{-1}$) and the signal saturates after
a few hrs. Reaction rates can then be derived from these growth
curves depending on temperature and other experimental parameters.
Reaction products can also be probed through temperature programmed
desorption experiments, in which the ice is heated up after a certain
reaction period and products are measured using mass spectroscopy as
they come off the ice.

Unlike gas phase experiments, however, the transfer of even such
quantitative laboratory data to interstellar conditions is not simple.
The difference in time scales, for example, between laboratory (hours)
and interstellar processes (up to 0.1 Myr) can lead to difficulties in
interpretation \citep{Cuppen09}.  Rather than reaction rates, full
microscopic modeling of the experiments is needed to infer basic
molecular parameters such as binding energies $E_D$ and diffusion
barriers $E_b$, but such modeling has been done only in a few cases
\citep{Ward11,Lamberts13}.  Diffusion barriers $E_b$ are particularly
difficult to determine, and so are usually simply taken to be a
fraction of $E_D$ in the models.
Thus, the rate coefficients for
surface/ice processes found in papers in the astrochemical literature
have considerable uncertainty.

\subsubsection{Water ice formation}

With this background, we can look more closely at the various
approaches to the formation of interstellar water ice
(Fig.~\ref{fig:icenetwork}).
The mechanisms discussed below have been recent subjects of
experimental study
\citep{Ioppolo08,Ioppolo10o2,Miyauchi08,Dulieu10,Romanzin11,Lamberts13,Accolla13}.
We maintain the assumption that the formation of water ice occurs via
diffusive (Langmuir-Hinshelwood) reactions, but are aware of the
possibility that the chemistry on an ice mantle is far more complex
than this simple diffusive approach, as noted above.  With these
reservations, we discuss the studied mechanisms below, originally put
forward by \citet{Tielens82} before much experimental evidence
existed.

\begin{itemize}

\item {\it Mechanism 1:   starting from O.}
In this series of reactions, both atomic H and atomic O accrete onto a
grain, diffuse towards one another, and form the radical OH.  A second
H atom lands on the grain and diffuses to the OH to form water: 

$s$-H + $s$-O $\to$ $s$-OH; \ \ $s$-OH + $s$-H $\to$ $s$-H$_2$O.

\item {\it Mechanism 2: starting from O$_2$.} This mechanism starts with the
diffusive formation of molecular oxygen: $s$-O + $s$-O $\to$ $s$-O$_2$.  The
molecular oxygen can then add atomic H twice to form the $s$-HO$_2$ radical
and then $s$-H$_2$O$_2$: 

$s$-O$_2$ + $s$-H $\to$ $s$-HO$_2$; \ \ $s$-HO$_2$ + $s$-H $\to$ $s$-H$_2$O$_2$.

It is also possible for the second step to occur
with molecular rather than atomic hydrogen, although the amount of H$_2$
on or in the ice mantle is poorly determined.  Moreover, the reaction
between $s$-HO$_2$ and $s$-H$_2$ may have an activation energy barrier.  Once
hydrogen peroxide is produced, it can react with atomic hydrogen to
form $s$-OH and $s$-H$_2$O: 

$s$-H$_2$O$_2$ + $s$-H $\to$ $s$-H$_2$O + $s$-OH.
  
The newly formed $s$-OH can react with $s$-H to once again form
$s$-H$_2$O. Similarly, $s$-HO$_2$ + $s$-H can lead to 2 $s$-OH, which
can react further to $s$-H$_2$O.

\item {\it Mechanism 3: starting from O$_3$.}  Once $s$-O$_2$ is produced,
addition of another surface/ice oxygen atom can produce ozone. The
ozone can then react with $s$-H to form $s$-OH + $s$-O$_2$, followed by
reaction of $s$-OH with $s$-H to form water, as already discussed.

\item {\it Mechanism 4: use of H$_2$.} The reaction between $s$-OH and
  molecular hydrogen to form water and H is controversial
  \citep{Dulieu11}. There is a newly measured rate coefficient which
  suggests that the reaction can occur via tunneling although this
  reaction possesses considerable activation energy in the gas-phase
  \citep{Oba12}.

\end{itemize}

The role of mechanisms 1,2, and 4 have been compared by
\citet{Cuppen07} under a variety of physical conditions in the
interstellar medium.  Under diffuse and translucent cloud conditions
($A_V \leq 3$ mag; $n_{\rm H} \leq 10^3$ cm$^{-3}$) , mechanism 1
dominates, although little water ice is produced due to efficient
photodesorption of all grain surface species.  The controversial
reaction between $s$-OH and $s$-H$_2$ (mechanism 4) becomes important
under the conditions of cold dense cores ($A_V \leq 10$ mag; $5 \times
10^3 \leq n_H \leq 5 \times 10^4$ cm$^{-3}$) if it occurs at all.  In
the models of \citet{Cuppen07}, mechanism 2 occurs at the 20\% level
for dense cores but \citet{Ioppolo08} and \citet{Du12} find mechanism
2 to be of at least comparable efficiency based on new experiments and
models.  Of course, these results are very model dependent, since, for
example, the amount of solid phase H$_2$ differs strongly from model
to model.  Also, the results depend on the ratio of gas-phase atomic O
and H arriving at the surface. Nevertheless, it is important for
modelers to include all of the mechanisms so that water can be
produced under diverse physical conditions.

As the dust temperature rises from 10 K, larger species than atoms
begin to diffuse both on and inside the ice mantles.  Most of these
species are not reactive unless activated by the formation of
radicals, which can occur via photodissociation or via energetic
particle bombardment such as cosmic rays.  Association reactions of
radicals can lead to complex organic species, as discussed by Garrod
and Widicus Weaver in this volume. 

\subsubsection{Water ice desorption}
\label{sect:desorption}

{\it Thermal sublimation.} As the dust temperature rises from 10 K to
about 30 K, sublimation of the more volatile molecules, such as CO,
begins. Once the dust temperature rises above 100 K, water ice and
complex organic species sublimate on very short time scales
under interstellar conditions \citep{Sandford90,Fraser01}, leading to
high gas-phase abundances of water and organics.  The binding energy
$E_D$ of water ice to be used in Equation (16) is 5600~K for pure
amorphous ice, with a slightly higher value of 5770 K found for
crystalline ice \citep{Fraser01}. Above $\sim$100 K, the desorption
rate of interstellar ice is so rapid due to the exponential dependence
that the half-life time of ice mantles (i.e., the time it takes for
the surface population of H$_2$O molecules on an interstellar grain to
decrease to half its initial value) becomes less a year (see Table~2
in Fraser et al.\citep{Fraser01}). As a result, the gas-phase
abundance of water can become temporarily as high as the original ice
abundance, on the order of $10^{-4}$ with respect to H$_{2}$.

In realistic astronomical ice mixtures, the sublimation pattern
can be rather complex, since a number of different effects can occur
in which the combined ices do not have independent desorption rates
\citep{Collings04,McCoustra05}, but since water is the principal ice
ingredient this does not affect the water chemistry. Note that the
precise dust temperature at which rapid desorption sets in is density
dependent through the Clausius-Clapeyron relation, and can be obtained
by solving the desorption/accretion balance: at densities of $10^{13}$
cm$^{-3}$ found in the inner regions of protoplanetary disks
\citep{Meijerink09}, the sublimation temperature is increased to
160~K. Similarly, planetary surfaces such as those of the Jovian moons
Europa and Ganymede in our own solar system are covered by water
ice even though their temperatures are somewhat above 100 K.

{\it Photodesorption.} At dust temperatures below $\sim$100 K, the
best-studied mechanism that can get some small fraction of water ice
returned to the gas phase is photodesorption. This process has been
studied in detail in the laboratory \citep{Westley95,Oberg09h2o} and
through classical molecular dynamics simulations
\citep{Andersson06,Andersson08,Arasa10,Arasa11,Koning13,Arasa13}. It
starts with photodissociation of a water molecule in the ice after
absorption of a UV photon, followed by a number of subsequent
processes involving the dissociation fragments:

$$ s{\rm -H_2O} \to s{\rm -H +} s{\rm -OH} \to {\rm products} \eqno(19) $$

\begin{figure}
\includegraphics[angle=0,width=0.6\textwidth]{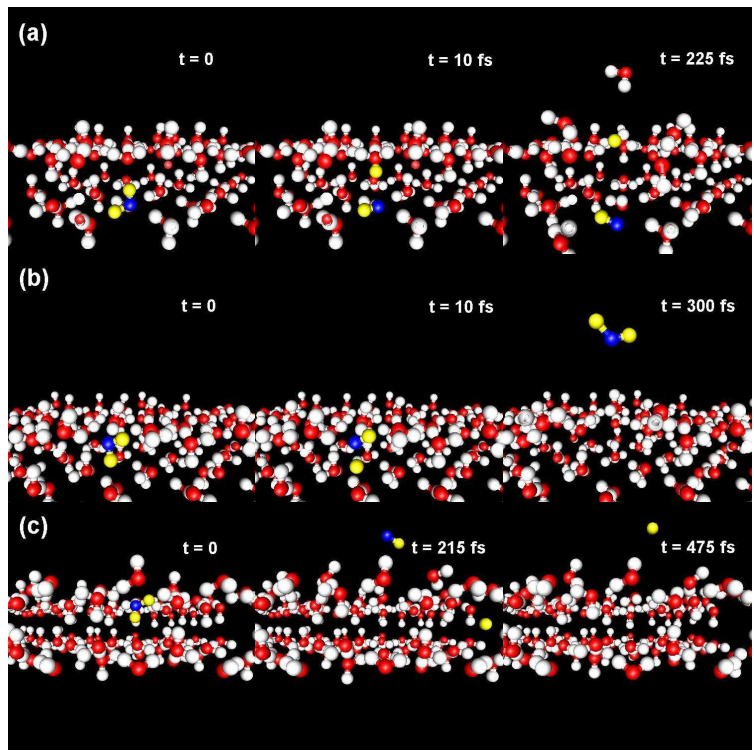}
\caption{ Snapshots of trajectories of mechanisms of H$_2$O photodesorption
  for a crystalline ice model. a) One of the surrounding molecules
  desorbs (`kick-out' mechanism).  b) The photofragments H and OH
  recombine and desorb as H$_2$O. c) The photofragments both desorb as
  separate species. The red and white atoms correspond to O and H
  atoms in the surrounding ice and the blue and yellow atoms
  correspond to the O and H atoms of the photodissociated H$_2$O
  molecule. Reprinted with permission from Reference 115. Copyright 2008
 European Southern Observatory.}
\label{fig:photodesorptionmech}
\end{figure}

Since both fragments have excess energy and are able to move through
the ice, there are a number of possible outcomes of the process
(Fig.~\ref{fig:photodesorptionmech}):

\begin{enumerate}

\item H desorbs while OH stays trapped in the ice

\item OH desorbs while H stays trapped in the ice

\item H and OH both desorb

\item H and OH are both trapped in the ice

\item H and OH recombine to H$_2$O which desorbs

\item H and OH recombine to H$_2$O which stays trapped in the ice

\end{enumerate} 

\begin{figure}
\begin{minipage}{0.475\textwidth}
\includegraphics[angle=-90,width=\textwidth]{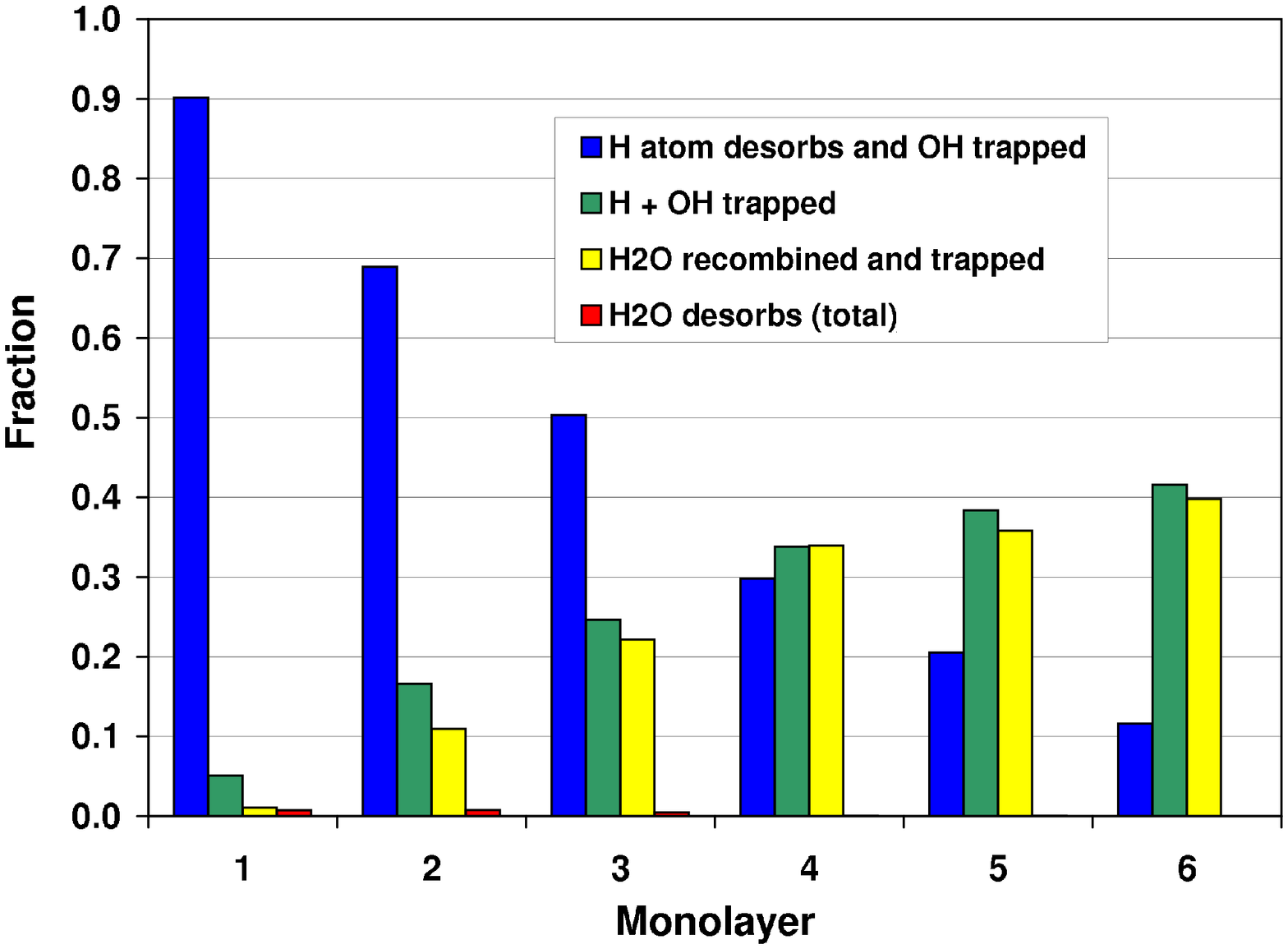}
\end{minipage}\hfill%
\begin{minipage}{0.475\textwidth}
\includegraphics[angle=-90,width=\textwidth]{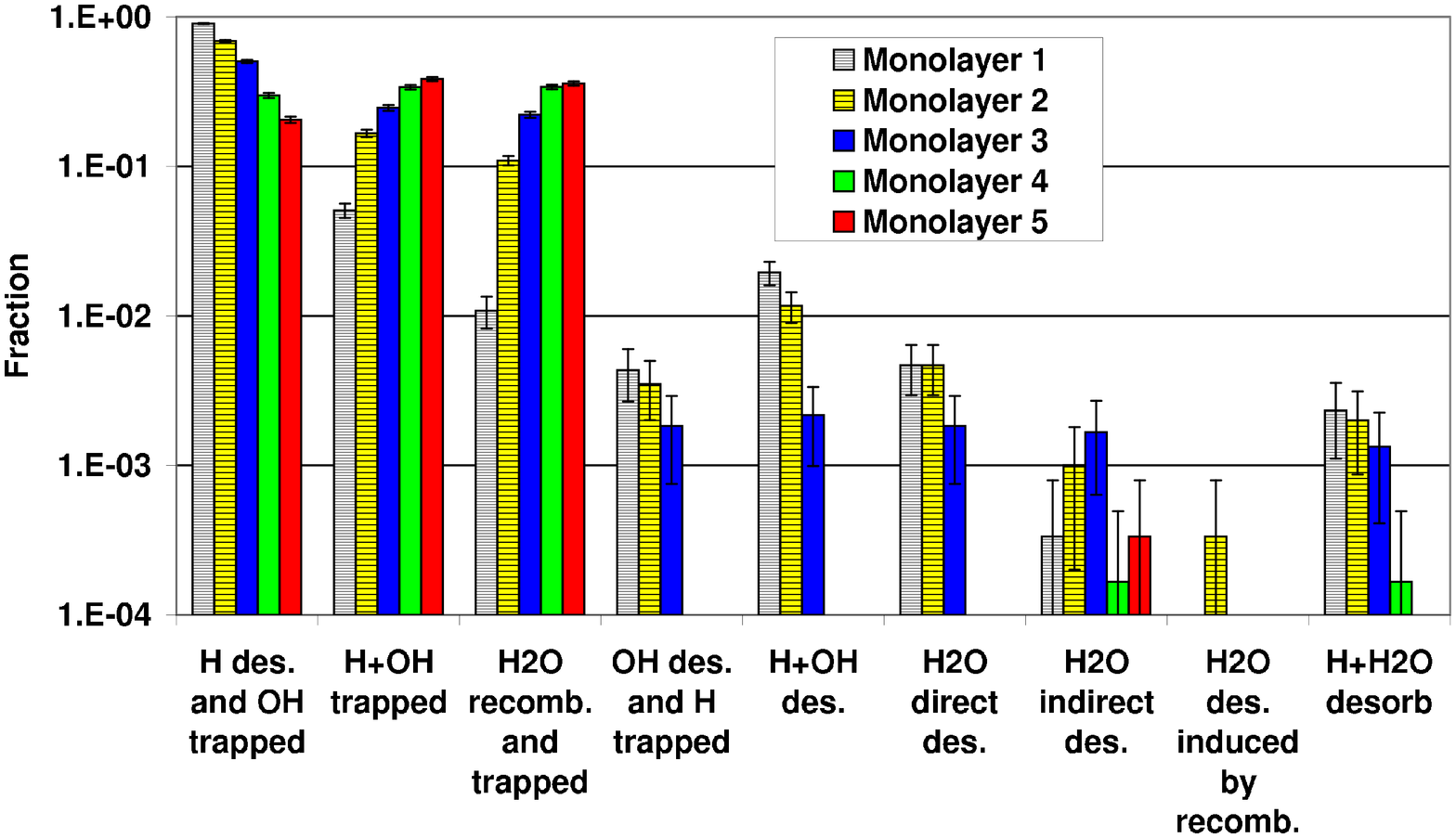}
\end{minipage}\hfill%
\caption{Left: Fractions of the main outcomes after H$_2$O ice
  photodissociation for the top six monolayers of amorphous ice on a
  linear scale. These probabilities are calculated from all
  trajectories in classical molecular dynamics simulations,
  irrespective of excitation energy. Note that H$_2$O desorbs in only
  a very small fraction of the outcomes.  Right: Fractions of the
  detailed outcomes after H$_2$O photodissociation for the top five
  monolayers of amorphous ice on a logarithmic scale. The error bars
  correspond to a 95\% confidence interval.  Reprinted with permission
  from Reference 115. Copyright 2008 European Southern Observatory.}
\label{fig:photodesorptionprob}
\end{figure}

The probabilities of each of these processes have been computed and
depend strongly on the depth into the
ice. Figure~\ref{fig:photodesorptionprob} summarizes the probabilities
for the first 6 monolayers for the case of a compact amorphous ice
structure \citep{Andersson08}. Consistent with experiments,
photodesorption of H, OH and H$_2$O (outcomes 1, 2, 3 and 5) only occurs
for the top few monolayers of the ice; the other outcomes (4 and 6)
are part of the normal photodissociation processes that also take
place deep inside the ice. The model finding that the OH desorption
probability is larger than that of H$_2$O is consistent with
experiments, although the theoretical OH/H$_2$O production ratio is
somewhat higher than the experimental value\citep{Oberg09h2o} of
$\sim$2. Interestingly, the probabilities for crystalline and compact
amorphous water ice are found to be very similar
\citep{Andersson06}. Also, variations with ice temperature in the
10--100 K range are small: there is a $\sim$30\% increase for the OH
and H$_2$O desorption probabilities when the ice temperature is
increased from 10 K up to 90 K \citep{Arasa10,Koning13,Oberg09h2o}.

In parallel with the six outcomes noted above, there is another
mechanism operating in the top few layers initiated by the same UV
photon: the energetic H atom can `kick-out' a neighboring H$_2$O
molecule before it desorbs, becomes trapped or recombines (see
Fig.~\ref{fig:photodesorptionmech}). Probabilities for this `kick-out'
mechanism and all other mechanisms are summarized for water and its
isotopologues in Arasa et al. \citep{Arasa10,Arasa11,Arasa13}.

There are two important caveats in these simulations. First, they only
consider excitation of H$_2$O ice into the first excited $\tilde A$
electronic state, but not into the higher $\tilde B$ or $\tilde C$
states. The probabilities for each of the outcomes may be different if
a higher electronic state is excited because the fragments will have
more translational energy. Also, the higher OH internal rotation
following the $\tilde B$ state dissociation (see \S
\ref{sect:photodissociation}) may change the relative outcomes. Since
the $\tilde A$ state is the dominating channel, the main conclusions
are unlikely to change, however. Second, the molecular dynamics
calculation only follows the processes on picosecond timescales. On the
much longer timescales in the laboratory experiments and in space,
slower diffusion processes may play a role. Wavelength dependent water
ice photodesorption experiments measuring both the OH and H$_2$O
channels are warranted.

\subsection{Solving the kinetic equations}

The chemistry that occurs in astronomical sources is rarely defined by
thermodynamics for a number of reasons.  In low temperature sources
such as cold interstellar clouds, most reactions are exceedingly slow
and do not occur significantly within the lifetimes of the sources.
Those reactions that do occur efficiently must be barrierless and
exothermic; typical destruction routes are not the reverse of
formation routes so that even though a steady-state can occur
eventually, the situation never reaches true thermodynamic
equilibrium, although astronomers often use the term inappropriately.
Also, the UV radiation field is characterized by a much higher
temperature than the kinetic temperature.  It is thus necessary to
solve the kinetics directly.  The reactions that directly lead to the
formation and destruction of water in various sources and varying
conditions depicted in Figure~\ref{fig:network} must therefore all be
considered.  A summary table of important gas-phase reactions leading
to water is given in Table~1 of \citet{Hollenbach12}. Networks of
chemical reactions to be included in general models of interstellar
chemistry can be found in assorted URLs such as the UDFA and KIDA
sites {\tt www.udfa.net/} and {\tt kida.obs.u-bordeaux1.fr/model};
these networks can contain up to 10,000 reactions involving 1000
species; many of the reactions included have not yet been studied in
the laboratory under the relevant physical conditions.
The KIDA database also contains codes for the solution of the
chemistry via rate equations.

\subsubsection{Rate equations and models}

The standard approach to the determination of time-dependent atomic
and molecular concentrations $n$(AB) in a system of coupled chemical
reactions is to use standard kinetic (rate) equations for each of the
species in the system, and to integrate them as a function of time
subject to constraints such as the total gas density and temperature,
the elemental abundances, and the initial abundances of each species.
Rather than concentrations, the modelers usually report `abundances' ,
which are referenced to the density of molecular hydrogen:
$x$(AB)=$n$(AB)/$n$(H$_2$) or to total hydrogen nuclei $n_{\rm
  H}$=$n$(H) + 2$n$(H$_2$): $x$(AB)= $n$(AB)/$n_{\rm H}$. Note the
  factor of at least 2 difference between these two definitions.  In
  addition to chemical reactions, processes such as photodissociation
  and photoionization need to be included, as well as accretion and
  desorption processes for models that include grain chemistry.

In the simplest models, the density and temperature remain homogeneous
and time independent.  These models are often labeled
`pseudo-time-dependent' because only the chemistry changes with time,
eventually reaching a steady state.  If only the temperature is
allowed to change with time and the density is constant and
homogeneous, the term used to describe the situation is `0-D', a terse
term meaning that there is no dependence on spatial parameters.  If
the physical conditions change along one particular axis, the term
used is `1-D'.  This term is often used to describe diffuse and
translucent clouds, and more generally the photon-dominated regions
(PDRs) in which a nearby star is the source of photons traveling into
a nearby cloud \citep{Hollenbach97}. The photons cause both the physical
diversity as well as the consequent chemical diversity along the axis.
Static 2-D models in which the physical parameters remain constant
with time also exist: examples include protoplanetary disks
\citep{Aikawa02} or protostellar envelopes in which an outflow cavity
has been carved out \citep{Bruderer10}. The most complex physical
models couple hydrodynamics (either semi-analytic or numerical, in 1D,
2D or 3D) with chemistry; these have been calculated primarily for
low-mass star-forming regions representative of the origins of our
solar system, in which material collapses inward to form a star
surrounded by a protoplanetary disk
\citep{Lee04,Aikawa08,Visser09,Aikawa12}.

Although most sources studied by astrochemists contain both a gas
phase and a condensed phase consisting of small dust particles, some
aspects of the chemistry can be handled by networks that are almost
exclusively gas-phase in nature, especially in low-density diffuse
clouds.  The one major exception is the formation of molecular
hydrogen from two neutral H atoms, which can only occur efficiently in
the interstellar medium on the surfaces of dust particles.  In such a
gas-phase model, the formation and destruction of gaseous water at low
temperature (Figure \ref{fig:network}) is incorporated through the
differential rate equation

$${d{\rm[ H_2O]}/dt}_{\rm LT} = k_{\rm dr} {\rm [H_3O^+][e]} - k_{\rm im}{\rm
  [H_2O][I^+]} - k_{\rm pd}{\rm [H_2O]} \eqno(20) $$ where LT stands
for low temperature, the symbols [...] for concentration with I=Ion,
and only the dominant reactions are included.  Formation occurs
through a dissociative recombination reaction, which also forms the
radical OH.  Such reactions tend to become more rapid as the
temperature is reduced; viz., $k_{\rm dr}\propto T^{-0.5}$
(\S~\ref{sect:ion}). It is easily seen that this equation is coupled
with differential equations for the abundances of other species, such
as that for the protonated water ion, H$_3$O$^+$, which itself is
formed by the series of ion-neutral reactions depicted in
Figure~\ref{fig:network}.  Destruction of water occurs through a
number of ion-neutral reactions involving positive ions $I^+$ such as
HCO$^+$ and H$_3^+$; photodissociation is also important under certain
conditions.

For sources or portions of sources in which the temperature exceeds
200 K, neutral-neutral reactions can become dominant even if they are
endothermic or possess barriers.  For water, in particular, equation
(20) has to be supplemented (or replaced) by the formation reaction
between OH and H$_2$ (reaction 13), which is exothermic but possesses
an activation energy barrier:

$$d{\rm [H_2O]}/dt =  {d{\rm [H_2O]}/dt}_{\rm LT}  +  k_{\rm OH-H_2} {\rm [OH][H_2]}.
\eqno(21)$$

The activation energy barrier is also counteracted by the high
abundance of H$_2$, which is 10$^4$ times greater than the second most
abundant molecule, CO.  At high temperatures, the radical OH is itself
formed mainly via the neutral-neutral between O and H$_2$ (reaction
(12)), as depicted in Figure~\ref{fig:network} and discussed above.
Up-to-date general networks for the gas-phase chemistry including
reactions and their rate coefficients as functions of temperature can
be found in the above mentioned KIDA and UDFA data bases; the former
contains a special high-temperature addendum \citep{Harada10} for
temperatures up to 800 K.

For gas-phase models, sensitivity analyses are available to determine
the relative importance of reactions for specific species under given
physical conditions, and to determine the uncertainty of the
calculated abundances based on the uncertainties, measured or
estimated, of the rate coefficients in the reaction network utilized
\citep{Wakelam10b}. For example, for the case of water vapor, the
cumulative uncertainties in all gas-phase reactions leading to water
result in a $\pm$0.5 dex spread in water abundances for a dark cloud
model \citep{Wakelam06}.  Such analyses provide useful information in
the ongoing collaborations between modelers of interstellar sources
and the laboratory and theoretical chemists who measure or calculate
rate coefficients for use in the chemical simulations.

\subsubsection{Gas-grain simulations and surface chemistry}

The solution of gas-grain chemical simulations of interstellar sources
is most easily accomplished by using rate equations for both gas-phase
and grain-surface reactions.  In the rate equations, both accretion
and desorption processes must be included.  Using water as an example,
a simplified version of the rate equation of water ice, denoted as
$s$-H$_2$O, can be written as

$$d {\rm [s-H_2O]}/dt = K_{\rm LH} {\rm [s-OH] [s-H]}  +  K_{\rm LH}{\rm [s-OH][s-H_2]}  
+ K_{\rm acc}{\rm [H_2O]}$$ 

$$ - K_{\rm subl} {\rm [s-H_2O]}
                         -  K_{\rm ntd} {\rm [s-H_2O]}	- K_{\rm pd} 
       {\rm [s-H_2O]}				\eqno(22) $$
where, as before, an upper case $K$ is used for the rate
coefficients on the grains to distinguish them from gas-phase
       rate coefficients, and no distinction is made between bulk and
       surface reactants.  Although chemical destruction routes for
       other surface molecules exist, for water the dominant
       destruction is via non-thermal desorption (ntd) at low
       temperatures and sublimation (subl) at temperatures when they
       rise to near 100 K.  Photodissociation (pd) in which the H
       and/or OH fragments remain in the ice or photodesorb can also
       be important (see \S~\ref{sect:desorption}).

\subsubsection{More advanced approaches}
\label{sect:advanced}

In the previous sections, the rate equation method for surface/ice
chemistry was discussed, with no distinction being made between the
surface and the interior layers of the ice mantle, other than for the
distinction between photodesorption and photodissociation.  It is
possible to use the rate equation approach and consider the surface
and interior layers separately; while the simple approach is known as
a two-phase approach (surface + interior and gas phase), the more
complex method is known as a three-phase approach (surface, interior,
gas phase).  Although this approach has been tried since the initial
work of \citet{Hasegawa93}, it does not solve all of the problems with
the rate equation approach.  One rather basic problem is that grains
are small and the flux of accreting species is also small, except in
the densest cores.  The result is that on some of the smaller grains,
with radii much less than the standardly used value of 0.1 $\mu$m, 
the average abundance of reactive species such as H
can be considerably less than 1 per grain.  In such a situation, one
must take account of both the discreteness of the rate problem, and
the fact that fluctuations are likely to be large.  For example,
consider the formation of molecular hydrogen under these
circumstances.  For the process to happen, there must be at least two
hydrogen atoms on a given grain at the same time, clearly a
fluctuation. To achieve these goals requires what is known as a
`stochastic' approach, in which the laws of probability are
utilized, although it is possible to mimic some of the stochastic
effects with an empirical modification of the rate equations \citep{Caselli98}, whether
the calculation be a two-phase or three-phase one.  This modification, known 
as the modified rate approach, has been improved by Garrod \citep{Garrod08} and
shown to be a good substitute for more time-consuming stochastic methods over certain 
ranges of physical conditions \citep{Garrod09}.

There are two principal stochastic approaches, which can be applied to
grain surface/ice chemistry --- one is known as the master equation
approach \citep{Stantcheva02} and the other is known as the Monte
Carlo approach \citep{Tielens82,Caselli02model,Chang05}.  In the
former approach, the concentrations are replaced by probabilities.
For example, if we consider a simple system in which only H atoms land
on grains, one would formulate individual deterministic equations for
the probabilities of 0 H atoms, 1 H atom, 2 H atoms, etc. on a grain;
their solution would then give a time-dependent distribution of H
atoms, which would allow one to determine both the average and the
standard deviation.  Unfortunately, if we expand the system from one
surface reactant to many, we have to compute joint probabilities;
e.g., the simultaneous probability of 1 H atom, 1 O atom, 10 CO
molecules, etc. on an individual grain, resulting in very large
numbers of equations to solve and necessitating some approximations.
At present, the most popular approximation to the master equation
method is known as the method of moments, and has been applied
successfully on a small number of attempts \citep{Du11}.  It is used
in a `hybrid' approach with rate equations depending upon whether the
abundances of reactants are large or small.  The strong point of a
method based on the master equation is that the differential equations
can be solved simultaneously using one integrator with the rate
equations used for the gas phase chemistry.

\begin{figure}[t]
  \includegraphics[angle=0,width=0.7\textwidth]{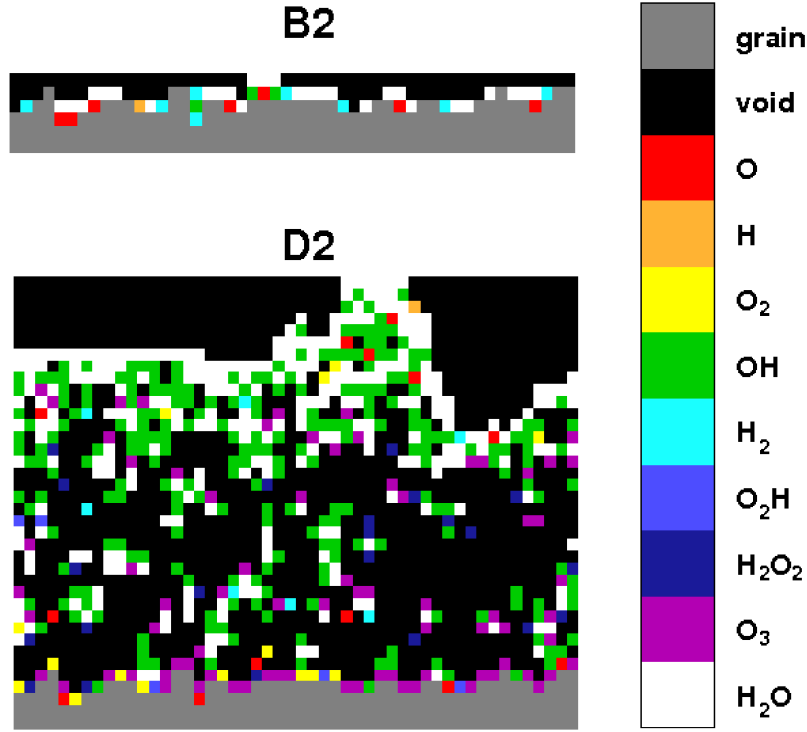}
  \caption{ Vertical cross section of the production of water ice
    after $3 \times 10^5$ yr in a translucent interstellar cloud from
    different mechanisms shown in Figure~\ref{fig:network}.  The CTRW
    Monte Carlo results are taken from the B2 and D2 models of
    \citet{Cuppen07} with $n_{\rm H}=250$ cm$^{-3}$, $T_d=16$ K and $A_V=1$
    mag and $n_{\rm H}=10^3$ cm$^{-3}$, $T_d=14$ K and $A_V=3$
    mag, respectively. Water molecules are shown in white, OH radicals in green,
    H$_2$O$_2$ in blue, H$_2$ in light blue, O$_2$ in yellow, O$_3$ in
    purple, and pores in black.  The topmost layers of the actual
    grain are seen to be rough. Reprinted with permission from Reference 252.
    Copyright American Astronomical Society.}
\label{fig:cuppen}
\end{figure}

The Monte Carlo approach is based on the choice of random numbers to
mimic the stochastic chemistry.  In a given time period, one compares
the rates of various processes; e.g., chemical reactions on the
surface, accretion, desorption, and links them to proportional ranges
of random numbers in the range 0-1.  Thus if there are two processes
to consider and one is twice as fast as the other, the faster one
could be assigned random numbers from 0-0.6666 and the slower from
this number to unity.  So, after many time periods of calling random
numbers, the comparison in rate between the two processes should
indeed be two.  The rates of the processes are still determined using
rate coefficients and concentrations of species.  One strong point of
the Monte Carlo approach is that it can be used in both a macroscopic
and microscopic sense, and anything in between.  In the macroscopic
sense, the only interest is in the number of atoms or molecules of a
given species on a grain and not where they are located.  In a
fully microscopic Monte Carlo treatment, on the other hand, the
interest is in exactly where on each monolayer individual species are
located as well as pores in the ices.

\begin{figure}[t]
  \includegraphics[angle=0,width=0.7\textwidth]{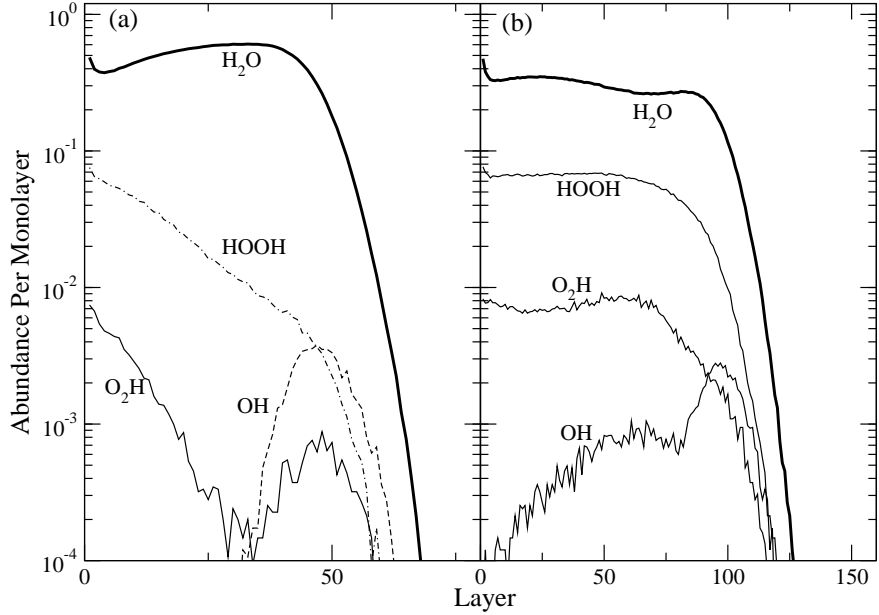}
  \caption{ The fraction of each monolayer occupied by assorted
    molecules including water using gas (macroscopic Monte
    Carlo)-grain (microscopic Monte Carlo) models for 10 K cold dense
    interstellar clouds. The layers increase from the innermost layer
    0 to the outermost layer. The gas temperature is taken to be the
    same as the dust temperature.  Model (a) (left) has $n_{\rm
      H}=2\times 10^4$ cm$^{-3}$, Model (b) (right) has $n_{\rm
      H}=10^5$ cm$^{-3}$. Adapted with permission from Reference 278.
    Copyright American Astronomical Society.}
\label{fig:chang}
\end{figure}

The details of the microscopic approach used for surface/ice
chemistry, known as the Continuous Time Random Walk (CTRW) approach
\citep{Chang05}, are discussed in the review by Cuppen et al., so will
not be repeated here in detail.  The basic idea is to consider a grain
as a square lattice of binding sites with periodic boundary
conditions.  One can start with either a `flat' surface of silicates
or amorphous carbon, in which all binding sites are the same, or with
a `rough' surface in which there are irregularities among binding
sites.  As accretion, chemistry, and desorption occur, monolayers of
ices are built up and the details of the ice on a specific grain
emerge.  Figure \ref{fig:cuppen} shows the formation of water ice and
other molecules such as OH from H, H$_2$, O, O$_2$, and O$_3$ for an
individual grain in a diffuse (top) and translucent (bottom)
interstellar cloud, with the latter having a visual extinction of
$A_V=3$ mag and a total hydrogen density of $10^3$ cm$^{-3}$.  The
gas-phase consists of atomic hydrogen and oxygen and is unchanged as
the grains evolve for $3 \times 10^5$ yr.  No gas-phase chemistry is
included.  In the diffuse cloud model, virtually no water ice is
formed because of rapid photodissociation and photodesorption of all
the reactants. In the translucent cloud model, water is formed on the
surface but there is a still a significant residual amount of OH in
the lower monolayers where H cannot penetrate to react with OH.
These models have a significant amount of empty space in the ice, but
the model porosity depends sensitively on the initial roughness of the
surface and associated diffusion rates, as well as the deposition
rate, and may be overestimated in these models.

The basic difficulty with the microscopic treatment is that it takes a
large amount of computer time even without coupling to the gas phase
chemistry, and the weakness of the Monte Carlo method as a whole is
that it cannot easily be coupled with normal rate equations for the
gas phase chemistry.  Rather the gas-phase chemistry must be treated
by a macroscopic Monte Carlo method whatever type of Monte Carlo
approach is used for the surface/ice chemistry.  The result is that
only a few stochastic calculations with the microscopic Monte Carlo
approach have been attempted, and these are mainly simulations of ice
build-up without gas phase chemistry, or simulations with gas-phase
chemistry and only a small number of surface reactions.  Figure
\ref{fig:chang} shows model molecular abundances\citep{Chang12} as a
function of monolayer for cold grains in dense clouds at 10 K.
Unlike the translucent case, the water is formed mainly in the inner
rather than outer monolayer.
It is still not yet
possible to compare rate-equation with Monte Carlo stochastic
approaches, although Chang et al.\ are working on an
approximation that should make large gas-grain simulations with many
surface reactions possible with the microscopic-macroscopic Monte
Carlo approach.  In addition, a method developed by
\citet{Vasyunin13}, in which there is somewhat less information about
the positions of species within individual monolayers, is now in the
public domain. 

A number of papers have shown that gas-grain chemical simulations with
macroscopic stochastic approaches to the surface chemistry can differ
significantly from those using rate equations under specific physical
conditions and times.  Detailed examples are shown by
\citet{Garrod09}, where, for example, under some conditions and times,
the CO gas phase and ice phase abundances calculated by rate equations
can differ by more than one order of magnitude from a stochastic
approach.  CO is one of the worst cases, however, and discrepancies
are much smaller when the rate equation approach is modified according
to the prescription of \citet{Garrod08}. For the case of H$_2$O,
agreement is generally within 10\% for the two approaches.  One major
advantage of the stochastic approaches is that they are correct
physically and should give the correct answer, if the chemical and
physical processes are correctly treated, whereas there is no security
in using the rate equation method, whether it be done in a two-phase
or a three-phase calculation.  Moreover, microscopic stochastic
approaches can take into account grains with roughness on the surface,
which are far more likely to occur in space than flat grains. The bulk
of the models described in \S 4 have been carried out with the two-phase
 rate equation method, however, with at most a partial modification.

\section{Comparison with observations}

In \S 3, three different routes to water in space have been described
in detail and important individual molecular processes leading to the
formation and destruction of water have been identified, both in the
gas-phase and in the solid-phase, as well as at the gas-solid
interface. In this section, we summarize the observations of water in
different astrophysical sources and explore to what extent the
observed abundances of water are consistent with the chemistry
described in \S 3. The discussion is ordered by the different types of
chemistry that dominate the formation and destruction of water in a
particular set of sources, rather than by type of astronomical
source. Thus, some types of sources can appear more than once in the
examples. A review of water from the astronomical perspective as it
cycles from cold clouds to disks and planets is given by
van Dishoeck et al.\citep{vanDishoeck14}.

As a general comment, it should be recalled that astronomers only observe
column densities $N$ in cm$^{-2}$ integrated along the line of sight
$L$ in cm rather than local concentrations $n$ in cm$^{-3}$, which are
usually obtained from models. 
As noted in \S~\ref{sect:radiation}, astronomers often adopt the
visual extinction $A_V$ as a measure of depth into a cloud, using the
empirical relation $N_{\rm H}=1.8\times 10^{21} A_V$ cm$^{-2}$.  For
dense clouds shielded from radiation there is little variation with
depth into a cloud and the fractional abundances are usually taken to
be the same as the column density ratios, i.e., $n$(X)/$n$(H$_2$)=
$N$(X)/$N$(H$_2$). For low and high density PDRs, proper comparison
requires the calculation of column densities integrated through the
model cloud. Unless stated otherwise, fractional abundances are quoted
relative to H$_2$, and are simply called `abundances'. To convert to
abundances with respect to total hydrogen nuclei, $n_{\rm H}$=$n$(H) +
2$n$(H$_2$), the abundances need to be divided by a factor of 2
(assuming the atomic hydrogen fraction to be negligible).

In many cases, the accuracy in observed abundances is limited to a
factor of a few, primarily by uncertainties in the {\it denominator},
i.e., the H$_2$ column.  Because H$_2$ can generally not be observed
directly, indirect tracers have to be used. Many different methods
exist, ranging from the use of simple molecules like CH in diffuse
clouds to measurements of CO and its isotopologues in dense
clouds. For dense star forming regions where freeze-out plays a role,
dust continuum observations often provide a more reliable tracer,
although this assumes a gas/dust mass ratio of typially 100.
Agreement between models and observations is usually considered to be
good if they are within one order of magnitude in observed
columns. Some of the chemistry tests described below are at the factor
of 2--3 level, which is considered excellent agreement.  Measured abundance
ratios of species, e.g., H$_2$O/OH, are often more accurate
than absolute abundances, H$_2$O/H$_2$.

\begin{figure}
\begin{minipage}{0.475\textwidth}
\includegraphics[angle=0,width=\textwidth]{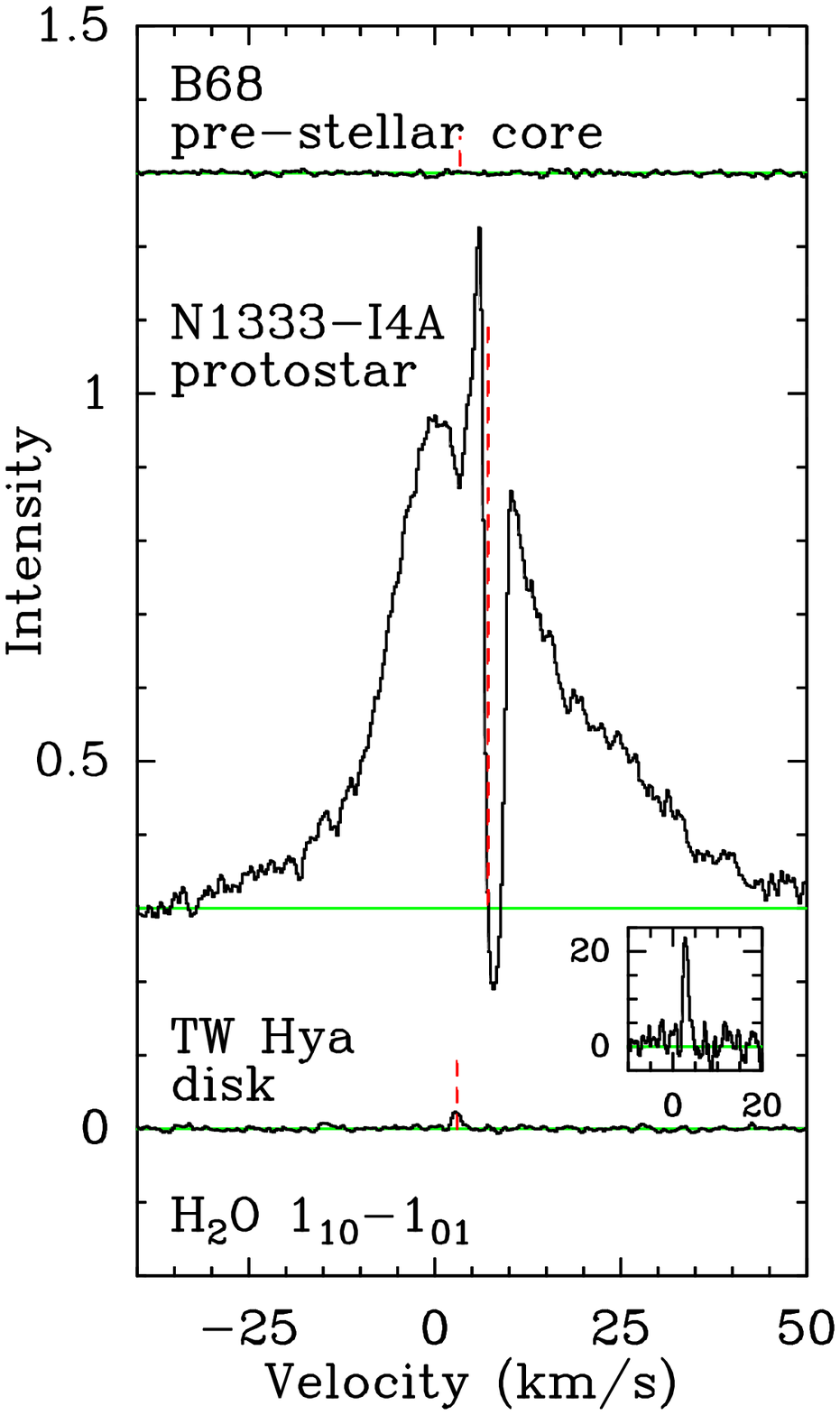}
\end{minipage}\hfill%
\begin{minipage}{0.475\textwidth}
\includegraphics[angle=0,width=0.8\textwidth]{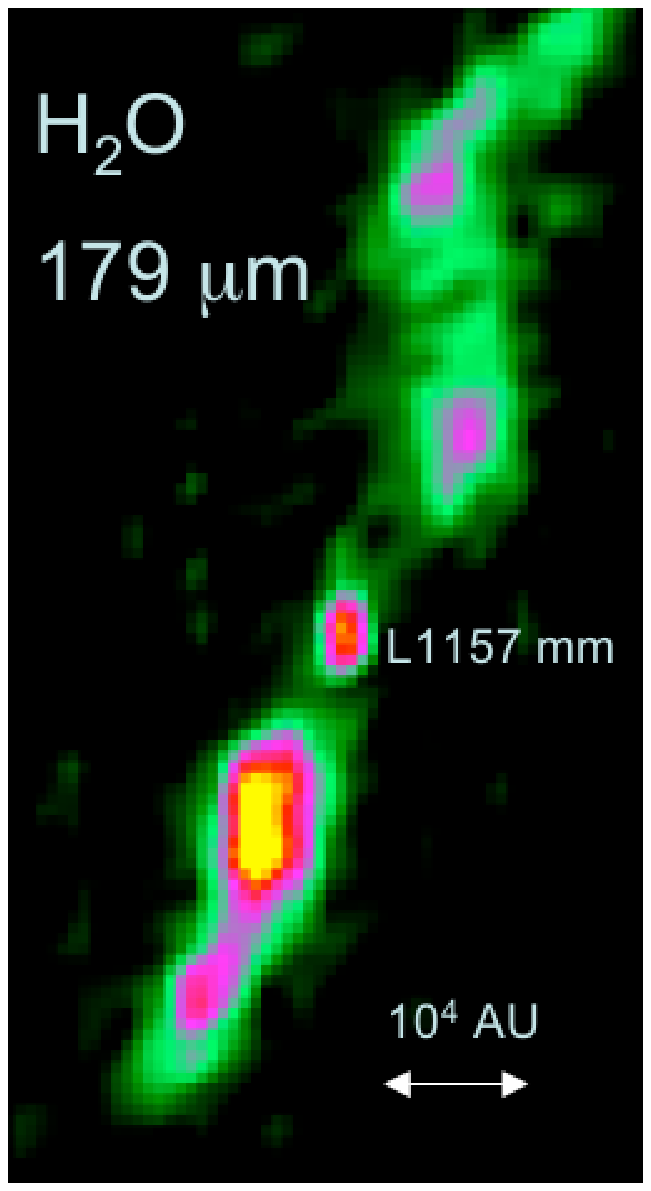}
\end{minipage}\hfill%
\caption{Left: {\it Herschel}-HIFI spectra of the ortho-H$_2$O
  $1_{10}-1_{01}$ 557 GHz line across the different evolutionary
  stages of a low-mass protostar.  From top to bottom: the pre-stellar
  core B68 \citep{Caselli10}, the low-mass protostar NGC 1333 IRAS4A,
  and the protoplanetary disk TW Hya \citep{Hogerheijde11}. Note that
  only the protostellar stage shows strong and broad water lines,
  coupled with narrow absorption lines. The intensity is in units of
  K, whereas the insert for TW Hya is in units of milli-K. Adapted
  with permission from Reference 281. Copyright 2012 Wiley. Right:
  {\it Herschel}-PACS image of the H$_2$O $2_{12}-1_{01}$ line at 179
  $\mu$m toward the L 1157 protostar\citep{Nisini10}.  Water emission
  is strong toward the protostar itself and also in `hot spots' along
  the outflow. Adapted with permission from Reference 40, copyright
  2011 The University of Chicago Press and from Reference 282,
  copyright 2010 European Southern Observatory.}
\label{fig:557spectra}
\end{figure}

Figure~\ref{fig:557spectra} (left) shows {\it Herschel-}HIFI spectra
of H$_2$O of astronomical sources at different stages of evolution
\cite{Kristensen11an}, whereas Figure~\ref{fig:557spectra} (right)
shows a PACS image of water in the protostellar phase
\cite{Nisini10}. The strength of the lines is related to the abundance
of water, whereas the line profiles provide information on the
kinematics of the gas containing water. Clearly, water emission is
weak in some sources and strong in others. The water line profiles are
complex with a mixture of narrow lines originating in cold quiescent
gas and very broad and strong lines due to fast outflow gas. These
observational characteristics point to different types of chemistry at
work in different regions.

\subsection{Testing ion-molecular chemistry}
\label{sect:diffuse}

Diffuse and translucent molecular clouds present a valuable laboratory
for testing the gas-phase chemical models described in
\S~\ref{sect:ion} above.  Such clouds may be observed in absorption
toward background sources of continuum radiation, providing robust
estimates of the molecular column densities.
Because the Milky Way galaxy is rotating differentially, with
an angular velocity that decreases with distance from the Galactic
center, multiple diffuse clouds along the sight-line to a given
continuum source may be distinguished by their Doppler shifts.  Thanks
to the relatively low densities within such clouds
(Table~\ref{tab:clouds}), and the weakness of the submillimeter
radiation to which they are typically exposed, most molecules are in
the ground rotational state; in this regime, the inferred 
column densities are very insensitive to the assumed physical
conditions.  
The main uncertainty in the abundances therefore comes from the
determination of $N$(H$_2$). In these clouds, the CH and HF molecules
are often used as surrogates \citep{Sonnentrucker10,Monje11}, although
 these two tracers do not always agree. The $N$(HF)/$N$(H$_2$)
conversion has recently been calibrated directly using near-infrared
observations of both species \citep{Indriolo13}.

\subsubsection{Water abundances in diffuse and translucent clouds }

While ultraviolet searches for water vapor in translucent molecular
clouds have not been successful\citep{Spaans98} (see \S 2.3
above), submillimeter absorption line observations of water have
proven to be a powerful tool for the study of interstellar water
vapor.  Detections of foreground absorption by water along the
sight-lines to several bright continuum sources were obtained by {\it
  ISO}, which was capable of detecting the 179 $\mu$m $2_{12} -
1_{01}$ transition of ortho-water\citep{Cernicharo97}, and by SWAS and
Odin, which targeted the fundamental $1_{10} - 1_{01}$ transition of
ortho-water near 557~GHz (Table~\ref{tab:telescopes}).

\begin{figure}
\includegraphics[angle=0,width=0.6\textwidth]{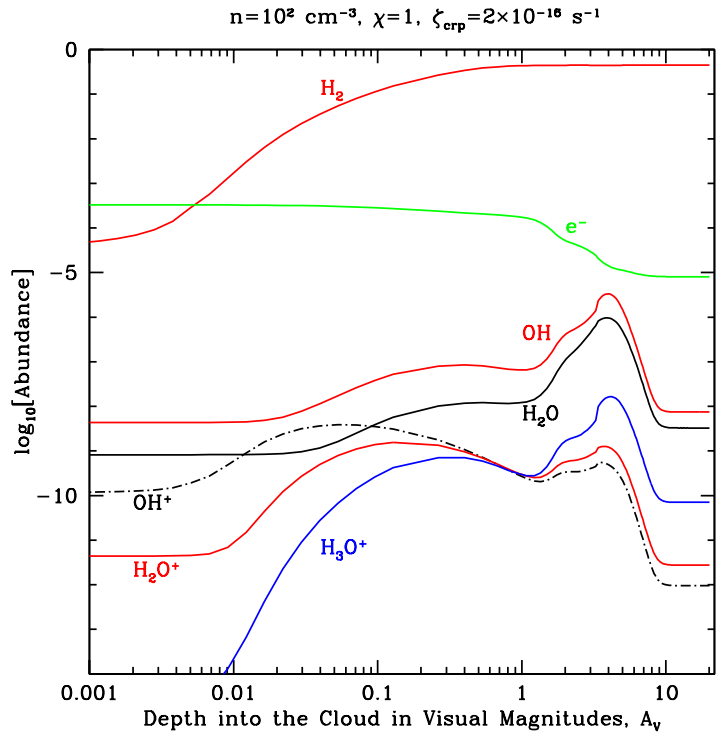}
\caption{Predicted abundances (relative to H nuclei) as a function of
  depth into a molecular cloud measured by the visual extinction (see
  \S~\ref{sect:radiation}).  The assumed cloud density $n_{\rm H}=100$
  cm$^{-3}$, the incident UV radiation field is equal to the average
  Galactic value ($\chi$=1), and the cosmic ray ionization rate
  $\zeta_{\rm H}=2 \times 10^{-16}\,\rm s^{-1}$. Reprinted with
  permission from Reference 262. Copyright 2012 American Astronomical
  Society.}\label{fig:Hollenbach12}
\end{figure}

Much more sensitive observations performed with {\it Herschel} have
greatly expanded the number of sight-lines along which water
absorption has been detected, and have permitted observations of the
$1_{11} - 0_{00}$ transition of para-water, near 1113 GHz, along with
the 557 GHz ortho-water transition\citep{Sonnentrucker10}.  These
observations have shown that the water abundance is remarkably
constant\citep{Flagey13} within the diffuse interstellar medium, with
the exception of the Galactic center region, which appears to contain
diffuse clouds with significantly higher water abundances by a factor
of three than those found elsewhere in the
Galaxy\citep{Monje11,Sonnentrucker13}.  The typical water abundance
relative to H nuclei is $\sim 0.5 - 1.5 \times 10^{-8}$, with no
apparent trend with H$_2$ column density over the range 3 -- 30
$\times 10^{20}\,\rm cm^{-2}$ or with Galactocentric distance over the
range 5.5 -- 8 kpc.  These values are broadly consistent within
factors of a few with model predictions for diffuse
clouds\citep{Hollenbach12}, providing support for the ion-molecule
production route discussed in \S 3 above.
Figure~\ref{fig:Hollenbach12} shows the abundances of water molecules
predicted by Hollenbach et
al.\citep{Hollenbach12}, as a function of depth into a molecular
cloud.

Absorption lines of H$_2$O and H$_2^{18}$O have also been seen with
{\it Herschel}-PACS toward the far-infrared continuum emission in
external galaxies \citep{Fischer10,Gonzalez12}. Part of this
absorption has been inferred to arise in a low density extended component
with water abundances of a few $\times 10^{-8}$, similar to those
found in galactic diffuse and translucent clouds.  Thus, ion-molecule
chemistry appears to widespread throughout galaxies.

\subsubsection{Related species in diffuse and translucent clouds}

In addition to water itself, {\it Herschel/HIFI} observations have
allowed the intermediaries OH$^+$ and H$_2$O$^+$ to be measured along
the same diffuse cloud sight-lines\citep{Gerin10, Neufeld10}.  OH$^+$
was first detected from the ground using the submillimeter APEX
telescope \citep{Wyrowski10oh+}. Once again, the observed abundances
are found to be in good agreement with the predictions of ion-neutral
chemistry, for reasonable estimates of the cosmic-ray ionization
rate\citep{Indriolo12}.  Because the OH$^+$ and H$_2$O$^+$ abundances
peak closer to the cloud surface than do OH and H$_2$O
(Figure~\ref{fig:Hollenbach12}), they can be detected in clouds with a
small molecular fraction \citep{Gerin10}.  By contrast, the lowest
frequency transition (at 2.5~THz) of the neutral OH radical in its
ground state lies outside the frequency range covered by the HIFI
instrument.  It can, however, be observed at high spectral resolution
using the GREAT instrument on the NASA/DLR airborne observatory SOFIA.
Recent absorption line observations of this transition have provided
robust estimates of the H$_2$O/OH ratio in diffuse clouds along the
sight-line to three bright continuum sources\citep{Wiesemeyer12};
observed values in the range 0.3 -- 1 are in good agreement with model
predictions (Figure~\ref{fig:Hollenbach12}). Similar H$_2$O/OH ratios
are found in diffuse extended regions of galaxies
\citep{Gonzalez12}. Previous observations of OH at UV wavelengths in a
different set of diffuse clouds toward bright
stars\citep{Roueff96,Weselak09,Weselak10} have also been reproduced
well with the basic ion-molecule network within a factor of two
\citep{vanDishoeck86}.

OH$^+$, H$_2$O$^+$ and H$_3$O$^+$ absorption has been detected as well
in some external galaxies, even from energy levels up to $\sim$200 K
using {\it Herschel-}PACS \citep{Gonzalez13}. The abundance ratios are
consistent with the ion-molecule chemistry in relatively low density
gas with a high atomic fraction, starting with H$^+$ and followed by
charge transfer to O$^+$ (Eq.\ (8)). The inferred cosmic ray
ionization rate $\zeta_{\rm H} >10^{-13}$ s$^{-1}$ for clouds in these
galaxies is at least two orders of magnitude higher than those found
in galactic sources, however.

\subsubsection{Dense PDRs}

Dense molecular clouds close to bright O or B stars are exposed to
much more intense UV radiation than diffuse and translucent clouds,
typically by a factor of $10^4$. An illustrative example is provided
by the Orion Bar PDR. The {\it Herschel}-SPIRE spectrum of this source
\citep{Habart10} reveals strong CO lines, but only weak low-$J$ H$_2$O
lines implying a water abundance $\leq 5 \times 10^{-7}$ . In
contrast, a number of OH lines are readily detected with {\it
  Herschel}-PACS\citep{Goicoechea11} indicating a high OH/H$_2$O
column density ratio $>$1. These findings are consistent with dense
PDR models \citep{Sternberg95}, where rapid photodissociation due to
the intense UV radiation limits the build up of water.  The strong OH
emission most likely originates in the warm ($\sim 200$~K) H/H$_2$
transition layer where vibrationally excited H$_2$ is produced, which
can react with atomic O without barrier to form OH (see Eq.\ (12) and
associated discussion).  The only other dense PDR for which water data
have been published is the Monoceros R2 region \citep{Pilleri12}. The
inferred water abundance in the quiescent part of the cloud is again
low, about $10^{-8}$.

In summary, observations of water and related species in diffuse and
translucent clouds as well as dense PDRs confirm the basic ion neutral
chemistry outlined in \S 3.1, within the factor of a few uncertainty
in both observations and models. The fact that all the intermediate
ions in the network leading to water have now been observed, both
within our own Milky Way and in external galaxies, leaves little doubt
about the main processes at work. Since the principal reactions are well
determined, observations of these species can now be used as
diagnostics of astrophysical parameters, such as the cosmic ray
ionization rate, UV radiation field and density.

\subsection{Testing solid-state chemistry: ice formation and photodesorption}

\subsubsection{Cold clouds: ice observations}

The most direct evidence for solid-state formation of interstellar
water comes from the detection of the 3 $\mu$m O-H stretching
vibration band of water ice toward numerous infrared sources
\citep{Whittet88,Boogert08}.  In many cases, the 6 $\mu$m bending and
11 $\mu$m librational modes have also been observed (see
Fig.~\ref{fig:waterice}).  The observations\citep{Whittet03} show that
water ice formation starts at a threshold extinction of $A_V\approx 5$
mag in clouds that have densities of at least 1000 cm$^{-3}$.
\citet{Cuppen07} have simulated the formation of water ice at the
microscopic level, layer by layer, for clouds with densities ranging
from $10^2-5\times 10^4$ cm$^{-3}$ and extinctions up to 10 mag
(Fig.~\ref{fig:cuppen}). Critical parameters in the models are the
binding energies and diffusion barriers, as discussed in
\S~\ref{sect:icechem}. 
Only the higher extinction models start to approach the observed
columns of water ice because of rapid photodissociation and
photodesorption of grain-surface species. The lifetime of the cloud is
also an important parameter\citep{Walmsley92}: densities need to be
larger than $\sim$10$^4$ cm$^{-3}$ in order for the timescale for an O
atom or molecule other than H$_2$ to collide with a grain and stick on
it to become shorter than the age of the cloud, the latter being at
least a few $\times 10^5$ yr.

\begin{figure}[t]
  \includegraphics[width=0.5\textwidth]{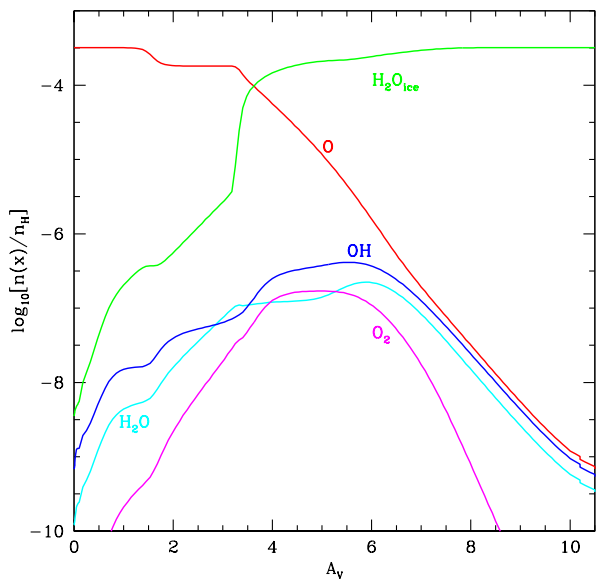}
  \caption{Oxygen chemistry for a standard steady-state cloud exposed
    to ultraviolet radiation including freeze-out, grain surface
    reactions, sublimation, photodesorption, and cosmic-ray
    desorption. Model abundances with respect to total hydrogen are
    shown as a function of visual extinction from the cloud surface,
    $A_V$. The density is $n_{\rm H}=10^4$ cm$^{-3}$ and the cloud is
    illuminated by 100 times the average interstellar radiation field.
    Reprinted with permission from Reference 9. Copyright 2009 American
    Astronomical Society.}
  \label{fig:Hollenbach}
\end{figure}

In dense clouds, the bulk of the water is in the form of ice
\citep{Whittet88,Smith89,Gibb04,Murakawa00,Pontoppidan04,Boogert11},
at levels H$_2$O ice/H$_2 \approx 10^{-4}$.  Such high ice abundances
are too large to result from freeze-out of gas-phase water produced by
ion-molecule reactions \citep[e.g.,][]{Lee96b}, so grain surface
reactions must surely happen. 
The overall abundance of elemental oxygen with respect to hydrogen nuclei in the
solar neighborhood\citep{Przybilla08} is estimated to be $5.75\times
10^{-4}$, of which 16--24\% is locked up in refractory silicate
material in diffuse clouds \citep{Whittet10}. If this silicate
fraction stays the same in dense clouds, and if the amount of oxygen
locked up in CO gas and ice ($\sim 1.4\times 10^{-4}$) is subtracted,
the maximum abundance of water ice in dense clouds would be
$(2.4-2.9)\times 10^{-4}$ with respect to hydrogen nuclei or
$(5-6)\times 10^{-4}$ with respect to H$_2$. Thus, the observed ice
abundances in dense clouds of $\sim 10^{-4}$ indicate that water ice
contains a significant fraction of the available oxygen, but 
not all \citep{Boonman03}. It is still unclear whether all oxygen is
fully accounted for and if not, in which form the missing oxygen is 
\citep{Whittet10}.

\subsubsection{Cold pre- and protostellar cores: gas-phase water}  

Another strong but more indirect argument in favor of ice chemistry
comes from the weak gas-phase H$_2$O lines and lack of O$_2$ lines
observed by SWAS and Odin in cold molecular clouds.  Most likely,
atomic O is converted rapidly on the grains into water ice, only a
small fraction of which is subsequently desorbed back into the gas by
non-thermal processes \citep{Bergin00,Roberts02,Hollenbach09}.  A
generic set of models for this situation has been developed by
\citet{Hollenbach09},  which is illustrated in
Fig.~\ref{fig:Hollenbach}. Because H$_2$O and O$_2$ are only abundant
at intermediate depth, their integrated column densities over depth
reach a maximum value that is much lower than if freeze-out were not
taking place.

\begin{figure}
\begin{minipage}{0.475\textwidth}
\includegraphics[angle=0,width=\textwidth]{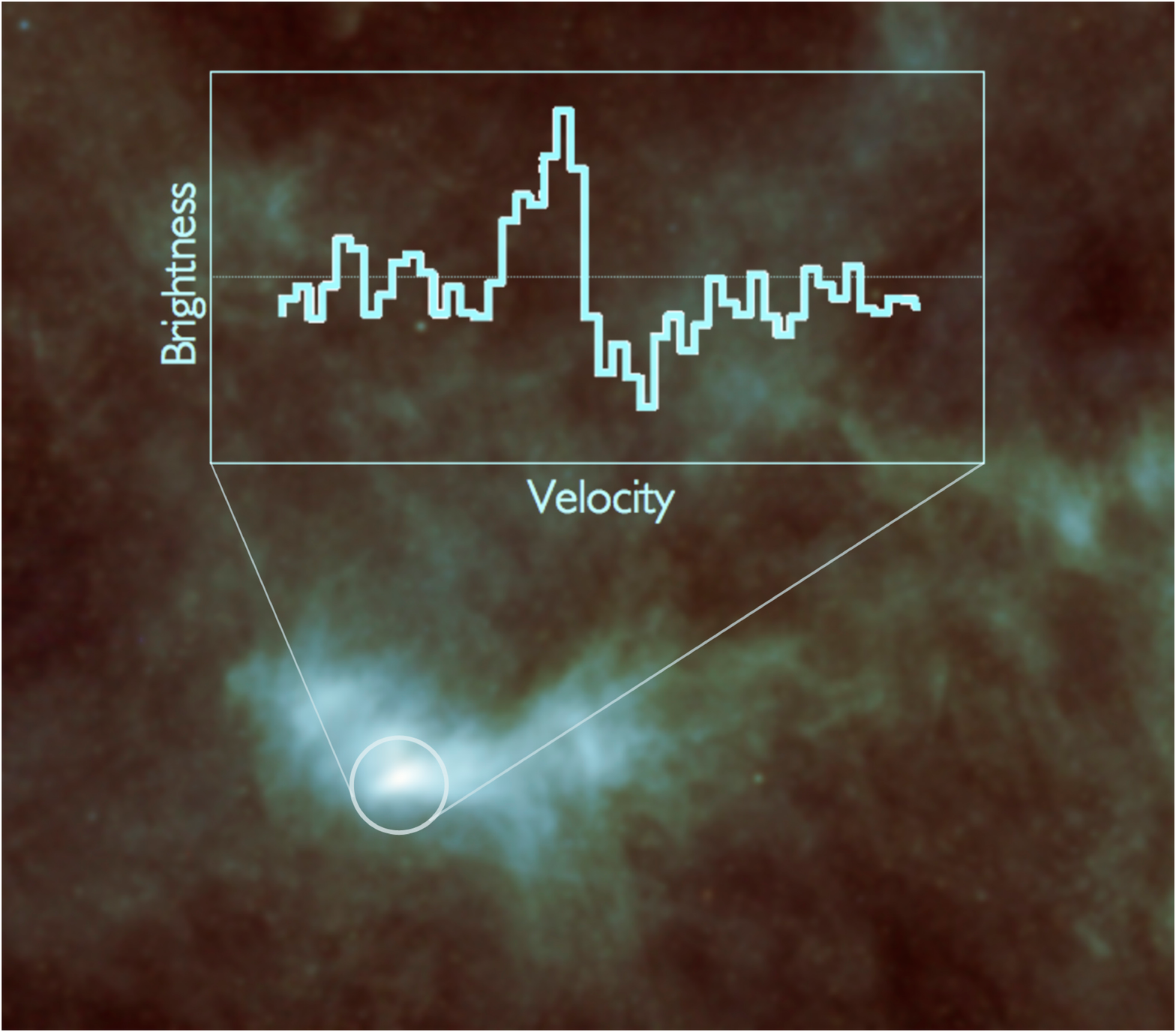}
\end{minipage}\hfill%
\begin{minipage}{0.475\textwidth}
\includegraphics[angle=0,width=\textwidth]{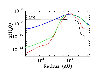}
\end{minipage}\hfill%
  \caption{Left: HIFI spectrum of o-H$_2$O 1$_{10}-1_{01}$ at 557 GHz 
    tracing the dense cold gas toward the pre-stellar core L1544
    superposed on a {\it Herschel} image of part of the Taurus cloud
    (Credit: ESA). The water line shows emission only at blue-shifted
    velocities coming from the densest gas at the backside of the
    core, whereas water on the near-side of the core absorbs the
    far-infrared dust continuum emission produced by the core central
    regions. Right: Best fitting water vapor abundance profile through
    the core indicated in blue, obtained using a simplified water
    chemistry with CR induced photodesorption of water ice
    included. The red and green lines indicate the results of a full
    gas-grain and simplified water chemistry model, respectively,
    without CR induced photodesorption. The water ice abundance (black
    line) has been divided by a factor of $10^4$ for clarity of
    presentation. Reprinted with permission from Reference 313.
    Copyright 2012 American Astronomical Society.}
  \label{fig:L1544}
\end{figure}

The most recent, quantitative test of this basic gas-grain chemistry comes
from the first detection of extremely weak water vapor emission and
absorption in the centrally concentrated pre-stellar core L1544, a
cloud which is likely on the verge of collapse to form a new star
\citep{Caselli12}.  The {\it Herschel}-HIFI line profile of the H$_2$O
$1_{10}-1_{01}$ 557 GHz transition toward L1544 is presented in
Figure \ref{fig:L1544} (left) and shows both emission and absorption. This
so-called `inverse P-Cygni profile' is indicative of inward motions in
the core: the infalling red-shifted gas on the near-side of the core
is partly absorbed against line and continuum emission produced deeper
in the cloud whereas the blue-shifted emission from the backside of
the cloud can proceed unhindered to the observer
\citep{Evans99}. Because the different parts of the line profile probe
different parts of the core, this can be used to reconstruct the water
vapor abundance as a function of position throughout the entire
core. Specifically, the detection of an emission feature signals the
presence of water vapor in the densest central part of the core,
whereas the absorption profile is sensitive to the water abundance in
the outer part of the core.

Figure~\ref{fig:L1544} (right) shows the best fitting water vapor
abundance for L1544, using a temperature and density structure that
has been determined independently from other data
\cite{Caselli12}. The main features of this figure are well described
by a highly simplified water chemistry that contains just the
freeze-out of gas-phase O and H$_2$O on the grains, the immediate
formation of water ice from adsorbed $s$-O, the photodesorption of
$s$-H$_2$O ice by the ambient and cosmic-ray induced UV radiation
fields back into the gas phase, and the photodissociation of gaseous
H$_2$O.  The critical parameters are the strength of the cosmic-ray
induced field combined with the photodesorption yield. Indeed, the
presence of the blue-shifted emission requires efficient cosmic-ray
induced photodesorption of water ice in the center of the core, which
was neglected in the models \citep{Hollenbach09} presented in
Fig.~\ref{fig:Hollenbach}.  Closer to the edge, at $A_V<$5 mag, the
interstellar radiation field takes over as the main photodesorption
mechanism and the gas-phase water abundance reaches its maximum of
$\sim 10^{-7}$. At the very edge of the cloud, $A_V<$2 mag, water gas
is photodissociated and its abundance drops. Thus, water gas has a
ring-like abundance structure with its peak at intermediate depth into
a core, and decreasing abundances towards the edge and the center of
the core that are measurable in the absorption and emission features.

\begin{figure}
\begin{minipage}{0.475\textwidth}
  \includegraphics[angle=0,width=\textwidth]{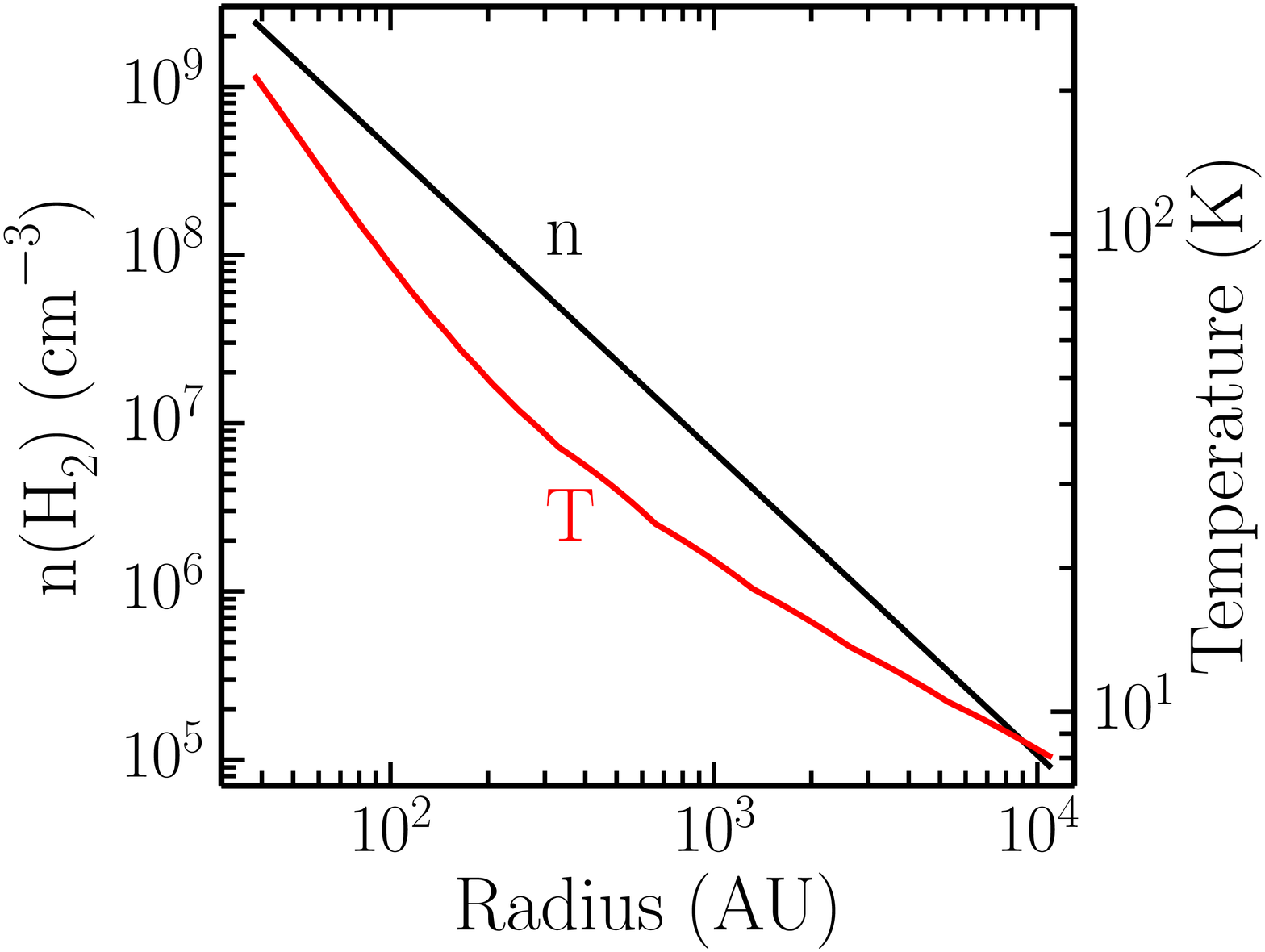}
\end{minipage}\hfill%
\begin{minipage}{0.475\textwidth}
  \includegraphics[angle=0,width=\textwidth]{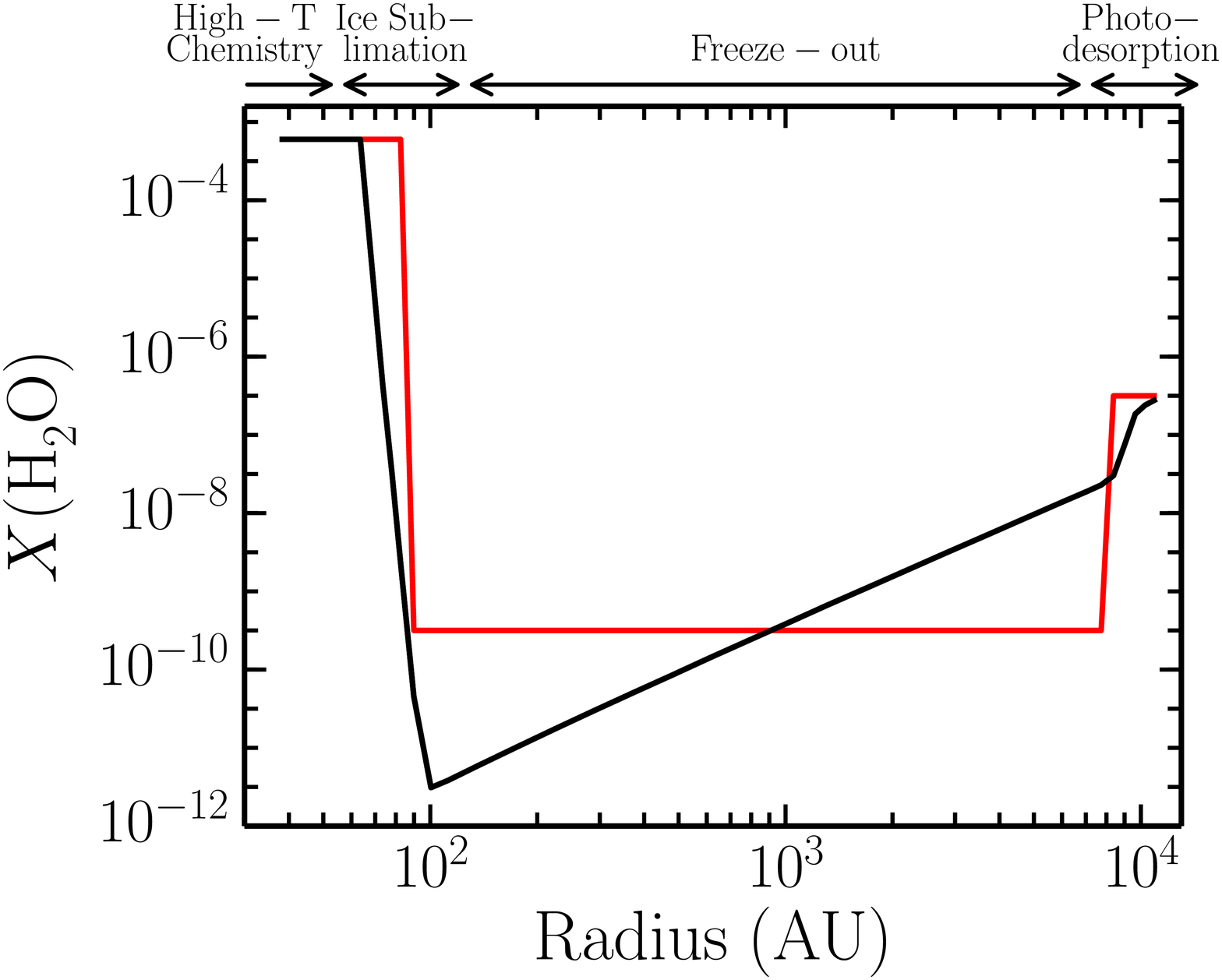}
\end{minipage}\hfill%
  \caption{Left: Density (black line) and dust temperature (red line)
    structure of the low-mass protostar NGC 1333 IRAS4A as a function of
    distance from the source. The gas temperature is
    taken to be equal to that of the dust. Right: The inferred 
     abundance profile of
    water (black) together with the step-function model 
   traditionally used to analyze observations 
    (red line). In the colder
    envelope ($T < 100$ K), the water molecules are trapped in the ice
    mantles, whereas in the inner part of the envelope, they are
    released into the gas phase by sublimation, leading to an
    enhancement of the water abundance. In the outer part of the cloud
    ($A_V \sim$ 1--4 mag), the water abundance increases due to
    ice photodesorption.
    Adapted with permission from Reference 316. Copyright 2013
    European Southern Observatory. Similar physical and water abundance
    structures apply to high-mass protostars, but scaled to larger
    sizes (see Figure 2 in \citealt{vanDishoeck11}).}
  \label{fig:Coutens}
\end{figure}

The same simple water chemistry has also been found to reproduce well
the water line profiles observed toward low-mass protostars, which
show similar absorption and emission features after the outflow
contributions have been removed
\citep{Kristensen12,Coutens12,Mottram13}. A prominent example is the
NGC 1333 IRAS4A protostar, for which the observed HIFI spectrum is
presented in Figure~\ref{fig:557spectra} and the best fitting model in
Figure~\ref{fig:Coutens}. The main difference with prestellar cores as
shown in Figure~\ref{fig:L1544} is that the dust temperature now
increases rather than decreases from the edge of the core inwards.
The two cold core studies differ slightly in the best fitting
parameters: the protostar study \citep{Mottram13} finds that the
standard CR-UV flux of $10^4$ photons cm$^{-2}$ s$^{-1}$ combined with
the measured photodesorption yield of $10^{-3}$ per incident photon
\citep{Oberg09h2o} fits the data well for several low-mass protostars,
whereas the prestellar core study \citep{Caselli12} uses the order of
magnitude higher CR-UV flux adopted in earlier models
\citep{Hollenbach09}.  Also, the ortho/para ratio of H$_2$ used for
the collisional excitation of H$_2$O plays a role \citep{Caselli12}.

Both studies, as well as \citet{Coutens12} and \citet{Mottram13},
agree on the need for a cloud layer in which water ice is
photodesorbed by external UV radiation in order to obtain deep enough
absorptions (Fig.~\ref{fig:Coutens}, adapted from
\citep{Mottram13}). The discussion centers on whether this is simply
the photodesorption layer at the illuminated edge around $A_V=2$ mag
of the envelope \citep{Mottram13} or whether it is a separate cloud
\citep{Coutens12}. More generally, the step-function abundance
structure used traditionally in the analysis of observations is now
being replaced by a more gradual power-law decrease of the abundance
inwards (Fig.~\ref{fig:Coutens}) until the ice sublimation radius is
reached \citep{Mottram13}.

\subsubsection{Outer protoplanetary disks}  
\label{sect:outerdisk}

Protoplanetary disks around young stars, representing a later stage in
the star formation cycle (Fig.~\ref{fig:cycle}), provide another
environment to test the water ice
chemistry. \citet{Hogerheijde11} detected the ground-state ortho- and
para-water lines toward the young star TW Hya with {\it Herschel}-HIFI
(Fig.~\ref{fig:557spectra}). Disks have typical sizes of 200 AU
diameter and are therefore small on the sky, of order 1$''$ in the
nearest star-forming regions. Thus, the signals from disks are
strongly diluted in the large {\it Herschel} beam. Moreover, disks
have typical masses of only 0.01 M$_\odot$, a factor of 100 less
than their parent clouds. Therefore, the signals of water from disks
are expected to be very weak, requiring long integration times of
$>$10 hours to detect them. The TW Hya disk has a favorable near
face-on geometry and is the closest known source of its kind,
facilitating the detection of the water emission lines. Deep searches
for water also exist for the outer portions of other disks, showing a
common pattern of very weak or no emission.

\begin{figure}
  \includegraphics[width=0.7\textwidth]{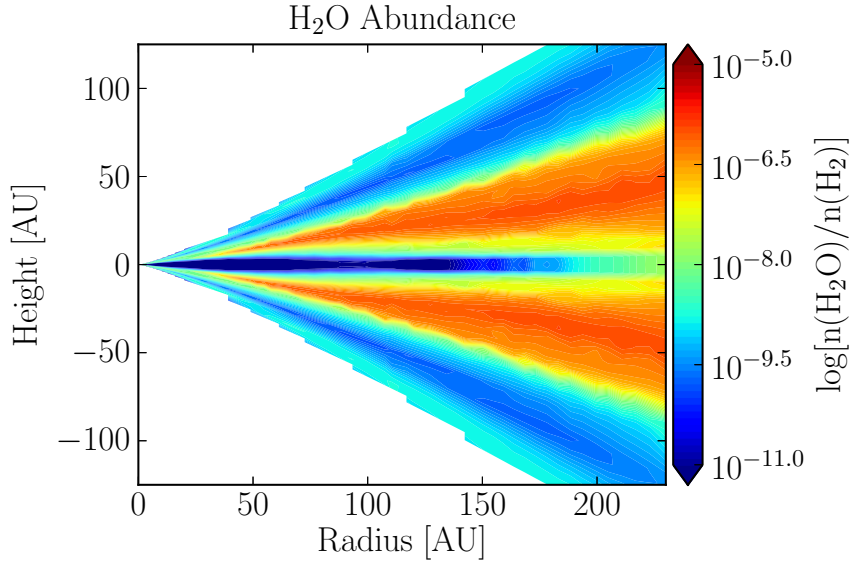}
  \caption{Abundance of water vapor relative to total hydrogen 
in the outer disk as
    function of both radial distance and vertical height. 
Adapted with permission from Reference 317.
Copyright 2010 European Southern Observatory.}
  \label{fig:Bergin}
\end{figure}

The chemical structure of water in disks obtained from models
\citep{Bergin10,Fogel11} is illustrated in Figure \ref{fig:Bergin} and
follows the typical layered `sandwich' structure \citep{Aikawa02} also
found for other molecules. This structure is similar to that in
Figures~\ref{fig:Hollenbach}, \ref{fig:L1544} and \ref{fig:Coutens},
but now in the vertical direction. The upper layers of the disk
contain very little water due to rapid photodissociation by the
stellar and interstellar UV radiation. In the cold midplane, where
densities reach values of $>10^{10}$ cm$^{-3}$, virtually all
molecules are frozen out. The bulk of the emission comes from the
intermediate layers of the disk where UV radiation can still penetrate
to photodesorb the icy grains, yet photodissociation is not too rapid
\citep{Dominik05}. Because of the large range of temperatures and
densities contained in the beam, the test of the water chemistry is
not as detailed as for the prestellar cores and protostellar envelopes
discussed above, but the observations are broadly consistent with the
models described here.

\subsubsection{Related ice species }

The above observational studies do not distinguish between the three
different channels postulated by \citet{Tielens82} to form water ice
and confirmed and extended by the laboratory studies presented in
\S~\ref{sect:icechem} (Fig.~\ref{fig:icenetwork}).  If the solid O and
O$_2$ pathways are indeed equally important in dense cores as
suggested by models \citep{Ioppolo08,Du11}, then the intermediate
products HO$_2$ and H$_2$O$_2$ should also be present in the gas at
low levels assuming they have similar photodesorption yields as water
ice. Indeed, a major confirmation of the ice chemistry comes from the
recent detections of hydrogen peroxide, H$_2$O$_2$, and the
hydroperoxyl radical, HO$_2$ toward $\rho$ OphA
\citep{Bergman11,Parise12}.

In summary, both the original ice detections as well as more recent
{\it Herschel}-HIFI and other studies validate the solid-state ice
chemistry described in \S~\ref{sect:icechem} at a quantitative level
to better than an order of magnitude.  The detection of HO$_2$ and
H$_2$O$_2$ demonstrates that gas-grain chemistry now has predictive
power for other molecules.  The era in which grain-surface chemistry
could be considered `the last refuge of a scoundrel'
\citep{Charnley92} is definitely over \citep{Parise12}.

\subsection{Testing gas-grain chemistry: water ice sublimation}

\subsubsection{Warm protostellar envelopes and hot cores}

Close to a protostar, the dust temperature exceeds 100~K, at which
point the sublimation of water ice becomes very fast under
interstellar conditions and the dust grains lose their icy mantles on
timescales short compared with the lifetimes of the sources (see
\S~\ref{sect:icechem}).  These regions of warm, dense gas are known as
`hot cores' (Fig.~\ref{fig:protostar} and \ref{fig:Coutens}), after
the prototype in the Orion molecular cloud. They often have large
abundances of complex organic molecules which are formed on the icy
surface, trapped in the water ice, and released together with the
water ice when the grains are heated. Thus, the water abundance in
these hot cores is expected to be equal to the original ice abundance
\citep{Boogert08}, which is of order 10$^{-4}$ within a factor of
2. This high water abundance will be maintained in the gas if the
water vapor destruction is slow compared with the lifetime of the
source or as long as there is continuous replenishment, e.g. by a
moving ice sublimation front.

\begin{figure}
\begin{minipage}{0.475\textwidth}
\includegraphics[angle=0,width=\textwidth]{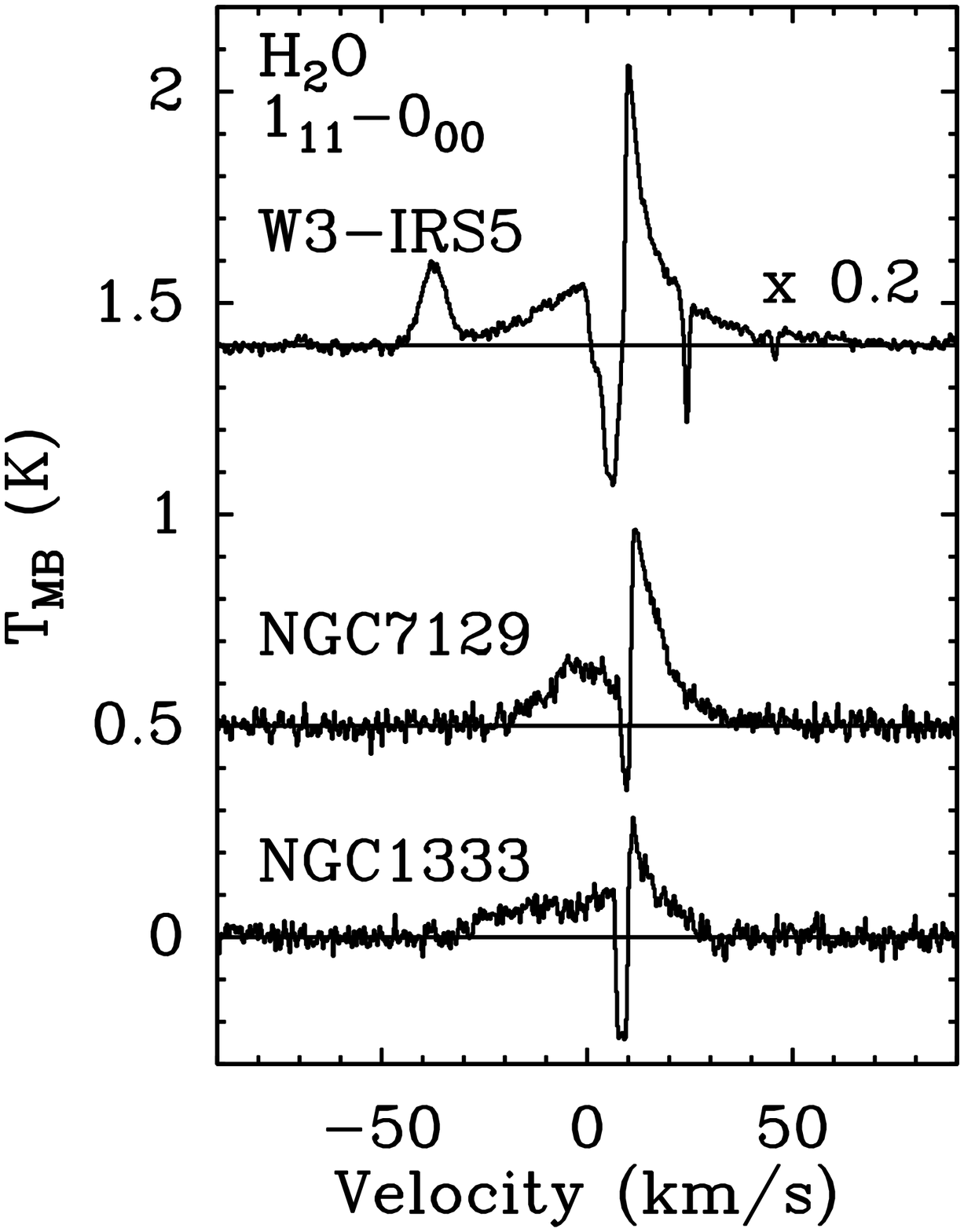}
\end{minipage}\hfill%
\begin{minipage}{0.475\textwidth}
\includegraphics[angle=0,width=\textwidth]{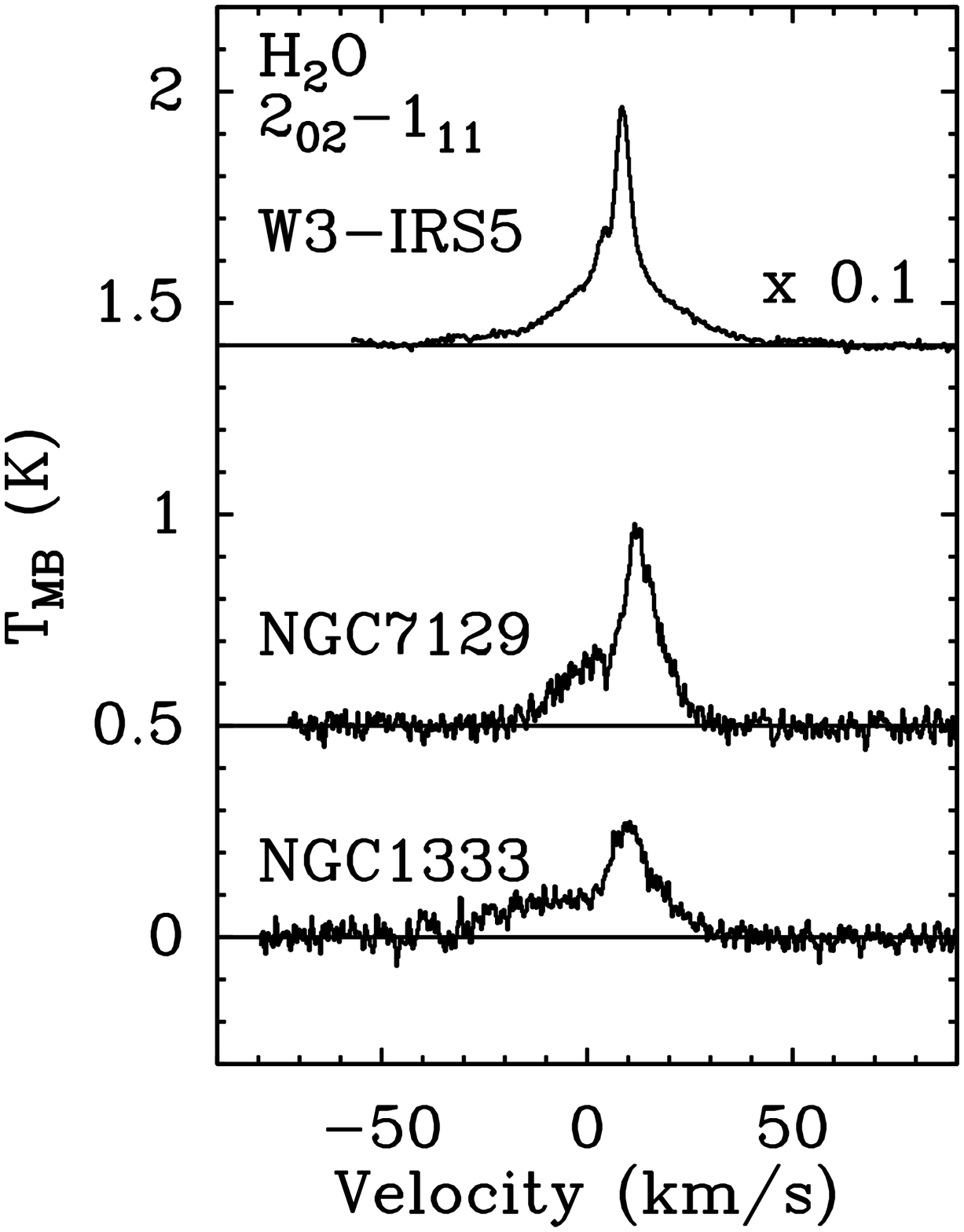}
\end{minipage}\hfill%
\caption{{\it Herschel}-HIFI spectra of the $p-$H$_2$O $1_{11}-0_{00}$
  1113 GHz (left) and $2_{02}-1_{11}$ 988 GHz (right) lines from low
  to high mass protostars.  From top to bottom: the high-mass
  protostar W3 IRS5 ($L=1.5\times 10^5$ L$_\odot$, $d=$2.0 kpc)
  \citep{Chavarria10}, the intermediate-mass protostar NGC 7129 FIRS2
  ($430$ L$_\odot$, 1260 pc) \citep{Johnstone10}, and the low-mass
  protostar NGC 1333 IRAS2A ($20$ L$_\odot$, 235 pc)
  \citep{Kristensen10}.  Note the complexity of the line profiles and
  the similarities from low to high mass. All spectra have been
  shifted to a central velocity of 0 km s$^{-1}$. The red-shifted
  absorption features seen in the $1_{11}-0_{00}$ spectrum toward W3
  IRS5 are due to water in foreground diffuse clouds whereas the small
  emission line on the blue side can be ascribed to the SO$_2$
  $13_{9,5}-12_{8,4}$ transition. Reprinted with permission from
  Reference 40. Copyright 2011 The
  University of Chicago press).}
\label{fig:paraspectra}
\end{figure}

Tests of this simple prediction are possible for high-mass
protostars. In particular, the ISO-SWS instrument was able to observe
both the water ice (through the 3 and 6 $\mu$m ice features) and gas
(through the 6 $\mu$m vibration-rotation band) in absorption along the
line of sight toward several high-mass protostars in a single
spectrum, thereby allowing an acccurate determination of their
relative column densities. A clear increase in the gas/ice ratio with
increasing (source-averaged) dust temperature has been found, reaching
values greater than unity
\citep{vanDishoeck96,vanDishoeck98,Boonman03h2o}. \citet{Boonman03}
coupled a simple gas-phase and ice chemistry with the physical
structure of these sources to model both the ISO-SWS, LWS and SWAS
lines from 3--540 $\mu$m. A clear conclusion from this work is that
scenarios without ice sublimation do not have large enough
columns of warm ($>$100 K) water to reproduce the ISO-SWS spectra.

The wealth of {\it Herschel}-HIFI lines of H$_2$O, H$_2^{18}$O and
H$_2^{17}$O toward high-mass protostars (Fig.~\ref{fig:paraspectra})
can be used to determine the gaseous water abundance in the hot cores
through emission lines.  Thanks to the high spectral resolution of
HIFI, the quiescent core contribution can be separated from the
outflow component and analyzed separately. Also, the H$_2^{17}$O and
H$_2^{18}$O lines have low optical depths facilitating accurate
determinations of the water abundances. Most models have used a `jump'
abundance profile in which the water abundance increases from a
constant low value in the outer envelope (of order 10$^{-8}$) to a
high abundance in the warm region, as illustrated in
Fig.~\ref{fig:Coutens}. The hot core abundances derived from
observations are found to range from values as low as $10^{-6}$ for
some sources \citep{Emprechtinger13} to the expected abundance of
$10^{-4}$ for other sources \citep{Chavarria10,Herpin12}. The low hot
core abundances are surprising, and are not yet understood. They
require a rapid gas-phase destruction mechanism such as UV
photodissociation or X-rays\citep{Stauber06} but it is not known
whether the fluxes are high enough in these inner shielded regions.

For low-mass protostars, the analysis is complicated by the fact that
ISO-SWS did not have the sensitivity to observe these low luminosity
sources whereas the observed {\it Herschel}-HIFI lines are dominated
by broad emission from outflows (Fig.~\ref{fig:557spectra} and
\ref{fig:paraspectra}), even for isotopologue lines
\citep{Kristensen10,Kristensen12}. Several H$_2^{18}$O and H$_2^{17}$O
lines toward the chemically rich low-mass protostar IRAS 16293 -2422
have been interpreted\citep{Coutens12} as originating in a hot core
with a water abundance of only $5\times 10^{-6}$. Deep integrations of
the excited H$_2^{18}$O $3_{12}-3_{03}$ line ($E_u$/k=249 K) near 1095
GHz toward a few other sources reveal a narrow feature that must come
from the hot core region \citep{Visser13}.  Also, the H$_2^{18}$O
$3_{13}-2_{20}$ line ($E_u/k$=204 K) at 203 GHz has been mapped
interferometrically in some of the same sources showing compact narrow
emission \citep{Jorgensen10,Persson12} .  While the determination of
the hot water column is relatively straightforward from such data if
the lines are optically thin (as in the case of the 203 GHz line, but
not the 1095 GHz line), the water abundance is much more uncertain due
to the uncertain warm H$_2$ column, which in turn is related to the
source structure on scales of 100 AU representative of the hot
core. Using independent data from C$^{18}$O 9--8 and 10--9, a high
water abundance of a few $\times 10^{-5} - 10^{-4}$ has been argued,
with values as low as $10^{-6}-10^{-5}$ clearly excluded
\citep{Visser13}.

In summary, there is considerable evidence for water ice sublimation
in hot cores, but warm water abundance determinations still differ by
up to two orders of magnitude. Thus, the question whether hot cores
are generally `dry' or `wet' due to ice sublimation is still open and
may differ from source to source.

\subsection{Testing high-temperature chemistry}

The reaction scheme discussed in \S~\ref{sect:hightemp} predicts that
almost all available oxygen not locked up into grains or CO is driven
into water at high temperatures \citep[$>$230
K][]{Draine83,Kaufman96}. Astronomical regions in which this chemistry
can be tested include shocks associated with the outflows from young
stars, supernova remnants expanding into the interstellar medium, the
warm envelopes around evolved stars, the hot cores around high-mass
protostars, the inner zones of protoplanetary disks and the nuclei of
external galaxies. The expected water vapor abundance in these regions
is H$_2$O/H$_2$$\approx (5-6)\times 10^{-4}$ (see above), unless
photodissociation or reactions with atomic H drive the water back to O
and OH (Fig.~\ref{fig:network}). If all non-refractory carbon is
locked up in CO, the expected H$_2$O/CO ratio is $\sim$1.3.

\subsubsection{Shocks}
\label{sect:shocks}

{\it Orion.}  The shocked regions in Orion-KL form one of the best
test cases of high temperature chemistry because of the wealth of
lines observed at mid- and far-infrared wavelengths. Using ISO-LWS,
\citet{Harwit98} detected 8 far-infrared water emission lines that
indicate H$_2$O/H$_2$$\approx 5\times 10^{-4}$, consistent with most
oxygen driven into H$_2$O. Other ISO studies of Orion based on many
additional absorption and emission lines with the SWS and LWS either
came to a similar conclusion \citep[2--5$\times 10^{-4}$][]{Wright00}
or found a lower abundance by a factor of 5--10
\citep{Gonzalez02,Cernicharo06}. Observations of the $1_{10}-1_{01}$
557 GHz line with SWAS and Odin give abundances of 3.5$\times 10^{-4}$
and $10^{-5}-10^{-4}$, respectively
\citep{Melnick00,Olofsson03}. Differences can be due to different
positions targeted in the various studies (central source IRc2 versus
outflow positions $\sim 30''$ offset from IRc2), different observing
beams (from $\sim$20$''$ to $>3'$) containing a range of physical
components with uncertain source sizes, different reference species
used to compute abundances (H$_2$ or CO), different conclusions on the
optical depths of the lines, and different excitation analyses with
infrared radiative pumping included or not.

Most recently, \citet{Melnick10} observed a plethora of
spectrally-resolved H$_2$O, H$_2^{18}$O and H$_2^{17}$O lines with
{\it Herschel-}HIFI (Fig.~\ref{fig:melnick}) to separate the different
physical components and infer abundances of $7.4\times 10^{-5}$,
$1.0\times 10^{-5}$ and $1.6\times 10^{-5}$ for the shock (`plateau'),
hot core and extended warm gas, respectively. The plateau value is
consistent with a significant fraction of the oxygen driven into
water, but not all. Either some fraction of oxygen is locked up in
unidentified refractory material, or photodissociation of H$_2$O by
the intense UV radiation pervading the Orion cloud reduces the water
abundance somewhat below its maximum value.

{\it Other shocks.}  Table~\ref{tab:shocks} contains a
(non-exhaustive) summary of H$_2$O abundances determined in shocks
associated with young stellar objects, using a variety of
instruments. The inferred values for sources other than Orion vary by
three orders of magnitude, with only a few sources consistent with all
oxygen driven into water. This has led to the suggestion based on SWAS
data that only a small fraction ($<$1\%) of the outflow gas has passed
through shocks strong (i.e., hot, $>$400 K) enough to convert oxygen
into water during the lifetime of the shock, primarily the gas at high
velocities \citep{Franklin08} . A similar conclusion is reached based
on {\it Herschel}-HIFI data at much higher spatial resolution
\citep{Lefloch10}, which suggests \citep{Kristensen12} that the water
abundance increases with velocity from $10^{-7}$ to $10^{-5}$.

As for Orion, part of the problem in this comparison of results stems
from the fact that different methods are used by the various teams to
infer the H$_2$ column.  Most studies use low-$J$ ($J_u \leq$3) CO
lines as their reference \citep{Franklin08,Kristensen12}, but these
line wings trace primarily the cool ($<$100 K) swept-up ambient gas
rather than the warm {\it currently shocked} gas with temperatures of
a few hundred up to few thousand K. If the H$_2$O line profiles are
compared with those of high-$J$ ($J_u>10$) CO, no increase in the
water abundance with velocity is seen and the higher abundance of a
few $\times 10^{-5}$ is found (Kristensen, priv.\
comm.). Alternatively, comparison with warm H$_2$ columns obtained
directly from {\it Spitzer} mid-IR data should result in reliable
water abundances.  Such studies point to high H$_2$O abundances of
$\sim$10$^{-4}$ in the densest, heavily extincted regions centered on
the deeply embedded protostars \citep[e.g.][]{Herczeg12,Goicoechea12},
but to lower H$_2$O abundances of $10^{-7}-10^{-6}$ at shock positions
offset from the protostars \citep{Tafalla13,Nisini13,Santangelo13}
(see Fig.~\ref{fig:557spectra} (right)). Thus, there do appear to be
real variations in the water abundance in high temperature gas
depending on the conditions.

The explanation for the low H$_2$O abundances centers again on the
presence of UV photons and on the type of shocks involved. Broadly
speaking, two classes of shocks occur in interstellar clouds
\citep{Draine03}: in $J$-shocks, there is sudden jump in the physical
variables (temperature, density) at the shock position and
temperatures often reach high enough values ($>$4000 K) to dissociate
most molecules, although non-dissociative $J$-shocks exist as
well. The gas is significantly compressed in the cooler post-shock
gas. In $C-$type shocks, the ions in a magnetic precursor are able to
communicate the upcoming shock to the neutrals so that temperature and
density vary much more smoothly through the shock and do not
dissociate the molecules. In dissociative $J$-type shocks, the H$_2$O
abundance is low \citep{Neufeld89b}, in $C-$type and non-dissociative
$J$ shocks it is high \citep{Kaufman96,Flower10}.  Analysis of the
H$_2$O excitation in several shock spots offset from the protostar
indicates that most of the water emission comes from gas that is
overpressured with respect to the surroundings by a factor of $\sim
10^4$, consistent with $J-$type shocks \citep{Tafalla13,Nisini13}.
Typical temperatures of the water emitting gas are around 500~K and
densities of order $10^7$ cm$^{-3}$, and the inferred water abundances
are indeed low, down to $10^{-7}$.

One further option to distinguish between the various models is to
look at the dissociation products, OH and [O I].  Non-dissociative
continuous shocks predict OH/H$_2$O $<<$1 if photodissociation is
negligible.  Spectrally resolved data on OH are still sparse, but for
one high-mass protostar, a ratio OH/H$_2$O $>$0.03 is found in the outflow
\citep{Wampfler11}. From spectrally unresolved data of the IRAS4B
low-mass protostar, OH/H$_2$O=0.2 is inferred \citep{Herczeg12}.
These high ratios, combined with strong emission from [O I] and
widespread detection of OH
\citep{vanKempen10,Wampfler13,Karska13,Green13}, also point to the presence
of dissociative shocks. On the other hand, the high abundance of
H$_2$O at the source position together with high$-J$ CO excitation
indicate the presence of some non-dissociative shocks as well
\citep{Karska13,Goicoechea12}.  Most likely, a combination of $J$-type
shocks and UV-irradiated $C$-type shocks is needed to explain the data
at the various positions. The UV radiation does not have to originate
externally from nearby bright stars as in the case of Orion, but can be
produced locally by fast $J-$ type shocks
\citep{Neufeld89a,Snell05}. One firm conclusion from all these studies
is that UV irradiated shocks are common near protostars and that a new
generation of shock models needs to be developed to interpret the
observed abundance ratios.

Alternative explanations for the low H$_2$O abundance include rapid
freeze-out onto grains in the cool post-shock gas. For a density of
$10^6$ cm$^{-3}$, the timescale is $\sim 10^3$ yr, longer than typical
shock timescales \citep{Bergin98}. Thus, freeze-out will not be
significant in the still warm post-shock gas probed by the {\it
  Herschel} data.  Other time-dependent effects can also play a role.
Slow $C-$ type shocks with velocities $\leq$15 km s$^{-1}$ reach
post-shock temperatures of only 300--400 K \citep{Kaufman96}. The
timescale to convert all oxygen into water is $>$400 yr at such low
temperatures, longer than the cooling time of the shock itself, which
is less than a hundred years.  Coupled with their ineffectiveness in
releasing water ice into the gas, slow shocks could have low water
abundances. The main argument against this option is that the water
line profiles \citep{Nisini13} indicate shock velocities higher than
15 km s$^{-1}$ (Fig.~\ref{fig:557spectra} and \ref{fig:paraspectra}).

\begin{table}
\caption{Summary of H$_2$O/H$_2$ abundance determinations in shocks}
\begin{tabular}{l l c l l}
\hline
\hline
Source & Instrument & H$_2$O/H$_2$ & H$_2$ from & Reference \\
       &            & (10$^{-4}$) & \\
\hline
Orion-KL & ISO-LWS & 5 & &\citet{Harwit98} \\
Orion-IRc2& ISO-SWS & 2--5 & Model & \citet{Wright00} \\
Orion peak 1 & ISO-SWS & 0.2 & & \citet{Gonzalez02} \\
Orion peak 1+2 & ISO-LWS & 0.2 & & \citet{Cernicharo06} \\
Orion-KL & SWAS & 3.5 & & \citet{Melnick00} \\
Orion-KL & Odin & 0.1--1 & & \citet{Olofsson03} \\
Orion-KL & Herschel-HIFI & 7.4 & low-$J$ CO & \citet{Melnick10} \\
\smallskip
Several  & ISO-LWS & 0.1--1 (Class 0) & high-$J$ CO & \citet{Nisini02} \\
         &  & $<$0.1 (Class I)& high-$J$ CO & \citet{Nisini02} \\
Several & SWAS & 0.001-0.01$^a$ & low-$J$ CO& \citet{Franklin08} \\
Several & Odin & 0.003--5$^a$ & low-$J$ CO & \citet{Bjerkeli09} \\
HH 54 & ISO-LWS &0.1 & high-$J$ CO & \citet{Liseau96} \\ 
HH 54 & Herschel-HIFI & 0.1$^a$ & mid-$J$ CO, H$_2$ & \citet{Bjerkeli11} \\
L1157 B1 & Herschel-HIFI & 0.008(LV) & mid-$J$ CO & \citet{Lefloch10} \\
     &     & 0.8(HV) & & \citet{Lefloch10} \\
L1157 B2/R & Herschel-HIFI/PACS & 0.01 & warm H$_2$ & \citet{Vasta12} \\
L1448 B2/R4 & Herschel-HIFI/PACS & 0.05-0.1 (HV) & warm H$_2$ &\citet{Santangelo12} \\
L1448  & Herschel-HIFI/PACS & 0.005-0.01  & warm H$_2$& \citet{Nisini13} \\
VLA1623 & Herschel-HIFI & $<$0.01$^a$ & warm H$_2$ & \citet{Bjerkeli12} \\
Several & Herschel-HIFI & 0.1 (HV) & low-$J$ CO & \citet{Kristensen12} \\
NGC1333 IRAS4B & Herschel-PACS & 1.0 & H$_2$ & \citet{Herczeg12}\\
Serpens SMM1 & Herschel-PACS & 0.4 & high-$J$ CO & \citet{Goicoechea12} \\ 
L1448
Several B/R & Herschel-PACS & 0.01 & warm H$_2$ & \citet{Tafalla13} \\
NGC 7129 IRS & Herschel-HIFI/PACS & 0.2--0.3 & high-$J$ CO& 
   \citet{Johnstone10} \\
             & & & & \citet{Fich10} \\
NGC 6334I & Herschel-HIFI & 0.4 & mid-$J$ CO & \citet{Emprechtinger10} \\
DR21(OH) & Herschel-HIFI & 0.32$^b$ & high-$J$ $^{13}$CO 
   & \citet{vanderTak10} \\
\hline
\end{tabular}
Note: uncertainties claimed by the authors range from a factor of 2 up to an
order of magnitude. 
HV=high-velocity; LV=low
velocity. In most cases, the abundance refers to the outflow at the
central source position. B and R indicated blue- and red-shifted outflow spots offset from the source.
\\ $^a$ ortho-H$_2$O abundance. \\ $^b$ from para-H$_2$O assuming $o/p$=3
\\
\label{tab:shocks}
\end{table}

{\it High temperature chemistry vs.\ ice sputtering}.  The above shock
models produce water through the high temperature neutral-neutral
reactions. However, sputtering of ice by high-velocity particles in
the shocks is another option, thereby releasing the ice mantle
material back into the gas.  Models \citep{Draine83,Jimenez08}
indicate that this becomes effective for shocks faster than 10--15 km
s$^{-1}$. One possibility to distinguish between these two options is
to compare observations of H$_2$O in shocks with those of other
grain-surface products. In particular, NH$_3$ and CH$_3$OH are also
known to be among the abundant ice mantle constituents, with
abundances of $\sim$5 \% and 1--20\% with respect to that of H$_2$O
ice, respectively \citep{Oberg11}.

High $S/N$ {\it Herschel} and ground-based data indicate that for most
sources the observed NH$_3$ and/or CH$_3$OH emission profiles lack the
broad line wings seen for the H$_2$O lines, implying an absence of
these molecules at the highest velocities \citep{Jimenez05,Codella10}.
This could be taken to imply a lack of ice mantle sputtering.
However, Viti et al.\ \citep{Viti11} note that NH$_3$ and CH$_3$OH can
actually be destroyed by atomic H in high temperature gas following
ice sputtering, whereas H$_2$O is not. This is because the energy
barrier for destroying H$_2$O by atomic H is much higher, $\sim$9000 K
versus $\sim$5000~K and $\sim$3000~K for NH$_3$ and CH$_3$OH,
respectively. Thus, ice mantle sputtering is not yet excluded.  If
this scenario is valid, the lack of NH$_3$ or CH$_3$OH would allow the
shock temperature to be constrained. Further observational tests of
this scenario are needed.

\subsubsection{Supernova remnants}

IC 443 is a prototypical example of high velocity gas ejected by a
supernova interacting with the interstellar medium. A molecular cloud
lies across the face of the supernova remnant and at locations where
the remnant encounters this cloud, shocked molecular gas is revealed
by broad asymmetric line profiles.  SWAS and ISO data point to a very
low abundance of H$_2$O in the shock, H$_2$O/CO=$2\times
10^{-4}-3\times 10^{-3}$ in the downstream gas \citep{Snell05} . Only
in the warmer part of the shock does the ratio reach
H$_2$O/CO$\approx$0.2. As for the shocks associated with protostars, a
combination of freeze-out of H$_2$O in the pre- and postshock cloud
(with slow shocks preventing complete ice sputtering) as well as
photodissociation of H$_2$O in postshock gas are invoked to explain
the low abundances.  The UV radiation is thought to be generated by a
preceeding fast $J-$type shock. Further evidence for enhanced UV
radiation comes from the high C/CO ratio in the preshock gas
\citep{Keene97}. Compared with dense star-forming clouds, the
molecular cloud encountered by IC 443 has a lower density, more like a
translucent cloud.  The higher H/H$_2$ ratio in a translucent region
could potentially drive reactions that convert H$_2$O back to
O. Similarly, an enhanced cosmic ray ionization rate could result in
higher H/H$_2$ and C/CO ratios in the preshock gas
\citep{Indriolo10}. However, even a very high H/H$_2$ ratio of 10
drives most oxygen into water since, as noted above, the energy
barrier for the H$_2$O + H back reaction is very high ($\sim$9000 K)
\citep{Snell05} .

Similar conclusions have been reached for other supernova remnants
interacting with molecular clouds based on high OH/H$_2$O ratios
\citep{Reach98,Lockett99}.

\subsubsection{External galaxies}

The central regions of galaxies contain dense and warm gas, which
allow the water chemistry to be tested on scales of a few hundred pc,
as opposed to the sub-pc scales in our own Galaxy. Following early
detections with ISO \citep{Gonzalez04,Gonzalez08}, a wealth of water
lines has now been seen in nearby galaxies ($z<0.3$) with all three
instruments on {\it Herschel}
\citep{vanderWerf10,Weiss10,Rangwala11,Gonzalez12,Spinoglio12,Kamenetzky12,Gonzalez13,Meijerink13,Yang13}. The
lines are sometimes seen in emission and in other cases in absorption
against the bright nuclear continuum, and arise from levels with
$E_u/k_{\rm B}$ up to 650 K. Water is now even seen at high redshifts up to
$z=6.3$ \citep{Riechers13} with ground based telescopes
\citep{Omont11,vanderWerf11,Combes12,Weiss13,Omont13} (see
  Fig.~\ref{fig:waterhighz}). The line luminosities of water are high, only
  slightly below those of CO.

  The water lines in these distant sources arise from a mix of
  physical components including the compact nuclear region but also
  more extended diffuse gas (\S~\ref{sect:diffuse}), so detailed
  modeling is needed to disentangle them.  The warm and dense nuclear
  regions are found to have high water abundances of order $10^{-5}$,
  consistent with high temperature chemistry and grain mantle
  sublimation \citep{Gonzalez12}.  In some galaxies, water outflows
  out to several hundred km s$^{-1}$ are seen,  so shocks can
  play a role as well \citep{Meijerink13}. The high OH/H$_2$O ratios
  of order unity again point to enhanced water dissociation; however,
  rather than UV radiation, X-rays and/or cosmic rays are invoked. 

  The ubiquitous detection of high excitation water lines such as
  $3_{21}-3_{12}$ and $4_{22}-4_{13}$ has been interpreted to imply
  that pumping by far-infrared continuum radiation due to warm dust
  plays an important role in the excitation of water in many, but not
  all \citep{Meijerink13}, of these sources
  \citep{vanderWerf11,Yang13}.

\subsubsection{Evolved stars}

{\it O-rich envelopes.}  The dense and warm envelopes around dying
stars (Fig.~\ref{fig:cycle}) provide another good laboratory to
investigate the high temperature water chemistry.  O-rich stars have
[O]/[C]$>$1, so at densities up to $10^{13}$ cm$^{-3}$ and
temperatures up to $1000$~K most oxygen should be driven into
water: since H$_2$O is the
thermodynamically most stable oxygen-containing molecule after CO, its
abundance is basically set by the available oxygen reservoir minus the
abundance of oxygen locked up in CO. 
H$_2$O and OH have been detected for decades in these sources
through their masing lines at radio wavelengths \citep{Habing96}.
These data have led to the development of photochemical models of
the envelopes in which thermal equilibrium chemistry dominates in
the innermost envelope but two-body reactions and photodissociation
due to UV radiation from the interstellar field become more
important in the outer envelope
\citep{Goldreich76,Glassgold96,Justtanont94,Willacy97}. In the later
stages leading up to the planetary nebulae stage, asymmetries and
bipolar outflows develop, the morphology of which is probed by H$_2$O
and OH masers \citep{Amiri11}.

\begin{figure}
  \includegraphics[width=0.7\textwidth]{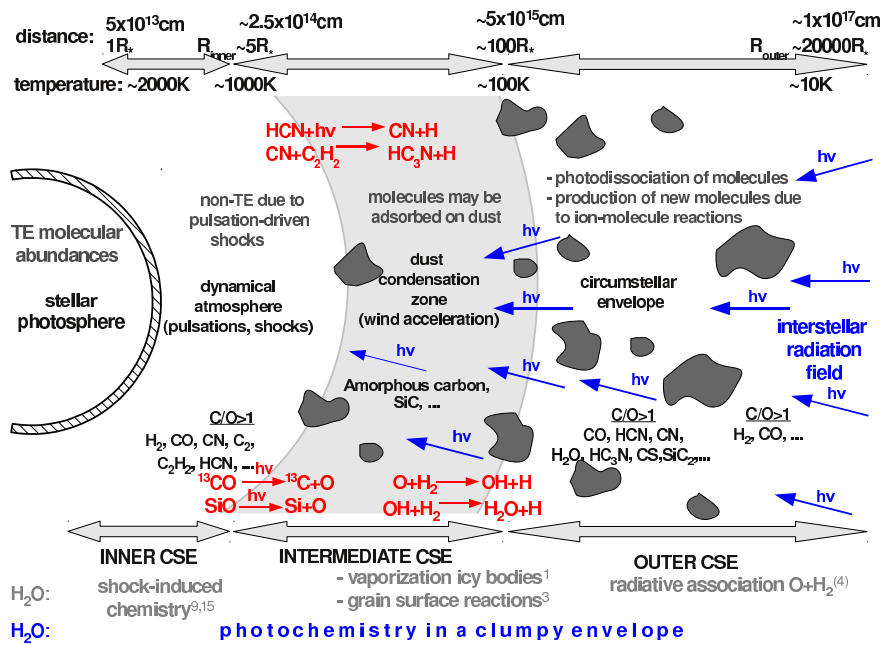}
  \caption{Schematic structure of the envelope of an evolved
    carbon-rich star.  Several chemical processes are indicated at
    their typical temperatures and radial distances from the star (not
    to scale). Thermodynamic equilibrium (TE) chemistry defines the
    abundances in the stellar atmosphere; shock-induced
    non-equilibrium chemistry can take place in the inner wind
    envelope; dust-gas and ion-molecule reactions alter the abundances
    in the intermediate wind zone; penetration of cosmic rays and UV
    photons dissociates the molecules and initiates an active
    photochemistry. If the medium is clumpy, the UV radiation can
    penetrate more deeply (blue arrows).  The different mechanisms
    proposed as origins of water in a carbon-rich environment are
    indicated at the bottom of the figure in gray at the typical
    distances where they occur. Reprinted with permission from
    Reference 405. Copyright 2010 Macmillan Publishers Ltd: Nature.  }
  \label{fig:Decin}
\end{figure}

To quantify the water chemistry, non-masing lines need to be observed.
A wealth of thermal water lines have been detected with ISO toward the
evolved stars W Hydrae \citep{Barlow96,Neufeld96} and VY CMa
\citep{Neufeld99}, showing that water is abundant and that the
emission originates primarily in the acceleration zone of the dense
outflows created by radiation pressure on the dust grains. The
presence of OH illustrates the importance of photodissociation at
larger distances from the star. A physical model of the W Hydrae
envelope has been used to infer H$_2$O/H$_2$=$8\times 10^{-4}$ in the
innermost layers and $3\times 10^{-4}$ at larger radii
\citep{Barlow96}. Comparable values have been found in other studies
based on ISO data
\citep{Justtanont05,Maercker08,Maercker09}. Alternatively, rather than
testing the oxygen chemistry, the water observations can be used to
make independent estimates of the temperature structure and mass loss
rates \citep{Neufeld96}.

Spectrally resolved lines of H$_2$O and its isotopologues with {\it
  Herschel}-HIFI provide the next step forward in the modeling of the
water chemistry in these sources. The best studied case so far is IK
Tau, for which an abundance ortho-H$_2$O/H$_2$=$5\times 10^{-5}$ 
has been derived \citep{Decin10b} . The
abundance is consistent with thermodynamic equilibrium chemistry and
does not require other processes that have been proposed such as
pulsationally induced non-equilibrium chemistry or grain surface
chemistry via Fischer-Tropsch catalysis \citep{Willacy04}.

For the star $\chi$ Cyg, which has [C]/[O]$\approx 1$, a water
abundance H$_2$O/H$_2$=$1.1\times 10^{-5}$ has been found with {\it
  Herschel}-HIFI, much lower than in O-rich stars
\citep{Justtanont10}. This implies that the bulk of the oxygen in this
source is locked up in CO rather than in H$_2$O and that [C]/[O] is
$\leq$0.98.

{\it C-rich envelopes.}  Water was not expected to be detected in the
envelopes of C-rich stars with [C]/[O]$>$1 since all free oxygen
should be locked up in the very stable CO molecule.  The detection of
the H$_2$O 557 GHz line toward the C-rich star IRC+10216 with SWAS
\citep{Melnick01} (and subsequently Odin \citep{Hasegawa06}) therefore
came as a surprise.  It has been speculated that the cold water could
result from sublimating icy bodies orbiting the star in an analog of
the Kuiper Belt in our own solar system
\citep{Melnick01,Ford01}. However, the subsequent {\it Herschel}
detection of a multitude of warm water lines \citep{Decin10} combined
with their line profiles \citep{Neufeld11a} argues against the
vaporization of ices and favors a scenario in which periodic high
temperature shocks\citep{Cherchneff11} and/or UV penetrating in a
clumpy medium\citep{Decin10} liberate oxygen from CO in the inner
envelope and drive it into water (Fig.~\ref{fig:Decin}).  Detection of
water in C-rich evolved stars is now found to be common
\citep{Neufeld11b}, with inferred H$_2$O/H$_2$ abundances ranging from
$8\times 10^{-8}$-$2\times 10^{-6}$.

\subsubsection{Inner protoplanetary disks}

\begin{figure}[t]
  \includegraphics[width=0.5\textwidth]{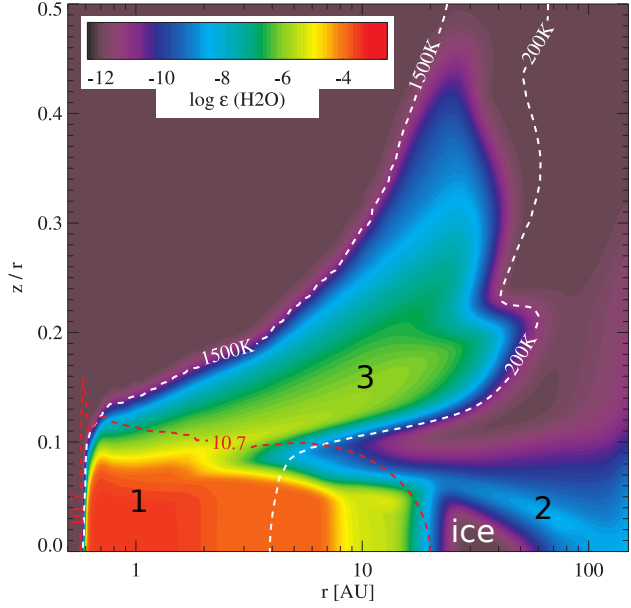}
  \caption{Abundance of water vapor relative to total hydrogen as
    function of both radial distance $r$ and relative height above the
    midplane $z/r$ for a disk around an A-type star ($T_*$=8600 K).
    Three regions with high H$_2$O abundance can be
    distinguished. Regions 1 and 3 involve high temperature chemistry,
    whereas region 2 lies beyond the ice line and involves
    photodesorption of water ice (see also Fig.~\ref{fig:Bergin}). The
    white contours indicate gas temperatures of 200 and 1500 K,
    whereas the red contour shows the $n_{\rm H}=5\times 10^{10}$
    cm$^{-3}$ density contour. Reprinted with permission from
    Reference 409. Copyright 2009 European Southern Observatory.}
  \label{fig:waterdisk}
\end{figure}

The inner few AU of protoplanetary disks around young stars have very
high temperatures ($>$ few hundred K) and densities ($> 10^{11}$
cm$^{-3}$) near the midplane.  Under these conditions, the chemistry
is close to thermodynamic equilibrium and most of the oxygen will be
driven into water.  This warm water reservoir is denoted as region 1
in the study by Woitke et al.~\citep{Woitke09h2o} and illustrated in
Figure~\ref{fig:waterdisk}. As pointed out by \citet{Bethell09}, the
water column in the upper layers is so large that the UV absorption
optical depth becomes greater than unity so that water can shield
itself from the dissociating stellar radiation and have an even higher
abundance at intermediate heights.

At somewhat larger distances (out to 30 AU, depending on the type of
star) and higher up in the disk atmosphere, the chemistry is no longer
in thermodynamic equilibrium but controlled by two body processes. In
these upper layers, the gas temperature is significantly higher than
the dust temperature, typically 300--2000 K. Thus the energy barriers
for the neutral-neutral reactions of O + H$_2$ and OH + H$_2$ are
readily overcome and H$_2$O is rapidly produced
\citep{Thi05,Woitke09,Glassgold09} (region 3 in
Fig.~\ref{fig:waterdisk}). In the uppermost surface layers, water is
even more rapidly destroyed by photodissociation and reactions with
atomic H than it is produced so there is little water at the very top
of the disk. The bulk of the warm water reservoir therefore lies at
intermediate heights in the disk where it is shielded from the most
intense stellar UV radiation. The cold water reservoir, region 2 in
Fig.~\ref{fig:waterdisk} is discussed in \S~\ref{sect:outerdisk}.

\begin{figure}
\includegraphics[angle=0,width=0.65\textwidth]{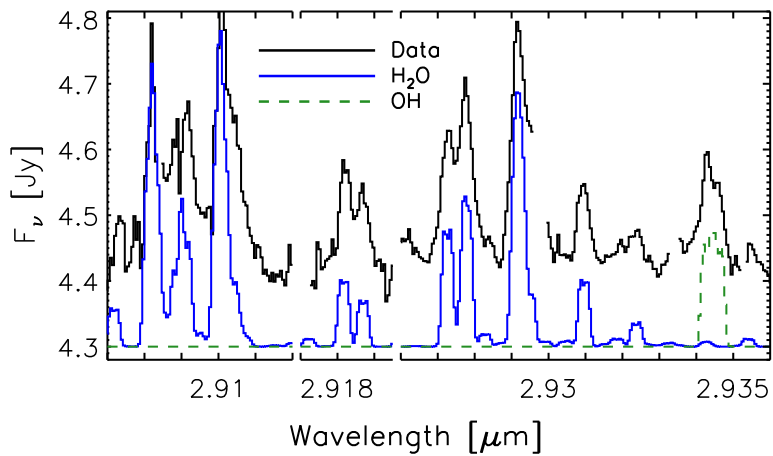}
\caption{Example of vibration-rotation lines of H$_2$O observed with
  Keck-NIRSPEC arising in the warm water reservoir in the inner AU of
  the AS 205 protoplanetary disk. Adapted with permission from
  Reference 73.  Copyright 2008 American Astronomical Society. }
\label{fig:salyk}
\end{figure}

The warm water reservoir in the inner regions of disks has been
detected using the {\it Spitzer}-IRS through highly-excited pure
rotational lines at mid-infrared wavelengths
\citep{Carr08,Salyk08,Salyk11,Pontoppidan10,Pontoppidan10visir}
(Fig.~\ref{fig:spitzer}) and through ground-based near-infrared
vibration rotation lines \citep{Salyk08,Mandell12}
(Fig.~\ref{fig:salyk}). Abundance ratios are difficult to extract from
the observations because the lines are highly saturated and in the
case of {\it Spitzer} data spectrally unresolved, but abundance ratios
of H$_2$O/CO$\sim$1--10 have been inferred for emitting radii up to a
few AU \citep{Salyk11,Mandell12}. Within the more than an order of
magnitude uncertainties, these values are consistent with the model
predictions for the warm layers (Fig.~\ref{fig:waterdisk}). The {\it
  Spitzer} and near-IR data show a clear dichotomy between disks
around the cooler T Tauri stars ($T_* \lapprox 5000$~K), where H$_2$O
is detected, and those around the hotter A-type stars ($T_*\sim
8000-10000$~K), where it is usually not, presumably due to the more
rapid photodissociation in the latter case producing the observed OH
(Fig.~\ref{fig:radfield}) \citep{Pontoppidan10,Fedele11}.

Longer wavelength observations with {\it Herschel}-PACS probe gas
further out into the disk. Far-infrared lines from warm water have
been detected in some sources, mostly around T Tauri stars
\citep{Zhang13,Riviere12} but also a few A-type stars
\citep{Fedele12,Meeus12,Fedele13}. Together with the shorter and
longer wavelength water data (see \S~\ref{sect:outerdisk}), these
observations can be used to constrain the water ice line in disks
\citep{Zhang13,Meijerink09}.

In summary, the observations of a wide range of sources confirm that
water can be produced in copious amounts by a combination of high
temperature chemistry and ice mantle sublimation and
sputtering. However, a common finding in all of these regions is that
the water abundance is lower than expected if all oxygen were driven
in to water, by up to 2--3 orders of magnitude, and that the OH/H$_2$O
ratios are correspondingly higher than expected. This indicates that
destructive radiation (UV, X-rays and/or cosmic rays) plays a
significant role in limiting the water abundance and must be included
in the next generation of models.

\subsection{Testing photodissociation models}
\label{sect:comets}

\subsubsection{Comets} 

\begin{figure}
\begin{minipage}{0.4\textwidth}
\includegraphics[angle=-90,width=\textwidth]{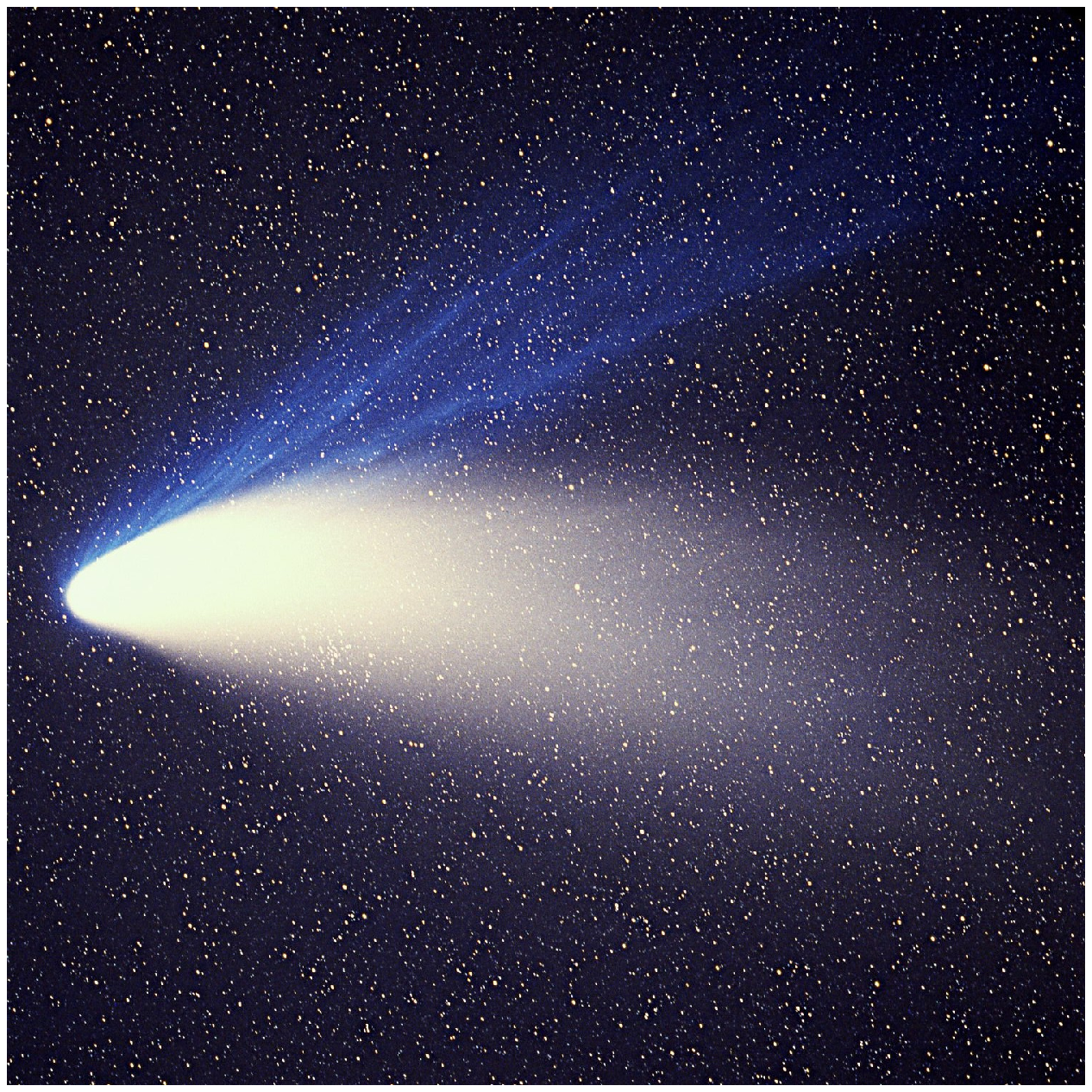}
\end{minipage}\hfill%
\begin{minipage}{0.52\textwidth}
\includegraphics[angle=0,width=\textwidth]{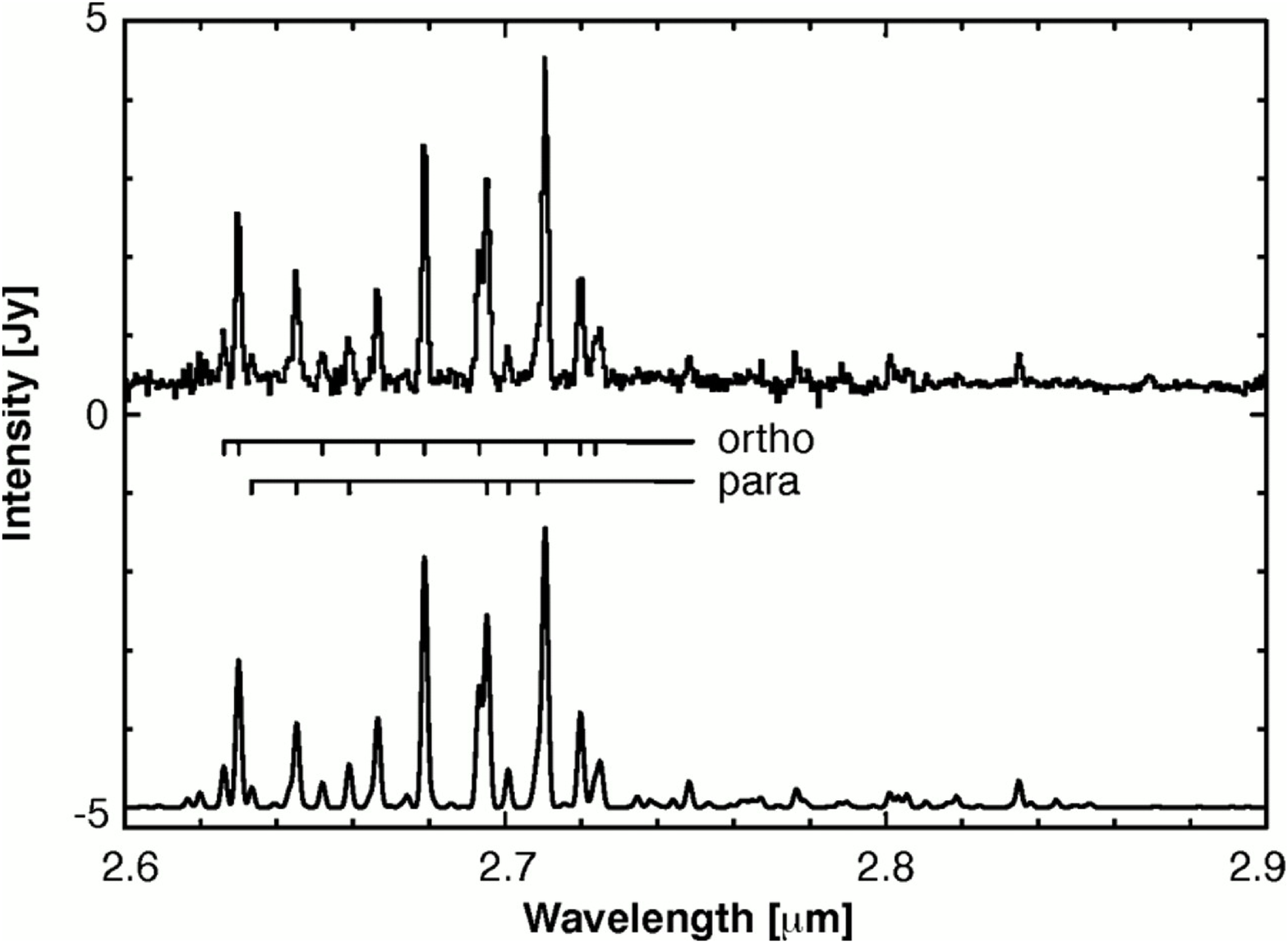}
\end{minipage}\hfill%
\caption{Left: Image of comet C/1995 O1 Hale-Bopp taken on April 4
  1997 by E. Kolmhofer and H. Raab ({\tt www.sternwarte.at}) and
  reproduced with their permission; Right: H$_2$O vibration-rotation
  lines of comet Hale-Bopp observed with the ISO-SWS. Reprinted with
  permission from Reference 429.  Copyright 1997 American
  Association for the Advancement of Science). }
  \label{fig:comet}
\end{figure}

Comets are small (few km diameter) solid bodies in the outer solar
system consisting of ice (mostly water), dust and small rocky
particles. When they come close to the Sun, the solids are heated up
and the ices sublimate into the gas, creating the coma and tails that
emit visible radiation due to the fluorescence of small molecules
(Figure~\ref{fig:comet}). The solar radiation also dissociates the
`parent' molecules that originate from the cometary ices into
`daughter' molecules \citep{Whipple51,Haser57}. Specifically, H$_2$O
is photodissociated into OH which in turn photodissociates into O. The
OH radical can readily be observed through its 18 cm radio
hyperfine/$\Lambda$-doubling transitions. The photodissociation model
is then used to infer the original H$_2$O production rate in the
cometary coma
\citep{Festou81,Crovisier89,Crovisier97,Mumma93,Bockelee04}. This
requires an accurate determination of the H$_2$O and OH
photodissociation rates by the solar radiation.

The radiation field of the Sun ($T_*=5780$~K) peaks at visible
wavelengths and has relatively few photons at UV wavelengths where
molecules like H$_2$O and OH can absorb and dissociate.  The main
photodissociation channel of H$_2$O is therefore through solar Lyman
$\alpha$ radiation which can excite H$_2$O into the $\tilde B$
electronic state. As discussed in \S~\ref{sect:photodissociation},
this process leads to OH in very high rotational levels as well as
some vibrational excitation.  This `prompt' emission of vibrationally
and rotationally excited OH produced directly from H$_2$O
photodissociation (in contrast to the fluorescently pumped OH
emission) has been detected in several comets
\citep{Brooke96,Magee02,Gibb03}. If the production rates into the
various OH states are well known, the observed emission can be
calibrated to provide an alternative method to determine the H$_2$O
release from the comet nucleus \citep{Bonev04,Bonev06}. Longer
wavelength radiation at 1570 \AA\ photodissociates H$_2$O through the
$\tilde A$ state and has been experimentally found to produce
population inversion in OH with a strong preference for populating the
upper $\Lambda$-doubling state \citep{Andresen83}. However, this
mechanism is unlikely to be responsible for the bulk of the observed
OH emission in comets.

\subsubsection{Bowshocks}  

\begin{figure}
\begin{minipage}{0.4\textwidth}
\includegraphics[angle=0,width=\textwidth]{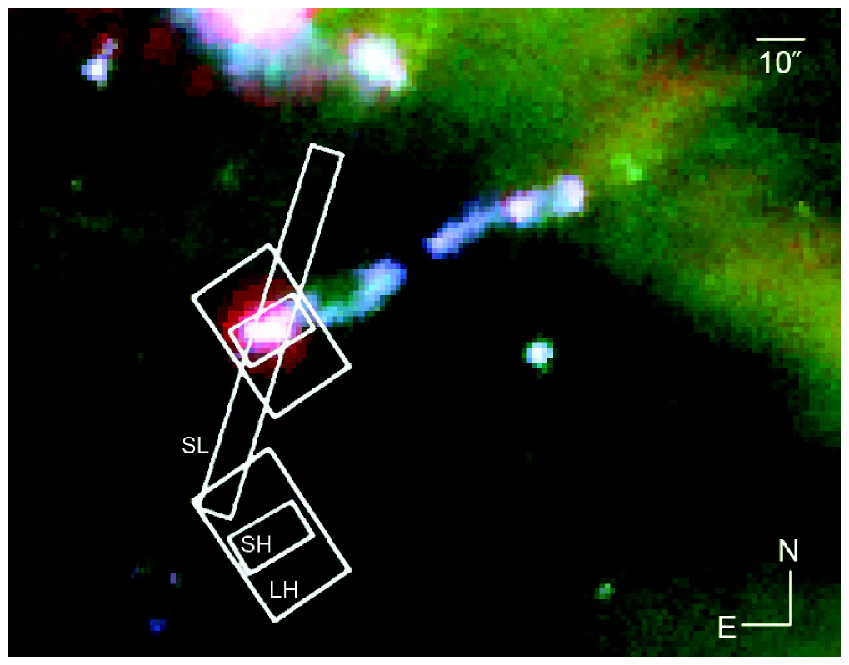}
\end{minipage}\hfill%
\begin{minipage}{0.55\textwidth}
\includegraphics[angle=0,width=\textwidth]{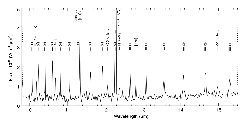}
\end{minipage}\hfill%
\caption{Left: {\it Spitzer} infrared image of the HH 211 jet and
  bowshock with the IRS spectrometer slits overlaid (blue=3.6 + 4.5
  $\mu$m; green=8 $\mu$m; red=24 $\mu$m).  Right: {\it Spitzer}-IRS
  spectrum of HH 211 at the bowshock (red tip). The OH lines originate
  from levels up to $J$=69/2. Reprinted with permission from
  References 437 and 438. Copyright 2008 and 2012 American
  Astronomical Society. }
  \label{fig:hh211}
\end{figure}

Another example of OH prompt emission following photodissociation is
found in some shocked regions. The most powerful shocks occur at the
tip of the outflows where the jet rams into the surrounding cloud with
velocities up to several hundred km s$^{-1}$. At these so-called bow
shocks, the gas temperature increases up to $10^5$ K resulting in
dissociation of molecules and ionization of atoms. As the gas cools,
the atomic ions recombine with electrons and emit UV radiation, specifically
Lyman $\alpha$ photons \citep{Raymond79,Neufeld89b}. These photons
dissociate H$_2$O produced further downstream in the cooling gas through the
$\tilde B$ electronic state, producing highly rotationally excited
OH. This scenario has been nicely confirmed by the detection of
OH pure rotational lines up to $J$=69/2 in the HH 211 outflow using
{\it Spitzer}-IRS \citep{Tappe08,Tappe12} (see Fig.~\ref{fig:hh211}).

In summary, the observed state-resolved OH product distribution leave
no doubt that photodissociation of water by Lyman $\alpha$ photons
occurs in these regions. This is a beautiful example of detailed
molecular physics and astronomy going hand-in-hand in enhancing our
understanding of the universe.

\subsection{Water ortho-to-para ratio}

Observations of water in space also provide constraints on the
ortho/para ratio (OPR) of water in the gas, which in turn could tell
astronomers something about the conditions, formation or thermal
history of water in specific regions.  An alternative way to describe
the OPR is through the `spin temperature', defined as the temperature
that characterizes the observed OPR if it were in thermal equilibrium.
The OPR becomes zero in the limit of low temperature and 3 in the
limit of high temperature.  For $T \ge 50$~K, the equilibrium OPR
exceeds 2.95, while for $T \le 25$~K, the OPR is well-approximated by
$9\,\exp(-\Delta E/kT)$ = $9\,\exp(-34.2 {\rm K}/T)$, where $\Delta E$
is the energy difference between the $1_{01}$ and $0_{00}$ rotational
states. This expression applies specifically to the gas-phase in which
the molecule is free to rotate; the relationship between the
temperature and the equilibrium OPR may be different if the molecule
is rotationally hindered \citep{Buntkowsky08}.

Accurate measurements of the OPR have been made for diffuse and
translucent clouds by combining absorption line observations of
optically thin 557~GHz and 1113~GHz transitions.  In a study of 13
translucent clouds observed with {\it Herschel}-HIFI, Flagey et
al.\citep{Flagey13} derived a mean OPR of 2.9, very close to the value
of 3 expected from nuclear spin statistics in thermal equilibrium in
the limit of high temperature.  While 10 of the 13 clouds exhibited an
OPR consistent with 3, one had a measured OPR significantly larger
than three ($4.3 \pm 0.2$), and two had OPRs significantly smaller
than 3 ($2.3 \pm 0.1$ and $2.4 \pm 0.2$). Gas-phase ion molecule
chemistry is expected to produce a ratio close to 3: the dissociative
recombination of H$_3$O$^+$ with electrons to produce H$_2$O, the
dominant channel leading to water in these clouds, has several eV of
excess energy and the nuclear spins should be produced in the
statistical high temperature ratio, although a detailed calculation
using the angular momentum approach of \citet{Oka04} should be used to
check this assumption.  Exchange reactions of H$_2$O with H$^+$
can further drive the OPR to the kinetic temperature if these reactions
are rapid enough compared with the lifetime of the cloud.

\begin{figure}
\includegraphics[angle=0,width=0.65\textwidth]{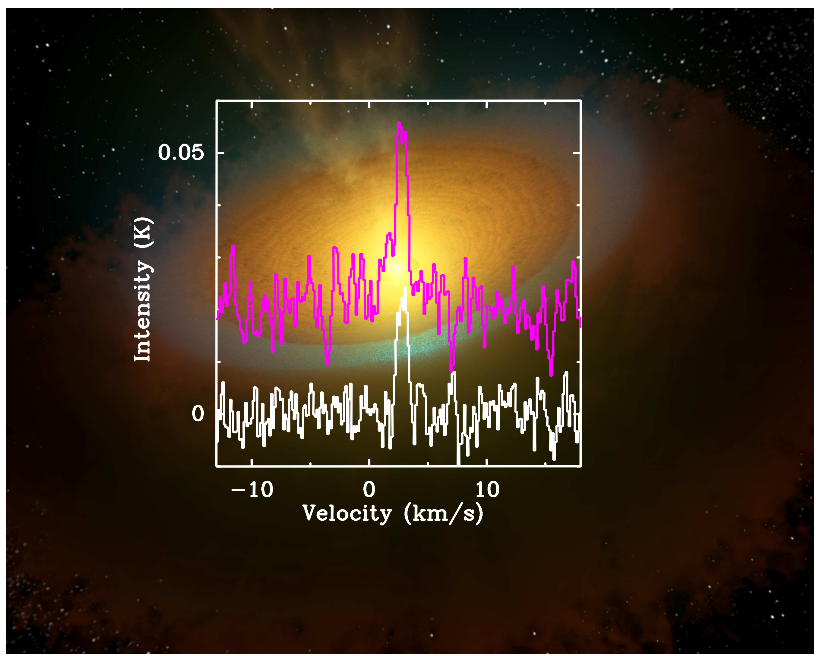}
\caption{Emisssion lines of ortho-H$_2$O ($1_{10}-1_{01}$ at 557 GHz)
  and para-H$_2$O ($1_{11}-0_{00}$ at 1113 GHz) observed with {\it
    Herschel}-HIFI toward the TW Hya disk. The para-H$_2$O line is
  offset for clarity. Adapted with permission from Reference
  284. Copyright 2011 American Association for the Advancement of
  Science. Background image: cartoon of the TW Hya disk, credit
  ESA/NASA/JPL-Caltech). }
\label{fig:waterop}
\end{figure}

In shocks and evolved stars where the water lines are seen in
emission, it is difficult to measure the OPR with the same precision
as for translucent clouds, even though many lines of both ortho and
para H$_2$O have been observed. Nevertheless, all analyses so
far\citep{Melnick10,Decin10b,Herczeg12,Emprechtinger13} are consistent
with an OPR of 3, as expected for hot water.  Similarly, the data on
warm water detected at mid infrared wavelengths in the inner regions
of protoplanetary disks\citep{Pontoppidan10} are consistent with
OPR=3.  In contrast, the OPR of cold water seen with {\it
  Herschel}-HIFI in the outer regions of disks \citep{Hogerheijde11}
imply an OPR$<1$ (Fig.~\ref{fig:waterop}, although smaller emitting
areas and correspondingly higher optical depths than assumed in the
original analysis could drive the observed ratios closer to 3.

Finally, near-infrared cometary spectra such as shown in
Fig.~\ref{fig:comet} allow an accurate determination of the OPR.  More
than a dozen comets have now been measured, with inferred OPRs ranging
from 2.4 to 3. A summary is presented in Fig.~10 of
\citealt{Mumma11}; the OPRs correspond to spin temperatures as low as
$\sim$20 K.  These spin temperatures have often been
interpreted as a measure of the temperature at which molecules formed,
or condensed, on the grain surfaces providing potentially a measure of
where the comets formed in the solar nebula. However, this
interpretation assumes that the nuclear spins of water equilibrate to
the grain temperature by some mechanism and that the OPR is preserved
upon sublimation.

These assumptions are discussed in the review by Watanabe et al. (this
volume) (see also \citealt{Dulieu11}) and are being heavily debated by
molecular physicists and astronomers.  
Recent laboratory measurements of water molecules
isolated in solid Ar matrices at 4 K, initially prepared as pure
para-H$_2$O \citep{Sliter11}, show a thermal OPR upon heating to
260--280 K. Similarly, studies of vapor deposited water at low
temperatures show a thermal OPR upon sublimation at 150 K, even if
 the water is kept at 8 K for several days \citep{Hama11}.  These
data suggests that spin conversion of water clusters and water ice is
fast upon heating and that the OPR retains no history of the formation
temperature.

In the outer regions of protoplanetary disks, the water vapor is
produced by photodesorption, not thermal sublimation. To what extent
does the process of photodesorption preserve the OPR in the ice?  As
discussed in \S~\ref{sect:desorption}, there are two possible
mechanisms for water produced by photodesorption, following
dissociation of H$_2$O ice into H + OH: (i) Recombination of H + OH
$\to$ H$_2$O in the ice, which then has enough energy to desorb; (ii)
Kick-out mechanism: the energtic H atom kicks out a neighboring H$_2$O
molecule on picosecond timescales.  In option (i), the H-OH bond is
broken and reformed, so the OPR should go to the statistical value of
3. In option (ii), no bond is broken and the original OPR in the ice
should be preserved.  The relative importance of option (i) and (ii)
depends on the ice monolayer and to a lesser extent on the ice
temperature, but are roughly of equal importance \citep{Arasa11}.
Thus, even if the water OPR in the ice were low, say $\leq$0.5, the
gas-phase OPR of the photodesorbed H$_2$O would be significantly
higher, about 2, but still lower than 3.

In summary, the molecular processes that lead to ortho/para
equilibration, and the timescale for those processes relative to
relevant astrophysical timescales, are still poorly understood. More
laboratory work on both the gas-phase and grain-surface processes that
convert nuclear spins is needed.  The results so far indicate that
water spin temperatures measured in comets and disks may tell
astronomers less about the water formation location than previously
thought (see also discussion in \citealt{Tielens13}).

\section{Concluding remarks}

In this review, we have illustrated the major facets of the
interdisciplinary science of astrochemistry by consideration of one
relatively small but important molecule: water.  A critical molecule
for life as we know it, water is found to a greater or lesser extent
in almost all environments studied in some detail by astrochemists, be
it in the gas or condensed phase.  Since it is made of two of the most
abundant elements, H and O, its formation is to some degree
inevitable, yet its chemistry is found to be varied, depending on
physical conditions.  Indeed, the water vapor abundance is found to be
high, about $10^{-4}$ with respect to H$_2$ in warm sources such as
dense shocks, oxygen rich circumstellar shells and inner disks. Its
abundance drops to $\sim 10^{-7}$ in diffuse clouds due to
photodissociation and is even lower, $\sim 10^{-9}$, in cold cores and
the outer regions of protoplanetary disks due to freeze-out. In
contrast, water ice is a major reservoir of oxygen in these latter
regions, at an abundance of $10^{-4}$.

The three main sets of chemical processes thought to form and destroy
water can be tested against observations these different sources.  In
some types of sources, agreement between models and observations is
good, at the level of a factor of 2--3, but in other regions there is
a difference of a few orders of magnitude. In the latter cases, the
large disagreement can point to missing chemical reactions in the
models but it can also be caused by observational difficulties in
determining the abundances.

The discussion of the three main sets of chemical processes -- low
temperature ion-molecule and grain-surface reactions, and
high-temperature neutral-neutral gas-phase reactions -- and their use
in chemical simulations should give the reader an understanding of the
role of these chemical processes.  Such simulations can be used to
confirm or identify not only the driving chemical processes but also
the physical conditions, homogeneous or heterogeneous, of the sources
where water is observed, the amount of UV radiation or cosmic rays,
and even tell us something of the history and/or fate of these
sources.  In this sense, chemistry helps astronomy, via spectroscopy
and chemical kinetics, and via theory and experiment, and it enhances
greatly the scientific return from the major investments made in
ground- and space-based telescopes.

Enormous progress has clearly been made in the last decades in
quantifying molecular processes involving water but there continues to
be further astronomical need for basic data.  Examples include
spectroscopy of water vapor and its isotopologues at high temperatures
for use in exoplanet and disk atmosphere models; collisional rate
coefficients for H$_2$O vibration-rotation transitions and for H$_2$O
with H for use in inner disk and dissociative shock models; an
independent check of the branching ratio for dissociative
recombination of H$_3$O$^+$ to water for benchmarking diffuse cloud
chemistry; state-resolved rate coefficients for high temperature
neutral-neutral reactions in shocks and in the surface layers of
disks; and determination of the OH vibration-rotation populations as
function of wavelength following H$_2$O photodissociation for
applications in comets, disks and shocks. In terms of water ice
formation, the main uncertainties are illustrated in
Figure~\ref{fig:icenetwork} and more generally further work is required
in the translation of laboratory data to quantities to be used in
gas-grain models. There is clearly a major need for wavelength
dependent photodesorption rates of H$_2$O and OH from water ice to
analyze water vapor observations of cold cores and the outer regions
of disks.

To audiences of chemists, it must be pointed out that astronomy has
and will continue to aid chemistry in making chemists aware that some
of the standard chemical curriculum is terrestrially biased, and that
`exotic' molecules (e.g. gaseous cations, anions, three-membered
rings, etc.) and `exotic' chemical processes (e.g., dissociative
recombination leading to water, radiative association) not normally
studied in the chemical laboratory are of great importance in the
universe as a whole. Indeed, the use of the word `exotic' may be more
applicable to our home planet than to the interstellar medium!
Similarly, there has been a surprising lack of chemical physics
studies of water ice formation until astronomers developed the field
of solid-state astrochemistry using surface science techniques.  Much
of the terrestrial chemistry had focused on thin, monolayer
experiments, often on metallic surfaces rather than on thick ices. Our
increase in knowledge of several of these fundamental chemical
processes has at least partially come about because chemists learned
of their likely importance to astronomy, and water is an excellent
case.  More examples are likely to follow.

Regarding the role of water in the new science of astrobiology,
astrochemistry has taught us much of relevance.  Astrobiology is the
study of the origin, evolution, distribution and future of life in the
universe. Water is commonly implicated as a prerequisite for
carbon-based life as found on Earth. We know that water is found in
many diverse sources, which certainly does not help to constrain the
location where life could form, but it does indicate that the
conditions for life are widespread throughout the universe. We also
know that in the colder regions that lie along the evolutionary trail
of star and planetary formation (Fig.~\ref{fig:cycle}), water is
mainly in its solid form, and is likely to be incorporated in this
form into the larger solid objects eventually formed from interstellar
grains, such as comets and asteroids \citep{vanDishoeck14}.  Indeed,
the terrestrial oceans have likely have come from cometary and
asteroidal bombardment, and solar system structures such as the Oort
cloud and Kuiper Belt may well be common.  Equally related to
astrobiology, the existence of gaseous water in exo-planets has
recently been observed (Fig.~\ref{fig:waterplanet}), and the first
generation of chemical simulations of this chemistry are already being
undertaken.  Eventually, we will have a much better understanding of
the trail of water from stellar atmospheres to diffuse clouds to dense
clouds, protostars, protoplanetary disks, and new generations of stars
and planets, all the way out to the edge of the universe.  By then,
the science of astrochemistry will have progressed from its essential
beginnings in the 1970's to a full-fledged science capable of helping
astronomers understand the universe to a degree not recognizable
today.

\acknowledgement

The authors thank many colleagues for fruitful discussions on water
chemistry and observations. We are particularly grateful to all the
physicists and chemists that have provided key molecular data to
analyze and interpret observations of water in space.  Magnus Persson
and Lars Kristensen provided support with figures, whereas Adwin
Boogert, Qiang Chang, Thanja Lamberts, Gary Melnick and Colette Salyk
kindly adapted their figures for this review. David Hollenbach, Alain
Baudry, Javier Goicoechea, Lars Kristensen, Geoff Mathews, Joe
Mottram, Amiel Sternberg, Ruud Visser, and three anonymous referees
provided detailed comments which have improved this paper.  EvD is
supported by the Netherlands Organization for Scientific Research
(NWO), the Netherlands Research School for Astronomy (NOVA), by a
Royal Netherlands Academy of Arts and Sciences (KNAW) professor prize,
and by a European Union A-ERC grant.  E.H. acknowledges the National
Science Foundation for its support of his program in astrochemistry
and NASA for support through its Exobiology and Evolutionary Biology
program and its program in Herschel science.

\bibliography{biblio_evd}

\providecommand*\mcitethebibliography{\thebibliography}
\csname @ifundefined\endcsname{endmcitethebibliography}
  {\let\endmcitethebibliography\endthebibliography}{}
\begin{mcitethebibliography}{444}
\providecommand*\natexlab[1]{#1}
\providecommand*\mciteSetBstSublistMode[1]{}
\providecommand*\mciteSetBstMaxWidthForm[2]{}
\providecommand*\mciteBstWouldAddEndPuncttrue
  {\def\EndOfBibitem{\unskip.}}
\providecommand*\mciteBstWouldAddEndPunctfalse
  {\let\EndOfBibitem\relax}
\providecommand*\mciteSetBstMidEndSepPunct[3]{}
\providecommand*\mciteSetBstSublistLabelBeginEnd[3]{}
\providecommand*\EndOfBibitem{}
\mciteSetBstSublistMode{f}
\mciteSetBstMaxWidthForm{subitem}{(\alph{mcitesubitemcount})}
\mciteSetBstSublistLabelBeginEnd
  {\mcitemaxwidthsubitemform\space}
  {\relax}
  {\relax}

\bibitem[{Steigman}(2007)]{Steigman07}
{Steigman},~G. \emph{Annu. Rev. Nucl. Part. Sci.} \textbf{2007}, \emph{57},
  463--491\relax
\mciteBstWouldAddEndPuncttrue
\mciteSetBstMidEndSepPunct{\mcitedefaultmidpunct}
{\mcitedefaultendpunct}{\mcitedefaultseppunct}\relax
\EndOfBibitem
\bibitem[{Hinshaw} et~al.(2013){Hinshaw}, {Larson}, {Komatsu}, {Spergel},
  {Bennett}, {Dunkley}, {Nolta}, {Halpern}, {Hill}, {Odegard}, {Page}, {Smith},
  {Weiland}, {Gold}, {Jarosik}, {Kogut}, {Limon}, {Meyer}, {Tucker}, {Wollack},
  and {Wright}]{WMAP13}
{Hinshaw},~G. et~al.  \emph{\apjs} \textbf{2013}, \emph{57}, 463--491\relax
\mciteBstWouldAddEndPuncttrue
\mciteSetBstMidEndSepPunct{\mcitedefaultmidpunct}
{\mcitedefaultendpunct}{\mcitedefaultseppunct}\relax
\EndOfBibitem
\bibitem[{Planck Collaboration} et~al.(2013){Planck Collaboration}, {Ade},
  {Aghanim}, {Armitage-Caplan}, {Arnaud}, {Ashdown}, {Atrio-Barandela},
  {Aumont}, {Baccigalupi}, {Banday}, and et~al.]{Planck13}
{Planck Collaboration},; {Ade},~P.~A.~R.; {Aghanim},~N.; {Armitage-Caplan},~C.;
  {Arnaud},~M.; {Ashdown},~M.; {Atrio-Barandela},~F.; {Aumont},~J.;
  {Baccigalupi},~C.; {Banday},~A.~J.; et~al., \emph{ArXiv e-prints 1303.5062}
  \textbf{2013}, \relax
\mciteBstWouldAddEndPunctfalse
\mciteSetBstMidEndSepPunct{\mcitedefaultmidpunct}
{}{\mcitedefaultseppunct}\relax
\EndOfBibitem
\bibitem[{Solomon} and {Klemperer}(1972){Solomon}, and {Klemperer}]{Solomon72}
{Solomon},~P.~M.; {Klemperer},~W. \emph{\apj} \textbf{1972}, \emph{178},
  389--422\relax
\mciteBstWouldAddEndPuncttrue
\mciteSetBstMidEndSepPunct{\mcitedefaultmidpunct}
{\mcitedefaultendpunct}{\mcitedefaultseppunct}\relax
\EndOfBibitem
\bibitem[{Herbst} and {Klemperer}(1973){Herbst}, and {Klemperer}]{Herbst73}
{Herbst},~E.; {Klemperer},~W. \emph{\apj} \textbf{1973}, \emph{185},
  505--534\relax
\mciteBstWouldAddEndPuncttrue
\mciteSetBstMidEndSepPunct{\mcitedefaultmidpunct}
{\mcitedefaultendpunct}{\mcitedefaultseppunct}\relax
\EndOfBibitem
\bibitem[{van de Hulst}(1996)]{VandeHulst96}
{van de Hulst},~H.~C. In \emph{Molecules in Astrophysics: Probes \& Processes};
  {van Dishoeck},~E.~F., Ed.; IAU Symposium 178, Kluwer: Dordrecht; 1996; Vol.
  178; p~13\relax
\mciteBstWouldAddEndPuncttrue
\mciteSetBstMidEndSepPunct{\mcitedefaultmidpunct}
{\mcitedefaultendpunct}{\mcitedefaultseppunct}\relax
\EndOfBibitem
\bibitem[{Allen} and {Robinson}(1977){Allen}, and {Robinson}]{Allen77}
{Allen},~M.; {Robinson},~G.~W. \emph{\apj} \textbf{1977}, \emph{212},
  396--415\relax
\mciteBstWouldAddEndPuncttrue
\mciteSetBstMidEndSepPunct{\mcitedefaultmidpunct}
{\mcitedefaultendpunct}{\mcitedefaultseppunct}\relax
\EndOfBibitem
\bibitem[{Tielens} and {Hagen}(1982){Tielens}, and {Hagen}]{Tielens82}
{Tielens},~A.~G.~G.~M.; {Hagen},~W. \emph{\aap} \textbf{1982}, \emph{114},
  245--260\relax
\mciteBstWouldAddEndPuncttrue
\mciteSetBstMidEndSepPunct{\mcitedefaultmidpunct}
{\mcitedefaultendpunct}{\mcitedefaultseppunct}\relax
\EndOfBibitem
\bibitem[{Cernicharo} and {Crovisier}(2005){Cernicharo}, and
  {Crovisier}]{Cernicharo05}
{Cernicharo},~J.; {Crovisier},~J. \emph{\ssr} \textbf{2005}, \emph{119},
  29--69\relax
\mciteBstWouldAddEndPuncttrue
\mciteSetBstMidEndSepPunct{\mcitedefaultmidpunct}
{\mcitedefaultendpunct}{\mcitedefaultseppunct}\relax
\EndOfBibitem
\bibitem[{Melnick}(2009)]{Melnick09}
{Melnick},~G.~J. In \emph{Submillimeter Astrophysics and Technology};
  {D.~C.~Lis, J.~E.~Vaillancourt, P.~F.~Goldsmith, T.~A.~Bell, N.~Z.~Scoville,
  \& J.~Zmuidzinas},, Ed.; Astron. Soc. Pac. Conf. Series: San Francisco; 2009;
  Vol. 417; p~59\relax
\mciteBstWouldAddEndPuncttrue
\mciteSetBstMidEndSepPunct{\mcitedefaultmidpunct}
{\mcitedefaultendpunct}{\mcitedefaultseppunct}\relax
\EndOfBibitem
\bibitem[{Hollenbach} et~al.(2009){Hollenbach}, {Kaufman}, {Bergin}, and
  {Melnick}]{Hollenbach09}
{Hollenbach},~D.; {Kaufman},~M.~J.; {Bergin},~E.~A.; {Melnick},~G.~J.
  \emph{\apj} \textbf{2009}, \emph{690}, 1497--1521\relax
\mciteBstWouldAddEndPuncttrue
\mciteSetBstMidEndSepPunct{\mcitedefaultmidpunct}
{\mcitedefaultendpunct}{\mcitedefaultseppunct}\relax
\EndOfBibitem
\bibitem[{Bergin} and {van Dishoeck}(2012){Bergin}, and {van
  Dishoeck}]{Bergin12}
{Bergin},~E.~A.; {van Dishoeck},~E.~F. \emph{\pta} \textbf{2012}, \emph{370},
  2778--2802\relax
\mciteBstWouldAddEndPuncttrue
\mciteSetBstMidEndSepPunct{\mcitedefaultmidpunct}
{\mcitedefaultendpunct}{\mcitedefaultseppunct}\relax
\EndOfBibitem
\bibitem[{Cheung} et~al.(1969){Cheung}, {Rank}, and {Townes}]{Cheung69}
{Cheung},~A.~C.; {Rank},~D.~M.; {Townes},~C.~H. \emph{\nat} \textbf{1969},
  \emph{221}, 626--628\relax
\mciteBstWouldAddEndPuncttrue
\mciteSetBstMidEndSepPunct{\mcitedefaultmidpunct}
{\mcitedefaultendpunct}{\mcitedefaultseppunct}\relax
\EndOfBibitem
\bibitem[{Miyoshi} et~al.(1995){Miyoshi}, {Moran}, {Herrnstein}, {Greenhill},
  {Nakai}, {Diamond}, and {Inoue}]{Miyoshi95}
{Miyoshi},~M.; {Moran},~J.; {Herrnstein},~J.; {Greenhill},~L.; {Nakai},~N.;
  {Diamond},~P.; {Inoue},~M. \emph{\nat} \textbf{1995}, \emph{373},
  127--129\relax
\mciteBstWouldAddEndPuncttrue
\mciteSetBstMidEndSepPunct{\mcitedefaultmidpunct}
{\mcitedefaultendpunct}{\mcitedefaultseppunct}\relax
\EndOfBibitem
\bibitem[{Birkby} et~al.(2013){Birkby}, {de Kok}, {Brogi}, {de Mooij},
  {Schwarz}, {Albrecht}, and {Snellen}]{Birkby13}
{Birkby},~J.~L.; {de Kok},~R.~J.; {Brogi},~M.; {de Mooij},~E.~J.~W.;
  {Schwarz},~H.; {Albrecht},~S.; {Snellen},~I.~A.~G. \emph{\mnras}
  \textbf{2013}, in press\relax
\mciteBstWouldAddEndPuncttrue
\mciteSetBstMidEndSepPunct{\mcitedefaultmidpunct}
{\mcitedefaultendpunct}{\mcitedefaultseppunct}\relax
\EndOfBibitem
\bibitem[{Wei{\ss}} et~al.(2013){Wei{\ss}}, {De Breuck}, {Marrone}, {Vieira},
  {Aguirre}, {Aird}, {Aravena}, {Ashby}, {Bayliss}, {Benson}, {B{\'e}thermin},
  {Biggs}, {Bleem}, {Bock}, {Bothwell}, {Bradford}, {Brodwin}, {Carlstrom},
  {Chang}, {Chapman}, {Crawford}, {Crites}, {de Haan}, {Dobbs}, {Downes},
  {Fassnacht}, {George}, {Gladders}, {Gonzalez}, {Greve}, {Halverson},
  {Hezaveh}, {High}, {Holder}, {Holzapfel}, {Hoover}, {Hrubes}, {Husband},
  {Keisler}, {Lee}, {Leitch}, {Lueker}, {Luong-Van}, {Malkan}, {McIntyre},
  {McMahon}, {Mehl}, {Menten}, {Meyer}, {Murphy}, {Padin}, {Plagge},
  {Reichardt}, {Rest}, {Rosenman}, {Ruel}, {Ruhl}, {Schaffer}, {Shirokoff},
  {Spilker}, {Stalder}, {Staniszewski}, {Stark}, {Story}, {Vanderlinde},
  {Welikala}, and {Williamson}]{Weiss13}
{Wei{\ss}},~A. et~al.  \emph{\apj} \textbf{2013}, \emph{767}, 88\relax
\mciteBstWouldAddEndPuncttrue
\mciteSetBstMidEndSepPunct{\mcitedefaultmidpunct}
{\mcitedefaultendpunct}{\mcitedefaultseppunct}\relax
\EndOfBibitem
\bibitem[{Seager} and {Deming}(2010){Seager}, and {Deming}]{Seager10}
{Seager},~S.; {Deming},~D. \emph{\araa} \textbf{2010}, \emph{48},
  631--672\relax
\mciteBstWouldAddEndPuncttrue
\mciteSetBstMidEndSepPunct{\mcitedefaultmidpunct}
{\mcitedefaultendpunct}{\mcitedefaultseppunct}\relax
\EndOfBibitem
\bibitem[{Madhusudhan} and {Seager}(2009){Madhusudhan}, and
  {Seager}]{Madhusudhan09}
{Madhusudhan},~N.; {Seager},~S. \emph{\apj} \textbf{2009}, \emph{707},
  24--39\relax
\mciteBstWouldAddEndPuncttrue
\mciteSetBstMidEndSepPunct{\mcitedefaultmidpunct}
{\mcitedefaultendpunct}{\mcitedefaultseppunct}\relax
\EndOfBibitem
\bibitem[{Fraser} et~al.(2001){Fraser}, {Collings}, {McCoustra}, and
  {Williams}]{Fraser01}
{Fraser},~H.~J.; {Collings},~M.~P.; {McCoustra},~M.~R.~S.; {Williams},~D.~A.
  \emph{\mnras} \textbf{2001}, \emph{327}, 1165--1172\relax
\mciteBstWouldAddEndPuncttrue
\mciteSetBstMidEndSepPunct{\mcitedefaultmidpunct}
{\mcitedefaultendpunct}{\mcitedefaultseppunct}\relax
\EndOfBibitem
\bibitem[{Gillett} and {Forrest}(1973){Gillett}, and {Forrest}]{Gillett73}
{Gillett},~F.~C.; {Forrest},~W.~J. \emph{\apj} \textbf{1973}, \emph{179},
  483--491\relax
\mciteBstWouldAddEndPuncttrue
\mciteSetBstMidEndSepPunct{\mcitedefaultmidpunct}
{\mcitedefaultendpunct}{\mcitedefaultseppunct}\relax
\EndOfBibitem
\bibitem[{Whittet}(2003)]{Whittet03}
{Whittet},~D.~C.~B. \emph{Dust in the galactic environment, 2nd ed.};
  {Institute of Physics Publishing, Bristol}, 2003\relax
\mciteBstWouldAddEndPuncttrue
\mciteSetBstMidEndSepPunct{\mcitedefaultmidpunct}
{\mcitedefaultendpunct}{\mcitedefaultseppunct}\relax
\EndOfBibitem
\bibitem[{Caselli} and {Ceccarelli}(2012){Caselli}, and
  {Ceccarelli}]{Caselli12aar}
{Caselli},~P.; {Ceccarelli},~C. \emph{\aapr} \textbf{2012}, \emph{20}, 56\relax
\mciteBstWouldAddEndPuncttrue
\mciteSetBstMidEndSepPunct{\mcitedefaultmidpunct}
{\mcitedefaultendpunct}{\mcitedefaultseppunct}\relax
\EndOfBibitem
\bibitem[{Ceccarelli} et~al.(2013){Ceccarelli}, {Caselli}, {Bockel\'ee-Morvan},
  {Mousis}, {Pizzarello}, {Robert}, and {Semenov}]{Ceccarelli14}
{Ceccarelli},~C.; {Caselli},~P.; {Bockel\'ee-Morvan},~D.; {Mousis},~O.;
  {Pizzarello},~F.; {Robert},~F.; {Semenov},~D. In \emph{Protostars \& Planets
  VI}; {Beuther, H., Klessen, R., Dullemond, K., Henning, Th.},, Ed.; Univ.
  Arizona Press: Tucson, 2013; p submitted\relax
\mciteBstWouldAddEndPuncttrue
\mciteSetBstMidEndSepPunct{\mcitedefaultmidpunct}
{\mcitedefaultendpunct}{\mcitedefaultseppunct}\relax
\EndOfBibitem
\bibitem[{de Graauw} et~al.(1996){de Graauw}, {Haser}, {Beintema}, {Roelfsema},
  {van Agthoven}, {Barl}, {Bauer}, {Bekenkamp}, {Boonstra}, {Boxhoorn}, {Cote},
  {de Groene}, {van Dijkhuizen}, {Drapatz}, {Evers}, {Feuchtgruber},
  {Frericks}, {Genzel}, {Haerendel}, {Heras}, {van der Hucht}, {van der Hulst},
  {Huygen}, {Jacobs}, {Jakob}, {Kamperman}, {Katterloher}, {Kester}, {Kunze},
  {Kussendrager}, {Lahuis}, {Lamers}, {Leech}, {van der Lei}, {van der Linden},
  {Luinge}, {Lutz}, {Melzner}, {Morris}, {van Nguyen}, {Ploeger}, {Price},
  {Salama}, {Schaeidt}, {Sijm}, {Smoorenburg}, {Spakman}, {Spoon},
  {Steinmayer}, {Stoecker}, {Valentijn}, {Vandenbussche}, {Visser}, {Waelkens},
  {Waters}, {Wensink}, {Wesselius}, {Wiezorrek}, {Wieprecht}, {Wijnbergen},
  {Wildeman}, and {Young}]{deGraauw96}
{de Graauw},~T. et~al.  \emph{\aap} \textbf{1996}, \emph{315}, L49--L54\relax
\mciteBstWouldAddEndPuncttrue
\mciteSetBstMidEndSepPunct{\mcitedefaultmidpunct}
{\mcitedefaultendpunct}{\mcitedefaultseppunct}\relax
\EndOfBibitem
\bibitem[{Clegg} et~al.(1996){Clegg}, {Ade}, {Armand}, {Baluteau}, {Barlow},
  {Buckley}, {Berges}, {Burgdorf}, {Caux}, {Ceccarelli}, {Cerulli}, {Church},
  {Cotin}, {Cox}, {Cruvellier}, {Culhane}, {Davis}, {di Giorgio}, {Diplock},
  {Drummond}, {Emery}, {Ewart}, {Fischer}, {Furniss}, {Glencross},
  {Greenhouse}, {Griffin}, {Gry}, {Harwood}, {Hazell}, {Joubert}, {King},
  {Lim}, {Liseau}, {Long}, {Lorenzetti}, {Molinari}, {Murray}, {Naylor},
  {Nisini}, {Norman}, {Omont}, {Orfei}, {Patrick}, {Pequignot}, {Pouliquen},
  {Price}, {Nguyen-Q-Rieu}, {Rogers}, {Robinson}, {Saisse}, {Saraceno},
  {Serra}, {Sidher}, {Smith}, {Smith}, {Spinoglio}, {Swinyard}, {Texier},
  {Towlson}, {Trams}, {Unger}, and {White}]{Clegg96}
{Clegg},~P.~E. et~al.  \emph{\aap} \textbf{1996}, \emph{315}, L38--L42\relax
\mciteBstWouldAddEndPuncttrue
\mciteSetBstMidEndSepPunct{\mcitedefaultmidpunct}
{\mcitedefaultendpunct}{\mcitedefaultseppunct}\relax
\EndOfBibitem
\bibitem[{van Dishoeck}(2004)]{vanDishoeck04}
{van Dishoeck},~E.~F. \emph{\araa} \textbf{2004}, \emph{42}, 119--167\relax
\mciteBstWouldAddEndPuncttrue
\mciteSetBstMidEndSepPunct{\mcitedefaultmidpunct}
{\mcitedefaultendpunct}{\mcitedefaultseppunct}\relax
\EndOfBibitem
\bibitem[{Melnick} et~al.(2000){Melnick}, {Stauffer}, {Ashby}, {Bergin},
  {Chin}, {Erickson}, {Goldsmith}, {Harwit}, {Howe}, {Kleiner}, {Koch},
  {Neufeld}, {Patten}, {Plume}, {Schieder}, {Snell}, {Tolls}, {Wang},
  {Winnewisser}, and {Zhang}]{Melnick00}
{Melnick},~G.~J. et~al.  \emph{\apjl} \textbf{2000}, \emph{539}, L77--L85\relax
\mciteBstWouldAddEndPuncttrue
\mciteSetBstMidEndSepPunct{\mcitedefaultmidpunct}
{\mcitedefaultendpunct}{\mcitedefaultseppunct}\relax
\EndOfBibitem
\bibitem[{Nordh} et~al.(2003){Nordh}, {von Sch{\'e}ele}, {Frisk}, {Ahola},
  {Booth}, {Encrenaz}, {Hjalmarson}, {Kendall}, {Kyr{\"o}l{\"a}}, {Kwok},
  {Lecacheux}, {Leppelmeier}, {Llewellyn}, {Mattila}, {M{\'e}gie}, {Murtagh},
  {Rougeron}, and {Witt}]{Nordh03}
{Nordh},~H.~L. et~al.  \emph{\aap} \textbf{2003}, \emph{402}, L21--L25\relax
\mciteBstWouldAddEndPuncttrue
\mciteSetBstMidEndSepPunct{\mcitedefaultmidpunct}
{\mcitedefaultendpunct}{\mcitedefaultseppunct}\relax
\EndOfBibitem
\bibitem[{Pilbratt} et~al.(2010){Pilbratt}, {Riedinger}, {Passvogel}, {Crone},
  {Doyle}, {Gageur}, {Heras}, {Jewell}, {Metcalfe}, {Ott}, and
  {Schmidt}]{Pilbratt10}
{Pilbratt},~G.~L.; {Riedinger},~J.~R.; {Passvogel},~T.; {Crone},~G.;
  {Doyle},~D.; {Gageur},~U.; {Heras},~A.~M.; {Jewell},~C.; {Metcalfe},~L.;
  {Ott},~S.; {Schmidt},~M. \emph{\aap} \textbf{2010}, \emph{518}, L1\relax
\mciteBstWouldAddEndPuncttrue
\mciteSetBstMidEndSepPunct{\mcitedefaultmidpunct}
{\mcitedefaultendpunct}{\mcitedefaultseppunct}\relax
\EndOfBibitem
\bibitem[{de Graauw} et~al.(2010){de Graauw}, {Helmich}, {Phillips}, {Stutzki},
  {Caux}, {Whyborn}, {Dieleman}, {Roelfsema}, {Aarts}, {Assendorp},
  {Bachiller}, {Baechtold}, {Barcia}, {Beintema}, {Belitsky}, {Benz}, {Bieber},
  {Boogert}, {Borys}, {Bumble}, {Ca{\"i}s}, {Caris}, {Cerulli-Irelli},
  {Chattopadhyay}, {Cherednichenko}, {Ciechanowicz}, {Coeur-Joly}, {Comito},
  {Cros}, {de Jonge}, {de Lange}, {Delforges}, {Delorme}, {den Boggende},
  {Desbat}, {Diez-Gonz{\'a}lez}, {di Giorgio}, {Dubbeldam}, {Edwards},
  {Eggens}, {Erickson}, {Evers}, {Fich}, {Finn}, {Franke}, {Gaier}, {Gal},
  {Gao}, {Gallego}, {Gauffre}, {Gill}, {Glenz}, {Golstein}, {Goulooze},
  {Gunsing}, {G{\"u}sten}, {Hartogh}, {Hatch}, {Higgins}, {Honingh}, {Huisman},
  {Jackson}, {Jacobs}, {Jacobs}, {Jarchow}, {Javadi}, {Jellema}, {Justen},
  {Karpov}, {Kasemann}, {Kawamura}, {Keizer}, {Kester}, {Klapwijk}, {Klein},
  {Kollberg}, {Kooi}, {Kooiman}, {Kopf}, {Krause}, {Krieg}, {Kramer},
  {Kruizenga}, {Kuhn}, {Laauwen}, {Lai}, {Larsson}, {Leduc}, {Leinz}, {Lin},
  {Liseau}, {Liu}, {Loose}, {L{\'o}pez-Fernandez}, {Lord}, {Luinge}, {Marston},
  {Mart{\'{\i}}n-Pintado}, {Maestrini}, {Maiwald}, {McCoey}, {Mehdi}, {Megej},
  {Melchior}, {Meinsma}, {Merkel}, {Michalska}, {Monstein}, {Moratschke},
  {Morris}, {Muller}, {Murphy}, {Naber}, {Natale}, {Nowosielski}, {Nuzzolo},
  {Olberg}, {Olbrich}, {Orfei}, {Orleanski}, {Ossenkopf}, {Peacock}, {Pearson},
  {Peron}, {Phillip-May}, {Piazzo}, {Planesas}, {Rataj}, {Ravera}, {Risacher},
  {Salez}, {Samoska}, {Saraceno}, {Schieder}, {Schlecht}, {Schl{\"o}der},
  {Schm{\"u}lling}, {Schultz}, {Schuster}, {Siebertz}, {Smit}, {Szczerba},
  {Shipman}, {Steinmetz}, {Stern}, {Stokroos}, {Teipen}, {Teyssier}, {Tils},
  {Trappe}, {van Baaren}, {van Leeuwen}, {van de Stadt}, {Visser}, {Wildeman},
  {Wafelbakker}, {Ward}, {Wesselius}, {Wild}, {Wulff}, {Wunsch}, {Tielens},
  {Zaal}, {Zirath}, {Zmuidzinas}, and {Zwart}]{deGraauw10}
{de Graauw},~T. et~al.  \emph{\aap} \textbf{2010}, \emph{518}, L6\relax
\mciteBstWouldAddEndPuncttrue
\mciteSetBstMidEndSepPunct{\mcitedefaultmidpunct}
{\mcitedefaultendpunct}{\mcitedefaultseppunct}\relax
\EndOfBibitem
\bibitem[{Poglitsch} et~al.(2010){Poglitsch}, {Waelkens}, {Geis},
  {Feuchtgruber}, {Vandenbussche}, {Rodriguez}, {Krause}, {Renotte}, {van
  Hoof}, {Saraceno}, {Cepa}, {Kerschbaum}, {Agn{\`e}se}, {Ali}, {Altieri},
  {Andreani}, {Augueres}, {Balog}, {Barl}, {Bauer}, {Belbachir}, {Benedettini},
  {Billot}, {Boulade}, {Bischof}, {Blommaert}, {Callut}, {Cara}, {Cerulli},
  {Cesarsky}, {Contursi}, {Creten}, {De Meester}, {Doublier}, {Doumayrou},
  {Duband}, {Exter}, {Genzel}, {Gillis}, {Gr{\"o}zinger}, {Henning},
  {Herreros}, {Huygen}, {Inguscio}, {Jakob}, {Jamar}, {Jean}, {de Jong},
  {Katterloher}, {Kiss}, {Klaas}, {Lemke}, {Lutz}, {Madden}, {Marquet},
  {Martignac}, {Mazy}, {Merken}, {Montfort}, {Morbidelli}, {M{\"u}ller},
  {Nielbock}, {Okumura}, {Orfei}, {Ottensamer}, {Pezzuto}, {Popesso},
  {Putzeys}, {Regibo}, {Reveret}, {Royer}, {Sauvage}, {Schreiber}, {Stegmaier},
  {Schmitt}, {Schubert}, {Sturm}, {Thiel}, {Tofani}, {Vavrek}, {Wetzstein},
  {Wieprecht}, and {Wiezorrek}]{Poglitsch10}
{Poglitsch},~A. et~al.  \emph{\aap} \textbf{2010}, \emph{518}, L2\relax
\mciteBstWouldAddEndPuncttrue
\mciteSetBstMidEndSepPunct{\mcitedefaultmidpunct}
{\mcitedefaultendpunct}{\mcitedefaultseppunct}\relax
\EndOfBibitem
\bibitem[{Cernicharo} et~al.(1990){Cernicharo}, {Thum}, {Hein}, {John},
  {Garcia}, and {Mattioco}]{Cernicharo90}
{Cernicharo},~J.; {Thum},~C.; {Hein},~H.; {John},~D.; {Garcia},~P.;
  {Mattioco},~F. \emph{\aap} \textbf{1990}, \emph{231}, L15--L18\relax
\mciteBstWouldAddEndPuncttrue
\mciteSetBstMidEndSepPunct{\mcitedefaultmidpunct}
{\mcitedefaultendpunct}{\mcitedefaultseppunct}\relax
\EndOfBibitem
\bibitem[{J{\o}rgensen} and {van Dishoeck}(2010){J{\o}rgensen}, and {van
  Dishoeck}]{Jorgensen10}
{J{\o}rgensen},~J.~K.; {van Dishoeck},~E.~F. \emph{\apjl} \textbf{2010},
  \emph{710}, L72--L76\relax
\mciteBstWouldAddEndPuncttrue
\mciteSetBstMidEndSepPunct{\mcitedefaultmidpunct}
{\mcitedefaultendpunct}{\mcitedefaultseppunct}\relax
\EndOfBibitem
\bibitem[{Houck} et~al.(2004){Houck}, {Roellig}, {van Cleve}, {Forrest},
  {Herter}, {Lawrence}, {Matthews}, {Reitsema}, {Soifer}, {Watson}, {Weedman},
  {Huisjen}, {Troeltzsch}, {Barry}, {Bernard-Salas}, {Blacken}, {Brandl},
  {Charmandaris}, {Devost}, {Gull}, {Hall}, {Henderson}, {Higdon}, {Pirger},
  {Schoenwald}, {Sloan}, {Uchida}, {Appleton}, {Armus}, {Burgdorf},
  {Fajardo-Acosta}, {Grillmair}, {Ingalls}, {Morris}, and {Teplitz}]{Houck04}
{Houck},~J.~R. et~al.  \emph{\apjs} \textbf{2004}, \emph{154}, 18--24\relax
\mciteBstWouldAddEndPuncttrue
\mciteSetBstMidEndSepPunct{\mcitedefaultmidpunct}
{\mcitedefaultendpunct}{\mcitedefaultseppunct}\relax
\EndOfBibitem
\bibitem[{Young} et~al.(2012){Young}, {Becklin}, {Marcum}, {Roellig}, {De
  Buizer}, {Herter}, {G{\"u}sten}, {Dunham}, {Temi}, {Andersson}, {Backman},
  {Burgdorf}, {Caroff}, {Casey}, {Davidson}, {Erickson}, {Gehrz}, {Harper},
  {Harvey}, {Helton}, {Horner}, {Howard}, {Klein}, {Krabbe}, {McLean}, {Meyer},
  {Miles}, {Morris}, {Reach}, {Rho}, {Richter}, {Roeser}, {Sandell}, {Sankrit},
  {Savage}, {Smith}, {Shuping}, {Vacca}, {Vaillancourt}, {Wolf}, and
  {Zinnecker}]{Sofia12}
{Young},~E.~T. et~al.  \emph{\apjl} \textbf{2012}, \emph{749}, L17\relax
\mciteBstWouldAddEndPuncttrue
\mciteSetBstMidEndSepPunct{\mcitedefaultmidpunct}
{\mcitedefaultendpunct}{\mcitedefaultseppunct}\relax
\EndOfBibitem
\bibitem[{Spaans} et~al.(1998){Spaans}, {Neufeld}, {Lepp}, {Melnick}, and
  {Stauffer}]{Spaans98}
{Spaans},~M.; {Neufeld},~D.; {Lepp},~S.; {Melnick},~G.~J.; {Stauffer},~J.
  \emph{\apj} \textbf{1998}, \emph{503}, 780\relax
\mciteBstWouldAddEndPuncttrue
\mciteSetBstMidEndSepPunct{\mcitedefaultmidpunct}
{\mcitedefaultendpunct}{\mcitedefaultseppunct}\relax
\EndOfBibitem
\bibitem[{Tielens}(2005)]{Tielens05}
{Tielens},~A.~G.~G.~M. \emph{The Physics and Chemistry of the Interstellar
  Medium}; Cambridge University Press, Cambridge, 2005\relax
\mciteBstWouldAddEndPuncttrue
\mciteSetBstMidEndSepPunct{\mcitedefaultmidpunct}
{\mcitedefaultendpunct}{\mcitedefaultseppunct}\relax
\EndOfBibitem
\bibitem[{Tielens}(2013)]{Tielens13}
{Tielens},~A.~G.~G.~M. \emph{\rmp} \textbf{2013}, \emph{85}, 1021--1081\relax
\mciteBstWouldAddEndPuncttrue
\mciteSetBstMidEndSepPunct{\mcitedefaultmidpunct}
{\mcitedefaultendpunct}{\mcitedefaultseppunct}\relax
\EndOfBibitem
\bibitem[{Tielens} and {Hollenbach}(1985){Tielens}, and
  {Hollenbach}]{Tielens85}
{Tielens},~A.~G.~G.~M.; {Hollenbach},~D. \emph{\apj} \textbf{1985}, \emph{291},
  722--754\relax
\mciteBstWouldAddEndPuncttrue
\mciteSetBstMidEndSepPunct{\mcitedefaultmidpunct}
{\mcitedefaultendpunct}{\mcitedefaultseppunct}\relax
\EndOfBibitem
\bibitem[{van Dishoeck} et~al.(2011){van Dishoeck}, {Kristensen}, {Benz},
  {Bergin}, {Caselli}, {Cernicharo}, {Herpin}, {Hogerheijde}, {Johnstone},
  {Liseau}, {Nisini}, {Shipman}, {Tafalla}, {van der Tak}, {Wyrowski},
  {Aikawa}, {Bachiller}, {Baudry}, {Benedettini}, {Bjerkeli}, {Blake},
  {Bontemps}, {Braine}, {Brinch}, {Bruderer}, {Chavarr{\'{\i}}a}, {Codella},
  {Daniel}, {de Graauw}, {Deul}, {di Giorgio}, {Dominik}, {Doty}, {Dubernet},
  {Encrenaz}, {Feuchtgruber}, {Fich}, {Frieswijk}, {Fuente}, {Giannini},
  {Goicoechea}, {Helmich}, {Herczeg}, {Jacq}, {J{\o}rgensen}, {Karska},
  {Kaufman}, {Keto}, {Larsson}, {Lefloch}, {Lis}, {Marseille}, {McCoey},
  {Melnick}, {Neufeld}, {Olberg}, {Pagani}, {Pani{\'c}}, {Parise}, {Pearson},
  {Plume}, {Risacher}, {Salter}, {Santiago-Garc{\'{\i}}a}, {Saraceno},
  {St{\"a}uber}, {van Kempen}, {Visser}, {Viti}, {Walmsley}, {Wampfler}, and
  {Y{\i}ld{\i}z}]{vanDishoeck11}
{van Dishoeck},~E.~F. et~al.  \emph{\pasp} \textbf{2011}, \emph{123},
  138--170\relax
\mciteBstWouldAddEndPuncttrue
\mciteSetBstMidEndSepPunct{\mcitedefaultmidpunct}
{\mcitedefaultendpunct}{\mcitedefaultseppunct}\relax
\EndOfBibitem
\bibitem[{Mottram} et~al.(2011){Mottram}, {Hoare}, {Davies}, {Lumsden},
  {Oudmaijer}, {Urquhart}, {Moore}, {Cooper}, and {Stead}]{Mottram11}
{Mottram},~J.~C.; {Hoare},~M.~G.; {Davies},~B.; {Lumsden},~S.~L.;
  {Oudmaijer},~R.~D.; {Urquhart},~J.~S.; {Moore},~T.~J.~T.; {Cooper},~H.~D.~B.;
  {Stead},~J.~J. \emph{\apjl} \textbf{2011}, \emph{730}, L33\relax
\mciteBstWouldAddEndPuncttrue
\mciteSetBstMidEndSepPunct{\mcitedefaultmidpunct}
{\mcitedefaultendpunct}{\mcitedefaultseppunct}\relax
\EndOfBibitem
\bibitem[{Habing}(1968)]{Habing68}
{Habing},~H.~J. \emph{\bain} \textbf{1968}, \emph{19}, 421\relax
\mciteBstWouldAddEndPuncttrue
\mciteSetBstMidEndSepPunct{\mcitedefaultmidpunct}
{\mcitedefaultendpunct}{\mcitedefaultseppunct}\relax
\EndOfBibitem
\bibitem[{Draine}(1978)]{Draine78}
{Draine},~B.~T. \emph{\apjs} \textbf{1978}, \emph{36}, 595--619\relax
\mciteBstWouldAddEndPuncttrue
\mciteSetBstMidEndSepPunct{\mcitedefaultmidpunct}
{\mcitedefaultendpunct}{\mcitedefaultseppunct}\relax
\EndOfBibitem
\bibitem[{Bergin} et~al.(2003){Bergin}, {Kaufman}, {Melnick}, {Snell}, and
  {Howe}]{Bergin03}
{Bergin},~E.~A.; {Kaufman},~M.~J.; {Melnick},~G.~J.; {Snell},~R.~L.;
  {Howe},~J.~E. \emph{\apj} \textbf{2003}, \emph{582}, 830--845\relax
\mciteBstWouldAddEndPuncttrue
\mciteSetBstMidEndSepPunct{\mcitedefaultmidpunct}
{\mcitedefaultendpunct}{\mcitedefaultseppunct}\relax
\EndOfBibitem
\bibitem[{van Dishoeck} et~al.(2006){van Dishoeck}, {Jonkheid}, and {van
  Hemert}]{vanDishoeck06photo}
{van Dishoeck},~E.~F.; {Jonkheid},~B.; {van Hemert},~M.~C. \emph{Faraday
  Discussions} \textbf{2006}, \emph{133}, 231\relax
\mciteBstWouldAddEndPuncttrue
\mciteSetBstMidEndSepPunct{\mcitedefaultmidpunct}
{\mcitedefaultendpunct}{\mcitedefaultseppunct}\relax
\EndOfBibitem
\bibitem[{van Dishoeck} and {Black}(1982){van Dishoeck}, and
  {Black}]{vanDishoeck82}
{van Dishoeck},~E.~F.; {Black},~J.~H. \emph{\apj} \textbf{1982}, \emph{258},
  533--547\relax
\mciteBstWouldAddEndPuncttrue
\mciteSetBstMidEndSepPunct{\mcitedefaultmidpunct}
{\mcitedefaultendpunct}{\mcitedefaultseppunct}\relax
\EndOfBibitem
\bibitem[{Hauschildt} et~al.(1999){Hauschildt}, {Allard}, {Ferguson}, {Baron},
  and {Alexander}]{Hauschildt99}
{Hauschildt},~P.~H.; {Allard},~F.; {Ferguson},~J.; {Baron},~E.;
  {Alexander},~D.~R. \emph{\apj} \textbf{1999}, \emph{525}, 871--880\relax
\mciteBstWouldAddEndPuncttrue
\mciteSetBstMidEndSepPunct{\mcitedefaultmidpunct}
{\mcitedefaultendpunct}{\mcitedefaultseppunct}\relax
\EndOfBibitem
\bibitem[{Roberge} et~al.(1991){Roberge}, {Jones}, {Lepp}, and
  {Dalgarno}]{Roberge91}
{Roberge},~W.~G.; {Jones},~D.; {Lepp},~S.; {Dalgarno},~A. \emph{\apjs}
  \textbf{1991}, \emph{77}, 287--297\relax
\mciteBstWouldAddEndPuncttrue
\mciteSetBstMidEndSepPunct{\mcitedefaultmidpunct}
{\mcitedefaultendpunct}{\mcitedefaultseppunct}\relax
\EndOfBibitem
\bibitem[{Bohlin} et~al.(1978){Bohlin}, {Savage}, and {Drake}]{Bohlin78}
{Bohlin},~R.~C.; {Savage},~B.~D.; {Drake},~J.~F. \emph{\apj} \textbf{1978},
  \emph{224}, 132--142\relax
\mciteBstWouldAddEndPuncttrue
\mciteSetBstMidEndSepPunct{\mcitedefaultmidpunct}
{\mcitedefaultendpunct}{\mcitedefaultseppunct}\relax
\EndOfBibitem
\bibitem[{Rachford} et~al.(2009){Rachford}, {Snow}, {Destree}, {Ross},
  {Ferlet}, {Friedman}, {Gry}, {Jenkins}, {Morton}, {Savage}, {Shull},
  {Sonnentrucker}, {Tumlinson}, {Vidal-Madjar}, {Welty}, and
  {York}]{Rachford09}
{Rachford},~B.~L. et~al.  \emph{\apjs} \textbf{2009}, \emph{180},
  125--137\relax
\mciteBstWouldAddEndPuncttrue
\mciteSetBstMidEndSepPunct{\mcitedefaultmidpunct}
{\mcitedefaultendpunct}{\mcitedefaultseppunct}\relax
\EndOfBibitem
\bibitem[{Prasad} and {Tarafdar}(1983){Prasad}, and {Tarafdar}]{Prasad83}
{Prasad},~S.~S.; {Tarafdar},~S.~P. \emph{\apj} \textbf{1983}, \emph{267},
  603--609\relax
\mciteBstWouldAddEndPuncttrue
\mciteSetBstMidEndSepPunct{\mcitedefaultmidpunct}
{\mcitedefaultendpunct}{\mcitedefaultseppunct}\relax
\EndOfBibitem
\bibitem[{Gredel} et~al.(1989){Gredel}, {Lepp}, {Dalgarno}, and
  {Herbst}]{Gredel89}
{Gredel},~R.; {Lepp},~S.; {Dalgarno},~A.; {Herbst},~E. \emph{\apj}
  \textbf{1989}, \emph{347}, 289--293\relax
\mciteBstWouldAddEndPuncttrue
\mciteSetBstMidEndSepPunct{\mcitedefaultmidpunct}
{\mcitedefaultendpunct}{\mcitedefaultseppunct}\relax
\EndOfBibitem
\bibitem[{Shen} et~al.(2004){Shen}, {Greenberg}, {Schutte}, and {van
  Dishoeck}]{Shen04}
{Shen},~C.~J.; {Greenberg},~J.~M.; {Schutte},~W.~A.; {van Dishoeck},~E.~F.
  \emph{\aap} \textbf{2004}, \emph{415}, 203--215\relax
\mciteBstWouldAddEndPuncttrue
\mciteSetBstMidEndSepPunct{\mcitedefaultmidpunct}
{\mcitedefaultendpunct}{\mcitedefaultseppunct}\relax
\EndOfBibitem
\bibitem[{Bruderer} et~al.(2009){Bruderer}, {Doty}, and {Benz}]{Bruderer09a}
{Bruderer},~S.; {Doty},~S.~D.; {Benz},~A.~O. \emph{\apjs} \textbf{2009},
  \emph{183}, 179--196\relax
\mciteBstWouldAddEndPuncttrue
\mciteSetBstMidEndSepPunct{\mcitedefaultmidpunct}
{\mcitedefaultendpunct}{\mcitedefaultseppunct}\relax
\EndOfBibitem
\bibitem[{Pickett} et~al.(1998){Pickett}, {Poynter}, {Cohen}, {Delitsky},
  {Pearson}, and {Muller}]{Pickett98}
{Pickett},~H.~M.; {Poynter},~I.~R.~L.; {Cohen},~E.~A.; {Delitsky},~M.~L.;
  {Pearson},~J.~C.; {Muller},~H.~S.~P. \emph{\jqsrt} \textbf{1998}, \emph{60},
  883--890\relax
\mciteBstWouldAddEndPuncttrue
\mciteSetBstMidEndSepPunct{\mcitedefaultmidpunct}
{\mcitedefaultendpunct}{\mcitedefaultseppunct}\relax
\EndOfBibitem
\bibitem[{M{\"u}ller} et~al.(2001){M{\"u}ller}, {Thorwirth}, {Roth}, and
  {Winnewisser}]{Muller01}
{M{\"u}ller},~H.~S.~P.; {Thorwirth},~S.; {Roth},~D.~A.; {Winnewisser},~G.
  \emph{\aap} \textbf{2001}, \emph{370}, L49--L52\relax
\mciteBstWouldAddEndPuncttrue
\mciteSetBstMidEndSepPunct{\mcitedefaultmidpunct}
{\mcitedefaultendpunct}{\mcitedefaultseppunct}\relax
\EndOfBibitem
\bibitem[{M{\"u}ller} et~al.(2005){M{\"u}ller}, {Schl{\"o}der}, {Stutzki}, and
  {Winnewisser}]{Muller05}
{M{\"u}ller},~H.~S.~P.; {Schl{\"o}der},~F.; {Stutzki},~J.; {Winnewisser},~G.
  \emph{\jmss} \textbf{2005}, \emph{742}, 215--227\relax
\mciteBstWouldAddEndPuncttrue
\mciteSetBstMidEndSepPunct{\mcitedefaultmidpunct}
{\mcitedefaultendpunct}{\mcitedefaultseppunct}\relax
\EndOfBibitem
\bibitem[De~Lucia et~al.({1972})De~Lucia, Gordy, Helminger, and
  Cook]{Delucia72}
De~Lucia,~F.; Gordy,~W.; Helminger,~P.; Cook,~R. \emph{{\pra}} \textbf{{1972}},
  \emph{{5}}, {487}\relax
\mciteBstWouldAddEndPuncttrue
\mciteSetBstMidEndSepPunct{\mcitedefaultmidpunct}
{\mcitedefaultendpunct}{\mcitedefaultseppunct}\relax
\EndOfBibitem
\bibitem[Polyansky et~al.({1997})Polyansky, Zobov, Viti, Tennyson, Bernath, and
  Wallace]{Polyansky97}
Polyansky,~O.; Zobov,~N.; Viti,~S.; Tennyson,~J.; Bernath,~P.; Wallace,~L.
  \emph{{\jms}} \textbf{{1997}}, \emph{{186}}, {422--447}\relax
\mciteBstWouldAddEndPuncttrue
\mciteSetBstMidEndSepPunct{\mcitedefaultmidpunct}
{\mcitedefaultendpunct}{\mcitedefaultseppunct}\relax
\EndOfBibitem
\bibitem[Toth({1999})]{Toth99}
Toth,~R. \emph{{\jms}} \textbf{{1999}}, \emph{{194}}, {28--42}\relax
\mciteBstWouldAddEndPuncttrue
\mciteSetBstMidEndSepPunct{\mcitedefaultmidpunct}
{\mcitedefaultendpunct}{\mcitedefaultseppunct}\relax
\EndOfBibitem
\bibitem[Yu et~al.({2012})Yu, Pearson, Drouin, Martin-Drumel, Pirali, Vervloet,
  Coudert, Mueller, and Bruenken]{Yu12}
Yu,~S.; Pearson,~J.; Drouin,~B.; Martin-Drumel,~M.-A.; Pirali,~O.;
  Vervloet,~M.; Coudert,~L.; Mueller,~H.; Bruenken,~S. \emph{{\jms}}
  \textbf{{2012}}, \emph{{279}}, {16--25}\relax
\mciteBstWouldAddEndPuncttrue
\mciteSetBstMidEndSepPunct{\mcitedefaultmidpunct}
{\mcitedefaultendpunct}{\mcitedefaultseppunct}\relax
\EndOfBibitem
\bibitem[Coudert et~al.({2008})Coudert, Wagner, Birk, Baranov, Lafferty, and
  Flaud]{Coudert08}
Coudert,~L.~H.; Wagner,~G.; Birk,~M.; Baranov,~Y.~I.; Lafferty,~W.~J.;
  Flaud,~J.~M. \emph{{\jms}} \textbf{{2008}}, \emph{{251}}, {339--357}\relax
\mciteBstWouldAddEndPuncttrue
\mciteSetBstMidEndSepPunct{\mcitedefaultmidpunct}
{\mcitedefaultendpunct}{\mcitedefaultseppunct}\relax
\EndOfBibitem
\bibitem[De~Lucia and Helminger({1975})De~Lucia, and Helminger]{Delucia75}
De~Lucia,~F.; Helminger,~P. \emph{{\jms}} \textbf{{1975}}, \emph{{56}},
  {138--145}\relax
\mciteBstWouldAddEndPuncttrue
\mciteSetBstMidEndSepPunct{\mcitedefaultmidpunct}
{\mcitedefaultendpunct}{\mcitedefaultseppunct}\relax
\EndOfBibitem
\bibitem[Steenbeck and Bellet({1971})Steenbeck, and Bellet]{Steenbeck71}
Steenbeck,~G.; Bellet,~J. \emph{{\crb}} \textbf{{1971}}, \emph{{273}},
  {471}\relax
\mciteBstWouldAddEndPuncttrue
\mciteSetBstMidEndSepPunct{\mcitedefaultmidpunct}
{\mcitedefaultendpunct}{\mcitedefaultseppunct}\relax
\EndOfBibitem
\bibitem[Johns({1985})]{Johns85}
Johns,~J. \emph{{\josab}} \textbf{{1985}}, \emph{{2}}, {1340--1354}\relax
\mciteBstWouldAddEndPuncttrue
\mciteSetBstMidEndSepPunct{\mcitedefaultmidpunct}
{\mcitedefaultendpunct}{\mcitedefaultseppunct}\relax
\EndOfBibitem
\bibitem[Guelachvili({1983})]{Guelachvili83}
Guelachvili,~G. \emph{{\josa}} \textbf{{1983}}, \emph{{73}}, {137--150}\relax
\mciteBstWouldAddEndPuncttrue
\mciteSetBstMidEndSepPunct{\mcitedefaultmidpunct}
{\mcitedefaultendpunct}{\mcitedefaultseppunct}\relax
\EndOfBibitem
\bibitem[{Melnick} et~al.(2010){Melnick}, {Tolls}, {Neufeld}, {Bergin},
  {Phillips}, {Wang}, {Crockett}, {Bell}, {Blake}, {Cabrit}, {Caux},
  {Ceccarelli}, {Cernicharo}, {Comito}, {Daniel}, {Dubernet}, {Emprechtinger},
  {Encrenaz}, {Falgarone}, {Gerin}, {Giesen}, {Goicoechea}, {Goldsmith},
  {Herbst}, {Joblin}, {Johnstone}, {Langer}, {Latter}, {Lis}, {Lord}, {Maret},
  {Martin}, {Menten}, {Morris}, {M{\"u}ller}, {Murphy}, {Ossenkopf}, {Pagani},
  {Pearson}, {P{\'e}rault}, {Plume}, {Qin}, {Salez}, {Schilke}, {Schlemmer},
  {Stutzki}, {Trappe}, {van der Tak}, {Vastel}, {Yorke}, {Yu}, and
  {Zmuidzinas}]{Melnick10}
{Melnick},~G.~J. et~al.  \emph{\aap} \textbf{2010}, \emph{521}, L27\relax
\mciteBstWouldAddEndPuncttrue
\mciteSetBstMidEndSepPunct{\mcitedefaultmidpunct}
{\mcitedefaultendpunct}{\mcitedefaultseppunct}\relax
\EndOfBibitem
\bibitem[{Herczeg} et~al.(2012){Herczeg}, {Karska}, {Bruderer}, {Kristensen},
  {van Dishoeck}, {J{\o}rgensen}, {Visser}, {Wampfler}, {Bergin},
  {Y{\i}ld{\i}z}, {Pontoppidan}, and {Gracia-Carpio}]{Herczeg12}
{Herczeg},~G.~J.; {Karska},~A.; {Bruderer},~S.; {Kristensen},~L.~E.; {van
  Dishoeck},~E.~F.; {J{\o}rgensen},~J.~K.; {Visser},~R.; {Wampfler},~S.~F.;
  {Bergin},~E.~A.; {Y{\i}ld{\i}z},~U.~A.; {Pontoppidan},~K.~M.;
  {Gracia-Carpio},~J. \emph{\aap} \textbf{2012}, \emph{540}, A84\relax
\mciteBstWouldAddEndPuncttrue
\mciteSetBstMidEndSepPunct{\mcitedefaultmidpunct}
{\mcitedefaultendpunct}{\mcitedefaultseppunct}\relax
\EndOfBibitem
\bibitem[{Coutens} et~al.(2012){Coutens}, {Vastel}, {Caux}, {Ceccarelli},
  {Bottinelli}, {Wiesenfeld}, {Faure}, {Scribano}, and {Kahane}]{Coutens12}
{Coutens},~A.; {Vastel},~C.; {Caux},~E.; {Ceccarelli},~C.; {Bottinelli},~S.;
  {Wiesenfeld},~L.; {Faure},~A.; {Scribano},~Y.; {Kahane},~C. \emph{\aap}
  \textbf{2012}, \emph{539}, A132\relax
\mciteBstWouldAddEndPuncttrue
\mciteSetBstMidEndSepPunct{\mcitedefaultmidpunct}
{\mcitedefaultendpunct}{\mcitedefaultseppunct}\relax
\EndOfBibitem
\bibitem[{Neill} et~al.(2013){Neill}, {Wang}, {Bergin}, {Crockett}, {Favre},
  {Plume}, and {Melnick}]{Neill13}
{Neill},~J.~L.; {Wang},~S.; {Bergin},~E.~A.; {Crockett},~N.~R.; {Favre},~C.;
  {Plume},~R.; {Melnick},~G.~J. \emph{\apj} \textbf{2013}, \emph{770},
  142\relax
\mciteBstWouldAddEndPuncttrue
\mciteSetBstMidEndSepPunct{\mcitedefaultmidpunct}
{\mcitedefaultendpunct}{\mcitedefaultseppunct}\relax
\EndOfBibitem
\bibitem[{Watson} et~al.(2007){Watson}, {Bohac}, {Hull}, {Forrest}, {Furlan},
  {Najita}, {Calvet}, {D'Alessio}, {Hartmann}, {Sargent}, {Green}, {Kim}, and
  {Houck}]{Watson07}
{Watson},~D.~M.; {Bohac},~C.~J.; {Hull},~C.; {Forrest},~W.~J.; {Furlan},~E.;
  {Najita},~J.; {Calvet},~N.; {D'Alessio},~P.; {Hartmann},~L.; {Sargent},~B.;
  {Green},~J.~D.; {Kim},~K.~H.; {Houck},~J.~R. \emph{\nat} \textbf{2007},
  \emph{448}, 1026--1028\relax
\mciteBstWouldAddEndPuncttrue
\mciteSetBstMidEndSepPunct{\mcitedefaultmidpunct}
{\mcitedefaultendpunct}{\mcitedefaultseppunct}\relax
\EndOfBibitem
\bibitem[{Carr} and {Najita}(2008){Carr}, and {Najita}]{Carr08}
{Carr},~J.~S.; {Najita},~J.~R. \emph{Science} \textbf{2008}, \emph{319},
  1504\relax
\mciteBstWouldAddEndPuncttrue
\mciteSetBstMidEndSepPunct{\mcitedefaultmidpunct}
{\mcitedefaultendpunct}{\mcitedefaultseppunct}\relax
\EndOfBibitem
\bibitem[{Salyk} et~al.(2008){Salyk}, {Pontoppidan}, {Blake}, {Lahuis}, {van
  Dishoeck}, and {Evans}]{Salyk08}
{Salyk},~C.; {Pontoppidan},~K.~M.; {Blake},~G.~A.; {Lahuis},~F.; {van
  Dishoeck},~E.~F.; {Evans},~N.~J.,~II \emph{\apjl} \textbf{2008}, \emph{676},
  L49--L52\relax
\mciteBstWouldAddEndPuncttrue
\mciteSetBstMidEndSepPunct{\mcitedefaultmidpunct}
{\mcitedefaultendpunct}{\mcitedefaultseppunct}\relax
\EndOfBibitem
\bibitem[{Pontoppidan} et~al.(2010){Pontoppidan}, {Salyk}, {Blake},
  {Meijerink}, {Carr}, and {Najita}]{Pontoppidan10}
{Pontoppidan},~K.~M.; {Salyk},~C.; {Blake},~G.~A.; {Meijerink},~R.;
  {Carr},~J.~S.; {Najita},~J. \emph{\apj} \textbf{2010}, \emph{720},
  887--903\relax
\mciteBstWouldAddEndPuncttrue
\mciteSetBstMidEndSepPunct{\mcitedefaultmidpunct}
{\mcitedefaultendpunct}{\mcitedefaultseppunct}\relax
\EndOfBibitem
\bibitem[Flaud and Camypeyret({1975})Flaud, and Camypeyret]{Flaud75}
Flaud,~J.; Camypeyret,~C. \emph{{\jms}} \textbf{{1975}}, \emph{{55}},
  {278--310}\relax
\mciteBstWouldAddEndPuncttrue
\mciteSetBstMidEndSepPunct{\mcitedefaultmidpunct}
{\mcitedefaultendpunct}{\mcitedefaultseppunct}\relax
\EndOfBibitem
\bibitem[Camypeyret and Flaud({1976})Camypeyret, and Flaud]{Camy76}
Camypeyret,~C.; Flaud,~J. \emph{{\molp}} \textbf{{1976}}, \emph{{32}},
  {523--537}\relax
\mciteBstWouldAddEndPuncttrue
\mciteSetBstMidEndSepPunct{\mcitedefaultmidpunct}
{\mcitedefaultendpunct}{\mcitedefaultseppunct}\relax
\EndOfBibitem
\bibitem[Rothman et~al.({2009})Rothman, Gordon, Barbe, Benner, Bernath, Birk,
  Boudon, Brown, Campargue, Champion, Chance, Coudert, Dana, Devi, Fally,
  Flaud, Gamache, Goldman, Jacquemart, Kleiner, Lacome, Lafferty, Mandin,
  Massie, Mikhailenko, Miller, Moazzen-Ahmadi, Naumenko, Nikitin, Orphal,
  Perevalov, Perrin, Predoi-Cross, Rinsland, Rotger, Simeckova, Smith, Sung,
  Tashkun, Tennyson, Toth, Vandaele, and Vander~Auwera]{Hitran09}
Rothman,~L.~S. et~al.  \emph{{\jqsrt}} \textbf{{2009}}, \emph{{110}},
  {533--572}\relax
\mciteBstWouldAddEndPuncttrue
\mciteSetBstMidEndSepPunct{\mcitedefaultmidpunct}
{\mcitedefaultendpunct}{\mcitedefaultseppunct}\relax
\EndOfBibitem
\bibitem[Barber et~al.({2006})Barber, Tennyson, Harris, and
  Tolchenov]{Barber06}
Barber,~R.; Tennyson,~J.; Harris,~G.; Tolchenov,~R. \emph{{\mnras}}
  \textbf{{2006}}, \emph{{368}}, {1087--1094}\relax
\mciteBstWouldAddEndPuncttrue
\mciteSetBstMidEndSepPunct{\mcitedefaultmidpunct}
{\mcitedefaultendpunct}{\mcitedefaultseppunct}\relax
\EndOfBibitem
\bibitem[Shirin et~al.({2006})Shirin, Polyansky, Zobov, Ovsyannikov, Csaszar,
  and Tennyson]{Shirin06}
Shirin,~S.; Polyansky,~O.; Zobov,~N.; Ovsyannikov,~R.; Csaszar,~A.;
  Tennyson,~J. \emph{{\jms}} \textbf{{2006}}, \emph{{236}}, {216--223}\relax
\mciteBstWouldAddEndPuncttrue
\mciteSetBstMidEndSepPunct{\mcitedefaultmidpunct}
{\mcitedefaultendpunct}{\mcitedefaultseppunct}\relax
\EndOfBibitem
\bibitem[Voronin et~al.({2010})Voronin, Tennyson, Tolchenov, Lugovskoy, and
  Yurchenko]{Voronin10}
Voronin,~B.~A.; Tennyson,~J.; Tolchenov,~R.~N.; Lugovskoy,~A.~A.;
  Yurchenko,~S.~N. \emph{{\mnras}} \textbf{{2010}}, \emph{{402}},
  {492--496}\relax
\mciteBstWouldAddEndPuncttrue
\mciteSetBstMidEndSepPunct{\mcitedefaultmidpunct}
{\mcitedefaultendpunct}{\mcitedefaultseppunct}\relax
\EndOfBibitem
\bibitem[Bruenken et~al.({2007})Bruenken, Mueller, Endres, Lewen, Giesen,
  Drouin, Pearson, and Maeder]{Brunken07}
Bruenken,~S.; Mueller,~H. S.~P.; Endres,~C.; Lewen,~F.; Giesen,~T.; Drouin,~B.;
  Pearson,~J.~C.; Maeder,~H. \emph{{\pccp}} \textbf{{2007}}, \emph{{9}},
  {2103--2112}\relax
\mciteBstWouldAddEndPuncttrue
\mciteSetBstMidEndSepPunct{\mcitedefaultmidpunct}
{\mcitedefaultendpunct}{\mcitedefaultseppunct}\relax
\EndOfBibitem
\bibitem[Tennyson and Shine({2012})Tennyson, and Shine]{Tennyson12}
Tennyson,~J.; Shine,~K.~P. \emph{{\pta}} \textbf{{2012}}, \emph{{370}},
  {2491--2494}\relax
\mciteBstWouldAddEndPuncttrue
\mciteSetBstMidEndSepPunct{\mcitedefaultmidpunct}
{\mcitedefaultendpunct}{\mcitedefaultseppunct}\relax
\EndOfBibitem
\bibitem[{Helmich} et~al.(1996){Helmich}, {van Dishoeck}, {Black}, {de Graauw},
  {Beintema}, {Heras}, {Lahuis}, {Morris}, and {Valentijn}]{Helmich96}
{Helmich},~F.~P.; {van Dishoeck},~E.~F.; {Black},~J.~H.; {de Graauw},~T.;
  {Beintema},~D.~A.; {Heras},~A.~M.; {Lahuis},~F.; {Morris},~P.~W.;
  {Valentijn},~E.~A. \emph{\aap} \textbf{1996}, \emph{315}, L173--L176\relax
\mciteBstWouldAddEndPuncttrue
\mciteSetBstMidEndSepPunct{\mcitedefaultmidpunct}
{\mcitedefaultendpunct}{\mcitedefaultseppunct}\relax
\EndOfBibitem
\bibitem[{D'Hendecourt} et~al.(1999){D'Hendecourt}, {Joblin}, and
  {Jones}]{DHendecourt99}
{D'Hendecourt},~L., {Joblin},~C., {Jones},~A., Eds. \emph{Solid Interstellar
  Matter: The ISO Revolution}; Springer, Berlin, 1999\relax
\mciteBstWouldAddEndPuncttrue
\mciteSetBstMidEndSepPunct{\mcitedefaultmidpunct}
{\mcitedefaultendpunct}{\mcitedefaultseppunct}\relax
\EndOfBibitem
\bibitem[{Gudipati} and {Castillo-Rogez}(2013){Gudipati}, and
  {Castillo-Rogez}]{Gudipati13}
{Gudipati},~M., {Castillo-Rogez},~J., Eds. \emph{The science of solar system
  ices}; Springer, New York, 2013\relax
\mciteBstWouldAddEndPuncttrue
\mciteSetBstMidEndSepPunct{\mcitedefaultmidpunct}
{\mcitedefaultendpunct}{\mcitedefaultseppunct}\relax
\EndOfBibitem
\bibitem[{Tielens}(1983)]{Tielens83}
{Tielens},~A.~G.~G.~M. \emph{\aap} \textbf{1983}, \emph{119}, 177--184\relax
\mciteBstWouldAddEndPuncttrue
\mciteSetBstMidEndSepPunct{\mcitedefaultmidpunct}
{\mcitedefaultendpunct}{\mcitedefaultseppunct}\relax
\EndOfBibitem
\bibitem[{Smith} et~al.(1989){Smith}, {Sellgren}, and {Tokunaga}]{Smith89}
{Smith},~R.~G.; {Sellgren},~K.; {Tokunaga},~A.~T. \emph{\apj} \textbf{1989},
  \emph{344}, 413--426\relax
\mciteBstWouldAddEndPuncttrue
\mciteSetBstMidEndSepPunct{\mcitedefaultmidpunct}
{\mcitedefaultendpunct}{\mcitedefaultseppunct}\relax
\EndOfBibitem
\bibitem[{Hudgins} et~al.(1993){Hudgins}, {Sandford}, {Allamandola}, and
  {Tielens}]{Hudgins93}
{Hudgins},~D.~M.; {Sandford},~S.~A.; {Allamandola},~L.~J.;
  {Tielens},~A.~G.~G.~M. \emph{\apjs} \textbf{1993}, \emph{86}, 713--870\relax
\mciteBstWouldAddEndPuncttrue
\mciteSetBstMidEndSepPunct{\mcitedefaultmidpunct}
{\mcitedefaultendpunct}{\mcitedefaultseppunct}\relax
\EndOfBibitem
\bibitem[Jenniskens et~al.({1995})Jenniskens, Blake, Wilson, and
  Pohorille]{Jenniskens95}
Jenniskens,~P.; Blake,~D.; Wilson,~M.; Pohorille,~A. \emph{{\apj}}
  \textbf{{1995}}, \emph{{455}}, {389--401}\relax
\mciteBstWouldAddEndPuncttrue
\mciteSetBstMidEndSepPunct{\mcitedefaultmidpunct}
{\mcitedefaultendpunct}{\mcitedefaultseppunct}\relax
\EndOfBibitem
\bibitem[Stevenson et~al.({1999})Stevenson, Kimmel, Dohnalek, Smith, and
  Kay]{Stevenson99}
Stevenson,~K.; Kimmel,~G.; Dohnalek,~Z.; Smith,~R.; Kay,~B. \emph{{\science}}
  \textbf{{1999}}, \emph{{283}}, {1505--1507}\relax
\mciteBstWouldAddEndPuncttrue
\mciteSetBstMidEndSepPunct{\mcitedefaultmidpunct}
{\mcitedefaultendpunct}{\mcitedefaultseppunct}\relax
\EndOfBibitem
\bibitem[Kimmel et~al.({2001})Kimmel, Stevenson, Dohnalek, Smith, and
  Kay]{Kimmel01}
Kimmel,~G.; Stevenson,~K.; Dohnalek,~Z.; Smith,~R.; Kay,~B. \emph{{\jcp}}
  \textbf{{2001}}, \emph{{114}}, {5284--5294}\relax
\mciteBstWouldAddEndPuncttrue
\mciteSetBstMidEndSepPunct{\mcitedefaultmidpunct}
{\mcitedefaultendpunct}{\mcitedefaultseppunct}\relax
\EndOfBibitem
\bibitem[{Schutte}(2002)]{Schutte02}
{Schutte},~W.~A. \emph{\aap} \textbf{2002}, \emph{386}, 1103--1105\relax
\mciteBstWouldAddEndPuncttrue
\mciteSetBstMidEndSepPunct{\mcitedefaultmidpunct}
{\mcitedefaultendpunct}{\mcitedefaultseppunct}\relax
\EndOfBibitem
\bibitem[Jenniskens and Blake({1994})Jenniskens, and Blake]{Jenniskens94}
Jenniskens,~P.; Blake,~D. \emph{{\science}} \textbf{{1994}}, \emph{{265}},
  {753--756}\relax
\mciteBstWouldAddEndPuncttrue
\mciteSetBstMidEndSepPunct{\mcitedefaultmidpunct}
{\mcitedefaultendpunct}{\mcitedefaultseppunct}\relax
\EndOfBibitem
\bibitem[{Ehrenfreund} et~al.(1996){Ehrenfreund}, {Boogert}, {Gerakines},
  {Jansen}, {Schutte}, {Tielens}, and {van Dishoeck}]{Ehrenfreund96}
{Ehrenfreund},~P.; {Boogert},~A.~C.~A.; {Gerakines},~P.~A.; {Jansen},~D.~J.;
  {Schutte},~W.~A.; {Tielens},~A.~G.~G.~M.; {van Dishoeck},~E.~F. \emph{\aap}
  \textbf{1996}, \emph{315}, L341--L344\relax
\mciteBstWouldAddEndPuncttrue
\mciteSetBstMidEndSepPunct{\mcitedefaultmidpunct}
{\mcitedefaultendpunct}{\mcitedefaultseppunct}\relax
\EndOfBibitem
\bibitem[{Bouwman} et~al.(2007){Bouwman}, {Ludwig}, {Awad}, {{\"O}berg},
  {Fuchs}, {van Dishoeck}, and {Linnartz}]{Bouwman07}
{Bouwman},~J.; {Ludwig},~W.; {Awad},~Z.; {{\"O}berg},~K.~I.; {Fuchs},~G.~W.;
  {van Dishoeck},~E.~F.; {Linnartz},~H. \emph{\aap} \textbf{2007}, \emph{476},
  995--1003\relax
\mciteBstWouldAddEndPuncttrue
\mciteSetBstMidEndSepPunct{\mcitedefaultmidpunct}
{\mcitedefaultendpunct}{\mcitedefaultseppunct}\relax
\EndOfBibitem
\bibitem[{Moore} and {Hudson}(1992){Moore}, and {Hudson}]{Moore92}
{Moore},~M.~H.; {Hudson},~R.~L. \emph{\apj} \textbf{1992}, \emph{401},
  353--360\relax
\mciteBstWouldAddEndPuncttrue
\mciteSetBstMidEndSepPunct{\mcitedefaultmidpunct}
{\mcitedefaultendpunct}{\mcitedefaultseppunct}\relax
\EndOfBibitem
\bibitem[{Moore} and {Hudson}(1994){Moore}, and {Hudson}]{Moore94}
{Moore},~M.~H.; {Hudson},~R.~L. \emph{\aaps} \textbf{1994}, \emph{103},
  45--56\relax
\mciteBstWouldAddEndPuncttrue
\mciteSetBstMidEndSepPunct{\mcitedefaultmidpunct}
{\mcitedefaultendpunct}{\mcitedefaultseppunct}\relax
\EndOfBibitem
\bibitem[{Boogert} et~al.(2000){Boogert}, {Tielens}, {Ceccarelli}, {Boonman},
  {van Dishoeck}, {Keane}, {Whittet}, and {de Graauw}]{Boogert00}
{Boogert},~A.~C.~A.; {Tielens},~A.~G.~G.~M.; {Ceccarelli},~C.;
  {Boonman},~A.~M.~S.; {van Dishoeck},~E.~F.; {Keane},~J.~V.;
  {Whittet},~D.~C.~B.; {de Graauw},~T. \emph{\aap} \textbf{2000}, \emph{360},
  683--698\relax
\mciteBstWouldAddEndPuncttrue
\mciteSetBstMidEndSepPunct{\mcitedefaultmidpunct}
{\mcitedefaultendpunct}{\mcitedefaultseppunct}\relax
\EndOfBibitem
\bibitem[{Gibb} et~al.(2004){Gibb}, {Whittet}, {Boogert}, and
  {Tielens}]{Gibb04}
{Gibb},~E.~L.; {Whittet},~D.~C.~B.; {Boogert},~A.~C.~A.; {Tielens},~A.~G.~G.~M.
  \emph{\apjs} \textbf{2004}, \emph{151}, 35--73\relax
\mciteBstWouldAddEndPuncttrue
\mciteSetBstMidEndSepPunct{\mcitedefaultmidpunct}
{\mcitedefaultendpunct}{\mcitedefaultseppunct}\relax
\EndOfBibitem
\bibitem[{Boogert} et~al.(2008){Boogert}, {Pontoppidan}, {Knez}, {Lahuis},
  {Kessler-Silacci}, {van Dishoeck}, {Blake}, {Augereau}, {Bisschop},
  {Bottinelli}, {Brooke}, {Brown}, {Crapsi}, {Evans}, {Fraser}, {Geers},
  {Huard}, {J{\o}rgensen}, {{\"O}berg}, {Allen}, {Harvey}, {Koerner}, {Mundy},
  {Padgett}, {Sargent}, and {Stapelfeldt}]{Boogert08}
{Boogert},~A.~C.~A. et~al.  \emph{\apj} \textbf{2008}, \emph{678},
  985--1004\relax
\mciteBstWouldAddEndPuncttrue
\mciteSetBstMidEndSepPunct{\mcitedefaultmidpunct}
{\mcitedefaultendpunct}{\mcitedefaultseppunct}\relax
\EndOfBibitem
\bibitem[{Whittet} et~al.(2007){Whittet}, {Shenoy}, {Bergin}, {Chiar},
  {Gerakines}, {Gibb}, {Melnick}, and {Neufeld}]{Whittet07}
{Whittet},~D.~C.~B.; {Shenoy},~S.~S.; {Bergin},~E.~A.; {Chiar},~J.~E.;
  {Gerakines},~P.~A.; {Gibb},~E.~L.; {Melnick},~G.~J.; {Neufeld},~D.~A.
  \emph{\apj} \textbf{2007}, \emph{655}, 332--341\relax
\mciteBstWouldAddEndPuncttrue
\mciteSetBstMidEndSepPunct{\mcitedefaultmidpunct}
{\mcitedefaultendpunct}{\mcitedefaultseppunct}\relax
\EndOfBibitem
\bibitem[{Boogert} et~al.(2011){Boogert}, {Huard}, {Cook}, {Chiar}, {Knez},
  {Decin}, {Blake}, {Tielens}, and {van Dishoeck}]{Boogert11}
{Boogert},~A.~C.~A.; {Huard},~T.~L.; {Cook},~A.~M.; {Chiar},~J.~E.; {Knez},~C.;
  {Decin},~L.; {Blake},~G.~A.; {Tielens},~A.~G.~G.~M.; {van Dishoeck},~E.~F.
  \emph{\apj} \textbf{2011}, \emph{729}, 92\relax
\mciteBstWouldAddEndPuncttrue
\mciteSetBstMidEndSepPunct{\mcitedefaultmidpunct}
{\mcitedefaultendpunct}{\mcitedefaultseppunct}\relax
\EndOfBibitem
\bibitem[{Soifer} et~al.(1981){Soifer}, {Willner}, {Rudy}, and
  {Capps}]{Soifer81}
{Soifer},~B.~T.; {Willner},~S.~P.; {Rudy},~R.~J.; {Capps},~R.~W. \emph{\apj}
  \textbf{1981}, \emph{250}, 631--635\relax
\mciteBstWouldAddEndPuncttrue
\mciteSetBstMidEndSepPunct{\mcitedefaultmidpunct}
{\mcitedefaultendpunct}{\mcitedefaultseppunct}\relax
\EndOfBibitem
\bibitem[{Sylvester} et~al.(1999){Sylvester}, {Kemper}, {Barlow}, {de Jong},
  {Waters}, {Tielens}, and {Omont}]{Sylvester99}
{Sylvester},~R.~J.; {Kemper},~F.; {Barlow},~M.~J.; {de Jong},~T.;
  {Waters},~L.~B.~F.~M.; {Tielens},~A.~G.~G.~M.; {Omont},~A. \emph{\aap}
  \textbf{1999}, \emph{352}, 587--599\relax
\mciteBstWouldAddEndPuncttrue
\mciteSetBstMidEndSepPunct{\mcitedefaultmidpunct}
{\mcitedefaultendpunct}{\mcitedefaultseppunct}\relax
\EndOfBibitem
\bibitem[{Spoon} et~al.(2004){Spoon}, {Armus}, {Cami}, {Tielens}, {Chiar},
  {Peeters}, {Keane}, {Charmandaris}, {Appleton}, {Teplitz}, and
  {Burgdorf}]{Spoon04}
{Spoon},~H.~W.~W.; {Armus},~L.; {Cami},~J.; {Tielens},~A.~G.~G.~M.;
  {Chiar},~J.~E.; {Peeters},~E.; {Keane},~J.~V.; {Charmandaris},~V.;
  {Appleton},~P.~N.; {Teplitz},~H.~I.; {Burgdorf},~M.~J. \emph{\apjs}
  \textbf{2004}, \emph{154}, 184--187\relax
\mciteBstWouldAddEndPuncttrue
\mciteSetBstMidEndSepPunct{\mcitedefaultmidpunct}
{\mcitedefaultendpunct}{\mcitedefaultseppunct}\relax
\EndOfBibitem
\bibitem[{Sajina} et~al.(2009){Sajina}, {Spoon}, {Yan}, {Imanishi}, {Fadda},
  and {Elitzur}]{Sajina09}
{Sajina},~A.; {Spoon},~H.; {Yan},~L.; {Imanishi},~M.; {Fadda},~D.;
  {Elitzur},~M. \emph{\apj} \textbf{2009}, \emph{703}, 270--284\relax
\mciteBstWouldAddEndPuncttrue
\mciteSetBstMidEndSepPunct{\mcitedefaultmidpunct}
{\mcitedefaultendpunct}{\mcitedefaultseppunct}\relax
\EndOfBibitem
\bibitem[{Shimonishi} et~al.(2010){Shimonishi}, {Onaka}, {Kato}, {Sakon},
  {Ita}, {Kawamura}, and {Kaneda}]{Shimonishi10}
{Shimonishi},~T.; {Onaka},~T.; {Kato},~D.; {Sakon},~I.; {Ita},~Y.;
  {Kawamura},~A.; {Kaneda},~H. \emph{\aap} \textbf{2010}, \emph{514}, A12\relax
\mciteBstWouldAddEndPuncttrue
\mciteSetBstMidEndSepPunct{\mcitedefaultmidpunct}
{\mcitedefaultendpunct}{\mcitedefaultseppunct}\relax
\EndOfBibitem
\bibitem[{Bossa} et~al.(2012){Bossa}, {Isokoski}, {de Valois}, and
  {Linnartz}]{Bossa12}
{Bossa},~J.-B.; {Isokoski},~K.; {de Valois},~M.~S.; {Linnartz},~H. \emph{\aap}
  \textbf{2012}, \emph{545}, A82\relax
\mciteBstWouldAddEndPuncttrue
\mciteSetBstMidEndSepPunct{\mcitedefaultmidpunct}
{\mcitedefaultendpunct}{\mcitedefaultseppunct}\relax
\EndOfBibitem
\bibitem[{Aitken} et~al.(1988){Aitken}, {Smith}, {James}, {Roche}, and
  {Hough}]{Aitken88}
{Aitken},~D.~K.; {Smith},~C.~H.; {James},~S.~D.; {Roche},~P.~F.; {Hough},~J.~H.
  \emph{\mnras} \textbf{1988}, \emph{230}, 629--638\relax
\mciteBstWouldAddEndPuncttrue
\mciteSetBstMidEndSepPunct{\mcitedefaultmidpunct}
{\mcitedefaultendpunct}{\mcitedefaultseppunct}\relax
\EndOfBibitem
\bibitem[{Schegerer} and {Wolf}(2010){Schegerer}, and {Wolf}]{Schegerer10}
{Schegerer},~A.~A.; {Wolf},~S. \emph{\aap} \textbf{2010}, \emph{517}, A87\relax
\mciteBstWouldAddEndPuncttrue
\mciteSetBstMidEndSepPunct{\mcitedefaultmidpunct}
{\mcitedefaultendpunct}{\mcitedefaultseppunct}\relax
\EndOfBibitem
\bibitem[Yoshino et~al.({1996})Yoshino, Esmond, Parkinson, Ito, and
  Matsui]{Yoshino96}
Yoshino,~K.; Esmond,~J.; Parkinson,~W.; Ito,~K.; Matsui,~T. \emph{{\cp}}
  \textbf{{1996}}, \emph{{211}}, {387--391}\relax
\mciteBstWouldAddEndPuncttrue
\mciteSetBstMidEndSepPunct{\mcitedefaultmidpunct}
{\mcitedefaultendpunct}{\mcitedefaultseppunct}\relax
\EndOfBibitem
\bibitem[Cheng et~al.({1999})Cheng, Chew, Liu, Bahou, Lee, Yung, and
  Gerstell]{Cheng99}
Cheng,~B.; Chew,~E.; Liu,~C.; Bahou,~M.; Lee,~Y.; Yung,~Y.; Gerstell,~M.
  \emph{{\grl}} \textbf{{1999}}, \emph{{26}}, {3657--3660}\relax
\mciteBstWouldAddEndPuncttrue
\mciteSetBstMidEndSepPunct{\mcitedefaultmidpunct}
{\mcitedefaultendpunct}{\mcitedefaultseppunct}\relax
\EndOfBibitem
\bibitem[Fillion et~al.({2001})Fillion, van Harrevelt, Ruiz, Castillejo,
  Zanganeh, Lemaire, van Hemert, and Rostas]{Fillion01}
Fillion,~J.; van Harrevelt,~R.; Ruiz,~J.; Castillejo,~N.; Zanganeh,~A.;
  Lemaire,~J.; van Hemert,~M.; Rostas,~F. \emph{{\jpca}} \textbf{{2001}},
  \emph{{105}}, {11414--11424}\relax
\mciteBstWouldAddEndPuncttrue
\mciteSetBstMidEndSepPunct{\mcitedefaultmidpunct}
{\mcitedefaultendpunct}{\mcitedefaultseppunct}\relax
\EndOfBibitem
\bibitem[{Keller-Rudek} et~al.(2013){Keller-Rudek}, {Moortgat}, {Sander}, and
  {S\"orensen}]{Mainz}
{Keller-Rudek},~H., {Moortgat},~G., {Sander},~R., {S\"orensen},~R., Eds.
  \emph{MPI-Mainz UV/VIS Spectral Atlas of Gaseous Molecules};
  www.uv-vis-spectral-atlas-mainz.org, 2013\relax
\mciteBstWouldAddEndPuncttrue
\mciteSetBstMidEndSepPunct{\mcitedefaultmidpunct}
{\mcitedefaultendpunct}{\mcitedefaultseppunct}\relax
\EndOfBibitem
\bibitem[{Andersson} et~al.(2006){Andersson}, {Al-Halabi}, {Kroes}, and {van
  Dishoeck}]{Andersson06}
{Andersson},~S.; {Al-Halabi},~A.; {Kroes},~G.-J.; {van Dishoeck},~E.~F.
  \emph{\jcp} \textbf{2006}, \emph{124}, 064715\relax
\mciteBstWouldAddEndPuncttrue
\mciteSetBstMidEndSepPunct{\mcitedefaultmidpunct}
{\mcitedefaultendpunct}{\mcitedefaultseppunct}\relax
\EndOfBibitem
\bibitem[{Andersson} and {van Dishoeck}(2008){Andersson}, and {van
  Dishoeck}]{Andersson08}
{Andersson},~S.; {van Dishoeck},~E.~F. \emph{\aap} \textbf{2008}, \emph{491},
  907--916\relax
\mciteBstWouldAddEndPuncttrue
\mciteSetBstMidEndSepPunct{\mcitedefaultmidpunct}
{\mcitedefaultendpunct}{\mcitedefaultseppunct}\relax
\EndOfBibitem
\bibitem[{Ajello}(1984)]{Ajello84}
{Ajello},~J.~M. \emph{\grl} \textbf{1984}, \emph{11}, 1195--1198\relax
\mciteBstWouldAddEndPuncttrue
\mciteSetBstMidEndSepPunct{\mcitedefaultmidpunct}
{\mcitedefaultendpunct}{\mcitedefaultseppunct}\relax
\EndOfBibitem
\bibitem[{Green} et~al.(1993){Green}, {Maluendes}, and {McLean}]{Green93}
{Green},~S.; {Maluendes},~S.; {McLean},~A.~D. \emph{\apjs} \textbf{1993},
  \emph{85}, 181--185\relax
\mciteBstWouldAddEndPuncttrue
\mciteSetBstMidEndSepPunct{\mcitedefaultmidpunct}
{\mcitedefaultendpunct}{\mcitedefaultseppunct}\relax
\EndOfBibitem
\bibitem[{Phillips} et~al.(1996){Phillips}, {Maluendes}, and
  {Green}]{Phillips96}
{Phillips},~T.~R.; {Maluendes},~S.; {Green},~S. \emph{\apjs} \textbf{1996},
  \emph{107}, 467\relax
\mciteBstWouldAddEndPuncttrue
\mciteSetBstMidEndSepPunct{\mcitedefaultmidpunct}
{\mcitedefaultendpunct}{\mcitedefaultseppunct}\relax
\EndOfBibitem
\bibitem[{Dubernet} et~al.(2009){Dubernet}, {Daniel}, {Grosjean}, and
  {Lin}]{Dubernet09}
{Dubernet},~M.; {Daniel},~F.; {Grosjean},~A.; {Lin},~C.~Y. \emph{\aap}
  \textbf{2009}, \emph{497}, 911--925\relax
\mciteBstWouldAddEndPuncttrue
\mciteSetBstMidEndSepPunct{\mcitedefaultmidpunct}
{\mcitedefaultendpunct}{\mcitedefaultseppunct}\relax
\EndOfBibitem
\bibitem[{Daniel} et~al.(2011){Daniel}, {Dubernet}, and {Grosjean}]{Daniel11}
{Daniel},~F.; {Dubernet},~M.-L.; {Grosjean},~A. \emph{\aap} \textbf{2011},
  \emph{536}, A76\relax
\mciteBstWouldAddEndPuncttrue
\mciteSetBstMidEndSepPunct{\mcitedefaultmidpunct}
{\mcitedefaultendpunct}{\mcitedefaultseppunct}\relax
\EndOfBibitem
\bibitem[Valiron et~al.(2008)Valiron, Wernli, Faure, Wiesenfeld, Rist, Kedzuch,
  and Noga]{Valiron08}
Valiron,~P.; Wernli,~M.; Faure,~A.; Wiesenfeld,~L.; Rist,~C.; Kedzuch,~S.;
  Noga,~J. \emph{J. Chem. Phys.} \textbf{2008}, \emph{129},
  134306--134306\relax
\mciteBstWouldAddEndPuncttrue
\mciteSetBstMidEndSepPunct{\mcitedefaultmidpunct}
{\mcitedefaultendpunct}{\mcitedefaultseppunct}\relax
\EndOfBibitem
\bibitem[{Indriolo} et~al.(2013){Indriolo}, {Neufeld}, {Seifahrt}, and
  {Richter}]{Indriolo13water}
{Indriolo},~N.; {Neufeld},~D.~A.; {Seifahrt},~A.; {Richter},~M.~J. \emph{\apj}
  \textbf{2013}, \emph{776}, 8\relax
\mciteBstWouldAddEndPuncttrue
\mciteSetBstMidEndSepPunct{\mcitedefaultmidpunct}
{\mcitedefaultendpunct}{\mcitedefaultseppunct}\relax
\EndOfBibitem
\bibitem[{Meijerink} et~al.(2009){Meijerink}, {Pontoppidan}, {Blake},
  {Poelman}, and {Dullemond}]{Meijerink09}
{Meijerink},~R.; {Pontoppidan},~K.~M.; {Blake},~G.~A.; {Poelman},~D.~R.;
  {Dullemond},~C.~P. \emph{\apj} \textbf{2009}, \emph{704}, 1471--1481\relax
\mciteBstWouldAddEndPuncttrue
\mciteSetBstMidEndSepPunct{\mcitedefaultmidpunct}
{\mcitedefaultendpunct}{\mcitedefaultseppunct}\relax
\EndOfBibitem
\bibitem[{Faure} and {Josselin}(2008){Faure}, and {Josselin}]{Faure08}
{Faure},~A.; {Josselin},~E. \emph{\aap} \textbf{2008}, \emph{492},
  257--264\relax
\mciteBstWouldAddEndPuncttrue
\mciteSetBstMidEndSepPunct{\mcitedefaultmidpunct}
{\mcitedefaultendpunct}{\mcitedefaultseppunct}\relax
\EndOfBibitem
\bibitem[{Neufeld}(2012)]{Neufeld12}
{Neufeld},~D.~A. \emph{\apj} \textbf{2012}, \emph{749}, 125\relax
\mciteBstWouldAddEndPuncttrue
\mciteSetBstMidEndSepPunct{\mcitedefaultmidpunct}
{\mcitedefaultendpunct}{\mcitedefaultseppunct}\relax
\EndOfBibitem
\bibitem[Wiesenfeld and Faure(2010)Wiesenfeld, and Faure]{Wiesenfeld10}
Wiesenfeld,~L.; Faure,~A. \emph{\pra} \textbf{2010}, \emph{82}, 040702\relax
\mciteBstWouldAddEndPuncttrue
\mciteSetBstMidEndSepPunct{\mcitedefaultmidpunct}
{\mcitedefaultendpunct}{\mcitedefaultseppunct}\relax
\EndOfBibitem
\bibitem[{Drouin} and {Wiesenfeld}(2012){Drouin}, and {Wiesenfeld}]{Drouin12}
{Drouin},~B.; {Wiesenfeld},~L. \emph{\pra} \textbf{2012}, \emph{86},
  022705\relax
\mciteBstWouldAddEndPuncttrue
\mciteSetBstMidEndSepPunct{\mcitedefaultmidpunct}
{\mcitedefaultendpunct}{\mcitedefaultseppunct}\relax
\EndOfBibitem
\bibitem[{Yang} et~al.(2010){Yang}, {Sarma}, {ter Meulen}, {Parker}, G.C.,
  {Wiesenfeld}, A., Y., and N.]{Yang10b}
{Yang},~C.-H.; {Sarma},~G.; {ter Meulen},~J.~J.; {Parker},~D.~H.; G.C.,~M.;
  {Wiesenfeld},~L.; A.,~F.; Y.,~S.; N.,~F. \emph{\jcp} \textbf{2010},
  \emph{122}, 131103\relax
\mciteBstWouldAddEndPuncttrue
\mciteSetBstMidEndSepPunct{\mcitedefaultmidpunct}
{\mcitedefaultendpunct}{\mcitedefaultseppunct}\relax
\EndOfBibitem
\bibitem[van~der Avoird and Nesbitt({2011})van~der Avoird, and
  Nesbitt]{Avoird11a}
van~der Avoird,~A.; Nesbitt,~D.~J. \emph{{\jcp}} \textbf{{2011}},
  \emph{{134}}\relax
\mciteBstWouldAddEndPuncttrue
\mciteSetBstMidEndSepPunct{\mcitedefaultmidpunct}
{\mcitedefaultendpunct}{\mcitedefaultseppunct}\relax
\EndOfBibitem
\bibitem[van~der Avoird et~al.({2012})van~der Avoird, Scribano, Faure, Weida,
  Fair, and Nesbitt]{Avoird11b}
van~der Avoird,~A.; Scribano,~Y.; Faure,~A.; Weida,~M.~J.; Fair,~J.~R.;
  Nesbitt,~D.~J. \emph{{\cp}} \textbf{{2012}}, \emph{{399}}, {28--38}\relax
\mciteBstWouldAddEndPuncttrue
\mciteSetBstMidEndSepPunct{\mcitedefaultmidpunct}
{\mcitedefaultendpunct}{\mcitedefaultseppunct}\relax
\EndOfBibitem
\bibitem[Ziemkiewicz et~al.({2012})Ziemkiewicz, Pluetzer, Nesbitt, Scribano,
  Faure, and van~der Avoird]{Avoird12}
Ziemkiewicz,~M.~P.; Pluetzer,~C.; Nesbitt,~D.~J.; Scribano,~Y.; Faure,~A.;
  van~der Avoird,~A. \emph{{\jcp}} \textbf{{2012}}, \emph{{137}}\relax
\mciteBstWouldAddEndPuncttrue
\mciteSetBstMidEndSepPunct{\mcitedefaultmidpunct}
{\mcitedefaultendpunct}{\mcitedefaultseppunct}\relax
\EndOfBibitem
\bibitem[{{Dick}, M. and {Drouin}, B. and {Pearson}, J.}(2010)]{Dick10}
{{Dick}, M. and {Drouin}, B. and {Pearson}, J.}, \emph{\pra} \textbf{2010},
  \emph{81}, 022706\relax
\mciteBstWouldAddEndPuncttrue
\mciteSetBstMidEndSepPunct{\mcitedefaultmidpunct}
{\mcitedefaultendpunct}{\mcitedefaultseppunct}\relax
\EndOfBibitem
\bibitem[{Daniel} et~al.(2012){Daniel}, {Goicoechea}, {Cernicharo}, {Dubernet},
  and {Faure}]{Daniel12}
{Daniel},~F.; {Goicoechea},~J.~R.; {Cernicharo},~J.; {Dubernet},~M.-L.;
  {Faure},~A. \emph{\aap} \textbf{2012}, \emph{547}, A81\relax
\mciteBstWouldAddEndPuncttrue
\mciteSetBstMidEndSepPunct{\mcitedefaultmidpunct}
{\mcitedefaultendpunct}{\mcitedefaultseppunct}\relax
\EndOfBibitem
\bibitem[{Crimier} et~al.(2007){Crimier}, {Faure}, {Ceccarelli}, {Valiron},
  {Wiesenfeld}, and {Dubernet}]{Crimier07}
{Crimier},~N.; {Faure},~A.; {Ceccarelli},~C.; {Valiron},~P.; {Wiesenfeld},~L.;
  {Dubernet},~M.~L. In \emph{Molecules in Space and Laboratory};
  J.L.~Lemaire,~F.~C., Ed.; S. Diana: Paris, 2007\relax
\mciteBstWouldAddEndPuncttrue
\mciteSetBstMidEndSepPunct{\mcitedefaultmidpunct}
{\mcitedefaultendpunct}{\mcitedefaultseppunct}\relax
\EndOfBibitem
\bibitem[{Crimier} et~al.(2007){Crimier}, {Faure}, {Ceccarelli}, {Valiron},
  {Wiesenfeld}, and {Dubernet}]{Crimier07b}
{Crimier},~C.; {Faure},~A.; {Ceccarelli},~C.; {Valiron},~P.; {Wiesenfeld},~L.;
  {Dubernet},~M.~L. In \emph{SF2A-2007: Proc. Annual meeting French Soc.
  Astron. Astrophys.}; {Bouvier},~J., {Chalabaev},~A., {Charbonnel},~C., Eds.;
  2007; p 241\relax
\mciteBstWouldAddEndPuncttrue
\mciteSetBstMidEndSepPunct{\mcitedefaultmidpunct}
{\mcitedefaultendpunct}{\mcitedefaultseppunct}\relax
\EndOfBibitem
\bibitem[{Grosjean} et~al.(2003){Grosjean}, {Dubernet}, and
  {Ceccarelli}]{Grosjean03}
{Grosjean},~A.; {Dubernet},~M.-L.; {Ceccarelli},~C. \emph{\aap} \textbf{2003},
  \emph{408}, 1197--1203\relax
\mciteBstWouldAddEndPuncttrue
\mciteSetBstMidEndSepPunct{\mcitedefaultmidpunct}
{\mcitedefaultendpunct}{\mcitedefaultseppunct}\relax
\EndOfBibitem
\bibitem[{Faure} et~al.(2007){Faure}, {Crimier}, {Ceccarelli}, {Valiron},
  {Wiesenfeld}, and {Dubernet}]{Faure07}
{Faure},~A.; {Crimier},~N.; {Ceccarelli},~C.; {Valiron},~P.; {Wiesenfeld},~L.;
  {Dubernet},~M.~L. \emph{\aap} \textbf{2007}, \emph{472}, 1029--1035\relax
\mciteBstWouldAddEndPuncttrue
\mciteSetBstMidEndSepPunct{\mcitedefaultmidpunct}
{\mcitedefaultendpunct}{\mcitedefaultseppunct}\relax
\EndOfBibitem
\bibitem[{Hollenbach} et~al.(2013){Hollenbach}, {Elitzur}, and
  {McKee}]{Hollenbach13}
{Hollenbach},~D.; {Elitzur},~M.; {McKee},~C.~F. \emph{\apj} \textbf{2013},
  \emph{773}, 70\relax
\mciteBstWouldAddEndPuncttrue
\mciteSetBstMidEndSepPunct{\mcitedefaultmidpunct}
{\mcitedefaultendpunct}{\mcitedefaultseppunct}\relax
\EndOfBibitem
\bibitem[{Dubernet} et~al.(2013){Dubernet}, {Alexander}, {Ba}, {Balakrishnan},
  {Balan{\c c}a}, {Ceccarelli}, {Cernicharo}, {Daniel}, {Dayou}, {Doronin},
  {Dumouchel}, {Faure}, {Feautrier}, {Flower}, {Grosjean}, {Halvick},
  {K{\l}os}, {Lique}, {McBane}, {Marinakis}, {Moreau}, {Moszynski}, {Neufeld},
  {Roueff}, {Schilke}, {Spielfiedel}, {Stancil}, {Stoecklin}, {Tennyson},
  {Yang}, {Vasserot}, and {Wiesenfeld}]{Dubernet13}
{Dubernet},~M.-L. et~al.  \emph{\aap} \textbf{2013}, \emph{553}, A50\relax
\mciteBstWouldAddEndPuncttrue
\mciteSetBstMidEndSepPunct{\mcitedefaultmidpunct}
{\mcitedefaultendpunct}{\mcitedefaultseppunct}\relax
\EndOfBibitem
\bibitem[{Sch{\"o}ier} et~al.(2005){Sch{\"o}ier}, {van der Tak}, {van
  Dishoeck}, and {Black}]{Schoier05}
{Sch{\"o}ier},~F.~L.; {van der Tak},~F.~F.~S.; {van Dishoeck},~E.~F.;
  {Black},~J.~H. \emph{\aap} \textbf{2005}, \emph{432}, 369--379\relax
\mciteBstWouldAddEndPuncttrue
\mciteSetBstMidEndSepPunct{\mcitedefaultmidpunct}
{\mcitedefaultendpunct}{\mcitedefaultseppunct}\relax
\EndOfBibitem
\bibitem[{Bernes}(1979)]{Bernes79}
{Bernes},~C. \emph{\aap} \textbf{1979}, \emph{73}, 67--73\relax
\mciteBstWouldAddEndPuncttrue
\mciteSetBstMidEndSepPunct{\mcitedefaultmidpunct}
{\mcitedefaultendpunct}{\mcitedefaultseppunct}\relax
\EndOfBibitem
\bibitem[{Rybicki} and {Hummer}(1991){Rybicki}, and {Hummer}]{Rybicki91}
{Rybicki},~G.~B.; {Hummer},~D.~G. \emph{\aap} \textbf{1991}, \emph{245},
  171--181\relax
\mciteBstWouldAddEndPuncttrue
\mciteSetBstMidEndSepPunct{\mcitedefaultmidpunct}
{\mcitedefaultendpunct}{\mcitedefaultseppunct}\relax
\EndOfBibitem
\bibitem[{Hummer} and {Rybicki}(1982){Hummer}, and {Rybicki}]{Hummer82}
{Hummer},~D.~G.; {Rybicki},~G.~B. \emph{\apj} \textbf{1982}, \emph{254},
  767--779\relax
\mciteBstWouldAddEndPuncttrue
\mciteSetBstMidEndSepPunct{\mcitedefaultmidpunct}
{\mcitedefaultendpunct}{\mcitedefaultseppunct}\relax
\EndOfBibitem
\bibitem[{Sobolev}(1960)]{Sobolev60}
{Sobolev},~V.~V. \emph{{Moving envelopes of stars}}; Harvard University Press,
  Cambridge, 1960\relax
\mciteBstWouldAddEndPuncttrue
\mciteSetBstMidEndSepPunct{\mcitedefaultmidpunct}
{\mcitedefaultendpunct}{\mcitedefaultseppunct}\relax
\EndOfBibitem
\bibitem[{Hogerheijde} and {van der Tak}(2000){Hogerheijde}, and {van der
  Tak}]{Hogerheijde00}
{Hogerheijde},~M.~R.; {van der Tak},~F.~F.~S. \emph{\aap} \textbf{2000},
  \emph{362}, 697--710\relax
\mciteBstWouldAddEndPuncttrue
\mciteSetBstMidEndSepPunct{\mcitedefaultmidpunct}
{\mcitedefaultendpunct}{\mcitedefaultseppunct}\relax
\EndOfBibitem
\bibitem[{Brinch} and {Hogerheijde}(2010){Brinch}, and {Hogerheijde}]{Brinch10}
{Brinch},~C.; {Hogerheijde},~M.~R. \emph{\aap} \textbf{2010}, \emph{523},
  A25\relax
\mciteBstWouldAddEndPuncttrue
\mciteSetBstMidEndSepPunct{\mcitedefaultmidpunct}
{\mcitedefaultendpunct}{\mcitedefaultseppunct}\relax
\EndOfBibitem
\bibitem[{van der Tak} et~al.(2007){van der Tak}, {Black}, {Sch{\"o}ier},
  {Jansen}, and {van Dishoeck}]{vanderTak07}
{van der Tak},~F.~F.~S.; {Black},~J.~H.; {Sch{\"o}ier},~F.~L.; {Jansen},~D.~J.;
  {van Dishoeck},~E.~F. \emph{\aap} \textbf{2007}, \emph{468}, 627--635\relax
\mciteBstWouldAddEndPuncttrue
\mciteSetBstMidEndSepPunct{\mcitedefaultmidpunct}
{\mcitedefaultendpunct}{\mcitedefaultseppunct}\relax
\EndOfBibitem
\bibitem[{de Jong}(1973)]{deJong73}
{de Jong},~T. \emph{\aap} \textbf{1973}, \emph{26}, 297\relax
\mciteBstWouldAddEndPuncttrue
\mciteSetBstMidEndSepPunct{\mcitedefaultmidpunct}
{\mcitedefaultendpunct}{\mcitedefaultseppunct}\relax
\EndOfBibitem
\bibitem[{Genzel}(1986)]{Genzel86}
{Genzel},~R. In \emph{Masers, Molecules, and Mass Outflows in Star Formation
  Regions}; {Haschick},~A.~D., {Moran},~J.~M., Eds.; Haystack: Westford, 1986;
  p 233\relax
\mciteBstWouldAddEndPuncttrue
\mciteSetBstMidEndSepPunct{\mcitedefaultmidpunct}
{\mcitedefaultendpunct}{\mcitedefaultseppunct}\relax
\EndOfBibitem
\bibitem[{Garay} et~al.(1989){Garay}, {Moran}, and {Haschick}]{Garay89}
{Garay},~G.; {Moran},~J.~M.; {Haschick},~A.~D. \emph{\apj} \textbf{1989},
  \emph{338}, 244--261\relax
\mciteBstWouldAddEndPuncttrue
\mciteSetBstMidEndSepPunct{\mcitedefaultmidpunct}
{\mcitedefaultendpunct}{\mcitedefaultseppunct}\relax
\EndOfBibitem
\bibitem[{Neufeld} et~al.(2013){Neufeld}, {Wu}, {Kraus}, {Menten}, {Tolls},
  {Melnick}, and {Nagy}]{Neufeld13maser}
{Neufeld},~D.~A.; {Wu},~Y.; {Kraus},~A.; {Menten},~K.~M.; {Tolls},~V.;
  {Melnick},~G.~J.; {Nagy},~Z. \emph{\apj} \textbf{2013}, \emph{769}, 48\relax
\mciteBstWouldAddEndPuncttrue
\mciteSetBstMidEndSepPunct{\mcitedefaultmidpunct}
{\mcitedefaultendpunct}{\mcitedefaultseppunct}\relax
\EndOfBibitem
\bibitem[{Neufeld} and {Melnick}(1991){Neufeld}, and {Melnick}]{Neufeld91}
{Neufeld},~D.~A.; {Melnick},~G.~J. \emph{\apj} \textbf{1991}, \emph{368},
  215--230\relax
\mciteBstWouldAddEndPuncttrue
\mciteSetBstMidEndSepPunct{\mcitedefaultmidpunct}
{\mcitedefaultendpunct}{\mcitedefaultseppunct}\relax
\EndOfBibitem
\bibitem[{Alcolea} and {Menten}(1993){Alcolea}, and {Menten}]{Alcolea93}
{Alcolea},~J.; {Menten},~K.~M. In \emph{Astrophysical Masers}; {Clegg},~A.~W.,
  {Nedoluha},~G.~E., Eds.; Lecture Notes in Physics; Springer: Berlin, 1993;
  Vol. 412; p 399\relax
\mciteBstWouldAddEndPuncttrue
\mciteSetBstMidEndSepPunct{\mcitedefaultmidpunct}
{\mcitedefaultendpunct}{\mcitedefaultseppunct}\relax
\EndOfBibitem
\bibitem[{Waters} et~al.(1980){Waters}, {Kakar}, {Kuiper}, {Roscoe}, {Swanson},
  {Rodriguez Kuiper}, {Kerr}, {Thaddeus}, and {Gustincic}]{Waters80}
{Waters},~J.~W.; {Kakar},~R.~K.; {Kuiper},~T.~B.~H.; {Roscoe},~H.~K.;
  {Swanson},~P.~N.; {Rodriguez Kuiper},~E.~N.; {Kerr},~A.~R.; {Thaddeus},~P.;
  {Gustincic},~J.~J. \emph{\apj} \textbf{1980}, \emph{235}, 57--62\relax
\mciteBstWouldAddEndPuncttrue
\mciteSetBstMidEndSepPunct{\mcitedefaultmidpunct}
{\mcitedefaultendpunct}{\mcitedefaultseppunct}\relax
\EndOfBibitem
\bibitem[{Menten} et~al.(1990){Menten}, {Melnick}, and {Phillips}]{Menten90a}
{Menten},~K.~M.; {Melnick},~G.~J.; {Phillips},~T.~G. \emph{\apjl}
  \textbf{1990}, \emph{350}, L41--L44\relax
\mciteBstWouldAddEndPuncttrue
\mciteSetBstMidEndSepPunct{\mcitedefaultmidpunct}
{\mcitedefaultendpunct}{\mcitedefaultseppunct}\relax
\EndOfBibitem
\bibitem[{Menten} et~al.(1990){Menten}, {Melnick}, {Phillips}, and
  {Neufeld}]{Menten90b}
{Menten},~K.~M.; {Melnick},~G.~J.; {Phillips},~T.~G.; {Neufeld},~D.~A.
  \emph{\apjl} \textbf{1990}, \emph{363}, L27--L31\relax
\mciteBstWouldAddEndPuncttrue
\mciteSetBstMidEndSepPunct{\mcitedefaultmidpunct}
{\mcitedefaultendpunct}{\mcitedefaultseppunct}\relax
\EndOfBibitem
\bibitem[{Phillips} et~al.(1980){Phillips}, {Kwan}, and {Huggins}]{Phillips80}
{Phillips},~T.~G.; {Kwan},~J.; {Huggins},~P.~J. In \emph{Interstellar
  Molecules}; {Andrew},~B.~H., Ed.; IAU Symposium; Kluwer: Dordrecht, 1980;
  Vol.~87; pp 21--24\relax
\mciteBstWouldAddEndPuncttrue
\mciteSetBstMidEndSepPunct{\mcitedefaultmidpunct}
{\mcitedefaultendpunct}{\mcitedefaultseppunct}\relax
\EndOfBibitem
\bibitem[{Melnick} et~al.(1993){Melnick}, {Menten}, {Phillips}, and
  {Hunter}]{Melnick93}
{Melnick},~G.~J.; {Menten},~K.~M.; {Phillips},~T.~G.; {Hunter},~T. \emph{\apjl}
  \textbf{1993}, \emph{416}, L37\relax
\mciteBstWouldAddEndPuncttrue
\mciteSetBstMidEndSepPunct{\mcitedefaultmidpunct}
{\mcitedefaultendpunct}{\mcitedefaultseppunct}\relax
\EndOfBibitem
\bibitem[{Menten} et~al.(2008){Menten}, {Lundgren}, {Belloche}, {Thorwirth},
  and {Reid}]{Menten08}
{Menten},~K.~M.; {Lundgren},~A.; {Belloche},~A.; {Thorwirth},~S.; {Reid},~M.~J.
  \emph{\aap} \textbf{2008}, \emph{477}, 185--192\relax
\mciteBstWouldAddEndPuncttrue
\mciteSetBstMidEndSepPunct{\mcitedefaultmidpunct}
{\mcitedefaultendpunct}{\mcitedefaultseppunct}\relax
\EndOfBibitem
\bibitem[{Justtanont} et~al.(2012){Justtanont}, {Khouri}, {Maercker},
  {Alcolea}, {Decin}, {Olofsson}, {Sch{\"o}ier}, {Bujarrabal}, {Marston},
  {Teyssier}, {Cernicharo}, {Dominik}, {de Koter}, {Melnick}, {Menten},
  {Neufeld}, {Planesas}, {Schmidt}, {Szczerba}, and {Waters}]{Justtanont12}
{Justtanont},~K. et~al.  \emph{\aap} \textbf{2012}, \emph{537}, A144\relax
\mciteBstWouldAddEndPuncttrue
\mciteSetBstMidEndSepPunct{\mcitedefaultmidpunct}
{\mcitedefaultendpunct}{\mcitedefaultseppunct}\relax
\EndOfBibitem
\bibitem[{Menten} and {Melnick}(1989){Menten}, and {Melnick}]{Menten89}
{Menten},~K.~M.; {Melnick},~G.~J. \emph{\apjl} \textbf{1989}, \emph{341},
  L91--L94\relax
\mciteBstWouldAddEndPuncttrue
\mciteSetBstMidEndSepPunct{\mcitedefaultmidpunct}
{\mcitedefaultendpunct}{\mcitedefaultseppunct}\relax
\EndOfBibitem
\bibitem[{Menten} et~al.(2006){Menten}, {Philipp}, {G{\"u}sten}, {Alcolea},
  {Polehampton}, and {Br{\"u}nken}]{Menten06}
{Menten},~K.~M.; {Philipp},~S.~D.; {G{\"u}sten},~R.; {Alcolea},~J.;
  {Polehampton},~E.~T.; {Br{\"u}nken},~S. \emph{\aap} \textbf{2006},
  \emph{454}, L107--L110\relax
\mciteBstWouldAddEndPuncttrue
\mciteSetBstMidEndSepPunct{\mcitedefaultmidpunct}
{\mcitedefaultendpunct}{\mcitedefaultseppunct}\relax
\EndOfBibitem
\bibitem[{Walsh} et~al.(2011){Walsh}, {Breen}, {Britton}, {Brooks}, {Burton},
  {Cunningham}, {Green}, {Harvey-Smith}, {Hindson}, {Hoare}, {Indermuehle},
  {Jones}, {Lo}, {Longmore}, {Lowe}, {Phillips}, {Purcell}, {Thompson},
  {Urquhart}, {Voronkov}, {White}, and {Whiting}]{WalshA11}
{Walsh},~A.~J. et~al.  \emph{\mnras} \textbf{2011}, \emph{416},
  1764--1821\relax
\mciteBstWouldAddEndPuncttrue
\mciteSetBstMidEndSepPunct{\mcitedefaultmidpunct}
{\mcitedefaultendpunct}{\mcitedefaultseppunct}\relax
\EndOfBibitem
\bibitem[{Lo}(2005)]{Lo05}
{Lo},~K.~Y. \emph{\araa} \textbf{2005}, \emph{43}, 625--676\relax
\mciteBstWouldAddEndPuncttrue
\mciteSetBstMidEndSepPunct{\mcitedefaultmidpunct}
{\mcitedefaultendpunct}{\mcitedefaultseppunct}\relax
\EndOfBibitem
\bibitem[{Surcis} et~al.(2009){Surcis}, {Tarchi}, {Henkel}, {Ott}, {Lovell},
  and {Castangia}]{Surcis09}
{Surcis},~G.; {Tarchi},~A.; {Henkel},~C.; {Ott},~J.; {Lovell},~J.;
  {Castangia},~P. \emph{\aap} \textbf{2009}, \emph{502}, 529--540\relax
\mciteBstWouldAddEndPuncttrue
\mciteSetBstMidEndSepPunct{\mcitedefaultmidpunct}
{\mcitedefaultendpunct}{\mcitedefaultseppunct}\relax
\EndOfBibitem
\bibitem[{Brunthaler} et~al.(2006){Brunthaler}, {Henkel}, {de Blok}, {Reid},
  {Greenhill}, and {Falcke}]{Brunthaler06}
{Brunthaler},~A.; {Henkel},~C.; {de Blok},~W.~J.~G.; {Reid},~M.~J.;
  {Greenhill},~L.~J.; {Falcke},~H. \emph{\aap} \textbf{2006}, \emph{457},
  109--114\relax
\mciteBstWouldAddEndPuncttrue
\mciteSetBstMidEndSepPunct{\mcitedefaultmidpunct}
{\mcitedefaultendpunct}{\mcitedefaultseppunct}\relax
\EndOfBibitem
\bibitem[{Braatz} et~al.(2010){Braatz}, {Reid}, {Humphreys}, {Henkel},
  {Condon}, and {Lo}]{Braatz10}
{Braatz},~J.~A.; {Reid},~M.~J.; {Humphreys},~E.~M.~L.; {Henkel},~C.;
  {Condon},~J.~J.; {Lo},~K.~Y. \emph{\apj} \textbf{2010}, \emph{718},
  657--665\relax
\mciteBstWouldAddEndPuncttrue
\mciteSetBstMidEndSepPunct{\mcitedefaultmidpunct}
{\mcitedefaultendpunct}{\mcitedefaultseppunct}\relax
\EndOfBibitem
\bibitem[{Reid} et~al.(2009){Reid}, {Menten}, {Zheng}, {Brunthaler}, and
  {Xu}]{Reid09}
{Reid},~M.~J.; {Menten},~K.~M.; {Zheng},~X.~W.; {Brunthaler},~A.; {Xu},~Y.
  \emph{\apj} \textbf{2009}, \emph{705}, 1548--1553\relax
\mciteBstWouldAddEndPuncttrue
\mciteSetBstMidEndSepPunct{\mcitedefaultmidpunct}
{\mcitedefaultendpunct}{\mcitedefaultseppunct}\relax
\EndOfBibitem
\bibitem[Wakelam et~al.({2010})Wakelam, Smith, Herbst, Troe, Geppert, Linnartz,
  Oeberg, Roueff, Agundez, Pernot, Cuppen, Loison, and Talbi]{Wakelam10ssr}
Wakelam,~V.; Smith,~I. W.~M.; Herbst,~E.; Troe,~J.; Geppert,~W.; Linnartz,~H.;
  Oeberg,~K.; Roueff,~E.; Agundez,~M.; Pernot,~P.; Cuppen,~H.~M.;
  Loison,~J.~C.; Talbi,~D. \emph{{\ssr}} \textbf{{2010}}, \emph{{156}},
  {13--72}\relax
\mciteBstWouldAddEndPuncttrue
\mciteSetBstMidEndSepPunct{\mcitedefaultmidpunct}
{\mcitedefaultendpunct}{\mcitedefaultseppunct}\relax
\EndOfBibitem
\bibitem[{Steinfeld} et~al.(1999){Steinfeld}, {Francisco}, and
  {Hase}]{Steinfeld99}
{Steinfeld},~J.; {Francisco},~J.; {Hase},~W. \emph{Chemical Kinetics and
  Dynamics}; Prentice Hall, Upper Saddle River, 1999\relax
\mciteBstWouldAddEndPuncttrue
\mciteSetBstMidEndSepPunct{\mcitedefaultmidpunct}
{\mcitedefaultendpunct}{\mcitedefaultseppunct}\relax
\EndOfBibitem
\bibitem[Su and Chesnavich({1982})Su, and Chesnavich]{Su82}
Su,~T.; Chesnavich,~W. \emph{{\jcp}} \textbf{{1982}}, \emph{{76}},
  {5183--5185}\relax
\mciteBstWouldAddEndPuncttrue
\mciteSetBstMidEndSepPunct{\mcitedefaultmidpunct}
{\mcitedefaultendpunct}{\mcitedefaultseppunct}\relax
\EndOfBibitem
\bibitem[Maergoiz et~al.({2009})Maergoiz, Nikitin, and Troe]{Maergoiz09}
Maergoiz,~A.~I.; Nikitin,~E.~E.; Troe,~J. \emph{{\ijms}} \textbf{{2009}},
  \emph{{280}}, {42--49}\relax
\mciteBstWouldAddEndPuncttrue
\mciteSetBstMidEndSepPunct{\mcitedefaultmidpunct}
{\mcitedefaultendpunct}{\mcitedefaultseppunct}\relax
\EndOfBibitem
\bibitem[{Woon} and {Herbst}(2009){Woon}, and {Herbst}]{Woon09}
{Woon},~D.~E.; {Herbst},~E. \emph{\apjs} \textbf{2009}, \emph{185},
  273--288\relax
\mciteBstWouldAddEndPuncttrue
\mciteSetBstMidEndSepPunct{\mcitedefaultmidpunct}
{\mcitedefaultendpunct}{\mcitedefaultseppunct}\relax
\EndOfBibitem
\bibitem[{Anicich}(2003)]{Anicich03}
{Anicich},~V. \emph{An Index of the Literature for Bimolecular Gas Phase
  Cation-Molecule Reaction Kinetics}; JPL Publication 03-19, Pasadena,
  2003\relax
\mciteBstWouldAddEndPuncttrue
\mciteSetBstMidEndSepPunct{\mcitedefaultmidpunct}
{\mcitedefaultendpunct}{\mcitedefaultseppunct}\relax
\EndOfBibitem
\bibitem[{Smith} et~al.(2004){Smith}, {Herbst}, and {Chang}]{Smith04}
{Smith},~I.~W.~M.; {Herbst},~E.; {Chang},~Q. \emph{\mnras} \textbf{2004},
  \emph{350}, 323--330\relax
\mciteBstWouldAddEndPuncttrue
\mciteSetBstMidEndSepPunct{\mcitedefaultmidpunct}
{\mcitedefaultendpunct}{\mcitedefaultseppunct}\relax
\EndOfBibitem
\bibitem[Mitchell and Florescu-Mitchell({2006})Mitchell, and
  Florescu-Mitchell]{Mitchell06}
Mitchell,~J.; Florescu-Mitchell,~A. \emph{{\pr}} \textbf{{2006}}, \emph{{430}},
  {277--374}\relax
\mciteBstWouldAddEndPuncttrue
\mciteSetBstMidEndSepPunct{\mcitedefaultmidpunct}
{\mcitedefaultendpunct}{\mcitedefaultseppunct}\relax
\EndOfBibitem
\bibitem[{Larsson} and {Orel}(2008){Larsson}, and {Orel}]{Larsson08}
{Larsson},~M.; {Orel},~A. In \emph{Dissociative Recombination of Molecular
  Ions}; {Larsson, M.},, Ed.; Cambridge Univ.\ Press, Cambridge, 2008\relax
\mciteBstWouldAddEndPuncttrue
\mciteSetBstMidEndSepPunct{\mcitedefaultmidpunct}
{\mcitedefaultendpunct}{\mcitedefaultseppunct}\relax
\EndOfBibitem
\bibitem[{Rimmer} et~al.(2012){Rimmer}, {Herbst}, {Morata}, and
  {Roueff}]{Rimmer12}
{Rimmer},~P.~B.; {Herbst},~E.; {Morata},~O.; {Roueff},~E. \emph{\aap}
  \textbf{2012}, \emph{537}, A7\relax
\mciteBstWouldAddEndPuncttrue
\mciteSetBstMidEndSepPunct{\mcitedefaultmidpunct}
{\mcitedefaultendpunct}{\mcitedefaultseppunct}\relax
\EndOfBibitem
\bibitem[{Padovani} et~al.(2009){Padovani}, {Galli}, and
  {Glassgold}]{Padovani09}
{Padovani},~M.; {Galli},~D.; {Glassgold},~A.~E. \emph{\aap} \textbf{2009},
  \emph{501}, 619--631, erratum 2013, 549, C3\relax
\mciteBstWouldAddEndPuncttrue
\mciteSetBstMidEndSepPunct{\mcitedefaultmidpunct}
{\mcitedefaultendpunct}{\mcitedefaultseppunct}\relax
\EndOfBibitem
\bibitem[{Padovani} et~al.(2013){Padovani}, {Galli}, and
  {Glassgold}]{Padovani13}
{Padovani},~M.; {Galli},~D.; {Glassgold},~A.~E. \emph{\aap} \textbf{2013},
  \emph{549}, C3\relax
\mciteBstWouldAddEndPuncttrue
\mciteSetBstMidEndSepPunct{\mcitedefaultmidpunct}
{\mcitedefaultendpunct}{\mcitedefaultseppunct}\relax
\EndOfBibitem
\bibitem[{McCall} et~al.(2003){McCall}, {Huneycutt}, {Saykally}, {Geballe},
  {Djuric}, {Dunn}, {Semaniak}, {Novotny}, {Al-Khalili}, {Ehlerding},
  {Hellberg}, {Kalhori}, {Neau}, {Thomas}, {{\"O}sterdahl}, and
  {Larsson}]{McCall03}
{McCall},~B.~J. et~al.  \emph{\nat} \textbf{2003}, \emph{422}, 500--502\relax
\mciteBstWouldAddEndPuncttrue
\mciteSetBstMidEndSepPunct{\mcitedefaultmidpunct}
{\mcitedefaultendpunct}{\mcitedefaultseppunct}\relax
\EndOfBibitem
\bibitem[{Indriolo} and {McCall}(2012){Indriolo}, and {McCall}]{Indriolo12}
{Indriolo},~N.; {McCall},~B.~J. \emph{\apj} \textbf{2012}, \emph{745}, 91\relax
\mciteBstWouldAddEndPuncttrue
\mciteSetBstMidEndSepPunct{\mcitedefaultmidpunct}
{\mcitedefaultendpunct}{\mcitedefaultseppunct}\relax
\EndOfBibitem
\bibitem[{Padovani} and {Galli}(2013){Padovani}, and {Galli}]{Padovani13b}
{Padovani},~M.; {Galli},~D. In \emph{Cosmic Rays in Star-Forming Environments};
  {Torres},~D.~F., {Reimer},~O., Eds.; Advances in Solid State Physics;
  Springer: Berlin, 2013; Vol.~34; p~61\relax
\mciteBstWouldAddEndPuncttrue
\mciteSetBstMidEndSepPunct{\mcitedefaultmidpunct}
{\mcitedefaultendpunct}{\mcitedefaultseppunct}\relax
\EndOfBibitem
\bibitem[{Hollenbach} and {Salpeter}(1971){Hollenbach}, and
  {Salpeter}]{Hollenbach71}
{Hollenbach},~D.; {Salpeter},~E.~E. \emph{\apj} \textbf{1971}, \emph{163},
  155\relax
\mciteBstWouldAddEndPuncttrue
\mciteSetBstMidEndSepPunct{\mcitedefaultmidpunct}
{\mcitedefaultendpunct}{\mcitedefaultseppunct}\relax
\EndOfBibitem
\bibitem[{Katz} et~al.(1999){Katz}, {Furman}, {Biham}, {Pirronello}, and
  {Vidali}]{Katz99}
{Katz},~N.; {Furman},~I.; {Biham},~O.; {Pirronello},~V.; {Vidali},~G.
  \emph{\apj} \textbf{1999}, \emph{522}, 305--312\relax
\mciteBstWouldAddEndPuncttrue
\mciteSetBstMidEndSepPunct{\mcitedefaultmidpunct}
{\mcitedefaultendpunct}{\mcitedefaultseppunct}\relax
\EndOfBibitem
\bibitem[{Chang} et~al.(2005){Chang}, {Cuppen}, and {Herbst}]{Chang05}
{Chang},~Q.; {Cuppen},~H.~M.; {Herbst},~E. \emph{\aap} \textbf{2005},
  \emph{434}, 599--611\relax
\mciteBstWouldAddEndPuncttrue
\mciteSetBstMidEndSepPunct{\mcitedefaultmidpunct}
{\mcitedefaultendpunct}{\mcitedefaultseppunct}\relax
\EndOfBibitem
\bibitem[{Iqbal} et~al.(2012){Iqbal}, {Acharyya}, and {Herbst}]{Iqbal12}
{Iqbal},~W.; {Acharyya},~K.; {Herbst},~E. \emph{\apj} \textbf{2012},
  \emph{751}, 58\relax
\mciteBstWouldAddEndPuncttrue
\mciteSetBstMidEndSepPunct{\mcitedefaultmidpunct}
{\mcitedefaultendpunct}{\mcitedefaultseppunct}\relax
\EndOfBibitem
\bibitem[{Cazaux} and {Tielens}(2002){Cazaux}, and {Tielens}]{Cazaux02}
{Cazaux},~S.; {Tielens},~A.~G.~G.~M. \emph{\apjl} \textbf{2002}, \emph{575},
  L29--L32\relax
\mciteBstWouldAddEndPuncttrue
\mciteSetBstMidEndSepPunct{\mcitedefaultmidpunct}
{\mcitedefaultendpunct}{\mcitedefaultseppunct}\relax
\EndOfBibitem
\bibitem[{Cazaux} and {Tielens}(2010){Cazaux}, and {Tielens}]{Cazaux10err}
{Cazaux},~S.; {Tielens},~A.~G.~G.~M. \emph{\apj} \textbf{2010}, \emph{715},
  698--699\relax
\mciteBstWouldAddEndPuncttrue
\mciteSetBstMidEndSepPunct{\mcitedefaultmidpunct}
{\mcitedefaultendpunct}{\mcitedefaultseppunct}\relax
\EndOfBibitem
\bibitem[{van Dishoeck} and {Black}(1986){van Dishoeck}, and
  {Black}]{vanDishoeck86}
{van Dishoeck},~E.~F.; {Black},~J.~H. \emph{\apjs} \textbf{1986}, \emph{62},
  109--145\relax
\mciteBstWouldAddEndPuncttrue
\mciteSetBstMidEndSepPunct{\mcitedefaultmidpunct}
{\mcitedefaultendpunct}{\mcitedefaultseppunct}\relax
\EndOfBibitem
\bibitem[{Cordiner} and {Millar}(2009){Cordiner}, and {Millar}]{Cordiner09}
{Cordiner},~M.~A.; {Millar},~T.~J. \emph{\apj} \textbf{2009}, \emph{697},
  68--78\relax
\mciteBstWouldAddEndPuncttrue
\mciteSetBstMidEndSepPunct{\mcitedefaultmidpunct}
{\mcitedefaultendpunct}{\mcitedefaultseppunct}\relax
\EndOfBibitem
\bibitem[{Wakelam} et~al.(2012){Wakelam}, {Herbst}, {Loison}, {Smith},
  {Chandrasekaran}, {Pavone}, {Adams}, {Bacchus-Montabonel}, {Bergeat},
  {B{\'e}roff}, {Bierbaum}, {Chabot}, {Dalgarno}, {van Dishoeck}, {Faure},
  {Geppert}, {Gerlich}, {Galli}, {H{\'e}brard}, {Hersant}, {Hickson},
  {Honvault}, {Klippenstein}, {Le Picard}, {Nyman}, {Pernot}, {Schlemmer},
  {Selsis}, {Sims}, {Talbi}, {Tennyson}, {Troe}, {Wester}, and
  {Wiesenfeld}]{Wakelam12}
{Wakelam},~V. et~al.  \emph{\apjs} \textbf{2012}, \emph{199}, 21\relax
\mciteBstWouldAddEndPuncttrue
\mciteSetBstMidEndSepPunct{\mcitedefaultmidpunct}
{\mcitedefaultendpunct}{\mcitedefaultseppunct}\relax
\EndOfBibitem
\bibitem[{Herbst} and {Millar}(2008){Herbst}, and {Millar}]{Herbst08}
{Herbst},~E.; {Millar},~T. In \emph{Low Temperatures and Cold Molecules};
  {I.W.M.~Smith},, Ed.; Imperial College: London, 2008; p~1\relax
\mciteBstWouldAddEndPuncttrue
\mciteSetBstMidEndSepPunct{\mcitedefaultmidpunct}
{\mcitedefaultendpunct}{\mcitedefaultseppunct}\relax
\EndOfBibitem
\bibitem[{Wyrowski} et~al.(2010){Wyrowski}, {Menten}, {G{\"u}sten}, and
  {Belloche}]{Wyrowski10oh+}
{Wyrowski},~F.; {Menten},~K.~M.; {G{\"u}sten},~R.; {Belloche},~A. \emph{\aap}
  \textbf{2010}, \emph{518}, A26\relax
\mciteBstWouldAddEndPuncttrue
\mciteSetBstMidEndSepPunct{\mcitedefaultmidpunct}
{\mcitedefaultendpunct}{\mcitedefaultseppunct}\relax
\EndOfBibitem
\bibitem[{Gerin} et~al.(2010){Gerin}, {de Luca}, {Black}, {Goicoechea},
  {Herbst}, {Neufeld}, {Falgarone}, {Godard}, {Pearson}, {Lis}, {Phillips},
  {Bell}, {Sonnentrucker}, {Boulanger}, {Cernicharo}, {Coutens}, {Dartois},
  {Encrenaz}, {Giesen}, {Goldsmith}, {Gupta}, {Gry}, {Hennebelle},
  {Hily-Blant}, {Joblin}, {Kazmierczak}, {Kolos}, {Krelowski},
  {Martin-Pintado}, {Monje}, {Mookerjea}, {Perault}, {Persson}, {Plume},
  {Rimmer}, {Salez}, {Schmidt}, {Stutzki}, {Teyssier}, {Vastel}, {Yu},
  {Contursi}, {Menten}, {Geballe}, {Schlemmer}, {Shipman}, {Tielens},
  {Philipp-May}, {Cros}, {Zmuidzinas}, {Samoska}, {Klein}, and
  {Lorenzani}]{Gerin10}
{Gerin},~M. et~al.  \emph{\aap} \textbf{2010}, \emph{518}, L110\relax
\mciteBstWouldAddEndPuncttrue
\mciteSetBstMidEndSepPunct{\mcitedefaultmidpunct}
{\mcitedefaultendpunct}{\mcitedefaultseppunct}\relax
\EndOfBibitem
\bibitem[{Benz} et~al.(2010){Benz}, {Bruderer}, {van Dishoeck}, {St{\"a}uber},
  {Wampfler}, {Melchior}, {Dedes}, {Wyrowski}, {Doty}, {van der Tak},
  {B{\"a}chtold}, {Csillaghy}, {Megej}, {Monstein}, {Soldati}, {Bachiller},
  {Baudry}, {Benedettini}, {Bergin}, {Bjerkeli}, {Blake}, {Bontemps}, {Braine},
  {Caselli}, {Cernicharo}, {Codella}, {Daniel}, {di Giorgio}, {Dieleman},
  {Dominik}, {Encrenaz}, {Fich}, {Fuente}, {Giannini}, {Goicoechea}, {de
  Graauw}, {Helmich}, {Herczeg}, {Herpin}, {Hogerheijde}, {Jacq}, {Jellema},
  {Johnstone}, {J{\o}rgensen}, {Kristensen}, {Larsson}, {Lis}, {Liseau},
  {Marseille}, {McCoey}, {Melnick}, {Neufeld}, {Nisini}, {Olberg}, {Ossenkopf},
  {Parise}, {Pearson}, {Plume}, {Risacher}, {Santiago-Garc{\'{\i}}a},
  {Saraceno}, {Schieder}, {Shipman}, {Stutzki}, {Tafalla}, {Tielens}, {van
  Kempen}, {Visser}, and {Y{\i}ld{\i}z}]{Benz10}
{Benz},~A.~O. et~al.  \emph{\aap} \textbf{2010}, \emph{521}, L35\relax
\mciteBstWouldAddEndPuncttrue
\mciteSetBstMidEndSepPunct{\mcitedefaultmidpunct}
{\mcitedefaultendpunct}{\mcitedefaultseppunct}\relax
\EndOfBibitem
\bibitem[{Bruderer} et~al.(2010){Bruderer}, {Benz}, {van Dishoeck}, {Melchior},
  {Doty}, {van der Tak}, {St{\"a}uber}, {Wampfler}, {Dedes}, {Y{\i}ld{\i}z},
  {Pagani}, {Giannini}, {de Graauw}, {Whyborn}, {Teyssier}, {Jellema},
  {Shipman}, {Schieder}, {Honingh}, {Caux}, {B{\"a}chtold}, {Csillaghy},
  {Monstein}, {Bachiller}, {Baudry}, {Benedettini}, {Bergin}, {Bjerkeli},
  {Blake}, {Bontemps}, {Braine}, {Caselli}, {Cernicharo}, {Codella}, {Daniel},
  {di Giorgio}, {Dominik}, {Encrenaz}, {Fich}, {Fuente}, {Goicoechea},
  {Helmich}, {Herczeg}, {Herpin}, {Hogerheijde}, {Jacq}, {Johnstone},
  {J{\o}rgensen}, {Kristensen}, {Larsson}, {Lis}, {Liseau}, {Marseille},
  {McCoey}, {Melnick}, {Neufeld}, {Nisini}, {Olberg}, {Parise}, {Pearson},
  {Plume}, {Risacher}, {Santiago-Garc{\'{\i}}a}, {Saraceno}, {Shipman},
  {Tafalla}, {van Kempen}, {Visser}, and {Wyrowski}]{Bruderer10}
{Bruderer},~S. et~al.  \emph{\aap} \textbf{2010}, \emph{521}, L44\relax
\mciteBstWouldAddEndPuncttrue
\mciteSetBstMidEndSepPunct{\mcitedefaultmidpunct}
{\mcitedefaultendpunct}{\mcitedefaultseppunct}\relax
\EndOfBibitem
\bibitem[{Gupta} et~al.(2010){Gupta}, {Rimmer}, {Pearson}, {Yu}, {Herbst},
  {Harada}, {Bergin}, {Neufeld}, {Melnick}, {Bachiller}, {Baechtold}, {Bell},
  {Blake}, {Caux}, {Ceccarelli}, {Cernicharo}, {Chattopadhyay}, {Comito},
  {Cabrit}, {Crockett}, {Daniel}, {Falgarone}, {Diez-Gonzalez}, {Dubernet},
  {Erickson}, {Emprechtinger}, {Encrenaz}, {Gerin}, {Gill}, {Giesen},
  {Goicoechea}, {Goldsmith}, {Joblin}, {Johnstone}, {Langer}, {Larsson},
  {Latter}, {Lin}, {Lis}, {Liseau}, {Lord}, {Maiwald}, {Maret}, {Martin},
  {Martin-Pintado}, {Menten}, {Morris}, {M{\"u}ller}, {Murphy}, {Nordh},
  {Olberg}, {Ossenkopf}, {Pagani}, {P{\'e}rault}, {Phillips}, {Plume}, {Qin},
  {Salez}, {Samoska}, {Schilke}, {Schlecht}, {Schlemmer}, {Szczerba},
  {Stutzki}, {Trappe}, {van der Tak}, {Vastel}, {Wang}, {Yorke}, {Zmuidzinas},
  {Boogert}, {G{\"u}sten}, {Hartogh}, {Honingh}, {Karpov}, {Kooi}, {Krieg},
  {Schieder}, and {Zaal}]{Gupta10}
{Gupta},~H. et~al.  \emph{\aap} \textbf{2010}, \emph{521}, L47\relax
\mciteBstWouldAddEndPuncttrue
\mciteSetBstMidEndSepPunct{\mcitedefaultmidpunct}
{\mcitedefaultendpunct}{\mcitedefaultseppunct}\relax
\EndOfBibitem
\bibitem[{Wyrowski} et~al.(2010){Wyrowski}, {van der Tak}, {Herpin}, {Baudry},
  {Bontemps}, {Chavarria}, {Frieswijk}, {Jacq}, {Marseille}, {Shipman}, {van
  Dishoeck}, {Benz}, {Caselli}, {Hogerheijde}, {Johnstone}, {Liseau},
  {Bachiller}, {Benedettini}, {Bergin}, {Bjerkeli}, {Blake}, {Braine},
  {Bruderer}, {Cernicharo}, {Codella}, {Daniel}, {di Giorgio}, {Dominik},
  {Doty}, {Encrenaz}, {Fich}, {Fuente}, {Giannini}, {Goicoechea}, {de Graauw},
  {Helmich}, {Herczeg}, {J{\o}rgensen}, {Kristensen}, {Larsson}, {Lis},
  {McCoey}, {Melnick}, {Nisini}, {Olberg}, {Parise}, {Pearson}, {Plume},
  {Risacher}, {Santiago}, {Saraceno}, {Tafalla}, {van Kempen}, {Visser},
  {Wampfler}, {Y{\i}ld{\i}z}, {Black}, {Falgarone}, {Gerin}, {Roelfsema},
  {Dieleman}, {Beintema}, {de Jonge}, {Whyborn}, {Stutzki}, and
  {Ossenkopf}]{Wyrowski10}
{Wyrowski},~F. et~al.  \emph{\aap} \textbf{2010}, \emph{521}, L34\relax
\mciteBstWouldAddEndPuncttrue
\mciteSetBstMidEndSepPunct{\mcitedefaultmidpunct}
{\mcitedefaultendpunct}{\mcitedefaultseppunct}\relax
\EndOfBibitem
\bibitem[{Menten} et~al.(2011){Menten}, {Wyrowski}, {Belloche}, {G{\"u}sten},
  {Dedes}, and {M{\"u}ller}]{Menten11}
{Menten},~K.~M.; {Wyrowski},~F.; {Belloche},~A.; {G{\"u}sten},~R.; {Dedes},~L.;
  {M{\"u}ller},~H.~S.~P. \emph{\aap} \textbf{2011}, \emph{525}, A77\relax
\mciteBstWouldAddEndPuncttrue
\mciteSetBstMidEndSepPunct{\mcitedefaultmidpunct}
{\mcitedefaultendpunct}{\mcitedefaultseppunct}\relax
\EndOfBibitem
\bibitem[{Field} and {Steigman}(1971){Field}, and {Steigman}]{Steigman71}
{Field},~G.~B.; {Steigman},~G. \emph{\apj} \textbf{1971}, \emph{166}, 59\relax
\mciteBstWouldAddEndPuncttrue
\mciteSetBstMidEndSepPunct{\mcitedefaultmidpunct}
{\mcitedefaultendpunct}{\mcitedefaultseppunct}\relax
\EndOfBibitem
\bibitem[{Spirko} et~al.(2003){Spirko}, {Zirbel}, and {Hickman}]{Spirko03}
{Spirko},~J.~A.; {Zirbel},~J.~J.; {Hickman},~A.~P. \emph{\jpb} \textbf{2003},
  \emph{36}, 1645--1662\relax
\mciteBstWouldAddEndPuncttrue
\mciteSetBstMidEndSepPunct{\mcitedefaultmidpunct}
{\mcitedefaultendpunct}{\mcitedefaultseppunct}\relax
\EndOfBibitem
\bibitem[{Jensen} et~al.(2000){Jensen}, {Bilodeau}, {Safvan}, {Seiersen},
  {Andersen}, {Pedersen}, and {Heber}]{Jensen00}
{Jensen},~M.~J.; {Bilodeau},~R.~C.; {Safvan},~C.~P.; {Seiersen},~K.;
  {Andersen},~L.~H.; {Pedersen},~H.~B.; {Heber},~O. \emph{\apj} \textbf{2000},
  \emph{543}, 764--774\relax
\mciteBstWouldAddEndPuncttrue
\mciteSetBstMidEndSepPunct{\mcitedefaultmidpunct}
{\mcitedefaultendpunct}{\mcitedefaultseppunct}\relax
\EndOfBibitem
\bibitem[Buhr et~al.({2010})Buhr, Stuetzel, Mendes, Novotny, Schwalm, Berg,
  Bing, Grieser, Heber, Krantz, Menk, Novotny, Orlov, Petrignani, Rappaport,
  Repnow, Zajfman, and Wolf]{Buhr10}
Buhr,~H. et~al.  \emph{{\prl}} \textbf{{2010}}, \emph{{105}}\relax
\mciteBstWouldAddEndPuncttrue
\mciteSetBstMidEndSepPunct{\mcitedefaultmidpunct}
{\mcitedefaultendpunct}{\mcitedefaultseppunct}\relax
\EndOfBibitem
\bibitem[Neau et~al.({2000})Neau, Al~Khalili, Rosen, Le~Padellec, Derkatch,
  Shi, Vikor, Larsson, Semaniak, Thomas, Nagard, Andersson, Danared, and
  af~Ugglas]{Neau00}
Neau,~A.; Al~Khalili,~A.; Rosen,~S.; Le~Padellec,~A.; Derkatch,~A.; Shi,~W.;
  Vikor,~L.; Larsson,~M.; Semaniak,~J.; Thomas,~R.; Nagard,~M.; Andersson,~K.;
  Danared,~H.; af~Ugglas,~M. \emph{{\jcp}} \textbf{{2000}}, \emph{{113}},
  {1762--1770}\relax
\mciteBstWouldAddEndPuncttrue
\mciteSetBstMidEndSepPunct{\mcitedefaultmidpunct}
{\mcitedefaultendpunct}{\mcitedefaultseppunct}\relax
\EndOfBibitem
\bibitem[{Herd} et~al.(1990){Herd}, {Adams}, and {Smith}]{Herd90}
{Herd},~C.~R.; {Adams},~N.~G.; {Smith},~D. \emph{\apj} \textbf{1990},
  \emph{349}, 388--392\relax
\mciteBstWouldAddEndPuncttrue
\mciteSetBstMidEndSepPunct{\mcitedefaultmidpunct}
{\mcitedefaultendpunct}{\mcitedefaultseppunct}\relax
\EndOfBibitem
\bibitem[{Okabe}(1978)]{Okabe78}
{Okabe},~H. \emph{Photochemistry of Small Molecules}; Wiley, New York,
  1978\relax
\mciteBstWouldAddEndPuncttrue
\mciteSetBstMidEndSepPunct{\mcitedefaultmidpunct}
{\mcitedefaultendpunct}{\mcitedefaultseppunct}\relax
\EndOfBibitem
\bibitem[van Hemert and van~der Avoird({1979})van Hemert, and van~der
  Avoird]{vanHemert79}
van Hemert,~M.; van~der Avoird,~A. \emph{{\jcp}} \textbf{{1979}}, \emph{{71}},
  {5310--5323}\relax
\mciteBstWouldAddEndPuncttrue
\mciteSetBstMidEndSepPunct{\mcitedefaultmidpunct}
{\mcitedefaultendpunct}{\mcitedefaultseppunct}\relax
\EndOfBibitem
\bibitem[Engel et~al.({1988})Engel, Schinke, and Staemmler]{Engel88}
Engel,~V.; Schinke,~R.; Staemmler,~V. \emph{{\jcp}} \textbf{{1988}},
  \emph{{88}}, {129--148}\relax
\mciteBstWouldAddEndPuncttrue
\mciteSetBstMidEndSepPunct{\mcitedefaultmidpunct}
{\mcitedefaultendpunct}{\mcitedefaultseppunct}\relax
\EndOfBibitem
\bibitem[Engel et~al.({1992})Engel, Staemmler, VanderWal, Crim, Sension,
  Hudson, Andresen, Hennig, Weide, and Schinke]{Engel92}
Engel,~V.; Staemmler,~V.; VanderWal,~R.; Crim,~F.; Sension,~R.; Hudson,~B.;
  Andresen,~P.; Hennig,~S.; Weide,~K.; Schinke,~R. \emph{{\jpca}}
  \textbf{{1992}}, \emph{{96}}, {3201--3213}\relax
\mciteBstWouldAddEndPuncttrue
\mciteSetBstMidEndSepPunct{\mcitedefaultmidpunct}
{\mcitedefaultendpunct}{\mcitedefaultseppunct}\relax
\EndOfBibitem
\bibitem[{Schinke}(1993)]{Schinke93}
{Schinke},~R. \emph{Photodissociation Dynamics}; Cambridge University Press,
  Cambridge, 1993\relax
\mciteBstWouldAddEndPuncttrue
\mciteSetBstMidEndSepPunct{\mcitedefaultmidpunct}
{\mcitedefaultendpunct}{\mcitedefaultseppunct}\relax
\EndOfBibitem
\bibitem[van Harrevelt and van Hemert({2001})van Harrevelt, and van
  Hemert]{vanHarrevelt01}
van Harrevelt,~R.; van Hemert,~M. \emph{{\jcp}} \textbf{{2001}}, \emph{{114}},
  {9453--9462}\relax
\mciteBstWouldAddEndPuncttrue
\mciteSetBstMidEndSepPunct{\mcitedefaultmidpunct}
{\mcitedefaultendpunct}{\mcitedefaultseppunct}\relax
\EndOfBibitem
\bibitem[Hwang et~al.({1999})Hwang, Yang, Harich, Lin, and Yang]{Hwang99}
Hwang,~D.; Yang,~X.; Harich,~S.; Lin,~J.; Yang,~X. \emph{{\jcp}}
  \textbf{{1999}}, \emph{{110}}, {4123--4126}\relax
\mciteBstWouldAddEndPuncttrue
\mciteSetBstMidEndSepPunct{\mcitedefaultmidpunct}
{\mcitedefaultendpunct}{\mcitedefaultseppunct}\relax
\EndOfBibitem
\bibitem[Yang et~al.({2000})Yang, Hwang, Lin, and Ying]{Yang00}
Yang,~X.; Hwang,~D.; Lin,~J.; Ying,~X. \emph{{\jcp}} \textbf{{2000}},
  \emph{{113}}, {10597--10604}\relax
\mciteBstWouldAddEndPuncttrue
\mciteSetBstMidEndSepPunct{\mcitedefaultmidpunct}
{\mcitedefaultendpunct}{\mcitedefaultseppunct}\relax
\EndOfBibitem
\bibitem[van Harrevelt and van Hemert({2000})van Harrevelt, and van
  Hemert]{vanHarrevelt00}
van Harrevelt,~R.; van Hemert,~M. \emph{{\jcp}} \textbf{{2000}}, \emph{{112}},
  {5787--5808}\relax
\mciteBstWouldAddEndPuncttrue
\mciteSetBstMidEndSepPunct{\mcitedefaultmidpunct}
{\mcitedefaultendpunct}{\mcitedefaultseppunct}\relax
\EndOfBibitem
\bibitem[Harich et~al.({2000})Harich, Hwang, Yang, Lin, Yang, and
  Dixon]{Harich00}
Harich,~S.; Hwang,~D.; Yang,~X.; Lin,~J.; Yang,~X.; Dixon,~R. \emph{{\jcp}}
  \textbf{{2000}}, \emph{{113}}, {10073--10090}\relax
\mciteBstWouldAddEndPuncttrue
\mciteSetBstMidEndSepPunct{\mcitedefaultmidpunct}
{\mcitedefaultendpunct}{\mcitedefaultseppunct}\relax
\EndOfBibitem
\bibitem[van Harrevelt et~al.({2001})van Harrevelt, van Hemert, and
  Schatz]{vanHarrevelt01comp}
van Harrevelt,~R.; van Hemert,~M.; Schatz,~G. \emph{{\jpca}} \textbf{{2001}},
  \emph{{105}}, {11480--11487}\relax
\mciteBstWouldAddEndPuncttrue
\mciteSetBstMidEndSepPunct{\mcitedefaultmidpunct}
{\mcitedefaultendpunct}{\mcitedefaultseppunct}\relax
\EndOfBibitem
\bibitem[{van Dishoeck}(1988)]{vanDishoeck88photo}
{van Dishoeck},~E.~F. In \emph{Rate Coefficients in Astrochemistry};
  {Millar},~T.~J., {Williams},~D.~A., Eds.; Astrophysics and Space Science
  Library; 1988; Vol. 146; pp 49--72\relax
\mciteBstWouldAddEndPuncttrue
\mciteSetBstMidEndSepPunct{\mcitedefaultmidpunct}
{\mcitedefaultendpunct}{\mcitedefaultseppunct}\relax
\EndOfBibitem
\bibitem[{van Dishoeck}(2006)]{vanDishoeck06}
{van Dishoeck},~E.~F. \emph{\pnas} \textbf{2006}, \emph{103},
  12249--12256\relax
\mciteBstWouldAddEndPuncttrue
\mciteSetBstMidEndSepPunct{\mcitedefaultmidpunct}
{\mcitedefaultendpunct}{\mcitedefaultseppunct}\relax
\EndOfBibitem
\bibitem[{Aikawa} and {Herbst}(1999){Aikawa}, and {Herbst}]{Aikawa99}
{Aikawa},~Y.; {Herbst},~E. \emph{\aap} \textbf{1999}, \emph{351},
  233--246\relax
\mciteBstWouldAddEndPuncttrue
\mciteSetBstMidEndSepPunct{\mcitedefaultmidpunct}
{\mcitedefaultendpunct}{\mcitedefaultseppunct}\relax
\EndOfBibitem
\bibitem[{St{\"a}uber} et~al.(2006){St{\"a}uber}, {J{\o}rgensen}, {van
  Dishoeck}, {Doty}, and {Benz}]{Stauber06}
{St{\"a}uber},~P.; {J{\o}rgensen},~J.~K.; {van Dishoeck},~E.~F.; {Doty},~S.~D.;
  {Benz},~A.~O. \emph{\aap} \textbf{2006}, \emph{453}, 555--565\relax
\mciteBstWouldAddEndPuncttrue
\mciteSetBstMidEndSepPunct{\mcitedefaultmidpunct}
{\mcitedefaultendpunct}{\mcitedefaultseppunct}\relax
\EndOfBibitem
\bibitem[{Harada} et~al.(2010){Harada}, {Herbst}, and {Wakelam}]{Harada10}
{Harada},~N.; {Herbst},~E.; {Wakelam},~V. \emph{\apj} \textbf{2010},
  \emph{721}, 1570--1578, erratum 2012, {\it Astrophys. J.} 756, 104\relax
\mciteBstWouldAddEndPuncttrue
\mciteSetBstMidEndSepPunct{\mcitedefaultmidpunct}
{\mcitedefaultendpunct}{\mcitedefaultseppunct}\relax
\EndOfBibitem
\bibitem[Baulch et~al.({1992})Baulch, Cobos, Cox, Esser, Frank, Just, Kerr,
  Pilling, Troe, Walker, and Warnatz]{Baulch92}
Baulch,~D.; Cobos,~C.; Cox,~R.; Esser,~C.; Frank,~P.; Just,~T.; Kerr,~J.;
  Pilling,~M.; Troe,~J.; Walker,~R.; Warnatz,~J. \emph{{\jpcrd}}
  \textbf{{1992}}, \emph{{21}}, {411--734}\relax
\mciteBstWouldAddEndPuncttrue
\mciteSetBstMidEndSepPunct{\mcitedefaultmidpunct}
{\mcitedefaultendpunct}{\mcitedefaultseppunct}\relax
\EndOfBibitem
\bibitem[Balakrishnan({2004})]{Balakrishnan04}
Balakrishnan,~N. \emph{{\jcp}} \textbf{{2004}}, \emph{{121}},
  {6346--6352}\relax
\mciteBstWouldAddEndPuncttrue
\mciteSetBstMidEndSepPunct{\mcitedefaultmidpunct}
{\mcitedefaultendpunct}{\mcitedefaultseppunct}\relax
\EndOfBibitem
\bibitem[{Charnley}(1997)]{Charnley97}
{Charnley},~S.~B. \emph{\apj} \textbf{1997}, \emph{481}, 396\relax
\mciteBstWouldAddEndPuncttrue
\mciteSetBstMidEndSepPunct{\mcitedefaultmidpunct}
{\mcitedefaultendpunct}{\mcitedefaultseppunct}\relax
\EndOfBibitem
\bibitem[Javoy et~al.({2003})Javoy, Naudet, Abid, and Paillard]{Javoy03}
Javoy,~S.; Naudet,~V.; Abid,~S.; Paillard,~C. \emph{{\etfs}} \textbf{{2003}},
  \emph{{27}}, {371}\relax
\mciteBstWouldAddEndPuncttrue
\mciteSetBstMidEndSepPunct{\mcitedefaultmidpunct}
{\mcitedefaultendpunct}{\mcitedefaultseppunct}\relax
\EndOfBibitem
\bibitem[{Atkinson} et~al.(2004){Atkinson}, {Baulch}, {Cox}, {Crowley},
  {Hampson}, and {Hynes}]{Atkinson04}
{Atkinson},~R.; {Baulch},~D.; {Cox},~R.; {Crowley},~J.; {Hampson},~R.;
  {Hynes},~R. \emph{\acp} \textbf{2004}, \emph{4}, 1461\relax
\mciteBstWouldAddEndPuncttrue
\mciteSetBstMidEndSepPunct{\mcitedefaultmidpunct}
{\mcitedefaultendpunct}{\mcitedefaultseppunct}\relax
\EndOfBibitem
\bibitem[Sultanov and Balakrishnan({2004})Sultanov, and
  Balakrishnan]{Sultanov04}
Sultanov,~R.; Balakrishnan,~N. \emph{{\jcp}} \textbf{{2004}}, \emph{{121}},
  {11038--11044}\relax
\mciteBstWouldAddEndPuncttrue
\mciteSetBstMidEndSepPunct{\mcitedefaultmidpunct}
{\mcitedefaultendpunct}{\mcitedefaultseppunct}\relax
\EndOfBibitem
\bibitem[Han et~al.({2000})Han, Chen, and Weiner]{Han00}
Han,~J.; Chen,~X.; Weiner,~B. \emph{{\cpl}} \textbf{{2000}}, \emph{{332}},
  {243--250}\relax
\mciteBstWouldAddEndPuncttrue
\mciteSetBstMidEndSepPunct{\mcitedefaultmidpunct}
{\mcitedefaultendpunct}{\mcitedefaultseppunct}\relax
\EndOfBibitem
\bibitem[{Draine} et~al.(1983){Draine}, {Roberge}, and {Dalgarno}]{Draine83}
{Draine},~B.~T.; {Roberge},~W.~G.; {Dalgarno},~A. \emph{\apj} \textbf{1983},
  \emph{264}, 485--507\relax
\mciteBstWouldAddEndPuncttrue
\mciteSetBstMidEndSepPunct{\mcitedefaultmidpunct}
{\mcitedefaultendpunct}{\mcitedefaultseppunct}\relax
\EndOfBibitem
\bibitem[{Bergin} et~al.(1998){Bergin}, {Neufeld}, and {Melnick}]{Bergin98}
{Bergin},~E.~A.; {Neufeld},~D.~A.; {Melnick},~G.~J. \emph{\apj} \textbf{1998},
  \emph{499}, 777\relax
\mciteBstWouldAddEndPuncttrue
\mciteSetBstMidEndSepPunct{\mcitedefaultmidpunct}
{\mcitedefaultendpunct}{\mcitedefaultseppunct}\relax
\EndOfBibitem
\bibitem[Cuppen et~al.(2010)Cuppen, Ioppolo, Romanzin, and Linnartz]{Cuppen10}
Cuppen,~H.~M.; Ioppolo,~S.; Romanzin,~C.; Linnartz,~H. \emph{Phys. Chem. Chem.
  Phys.} \textbf{2010}, \emph{12}, 12077--12088\relax
\mciteBstWouldAddEndPuncttrue
\mciteSetBstMidEndSepPunct{\mcitedefaultmidpunct}
{\mcitedefaultendpunct}{\mcitedefaultseppunct}\relax
\EndOfBibitem
\bibitem[Lamberts et~al.(2013)Lamberts, Cuppen, Ioppolo, and
  Linnartz]{Lamberts13}
Lamberts,~T.; Cuppen,~H.~M.; Ioppolo,~S.; Linnartz,~H. \emph{{\pccp}}
  \textbf{2013}, \emph{15}, 8287--8302\relax
\mciteBstWouldAddEndPuncttrue
\mciteSetBstMidEndSepPunct{\mcitedefaultmidpunct}
{\mcitedefaultendpunct}{\mcitedefaultseppunct}\relax
\EndOfBibitem
\bibitem[{{\"O}berg} et~al.(2011){{\"O}berg}, {Boogert}, {Pontoppidan}, {van
  den Broek}, {van Dishoeck}, {Bottinelli}, {Blake}, and {Evans}]{Oberg11}
{{\"O}berg},~K.~I.; {Boogert},~A.~C.~A.; {Pontoppidan},~K.~M.; {van den
  Broek},~S.; {van Dishoeck},~E.~F.; {Bottinelli},~S.; {Blake},~G.~A.;
  {Evans},~N.~J.,~II \emph{\apj} \textbf{2011}, \emph{740}, 109\relax
\mciteBstWouldAddEndPuncttrue
\mciteSetBstMidEndSepPunct{\mcitedefaultmidpunct}
{\mcitedefaultendpunct}{\mcitedefaultseppunct}\relax
\EndOfBibitem
\bibitem[{Oba} et~al.(2012){Oba}, {Watanabe}, {Hama}, {Kuwahata}, {Hidaka}, and
  {Kouchi}]{Oba12}
{Oba},~Y.; {Watanabe},~N.; {Hama},~T.; {Kuwahata},~K.; {Hidaka},~H.;
  {Kouchi},~A. \emph{\apj} \textbf{2012}, \emph{749}, 67\relax
\mciteBstWouldAddEndPuncttrue
\mciteSetBstMidEndSepPunct{\mcitedefaultmidpunct}
{\mcitedefaultendpunct}{\mcitedefaultseppunct}\relax
\EndOfBibitem
\bibitem[{Fayolle} et~al.(2011){Fayolle}, {{\"O}berg}, {Cuppen}, {Visser}, and
  {Linnartz}]{Fayolle11co2}
{Fayolle},~E.~C.; {{\"O}berg},~K.~I.; {Cuppen},~H.~M.; {Visser},~R.;
  {Linnartz},~H. \emph{\aap} \textbf{2011}, \emph{529}, A74\relax
\mciteBstWouldAddEndPuncttrue
\mciteSetBstMidEndSepPunct{\mcitedefaultmidpunct}
{\mcitedefaultendpunct}{\mcitedefaultseppunct}\relax
\EndOfBibitem
\bibitem[{Buch} and {Zhang}(1991){Buch}, and {Zhang}]{Buch91}
{Buch},~V.; {Zhang},~Q. \emph{\apj} \textbf{1991}, \emph{379}, 647--652\relax
\mciteBstWouldAddEndPuncttrue
\mciteSetBstMidEndSepPunct{\mcitedefaultmidpunct}
{\mcitedefaultendpunct}{\mcitedefaultseppunct}\relax
\EndOfBibitem
\bibitem[{Al-Halabi} and {van Dishoeck}(2007){Al-Halabi}, and {van
  Dishoeck}]{Alhalabi07}
{Al-Halabi},~A.; {van Dishoeck},~E.~F. \emph{\mnras} \textbf{2007}, \emph{382},
  1648--1656\relax
\mciteBstWouldAddEndPuncttrue
\mciteSetBstMidEndSepPunct{\mcitedefaultmidpunct}
{\mcitedefaultendpunct}{\mcitedefaultseppunct}\relax
\EndOfBibitem
\bibitem[{{\"O}berg} et~al.(2009){{\"O}berg}, {Linnartz}, {Visser}, and {van
  Dishoeck}]{Oberg09h2o}
{{\"O}berg},~K.~I.; {Linnartz},~H.; {Visser},~R.; {van Dishoeck},~E.~F.
  \emph{\apj} \textbf{2009}, \emph{693}, 1209--1218\relax
\mciteBstWouldAddEndPuncttrue
\mciteSetBstMidEndSepPunct{\mcitedefaultmidpunct}
{\mcitedefaultendpunct}{\mcitedefaultseppunct}\relax
\EndOfBibitem
\bibitem[{Fayolle} et~al.(2011){Fayolle}, {Bertin}, {Romanzin}, {Michaut},
  {{\"O}berg}, {Linnartz}, and {Fillion}]{Fayolle11}
{Fayolle},~E.~C.; {Bertin},~M.; {Romanzin},~C.; {Michaut},~X.;
  {{\"O}berg},~K.~I.; {Linnartz},~H.; {Fillion},~J.-H. \emph{\apjl}
  \textbf{2011}, \emph{739}, L36\relax
\mciteBstWouldAddEndPuncttrue
\mciteSetBstMidEndSepPunct{\mcitedefaultmidpunct}
{\mcitedefaultendpunct}{\mcitedefaultseppunct}\relax
\EndOfBibitem
\bibitem[{Dulieu} et~al.(2013){Dulieu}, {Congiu}, {Noble}, {Baouche},
  {Chaabouni}, {Moudens}, {Minissale}, and {Cazaux}]{Dulieu13}
{Dulieu},~F.; {Congiu},~E.; {Noble},~J.; {Baouche},~S.; {Chaabouni},~H.;
  {Moudens},~A.; {Minissale},~M.; {Cazaux},~S. \emph{\natr} \textbf{2013},
  \emph{3}\relax
\mciteBstWouldAddEndPuncttrue
\mciteSetBstMidEndSepPunct{\mcitedefaultmidpunct}
{\mcitedefaultendpunct}{\mcitedefaultseppunct}\relax
\EndOfBibitem
\bibitem[{Ioppolo} et~al.(2008){Ioppolo}, {Cuppen}, {Romanzin}, {van Dishoeck},
  and {Linnartz}]{Ioppolo08}
{Ioppolo},~S.; {Cuppen},~H.~M.; {Romanzin},~C.; {van Dishoeck},~E.~F.;
  {Linnartz},~H. \emph{\apj} \textbf{2008}, \emph{686}, 1474--1479\relax
\mciteBstWouldAddEndPuncttrue
\mciteSetBstMidEndSepPunct{\mcitedefaultmidpunct}
{\mcitedefaultendpunct}{\mcitedefaultseppunct}\relax
\EndOfBibitem
\bibitem[{Cuppen} et~al.(2009){Cuppen}, {van Dishoeck}, {Herbst}, and
  {Tielens}]{Cuppen09}
{Cuppen},~H.~M.; {van Dishoeck},~E.~F.; {Herbst},~E.; {Tielens},~A.~G.~G.~M.
  \emph{\aap} \textbf{2009}, \emph{508}, 275--287\relax
\mciteBstWouldAddEndPuncttrue
\mciteSetBstMidEndSepPunct{\mcitedefaultmidpunct}
{\mcitedefaultendpunct}{\mcitedefaultseppunct}\relax
\EndOfBibitem
\bibitem[{Ward} and {Price}(2011){Ward}, and {Price}]{Ward11}
{Ward},~M.~D.; {Price},~S.~D. \emph{\apj} \textbf{2011}, \emph{741}, 121\relax
\mciteBstWouldAddEndPuncttrue
\mciteSetBstMidEndSepPunct{\mcitedefaultmidpunct}
{\mcitedefaultendpunct}{\mcitedefaultseppunct}\relax
\EndOfBibitem
\bibitem[Ioppolo et~al.(2010)Ioppolo, Cuppen, Romanzin, van Dishoeck, and
  Linnartz]{Ioppolo10o2}
Ioppolo,~S.; Cuppen,~H.~M.; Romanzin,~C.; van Dishoeck,~E.~F.; Linnartz,~H.
  \emph{Phys. Chem. Chem. Phys} \textbf{2010}, \emph{12}, 12065--12076\relax
\mciteBstWouldAddEndPuncttrue
\mciteSetBstMidEndSepPunct{\mcitedefaultmidpunct}
{\mcitedefaultendpunct}{\mcitedefaultseppunct}\relax
\EndOfBibitem
\bibitem[{Miyauchi} et~al.(2008){Miyauchi}, {Hidaka}, {Chigai}, {Nagaoka},
  {Watanabe}, and {Kouchi}]{Miyauchi08}
{Miyauchi},~N.; {Hidaka},~H.; {Chigai},~T.; {Nagaoka},~A.; {Watanabe},~N.;
  {Kouchi},~A. \emph{Chem. Phys. Lett.} \textbf{2008}, \emph{456}, 27--30\relax
\mciteBstWouldAddEndPuncttrue
\mciteSetBstMidEndSepPunct{\mcitedefaultmidpunct}
{\mcitedefaultendpunct}{\mcitedefaultseppunct}\relax
\EndOfBibitem
\bibitem[{Dulieu} et~al.(2010){Dulieu}, {Amiaud}, {Congiu}, {Fillion}, {Matar},
  {Momeni}, {Pirronello}, and {Lemaire}]{Dulieu10}
{Dulieu},~F.; {Amiaud},~L.; {Congiu},~E.; {Fillion},~J.; {Matar},~E.;
  {Momeni},~A.; {Pirronello},~V.; {Lemaire},~J.~L. \emph{\aap} \textbf{2010},
  \emph{512}, A30\relax
\mciteBstWouldAddEndPuncttrue
\mciteSetBstMidEndSepPunct{\mcitedefaultmidpunct}
{\mcitedefaultendpunct}{\mcitedefaultseppunct}\relax
\EndOfBibitem
\bibitem[Romanzin et~al.({2011})Romanzin, Ioppolo, Cuppen, van Dishoeck, and
  Linnartz]{Romanzin11}
Romanzin,~C.; Ioppolo,~S.; Cuppen,~H.~M.; van Dishoeck,~E.~F.; Linnartz,~H.
  \emph{{\jcp}} \textbf{{2011}}, \emph{{134}}\relax
\mciteBstWouldAddEndPuncttrue
\mciteSetBstMidEndSepPunct{\mcitedefaultmidpunct}
{\mcitedefaultendpunct}{\mcitedefaultseppunct}\relax
\EndOfBibitem
\bibitem[Accolla et~al.({2013})Accolla, Congiu, Manico, Dulieu, Chaabouni,
  Lemaire, and Pirronello]{Accolla13}
Accolla,~M.; Congiu,~E.; Manico,~G.; Dulieu,~F.; Chaabouni,~H.; Lemaire,~J.~L.;
  Pirronello,~V. \emph{{\mnras}} \textbf{{2013}}, \emph{{429}},
  {3200--3206}\relax
\mciteBstWouldAddEndPuncttrue
\mciteSetBstMidEndSepPunct{\mcitedefaultmidpunct}
{\mcitedefaultendpunct}{\mcitedefaultseppunct}\relax
\EndOfBibitem
\bibitem[{Dulieu}(2011)]{Dulieu11}
{Dulieu},~F. In \emph{IAU Symposium}; {Cernicharo},~J., {Bachiller},~R., Eds.;
  IAU Symposium; Cambridge University Press: Cambridge, 2011; Vol. 280; p
  405\relax
\mciteBstWouldAddEndPuncttrue
\mciteSetBstMidEndSepPunct{\mcitedefaultmidpunct}
{\mcitedefaultendpunct}{\mcitedefaultseppunct}\relax
\EndOfBibitem
\bibitem[{Cuppen} and {Herbst}(2007){Cuppen}, and {Herbst}]{Cuppen07}
{Cuppen},~H.~M.; {Herbst},~E. \emph{\apj} \textbf{2007}, \emph{668},
  294--309\relax
\mciteBstWouldAddEndPuncttrue
\mciteSetBstMidEndSepPunct{\mcitedefaultmidpunct}
{\mcitedefaultendpunct}{\mcitedefaultseppunct}\relax
\EndOfBibitem
\bibitem[{Du} et~al.(2012){Du}, {Parise}, and {Bergman}]{Du12}
{Du},~F.; {Parise},~B.; {Bergman},~P. \emph{\aap} \textbf{2012}, \emph{538},
  A91\relax
\mciteBstWouldAddEndPuncttrue
\mciteSetBstMidEndSepPunct{\mcitedefaultmidpunct}
{\mcitedefaultendpunct}{\mcitedefaultseppunct}\relax
\EndOfBibitem
\bibitem[{Sandford} and {Allamandola}(1990){Sandford}, and
  {Allamandola}]{Sandford90}
{Sandford},~S.~A.; {Allamandola},~L.~J. \emph{\icarus} \textbf{1990},
  \emph{87}, 188--192\relax
\mciteBstWouldAddEndPuncttrue
\mciteSetBstMidEndSepPunct{\mcitedefaultmidpunct}
{\mcitedefaultendpunct}{\mcitedefaultseppunct}\relax
\EndOfBibitem
\bibitem[{Collings} et~al.(2004){Collings}, {Anderson}, {Chen}, {Dever},
  {Viti}, {Williams}, and {McCoustra}]{Collings04}
{Collings},~M.~P.; {Anderson},~M.~A.; {Chen},~R.; {Dever},~J.~W.; {Viti},~S.;
  {Williams},~D.~A.; {McCoustra},~M.~R.~S. \emph{\mnras} \textbf{2004},
  \emph{354}, 1133--1140\relax
\mciteBstWouldAddEndPuncttrue
\mciteSetBstMidEndSepPunct{\mcitedefaultmidpunct}
{\mcitedefaultendpunct}{\mcitedefaultseppunct}\relax
\EndOfBibitem
\bibitem[{Collings} and {McCoustra}(2005){Collings}, and
  {McCoustra}]{McCoustra05}
{Collings},~M.~P.; {McCoustra},~M.~R.~S. In \emph{Astrochemistry: Recent
  Successes and Current Challenges}; {Lis},~D.~C., {Blake},~G.~A.,
  {Herbst},~E., Eds.; IAU Symposium; Cambridge Univ. Press: Cambridge, 2005;
  Vol. 231; pp 405--414\relax
\mciteBstWouldAddEndPuncttrue
\mciteSetBstMidEndSepPunct{\mcitedefaultmidpunct}
{\mcitedefaultendpunct}{\mcitedefaultseppunct}\relax
\EndOfBibitem
\bibitem[{Westley} et~al.(1995){Westley}, {Baragiola}, {Johnson}, and
  {Baratta}]{Westley95}
{Westley},~M.~S.; {Baragiola},~R.~A.; {Johnson},~R.~E.; {Baratta},~G.~A.
  \emph{\nat} \textbf{1995}, \emph{373}, 405--407\relax
\mciteBstWouldAddEndPuncttrue
\mciteSetBstMidEndSepPunct{\mcitedefaultmidpunct}
{\mcitedefaultendpunct}{\mcitedefaultseppunct}\relax
\EndOfBibitem
\bibitem[{Arasa} et~al.(2010){Arasa}, {Andersson}, {Cuppen}, {van Dishoeck},
  and {Kroes}]{Arasa10}
{Arasa},~C.; {Andersson},~S.; {Cuppen},~H.~M.; {van Dishoeck},~E.~F.;
  {Kroes},~G.-J. \emph{\jcp} \textbf{2010}, \emph{132}, 184510\relax
\mciteBstWouldAddEndPuncttrue
\mciteSetBstMidEndSepPunct{\mcitedefaultmidpunct}
{\mcitedefaultendpunct}{\mcitedefaultseppunct}\relax
\EndOfBibitem
\bibitem[{Arasa} et~al.(2011){Arasa}, {Andersson}, {Cuppen}, {van Dishoeck},
  and {Kroes}]{Arasa11}
{Arasa},~C.; {Andersson},~S.; {Cuppen},~H.~M.; {van Dishoeck},~E.~F.;
  {Kroes},~G.~J. \emph{\jcp} \textbf{2011}, \emph{134}, 164503\relax
\mciteBstWouldAddEndPuncttrue
\mciteSetBstMidEndSepPunct{\mcitedefaultmidpunct}
{\mcitedefaultendpunct}{\mcitedefaultseppunct}\relax
\EndOfBibitem
\bibitem[Koning et~al.({2013})Koning, Kroes, and Arasa]{Koning13}
Koning,~J.; Kroes,~G.~J.; Arasa,~C. \emph{{\jcp}} \textbf{{2013}},
  \emph{{138}}\relax
\mciteBstWouldAddEndPuncttrue
\mciteSetBstMidEndSepPunct{\mcitedefaultmidpunct}
{\mcitedefaultendpunct}{\mcitedefaultseppunct}\relax
\EndOfBibitem
\bibitem[{Arasa} et~al.(2013){Arasa}, {Koning}, {Kroes}, {Walsh}, and {van
  Dishoeck}]{Arasa13}
{Arasa},~C.; {Koning},~J.; {Kroes},~G.; {Walsh},~C.; {van Dishoeck},~E.
  \emph{\aap} \textbf{2013}, submitted\relax
\mciteBstWouldAddEndPuncttrue
\mciteSetBstMidEndSepPunct{\mcitedefaultmidpunct}
{\mcitedefaultendpunct}{\mcitedefaultseppunct}\relax
\EndOfBibitem
\bibitem[{Hollenbach} et~al.(2012){Hollenbach}, {Kaufman}, {Neufeld},
  {Wolfire}, and {Goicoechea}]{Hollenbach12}
{Hollenbach},~D.; {Kaufman},~M.~J.; {Neufeld},~D.; {Wolfire},~M.;
  {Goicoechea},~J.~R. \emph{\apj} \textbf{2012}, \emph{754}, 105\relax
\mciteBstWouldAddEndPuncttrue
\mciteSetBstMidEndSepPunct{\mcitedefaultmidpunct}
{\mcitedefaultendpunct}{\mcitedefaultseppunct}\relax
\EndOfBibitem
\bibitem[{Hollenbach} and {Tielens}(1997){Hollenbach}, and
  {Tielens}]{Hollenbach97}
{Hollenbach},~D.~J.; {Tielens},~A.~G.~G.~M. \emph{\araa} \textbf{1997},
  \emph{35}, 179--216\relax
\mciteBstWouldAddEndPuncttrue
\mciteSetBstMidEndSepPunct{\mcitedefaultmidpunct}
{\mcitedefaultendpunct}{\mcitedefaultseppunct}\relax
\EndOfBibitem
\bibitem[{Aikawa} et~al.(2002){Aikawa}, {van Zadelhoff}, {van Dishoeck}, and
  {Herbst}]{Aikawa02}
{Aikawa},~Y.; {van Zadelhoff},~G.~J.; {van Dishoeck},~E.~F.; {Herbst},~E.
  \emph{\aap} \textbf{2002}, \emph{386}, 622--632\relax
\mciteBstWouldAddEndPuncttrue
\mciteSetBstMidEndSepPunct{\mcitedefaultmidpunct}
{\mcitedefaultendpunct}{\mcitedefaultseppunct}\relax
\EndOfBibitem
\bibitem[{Lee} et~al.(2004){Lee}, {Bergin}, and {Evans}]{Lee04}
{Lee},~J.; {Bergin},~E.~A.; {Evans},~N.~J.,~II \emph{\apj} \textbf{2004},
  \emph{617}, 360--383\relax
\mciteBstWouldAddEndPuncttrue
\mciteSetBstMidEndSepPunct{\mcitedefaultmidpunct}
{\mcitedefaultendpunct}{\mcitedefaultseppunct}\relax
\EndOfBibitem
\bibitem[{Aikawa} et~al.(2008){Aikawa}, {Wakelam}, {Garrod}, and
  {Herbst}]{Aikawa08}
{Aikawa},~Y.; {Wakelam},~V.; {Garrod},~R.~T.; {Herbst},~E. \emph{\apj}
  \textbf{2008}, \emph{674}, 984--996\relax
\mciteBstWouldAddEndPuncttrue
\mciteSetBstMidEndSepPunct{\mcitedefaultmidpunct}
{\mcitedefaultendpunct}{\mcitedefaultseppunct}\relax
\EndOfBibitem
\bibitem[{Visser} et~al.(2009){Visser}, {van Dishoeck}, {Doty}, and
  {Dullemond}]{Visser09}
{Visser},~R.; {van Dishoeck},~E.~F.; {Doty},~S.~D.; {Dullemond},~C.~P.
  \emph{\aap} \textbf{2009}, \emph{495}, 881--897\relax
\mciteBstWouldAddEndPuncttrue
\mciteSetBstMidEndSepPunct{\mcitedefaultmidpunct}
{\mcitedefaultendpunct}{\mcitedefaultseppunct}\relax
\EndOfBibitem
\bibitem[{Aikawa} et~al.(2012){Aikawa}, {Wakelam}, {Hersant}, {Garrod}, and
  {Herbst}]{Aikawa12}
{Aikawa},~Y.; {Wakelam},~V.; {Hersant},~F.; {Garrod},~R.~T.; {Herbst},~E.
  \emph{\apj} \textbf{2012}, \emph{760}, 40\relax
\mciteBstWouldAddEndPuncttrue
\mciteSetBstMidEndSepPunct{\mcitedefaultmidpunct}
{\mcitedefaultendpunct}{\mcitedefaultseppunct}\relax
\EndOfBibitem
\bibitem[{Wakelam} et~al.(2010){Wakelam}, {Herbst}, {Le Bourlot}, {Hersant},
  {Selsis}, and {Guilloteau}]{Wakelam10b}
{Wakelam},~V.; {Herbst},~E.; {Le Bourlot},~J.; {Hersant},~F.; {Selsis},~F.;
  {Guilloteau},~S. \emph{\aap} \textbf{2010}, \emph{517}, A21\relax
\mciteBstWouldAddEndPuncttrue
\mciteSetBstMidEndSepPunct{\mcitedefaultmidpunct}
{\mcitedefaultendpunct}{\mcitedefaultseppunct}\relax
\EndOfBibitem
\bibitem[{Wakelam} et~al.(2006){Wakelam}, {Herbst}, and {Selsis}]{Wakelam06}
{Wakelam},~V.; {Herbst},~E.; {Selsis},~F. \emph{\aap} \textbf{2006},
  \emph{451}, 551--562\relax
\mciteBstWouldAddEndPuncttrue
\mciteSetBstMidEndSepPunct{\mcitedefaultmidpunct}
{\mcitedefaultendpunct}{\mcitedefaultseppunct}\relax
\EndOfBibitem
\bibitem[{Hasegawa} and {Herbst}(1993){Hasegawa}, and {Herbst}]{Hasegawa93}
{Hasegawa},~T.~I.; {Herbst},~E. \emph{\mnras} \textbf{1993}, \emph{263},
  589\relax
\mciteBstWouldAddEndPuncttrue
\mciteSetBstMidEndSepPunct{\mcitedefaultmidpunct}
{\mcitedefaultendpunct}{\mcitedefaultseppunct}\relax
\EndOfBibitem
\bibitem[{Caselli} et~al.(1998){Caselli}, {Hasegawa}, and {Herbst}]{Caselli98}
{Caselli},~P.; {Hasegawa},~T.~I.; {Herbst},~E. \emph{\apj} \textbf{1998},
  \emph{495}, 309\relax
\mciteBstWouldAddEndPuncttrue
\mciteSetBstMidEndSepPunct{\mcitedefaultmidpunct}
{\mcitedefaultendpunct}{\mcitedefaultseppunct}\relax
\EndOfBibitem
\bibitem[{Garrod}(2008)]{Garrod08}
{Garrod},~R.~T. \emph{\aap} \textbf{2008}, \emph{491}, 239--251\relax
\mciteBstWouldAddEndPuncttrue
\mciteSetBstMidEndSepPunct{\mcitedefaultmidpunct}
{\mcitedefaultendpunct}{\mcitedefaultseppunct}\relax
\EndOfBibitem
\bibitem[{Garrod} et~al.(2009){Garrod}, {Vasyunin}, {Semenov}, {Wiebe}, and
  {Henning}]{Garrod09}
{Garrod},~R.~T.; {Vasyunin},~A.~I.; {Semenov},~D.~A.; {Wiebe},~D.~S.;
  {Henning},~T. \emph{\apjl} \textbf{2009}, \emph{700}, L43--L46\relax
\mciteBstWouldAddEndPuncttrue
\mciteSetBstMidEndSepPunct{\mcitedefaultmidpunct}
{\mcitedefaultendpunct}{\mcitedefaultseppunct}\relax
\EndOfBibitem
\bibitem[{Stantcheva} et~al.(2002){Stantcheva}, {Shematovich}, and
  {Herbst}]{Stantcheva02}
{Stantcheva},~T.; {Shematovich},~V.~I.; {Herbst},~E. \emph{\aap} \textbf{2002},
  \emph{391}, 1069--1080\relax
\mciteBstWouldAddEndPuncttrue
\mciteSetBstMidEndSepPunct{\mcitedefaultmidpunct}
{\mcitedefaultendpunct}{\mcitedefaultseppunct}\relax
\EndOfBibitem
\bibitem[{Caselli} et~al.(2002){Caselli}, {Stantcheva}, {Shalabiea},
  {Shematovich}, and {Herbst}]{Caselli02model}
{Caselli},~P.; {Stantcheva},~T.; {Shalabiea},~O.; {Shematovich},~V.~I.;
  {Herbst},~E. \emph{\planss} \textbf{2002}, \emph{50}, 1257--1266\relax
\mciteBstWouldAddEndPuncttrue
\mciteSetBstMidEndSepPunct{\mcitedefaultmidpunct}
{\mcitedefaultendpunct}{\mcitedefaultseppunct}\relax
\EndOfBibitem
\bibitem[{Du} and {Parise}(2011){Du}, and {Parise}]{Du11}
{Du},~F.; {Parise},~B. \emph{\aap} \textbf{2011}, \emph{530}, A131\relax
\mciteBstWouldAddEndPuncttrue
\mciteSetBstMidEndSepPunct{\mcitedefaultmidpunct}
{\mcitedefaultendpunct}{\mcitedefaultseppunct}\relax
\EndOfBibitem
\bibitem[{Chang} and {Herbst}(2012){Chang}, and {Herbst}]{Chang12}
{Chang},~Q.; {Herbst},~E. \emph{\apj} \textbf{2012}, \emph{759}, 147\relax
\mciteBstWouldAddEndPuncttrue
\mciteSetBstMidEndSepPunct{\mcitedefaultmidpunct}
{\mcitedefaultendpunct}{\mcitedefaultseppunct}\relax
\EndOfBibitem
\bibitem[{Vasyunin} and {Herbst}(2013){Vasyunin}, and {Herbst}]{Vasyunin13}
{Vasyunin},~A.~I.; {Herbst},~E. \emph{\apj} \textbf{2013}, \emph{762}, 86\relax
\mciteBstWouldAddEndPuncttrue
\mciteSetBstMidEndSepPunct{\mcitedefaultmidpunct}
{\mcitedefaultendpunct}{\mcitedefaultseppunct}\relax
\EndOfBibitem
\bibitem[{van Dishoeck} et~al.(2013){van Dishoeck}, {Bergin}, {Lis}, and
  {Lunine}]{vanDishoeck14}
{van Dishoeck},~E.; {Bergin},~E.; {Lis},~D.; {Lunine},~J. In \emph{Protostars
  \& Planets VI}; {Beuther, H., Klessen, R., Dullemond, K., Henning, Th.},,
  Ed.; Univ. Arizona Press: Tucson, 2013\relax
\mciteBstWouldAddEndPuncttrue
\mciteSetBstMidEndSepPunct{\mcitedefaultmidpunct}
{\mcitedefaultendpunct}{\mcitedefaultseppunct}\relax
\EndOfBibitem
\bibitem[{Kristensen} and {van Dishoeck}(2011){Kristensen}, and {van
  Dishoeck}]{Kristensen11an}
{Kristensen},~L.~E.; {van Dishoeck},~E.~F. \emph{Astronomische Nachrichten}
  \textbf{2011}, \emph{332}, 475\relax
\mciteBstWouldAddEndPuncttrue
\mciteSetBstMidEndSepPunct{\mcitedefaultmidpunct}
{\mcitedefaultendpunct}{\mcitedefaultseppunct}\relax
\EndOfBibitem
\bibitem[{Nisini} et~al.(2010){Nisini}, {Benedettini}, {Codella}, {Giannini},
  {Liseau}, {Neufeld}, {Tafalla}, {van Dishoeck}, {Bachiller}, {Baudry},
  {Benz}, {Bergin}, {Bjerkeli}, {Blake}, {Bontemps}, {Braine}, {Bruderer},
  {Caselli}, {Cernicharo}, {Daniel}, {Encrenaz}, {di Giorgio}, {Dominik},
  {Doty}, {Fich}, {Fuente}, {Goicoechea}, {de Graauw}, {Helmich}, {Herczeg},
  {Herpin}, {Hogerheijde}, {Jacq}, {Johnstone}, {J{\o}rgensen}, {Kaufman},
  {Kristensen}, {Larsson}, {Lis}, {Marseille}, {McCoey}, {Melnick}, {Olberg},
  {Parise}, {Pearson}, {Plume}, {Risacher}, {Santiago}, {Saraceno}, {Shipman},
  {van Kempen}, {Visser}, {Viti}, {Wampfler}, {Wyrowski}, {van der Tak},
  {Y{\i}ld{\i}z}, {Delforge}, {Desbat}, {Hatch}, {P{\'e}ron}, {Schieder},
  {Stern}, {Teyssier}, and {Whyborn}]{Nisini10}
{Nisini},~B. et~al.  \emph{\aap} \textbf{2010}, \emph{518}, L120\relax
\mciteBstWouldAddEndPuncttrue
\mciteSetBstMidEndSepPunct{\mcitedefaultmidpunct}
{\mcitedefaultendpunct}{\mcitedefaultseppunct}\relax
\EndOfBibitem
\bibitem[{Caselli} et~al.(2010){Caselli}, {Keto}, {Pagani}, {Aikawa},
  {Y{\i}ld{\i}z}, {van der Tak}, {Tafalla}, {Bergin}, {Nisini}, {Codella}, {van
  Dishoeck}, {Bachiller}, {Baudry}, {Benedettini}, {Benz}, {Bjerkeli}, {Blake},
  {Bontemps}, {Braine}, {Bruderer}, {Cernicharo}, {Daniel}, {di Giorgio},
  {Dominik}, {Doty}, {Encrenaz}, {Fich}, {Fuente}, {Gaier}, {Giannini},
  {Goicoechea}, {de Graauw}, {Helmich}, {Herczeg}, {Herpin}, {Hogerheijde},
  {Jackson}, {Jacq}, {Javadi}, {Johnstone}, {J{\o}rgensen}, {Kester},
  {Kristensen}, {Laauwen}, {Larsson}, {Lis}, {Liseau}, {Luinge}, {Marseille},
  {McCoey}, {Megej}, {Melnick}, {Neufeld}, {Olberg}, {Parise}, {Pearson},
  {Plume}, {Risacher}, {Santiago-Garc{\'{\i}}a}, {Saraceno}, {Shipman},
  {Siegel}, {van Kempen}, {Visser}, {Wampfler}, and {Wyrowski}]{Caselli10}
{Caselli},~P. et~al.  \emph{\aap} \textbf{2010}, \emph{521}, L29\relax
\mciteBstWouldAddEndPuncttrue
\mciteSetBstMidEndSepPunct{\mcitedefaultmidpunct}
{\mcitedefaultendpunct}{\mcitedefaultseppunct}\relax
\EndOfBibitem
\bibitem[{Hogerheijde} et~al.(2011){Hogerheijde}, {Bergin}, {Brinch},
  {Cleeves}, {Fogel}, {Blake}, {Dominik}, {Lis}, {Melnick}, {Neufeld},
  {Pani{\'c}}, {Pearson}, {Kristensen}, {Y{\i}ld{\i}z}, and {van
  Dishoeck}]{Hogerheijde11}
{Hogerheijde},~M.~R.; {Bergin},~E.~A.; {Brinch},~C.; {Cleeves},~L.~I.;
  {Fogel},~J.~K.~J.; {Blake},~G.~A.; {Dominik},~C.; {Lis},~D.~C.;
  {Melnick},~G.; {Neufeld},~D.; {Pani{\'c}},~O.; {Pearson},~J.~C.;
  {Kristensen},~L.; {Y{\i}ld{\i}z},~U.~A.; {van Dishoeck},~E.~F. \emph{Science}
  \textbf{2011}, \emph{334}, 338\relax
\mciteBstWouldAddEndPuncttrue
\mciteSetBstMidEndSepPunct{\mcitedefaultmidpunct}
{\mcitedefaultendpunct}{\mcitedefaultseppunct}\relax
\EndOfBibitem
\bibitem[{Sonnentrucker} et~al.(2010){Sonnentrucker}, {Neufeld}, {Phillips},
  {Gerin}, {Lis}, {de Luca}, {Goicoechea}, {Black}, {Bell}, {Boulanger},
  {Cernicharo}, {Coutens}, {Dartois}, {Ka{\'z}mierczak}, {Encrenaz},
  {Falgarone}, {Geballe}, {Giesen}, {Godard}, {Goldsmith}, {Gry}, {Gupta},
  {Hennebelle}, {Herbst}, {Hily-Blant}, {Joblin}, {Ko{\l}os}, {Kre{\l}owski},
  {Mart{\'{\i}}n-Pintado}, {Menten}, {Monje}, {Mookerjea}, {Pearson},
  {Perault}, {Persson}, {Plume}, {Salez}, {Schlemmer}, {Schmidt}, {Stutzki},
  {Teyssier}, {Vastel}, {Yu}, {Caux}, {G{\"u}sten}, {Hatch}, {Klein}, {Mehdi},
  {Morris}, and {Ward}]{Sonnentrucker10}
{Sonnentrucker},~P. et~al.  \emph{\aap} \textbf{2010}, \emph{521}, L12\relax
\mciteBstWouldAddEndPuncttrue
\mciteSetBstMidEndSepPunct{\mcitedefaultmidpunct}
{\mcitedefaultendpunct}{\mcitedefaultseppunct}\relax
\EndOfBibitem
\bibitem[{Monje} et~al.(2011){Monje}, {Emprechtinger}, {Phillips}, {Lis},
  {Goldsmith}, {Bergin}, {Bell}, {Neufeld}, and {Sonnentrucker}]{Monje11}
{Monje},~R.~R.; {Emprechtinger},~M.; {Phillips},~T.~G.; {Lis},~D.~C.;
  {Goldsmith},~P.~F.; {Bergin},~E.~A.; {Bell},~T.~A.; {Neufeld},~D.~A.;
  {Sonnentrucker},~P. \emph{\apjl} \textbf{2011}, \emph{734}, L23\relax
\mciteBstWouldAddEndPuncttrue
\mciteSetBstMidEndSepPunct{\mcitedefaultmidpunct}
{\mcitedefaultendpunct}{\mcitedefaultseppunct}\relax
\EndOfBibitem
\bibitem[{Indriolo} et~al.(2013){Indriolo}, {Neufeld}, {Seifahrt}, and
  {Richter}]{Indriolo13}
{Indriolo},~N.; {Neufeld},~D.~A.; {Seifahrt},~A.; {Richter},~M.~J. \emph{\apj}
  \textbf{2013}, \emph{764}, 188\relax
\mciteBstWouldAddEndPuncttrue
\mciteSetBstMidEndSepPunct{\mcitedefaultmidpunct}
{\mcitedefaultendpunct}{\mcitedefaultseppunct}\relax
\EndOfBibitem
\bibitem[{Cernicharo} et~al.(1997){Cernicharo}, {Lim}, {Cox},
  {Gonzalez-Alfonso}, {Caux}, {Swinyard}, {Martin-Pintado}, {Baluteau}, and
  {Clegg}]{Cernicharo97}
{Cernicharo},~J.; {Lim},~T.; {Cox},~P.; {Gonzalez-Alfonso},~E.; {Caux},~E.;
  {Swinyard},~B.~M.; {Martin-Pintado},~J.; {Baluteau},~J.~P.; {Clegg},~P.
  \emph{\aap} \textbf{1997}, \emph{323}, L25--L28\relax
\mciteBstWouldAddEndPuncttrue
\mciteSetBstMidEndSepPunct{\mcitedefaultmidpunct}
{\mcitedefaultendpunct}{\mcitedefaultseppunct}\relax
\EndOfBibitem
\bibitem[{Flagey} et~al.(2013){Flagey}, {Goldsmith}, {Lis}, {Gerin}, {Neufeld},
  {Sonnentrucker}, {De Luca}, {Godard}, {Goicoechea}, {Monje}, and
  {Phillips}]{Flagey13}
{Flagey},~N.; {Goldsmith},~P.~F.; {Lis},~D.~C.; {Gerin},~M.; {Neufeld},~D.;
  {Sonnentrucker},~P.; {De Luca},~M.; {Godard},~B.; {Goicoechea},~J.~R.;
  {Monje},~R.; {Phillips},~T.~G. \emph{\apj} \textbf{2013}, \emph{762},
  11\relax
\mciteBstWouldAddEndPuncttrue
\mciteSetBstMidEndSepPunct{\mcitedefaultmidpunct}
{\mcitedefaultendpunct}{\mcitedefaultseppunct}\relax
\EndOfBibitem
\bibitem[{Sonnentrucker} et~al.(2013){Sonnentrucker}, {Neufeld}, {Gerin}, {De
  Luca}, {Indriolo}, {Lis}, and {Goicoechea}]{Sonnentrucker13}
{Sonnentrucker},~P.; {Neufeld},~D.~A.; {Gerin},~M.; {De Luca},~M.;
  {Indriolo},~N.; {Lis},~D.~C.; {Goicoechea},~J.~R. \emph{\apjl} \textbf{2013},
  \emph{763}, L19\relax
\mciteBstWouldAddEndPuncttrue
\mciteSetBstMidEndSepPunct{\mcitedefaultmidpunct}
{\mcitedefaultendpunct}{\mcitedefaultseppunct}\relax
\EndOfBibitem
\bibitem[{Fischer} et~al.(2010){Fischer}, {Sturm}, {Gonz{\'a}lez-Alfonso},
  {Graci{\'a}-Carpio}, {Hailey-Dunsheath}, {Poglitsch}, {Contursi}, {Lutz},
  {Genzel}, {Sternberg}, {Verma}, and {Tacconi}]{Fischer10}
{Fischer},~J.; {Sturm},~E.; {Gonz{\'a}lez-Alfonso},~E.;
  {Graci{\'a}-Carpio},~J.; {Hailey-Dunsheath},~S.; {Poglitsch},~A.;
  {Contursi},~A.; {Lutz},~D.; {Genzel},~R.; {Sternberg},~A.; {Verma},~A.;
  {Tacconi},~L. \emph{\aap} \textbf{2010}, \emph{518}, L41\relax
\mciteBstWouldAddEndPuncttrue
\mciteSetBstMidEndSepPunct{\mcitedefaultmidpunct}
{\mcitedefaultendpunct}{\mcitedefaultseppunct}\relax
\EndOfBibitem
\bibitem[{Gonz{\'a}lez-Alfonso} et~al.(2012){Gonz{\'a}lez-Alfonso}, {Fischer},
  {Graci{\'a}-Carpio}, {Sturm}, {Hailey-Dunsheath}, {Lutz}, {Poglitsch},
  {Contursi}, {Feuchtgruber}, {Veilleux}, {Spoon}, {Verma}, {Christopher},
  {Davies}, {Sternberg}, {Genzel}, and {Tacconi}]{Gonzalez12}
{Gonz{\'a}lez-Alfonso},~E. et~al.  \emph{\aap} \textbf{2012}, \emph{541},
  A4\relax
\mciteBstWouldAddEndPuncttrue
\mciteSetBstMidEndSepPunct{\mcitedefaultmidpunct}
{\mcitedefaultendpunct}{\mcitedefaultseppunct}\relax
\EndOfBibitem
\bibitem[{Neufeld} et~al.(2010){Neufeld}, {Gonz{\'a}lez-Alfonso}, {Melnick},
  {Pu{\l}ecka}, {Schmidt}, {Szczerba}, {Bujarrabal}, {Alcolea}, {Cernicharo},
  {Decin}, {Dominik}, {Justtanont}, {de Koter}, {Marston}, {Menten},
  {Olofsson}, {Planesas}, {Sch{\"o}ier}, {Teyssier}, {Waters}, {Edwards},
  {McCoey}, {Shipman}, {Jellema}, {de Graauw}, {Ossenkopf}, {Schieder}, and
  {Philipp}]{Neufeld10}
{Neufeld},~D.~A. et~al.  \emph{\aap} \textbf{2010}, \emph{521}, L5\relax
\mciteBstWouldAddEndPuncttrue
\mciteSetBstMidEndSepPunct{\mcitedefaultmidpunct}
{\mcitedefaultendpunct}{\mcitedefaultseppunct}\relax
\EndOfBibitem
\bibitem[{Wiesemeyer} et~al.(2012){Wiesemeyer}, {G{\"u}sten}, {Heyminck},
  {Jacobs}, {Menten}, {Neufeld}, {Requena-Torres}, and {Stutzki}]{Wiesemeyer12}
{Wiesemeyer},~H.; {G{\"u}sten},~R.; {Heyminck},~S.; {Jacobs},~K.;
  {Menten},~K.~M.; {Neufeld},~D.~A.; {Requena-Torres},~M.~A.; {Stutzki},~J.
  \emph{\aap} \textbf{2012}, \emph{542}, L7\relax
\mciteBstWouldAddEndPuncttrue
\mciteSetBstMidEndSepPunct{\mcitedefaultmidpunct}
{\mcitedefaultendpunct}{\mcitedefaultseppunct}\relax
\EndOfBibitem
\bibitem[{Roueff}(1996)]{Roueff96}
{Roueff},~E. \emph{\mnras} \textbf{1996}, \emph{279}, L37--L40\relax
\mciteBstWouldAddEndPuncttrue
\mciteSetBstMidEndSepPunct{\mcitedefaultmidpunct}
{\mcitedefaultendpunct}{\mcitedefaultseppunct}\relax
\EndOfBibitem
\bibitem[{Weselak} et~al.(2009){Weselak}, {Galazutdinov}, {Beletsky}, and
  {Kre{\l}owski}]{Weselak09}
{Weselak},~T.; {Galazutdinov},~G.; {Beletsky},~Y.; {Kre{\l}owski},~J.
  \emph{\aap} \textbf{2009}, \emph{499}, 783--787\relax
\mciteBstWouldAddEndPuncttrue
\mciteSetBstMidEndSepPunct{\mcitedefaultmidpunct}
{\mcitedefaultendpunct}{\mcitedefaultseppunct}\relax
\EndOfBibitem
\bibitem[{Weselak} et~al.(2010){Weselak}, {Galazutdinov}, {Beletsky}, and
  {Kre{\l}owski}]{Weselak10}
{Weselak},~T.; {Galazutdinov},~G.~A.; {Beletsky},~Y.; {Kre{\l}owski},~J.
  \emph{\mnras} \textbf{2010}, \emph{402}, 1991--1994\relax
\mciteBstWouldAddEndPuncttrue
\mciteSetBstMidEndSepPunct{\mcitedefaultmidpunct}
{\mcitedefaultendpunct}{\mcitedefaultseppunct}\relax
\EndOfBibitem
\bibitem[{Gonz{\'a}lez-Alfonso} et~al.(2013){Gonz{\'a}lez-Alfonso}, {Fischer},
  {Bruderer}, {M{\"u}ller}, {Graci{\'a}-Carpio}, {Sturm}, {Lutz}, {Poglitsch},
  {Feuchtgruber}, {Veilleux}, {Contursi}, {Sternberg}, {Hailey-Dunsheath},
  {Verma}, {Christopher}, {Davies}, {Genzel}, and {Tacconi}]{Gonzalez13}
{Gonz{\'a}lez-Alfonso},~E. et~al.  \emph{\aap} \textbf{2013}, \emph{550},
  A25\relax
\mciteBstWouldAddEndPuncttrue
\mciteSetBstMidEndSepPunct{\mcitedefaultmidpunct}
{\mcitedefaultendpunct}{\mcitedefaultseppunct}\relax
\EndOfBibitem
\bibitem[{Habart} et~al.(2010){Habart}, {Dartois}, {Abergel}, {Baluteau},
  {Naylor}, {Polehampton}, {Joblin}, {Ade}, {Anderson}, {Andr{\'e}}, {Arab},
  {Bernard}, {Blagrave}, {Bontemps}, {Boulanger}, {Cohen}, {Compiegne}, {Cox},
  {Davis}, {Emery}, {Fulton}, {Gry}, {Huang}, {Jones}, {Kirk}, {Lagache},
  {Lim}, {Madden}, {Makiwa}, {Martin}, {Miville-Desch{\^e}nes}, {Molinari},
  {Moseley}, {Motte}, {Okumura}, {Pinheiro Gon{\c c}alves}, {Rodon}, {Russeil},
  {Saraceno}, {Sidher}, {Spencer}, {Swinyard}, {Ward-Thompson}, {White}, and
  {Zavagno}]{Habart10}
{Habart},~E. et~al.  \emph{\aap} \textbf{2010}, \emph{518}, L116\relax
\mciteBstWouldAddEndPuncttrue
\mciteSetBstMidEndSepPunct{\mcitedefaultmidpunct}
{\mcitedefaultendpunct}{\mcitedefaultseppunct}\relax
\EndOfBibitem
\bibitem[{Goicoechea} et~al.(2011){Goicoechea}, {Joblin}, {Contursi},
  {Bern{\'e}}, {Cernicharo}, {Gerin}, {Le Bourlot}, {Bergin}, {Bell}, and
  {R{\"o}llig}]{Goicoechea11}
{Goicoechea},~J.~R.; {Joblin},~C.; {Contursi},~A.; {Bern{\'e}},~O.;
  {Cernicharo},~J.; {Gerin},~M.; {Le Bourlot},~J.; {Bergin},~E.~A.;
  {Bell},~T.~A.; {R{\"o}llig},~M. \emph{\aap} \textbf{2011}, \emph{530},
  L16\relax
\mciteBstWouldAddEndPuncttrue
\mciteSetBstMidEndSepPunct{\mcitedefaultmidpunct}
{\mcitedefaultendpunct}{\mcitedefaultseppunct}\relax
\EndOfBibitem
\bibitem[{Sternberg} and {Dalgarno}(1995){Sternberg}, and
  {Dalgarno}]{Sternberg95}
{Sternberg},~A.; {Dalgarno},~A. \emph{\apjs} \textbf{1995}, \emph{99},
  565\relax
\mciteBstWouldAddEndPuncttrue
\mciteSetBstMidEndSepPunct{\mcitedefaultmidpunct}
{\mcitedefaultendpunct}{\mcitedefaultseppunct}\relax
\EndOfBibitem
\bibitem[{Pilleri} et~al.(2012){Pilleri}, {Fuente}, {Cernicharo}, {Ossenkopf},
  {Bern{\'e}}, {Gerin}, {Pety}, {Goicoechea}, {Rizzo}, {Montillaud},
  {Gonz{\'a}lez-Garc{\'{\i}}a}, {Joblin}, {Le Bourlot}, {Le Petit}, and
  {Kramer}]{Pilleri12}
{Pilleri},~P.; {Fuente},~A.; {Cernicharo},~J.; {Ossenkopf},~V.;
  {Bern{\'e}},~O.; {Gerin},~M.; {Pety},~J.; {Goicoechea},~J.~R.;
  {Rizzo},~J.~R.; {Montillaud},~J.; {Gonz{\'a}lez-Garc{\'{\i}}a},~M.;
  {Joblin},~C.; {Le Bourlot},~J.; {Le Petit},~F.; {Kramer},~C. \emph{\aap}
  \textbf{2012}, \emph{544}, A110\relax
\mciteBstWouldAddEndPuncttrue
\mciteSetBstMidEndSepPunct{\mcitedefaultmidpunct}
{\mcitedefaultendpunct}{\mcitedefaultseppunct}\relax
\EndOfBibitem
\bibitem[{Whittet} et~al.(1988){Whittet}, {Bode}, {Longmore}, {Adamson},
  {McFadzean}, {Aitken}, and {Roche}]{Whittet88}
{Whittet},~D.~C.~B.; {Bode},~M.~F.; {Longmore},~A.~J.; {Adamson},~A.~J.;
  {McFadzean},~A.~D.; {Aitken},~D.~K.; {Roche},~P.~F. \emph{\mnras}
  \textbf{1988}, \emph{233}, 321--336\relax
\mciteBstWouldAddEndPuncttrue
\mciteSetBstMidEndSepPunct{\mcitedefaultmidpunct}
{\mcitedefaultendpunct}{\mcitedefaultseppunct}\relax
\EndOfBibitem
\bibitem[{Walmsley}(1992)]{Walmsley92}
{Walmsley},~C.~M. In \emph{Chemistry and Spectroscopy of Interstellar
  Molecules}; {Bohme},~D.~K., Ed.; Univ. of Tokyo Press: Tokyo, 1992; p
  267\relax
\mciteBstWouldAddEndPuncttrue
\mciteSetBstMidEndSepPunct{\mcitedefaultmidpunct}
{\mcitedefaultendpunct}{\mcitedefaultseppunct}\relax
\EndOfBibitem
\bibitem[{Murakawa} et~al.(2000){Murakawa}, {Tamura}, and {Nagata}]{Murakawa00}
{Murakawa},~K.; {Tamura},~M.; {Nagata},~T. \emph{\apjs} \textbf{2000},
  \emph{128}, 603--613\relax
\mciteBstWouldAddEndPuncttrue
\mciteSetBstMidEndSepPunct{\mcitedefaultmidpunct}
{\mcitedefaultendpunct}{\mcitedefaultseppunct}\relax
\EndOfBibitem
\bibitem[{Pontoppidan} et~al.(2004){Pontoppidan}, {van Dishoeck}, and
  {Dartois}]{Pontoppidan04}
{Pontoppidan},~K.~M.; {van Dishoeck},~E.~F.; {Dartois},~E. \emph{\aap}
  \textbf{2004}, \emph{426}, 925--940\relax
\mciteBstWouldAddEndPuncttrue
\mciteSetBstMidEndSepPunct{\mcitedefaultmidpunct}
{\mcitedefaultendpunct}{\mcitedefaultseppunct}\relax
\EndOfBibitem
\bibitem[{Lee} et~al.(1996){Lee}, {Bettens}, and {Herbst}]{Lee96b}
{Lee},~H.-H.; {Bettens},~R.~P.~A.; {Herbst},~E. \emph{\aaps} \textbf{1996},
  \emph{119}, 111--114\relax
\mciteBstWouldAddEndPuncttrue
\mciteSetBstMidEndSepPunct{\mcitedefaultmidpunct}
{\mcitedefaultendpunct}{\mcitedefaultseppunct}\relax
\EndOfBibitem
\bibitem[{Przybilla} et~al.(2008){Przybilla}, {Nieva}, and
  {Butler}]{Przybilla08}
{Przybilla},~N.; {Nieva},~M.-F.; {Butler},~K. \emph{\apjl} \textbf{2008},
  \emph{688}, L103--L106\relax
\mciteBstWouldAddEndPuncttrue
\mciteSetBstMidEndSepPunct{\mcitedefaultmidpunct}
{\mcitedefaultendpunct}{\mcitedefaultseppunct}\relax
\EndOfBibitem
\bibitem[{Whittet}(2010)]{Whittet10}
{Whittet},~D.~C.~B. \emph{\apj} \textbf{2010}, \emph{710}, 1009--1016\relax
\mciteBstWouldAddEndPuncttrue
\mciteSetBstMidEndSepPunct{\mcitedefaultmidpunct}
{\mcitedefaultendpunct}{\mcitedefaultseppunct}\relax
\EndOfBibitem
\bibitem[{Boonman} et~al.(2003){Boonman}, {Doty}, {van Dishoeck}, {Bergin},
  {Melnick}, {Wright}, and {Stark}]{Boonman03}
{Boonman},~A.~M.~S.; {Doty},~S.~D.; {van Dishoeck},~E.~F.; {Bergin},~E.~A.;
  {Melnick},~G.~J.; {Wright},~C.~M.; {Stark},~R. \emph{\aap} \textbf{2003},
  \emph{406}, 937--955\relax
\mciteBstWouldAddEndPuncttrue
\mciteSetBstMidEndSepPunct{\mcitedefaultmidpunct}
{\mcitedefaultendpunct}{\mcitedefaultseppunct}\relax
\EndOfBibitem
\bibitem[{Bergin} et~al.(2000){Bergin}, {Melnick}, {Stauffer}, {Ashby}, {Chin},
  {Erickson}, {Goldsmith}, {Harwit}, {Howe}, {Kleiner}, {Koch}, {Neufeld},
  {Patten}, {Plume}, {Schieder}, {Snell}, {Tolls}, {Wang}, {Winnewisser}, and
  {Zhang}]{Bergin00}
{Bergin},~E.~A. et~al.  \emph{\apjl} \textbf{2000}, \emph{539},
  L129--L132\relax
\mciteBstWouldAddEndPuncttrue
\mciteSetBstMidEndSepPunct{\mcitedefaultmidpunct}
{\mcitedefaultendpunct}{\mcitedefaultseppunct}\relax
\EndOfBibitem
\bibitem[{Roberts} and {Herbst}(2002){Roberts}, and {Herbst}]{Roberts02}
{Roberts},~H.; {Herbst},~E. \emph{\aap} \textbf{2002}, \emph{395},
  233--242\relax
\mciteBstWouldAddEndPuncttrue
\mciteSetBstMidEndSepPunct{\mcitedefaultmidpunct}
{\mcitedefaultendpunct}{\mcitedefaultseppunct}\relax
\EndOfBibitem
\bibitem[{Caselli} et~al.(2012){Caselli}, {Keto}, {Bergin}, {Tafalla},
  {Aikawa}, {Douglas}, {Pagani}, {Y{\'{\i}}ld{\'{\i}}z}, {van der Tak},
  {Walmsley}, {Codella}, {Nisini}, {Kristensen}, and {van Dishoeck}]{Caselli12}
{Caselli},~P.; {Keto},~E.; {Bergin},~E.~A.; {Tafalla},~M.; {Aikawa},~Y.;
  {Douglas},~T.; {Pagani},~L.; {Y{\'{\i}}ld{\'{\i}}z},~U.~A.; {van der
  Tak},~F.~F.~S.; {Walmsley},~C.~M.; {Codella},~C.; {Nisini},~B.;
  {Kristensen},~L.~E.; {van Dishoeck},~E.~F. \emph{\apjl} \textbf{2012},
  \emph{759}, L37\relax
\mciteBstWouldAddEndPuncttrue
\mciteSetBstMidEndSepPunct{\mcitedefaultmidpunct}
{\mcitedefaultendpunct}{\mcitedefaultseppunct}\relax
\EndOfBibitem
\bibitem[{Evans}(1999)]{Evans99}
{Evans},~N.~J.,~II \emph{\araa} \textbf{1999}, \emph{37}, 311--362\relax
\mciteBstWouldAddEndPuncttrue
\mciteSetBstMidEndSepPunct{\mcitedefaultmidpunct}
{\mcitedefaultendpunct}{\mcitedefaultseppunct}\relax
\EndOfBibitem
\bibitem[{Kristensen} et~al.(2012){Kristensen}, {van Dishoeck}, {Bergin},
  {Visser}, {Y{\i}ld{\i}z}, {San Jose-Garcia}, {J{\o}rgensen}, {Herczeg},
  {Johnstone}, {Wampfler}, {Benz}, {Bruderer}, {Cabrit}, {Caselli}, {Doty},
  {Harsono}, {Herpin}, {Hogerheijde}, {Karska}, {van Kempen}, {Liseau},
  {Nisini}, {Tafalla}, {van der Tak}, and {Wyrowski}]{Kristensen12}
{Kristensen},~L.~E. et~al.  \emph{\aap} \textbf{2012}, \emph{542}, A8\relax
\mciteBstWouldAddEndPuncttrue
\mciteSetBstMidEndSepPunct{\mcitedefaultmidpunct}
{\mcitedefaultendpunct}{\mcitedefaultseppunct}\relax
\EndOfBibitem
\bibitem[{Mottram} et~al.(2013){Mottram}, {van Dishoeck}, {Schmalzl},
  {Kristensen}, {Visser}, {Hogerheijde}, and {Bruderer}]{Mottram13}
{Mottram},~J.~C.; {van Dishoeck},~E.~F.; {Schmalzl},~M.; {Kristensen},~L.~E.;
  {Visser},~R.; {Hogerheijde},~M.~R.; {Bruderer},~S. \emph{\aap} \textbf{2013},
  in press\relax
\mciteBstWouldAddEndPuncttrue
\mciteSetBstMidEndSepPunct{\mcitedefaultmidpunct}
{\mcitedefaultendpunct}{\mcitedefaultseppunct}\relax
\EndOfBibitem
\bibitem[{Bergin} et~al.(2010){Bergin}, {Hogerheijde}, {Brinch}, {Fogel},
  {Y{\i}ld{\i}z}, {Kristensen}, {van Dishoeck}, {Bell}, {Blake}, {Cernicharo},
  {Dominik}, {Lis}, {Melnick}, {Neufeld}, {Pani{\'c}}, {Pearson}, {Bachiller},
  {Baudry}, {Benedettini}, {Benz}, {Bjerkeli}, {Bontemps}, {Braine},
  {Bruderer}, {Caselli}, {Codella}, {Daniel}, {di Giorgio}, {Doty}, {Encrenaz},
  {Fich}, {Fuente}, {Giannini}, {Goicoechea}, {de Graauw}, {Helmich},
  {Herczeg}, {Herpin}, {Jacq}, {Johnstone}, {J{\o}rgensen}, {Larsson},
  {Liseau}, {Marseille}, {Mc Coey}, {Nisini}, {Olberg}, {Parise}, {Plume},
  {Risacher}, {Santiago-Garc{\'{\i}}a}, {Saraceno}, {Shipman}, {Tafalla}, {van
  Kempen}, {Visser}, {Wampfler}, {Wyrowski}, {van der Tak}, {Jellema},
  {Tielens}, {Hartogh}, {St{\"u}tzki}, and {Szczerba}]{Bergin10}
{Bergin},~E.~A. et~al.  \emph{\aap} \textbf{2010}, \emph{521}, L33\relax
\mciteBstWouldAddEndPuncttrue
\mciteSetBstMidEndSepPunct{\mcitedefaultmidpunct}
{\mcitedefaultendpunct}{\mcitedefaultseppunct}\relax
\EndOfBibitem
\bibitem[{Fogel} et~al.(2011){Fogel}, {Bethell}, {Bergin}, {Calvet}, and
  {Semenov}]{Fogel11}
{Fogel},~J.~K.~J.; {Bethell},~T.~J.; {Bergin},~E.~A.; {Calvet},~N.;
  {Semenov},~D. \emph{\apj} \textbf{2011}, \emph{726}, 29\relax
\mciteBstWouldAddEndPuncttrue
\mciteSetBstMidEndSepPunct{\mcitedefaultmidpunct}
{\mcitedefaultendpunct}{\mcitedefaultseppunct}\relax
\EndOfBibitem
\bibitem[{Dominik} et~al.(2005){Dominik}, {Ceccarelli}, {Hollenbach}, and
  {Kaufman}]{Dominik05}
{Dominik},~C.; {Ceccarelli},~C.; {Hollenbach},~D.; {Kaufman},~M. \emph{\apjl}
  \textbf{2005}, \emph{635}, L85--L88\relax
\mciteBstWouldAddEndPuncttrue
\mciteSetBstMidEndSepPunct{\mcitedefaultmidpunct}
{\mcitedefaultendpunct}{\mcitedefaultseppunct}\relax
\EndOfBibitem
\bibitem[{Bergman} et~al.(2011){Bergman}, {Parise}, {Liseau}, {Larsson},
  {Olofsson}, {Menten}, and {G{\"u}sten}]{Bergman11}
{Bergman},~P.; {Parise},~B.; {Liseau},~R.; {Larsson},~B.; {Olofsson},~H.;
  {Menten},~K.~M.; {G{\"u}sten},~R. \emph{\aap} \textbf{2011}, \emph{531},
  L8\relax
\mciteBstWouldAddEndPuncttrue
\mciteSetBstMidEndSepPunct{\mcitedefaultmidpunct}
{\mcitedefaultendpunct}{\mcitedefaultseppunct}\relax
\EndOfBibitem
\bibitem[{Parise} et~al.(2012){Parise}, {Bergman}, and {Du}]{Parise12}
{Parise},~B.; {Bergman},~P.; {Du},~F. \emph{\aap} \textbf{2012}, \emph{541},
  L11\relax
\mciteBstWouldAddEndPuncttrue
\mciteSetBstMidEndSepPunct{\mcitedefaultmidpunct}
{\mcitedefaultendpunct}{\mcitedefaultseppunct}\relax
\EndOfBibitem
\bibitem[{Charnley} et~al.(1992){Charnley}, {Tielens}, and
  {Millar}]{Charnley92}
{Charnley},~S.~B.; {Tielens},~A.~G.~G.~M.; {Millar},~T.~J. \emph{\apjl}
  \textbf{1992}, \emph{399}, L71--L74\relax
\mciteBstWouldAddEndPuncttrue
\mciteSetBstMidEndSepPunct{\mcitedefaultmidpunct}
{\mcitedefaultendpunct}{\mcitedefaultseppunct}\relax
\EndOfBibitem
\bibitem[{Chavarr{\'{\i}}a} et~al.(2010){Chavarr{\'{\i}}a}, {Herpin}, {Jacq},
  {Braine}, {Bontemps}, {Baudry}, {Marseille}, {van der Tak}, {Pietropaoli},
  {Wyrowski}, {Shipman}, {Frieswijk}, {van Dishoeck}, {Cernicharo},
  {Bachiller}, {Benedettini}, {Benz}, {Bergin}, {Bjerkeli}, {Blake},
  {Bruderer}, {Caselli}, {Codella}, {Daniel}, {di Giorgio}, {Dominik}, {Doty},
  {Encrenaz}, {Fich}, {Fuente}, {Giannini}, {Goicoechea}, {de Graauw},
  {Hartogh}, {Helmich}, {Herczeg}, {Hogerheijde}, {Johnstone}, {J{\o}rgensen},
  {Kristensen}, {Larsson}, {Lis}, {Liseau}, {McCoey}, {Melnick}, {Nisini},
  {Olberg}, {Parise}, {Pearson}, {Plume}, {Risacher}, {Santiago-Garc{\'{\i}}a},
  {Saraceno}, {Stutzki}, {Szczerba}, {Tafalla}, {Tielens}, {van Kempen},
  {Visser}, {Wampfler}, {Willem}, and {Y{\i}ld{\i}z}]{Chavarria10}
{Chavarr{\'{\i}}a},~L. et~al.  \emph{\aap} \textbf{2010}, \emph{521}, L37\relax
\mciteBstWouldAddEndPuncttrue
\mciteSetBstMidEndSepPunct{\mcitedefaultmidpunct}
{\mcitedefaultendpunct}{\mcitedefaultseppunct}\relax
\EndOfBibitem
\bibitem[{Johnstone} et~al.(2010){Johnstone}, {Fich}, {McCoey}, {van Kempen},
  {Fuente}, {Kristensen}, {Cernicharo}, {Caselli}, {Visser}, {Plume},
  {Herczeg}, {van Dishoeck}, {Wampfler}, {Bachiller}, {Baudry}, {Benedettini},
  {Bergin}, {Benz}, {Bjerkeli}, {Blake}, {Bontemps}, {Braine}, {Bruderer},
  {Codella}, {Daniel}, {di Giorgio}, {Dominik}, {Doty}, {Encrenaz}, {Giannini},
  {Goicoechea}, {de Graauw}, {Helmich}, {Herpin}, {Hogerheijde}, {Jacq},
  {J{\o}rgensen}, {Larsson}, {Lis}, {Liseau}, {Marseille}, {Melnick},
  {Neufeld}, {Nisini}, {Olberg}, {Parise}, {Pearson}, {Risacher},
  {Santiago-Garc{\'{\i}}a}, {Saraceno}, {Shipman}, {Tafalla}, {van der Tak},
  {Wyrowski}, {Y{\i}ld{\i}z}, {Caux}, {Honingh}, {Jellema}, {Schieder},
  {Teyssier}, and {Whyborn}]{Johnstone10}
{Johnstone},~D. et~al.  \emph{\aap} \textbf{2010}, \emph{521}, L41\relax
\mciteBstWouldAddEndPuncttrue
\mciteSetBstMidEndSepPunct{\mcitedefaultmidpunct}
{\mcitedefaultendpunct}{\mcitedefaultseppunct}\relax
\EndOfBibitem
\bibitem[{Kristensen} et~al.(2010){Kristensen}, {Visser}, {van Dishoeck},
  {Y{\i}ld{\i}z}, {Doty}, {Herczeg}, {Liu}, {Parise}, {J{\o}rgensen}, {van
  Kempen}, {Brinch}, {Wampfler}, {Bruderer}, {Benz}, {Hogerheijde}, {Deul},
  {Bachiller}, {Baudry}, {Benedettini}, {Bergin}, {Bjerkeli}, {Blake},
  {Bontemps}, {Braine}, {Caselli}, {Cernicharo}, {Codella}, {Daniel}, {de
  Graauw}, {di Giorgio}, {Dominik}, {Encrenaz}, {Fich}, {Fuente}, {Giannini},
  {Goicoechea}, {Helmich}, {Herpin}, {Jacq}, {Johnstone}, {Kaufman}, {Larsson},
  {Lis}, {Liseau}, {Marseille}, {McCoey}, {Melnick}, {Neufeld}, {Nisini},
  {Olberg}, {Pearson}, {Plume}, {Risacher}, {Santiago-Garc{\'{\i}}a},
  {Saraceno}, {Shipman}, {Tafalla}, {Tielens}, {van der Tak}, {Wyrowski},
  {Beintema}, {de Jonge}, {Dieleman}, {Ossenkopf}, {Roelfsema}, {Stutzki}, and
  {Whyborn}]{Kristensen10}
{Kristensen},~L.~E. et~al.  \emph{\aap} \textbf{2010}, \emph{521}, L30\relax
\mciteBstWouldAddEndPuncttrue
\mciteSetBstMidEndSepPunct{\mcitedefaultmidpunct}
{\mcitedefaultendpunct}{\mcitedefaultseppunct}\relax
\EndOfBibitem
\bibitem[{van Dishoeck} and {Helmich}(1996){van Dishoeck}, and
  {Helmich}]{vanDishoeck96}
{van Dishoeck},~E.~F.; {Helmich},~F.~P. \emph{\aap} \textbf{1996}, \emph{315},
  L177--L180\relax
\mciteBstWouldAddEndPuncttrue
\mciteSetBstMidEndSepPunct{\mcitedefaultmidpunct}
{\mcitedefaultendpunct}{\mcitedefaultseppunct}\relax
\EndOfBibitem
\bibitem[{van Dishoeck} and {Blake}(1998){van Dishoeck}, and
  {Blake}]{vanDishoeck98}
{van Dishoeck},~E.~F.; {Blake},~G.~A. \emph{\araa} \textbf{1998}, \emph{36},
  317--368\relax
\mciteBstWouldAddEndPuncttrue
\mciteSetBstMidEndSepPunct{\mcitedefaultmidpunct}
{\mcitedefaultendpunct}{\mcitedefaultseppunct}\relax
\EndOfBibitem
\bibitem[{Boonman} and {van Dishoeck}(2003){Boonman}, and {van
  Dishoeck}]{Boonman03h2o}
{Boonman},~A.~M.~S.; {van Dishoeck},~E.~F. \emph{\aap} \textbf{2003},
  \emph{403}, 1003--1010\relax
\mciteBstWouldAddEndPuncttrue
\mciteSetBstMidEndSepPunct{\mcitedefaultmidpunct}
{\mcitedefaultendpunct}{\mcitedefaultseppunct}\relax
\EndOfBibitem
\bibitem[{Emprechtinger} et~al.(2013){Emprechtinger}, {Lis}, {Rolffs},
  {Schilke}, {Monje}, {Comito}, {Ceccarelli}, {Neufeld}, and {van der
  Tak}]{Emprechtinger13}
{Emprechtinger},~M.; {Lis},~D.~C.; {Rolffs},~R.; {Schilke},~P.; {Monje},~R.~R.;
  {Comito},~C.; {Ceccarelli},~C.; {Neufeld},~D.~A.; {van der Tak},~F.~F.~S.
  \emph{\apj} \textbf{2013}, \emph{765}, 61\relax
\mciteBstWouldAddEndPuncttrue
\mciteSetBstMidEndSepPunct{\mcitedefaultmidpunct}
{\mcitedefaultendpunct}{\mcitedefaultseppunct}\relax
\EndOfBibitem
\bibitem[{Herpin} et~al.(2012){Herpin}, {Chavarr{\'{\i}}a}, {van der Tak},
  {Wyrowski}, {van Dishoeck}, {Jacq}, {Braine}, {Baudry}, {Bontemps}, and
  {Kristensen}]{Herpin12}
{Herpin},~F.; {Chavarr{\'{\i}}a},~L.; {van der Tak},~F.; {Wyrowski},~F.; {van
  Dishoeck},~E.~F.; {Jacq},~T.; {Braine},~J.; {Baudry},~A.; {Bontemps},~S.;
  {Kristensen},~L. \emph{\aap} \textbf{2012}, \emph{542}, A76\relax
\mciteBstWouldAddEndPuncttrue
\mciteSetBstMidEndSepPunct{\mcitedefaultmidpunct}
{\mcitedefaultendpunct}{\mcitedefaultseppunct}\relax
\EndOfBibitem
\bibitem[{Visser} et~al.(2013){Visser}, {J{\o}rgensen}, {Kristensen}, {van
  Dishoeck}, and {Bergin}]{Visser13}
{Visser},~R.; {J{\o}rgensen},~J.~K.; {Kristensen},~L.~E.; {van
  Dishoeck},~E.~F.; {Bergin},~E.~A. \emph{\apj} \textbf{2013}, \emph{769},
  19\relax
\mciteBstWouldAddEndPuncttrue
\mciteSetBstMidEndSepPunct{\mcitedefaultmidpunct}
{\mcitedefaultendpunct}{\mcitedefaultseppunct}\relax
\EndOfBibitem
\bibitem[{Persson} et~al.(2012){Persson}, {J{\o}rgensen}, and {van
  Dishoeck}]{Persson12}
{Persson},~M.~V.; {J{\o}rgensen},~J.~K.; {van Dishoeck},~E.~F. \emph{\aap}
  \textbf{2012}, \emph{541}, A39\relax
\mciteBstWouldAddEndPuncttrue
\mciteSetBstMidEndSepPunct{\mcitedefaultmidpunct}
{\mcitedefaultendpunct}{\mcitedefaultseppunct}\relax
\EndOfBibitem
\bibitem[{Kaufman} and {Neufeld}(1996){Kaufman}, and {Neufeld}]{Kaufman96}
{Kaufman},~M.~J.; {Neufeld},~D.~A. \emph{\apj} \textbf{1996}, \emph{456},
  611\relax
\mciteBstWouldAddEndPuncttrue
\mciteSetBstMidEndSepPunct{\mcitedefaultmidpunct}
{\mcitedefaultendpunct}{\mcitedefaultseppunct}\relax
\EndOfBibitem
\bibitem[{Harwit} et~al.(1998){Harwit}, {Neufeld}, {Melnick}, and
  {Kaufman}]{Harwit98}
{Harwit},~M.; {Neufeld},~D.~A.; {Melnick},~G.~J.; {Kaufman},~M.~J. \emph{\apjl}
  \textbf{1998}, \emph{497}, L105\relax
\mciteBstWouldAddEndPuncttrue
\mciteSetBstMidEndSepPunct{\mcitedefaultmidpunct}
{\mcitedefaultendpunct}{\mcitedefaultseppunct}\relax
\EndOfBibitem
\bibitem[{Wright} et~al.(2000){Wright}, {van Dishoeck}, {Black},
  {Feuchtgruber}, {Cernicharo}, {Gonz{\'a}lez-Alfonso}, and {de
  Graauw}]{Wright00}
{Wright},~C.~M.; {van Dishoeck},~E.~F.; {Black},~J.~H.; {Feuchtgruber},~H.;
  {Cernicharo},~J.; {Gonz{\'a}lez-Alfonso},~E.; {de Graauw},~T. \emph{\aap}
  \textbf{2000}, \emph{358}, 689--700\relax
\mciteBstWouldAddEndPuncttrue
\mciteSetBstMidEndSepPunct{\mcitedefaultmidpunct}
{\mcitedefaultendpunct}{\mcitedefaultseppunct}\relax
\EndOfBibitem
\bibitem[{Gonz{\'a}lez-Alfonso} et~al.(2002){Gonz{\'a}lez-Alfonso}, {Wright},
  {Cernicharo}, {Rosenthal}, {Boonman}, and {van Dishoeck}]{Gonzalez02}
{Gonz{\'a}lez-Alfonso},~E.; {Wright},~C.~M.; {Cernicharo},~J.; {Rosenthal},~D.;
  {Boonman},~A.~M.~S.; {van Dishoeck},~E.~F. \emph{\aap} \textbf{2002},
  \emph{386}, 1074--1102\relax
\mciteBstWouldAddEndPuncttrue
\mciteSetBstMidEndSepPunct{\mcitedefaultmidpunct}
{\mcitedefaultendpunct}{\mcitedefaultseppunct}\relax
\EndOfBibitem
\bibitem[{Cernicharo} et~al.(2006){Cernicharo}, {Goicoechea}, {Daniel},
  {Lerate}, {Barlow}, {Swinyard}, {van Dishoeck}, {Lim}, {Viti}, and
  {Yates}]{Cernicharo06}
{Cernicharo},~J.; {Goicoechea},~J.~R.; {Daniel},~F.; {Lerate},~M.~R.;
  {Barlow},~M.~J.; {Swinyard},~B.~M.; {van Dishoeck},~E.~F.; {Lim},~T.~L.;
  {Viti},~S.; {Yates},~J. \emph{\apjl} \textbf{2006}, \emph{649},
  L33--L36\relax
\mciteBstWouldAddEndPuncttrue
\mciteSetBstMidEndSepPunct{\mcitedefaultmidpunct}
{\mcitedefaultendpunct}{\mcitedefaultseppunct}\relax
\EndOfBibitem
\bibitem[{Olofsson} et~al.(2003){Olofsson}, {Olofsson}, {Hjalmarson},
  {Bergman}, {Black}, {Booth}, {Buat}, {Curry}, {Encrenaz}, {Falgarone},
  {Feldman}, {Fich}, {Flor{\'e}n}, {Frisk}, {Gerin}, {Gregersen}, {Harju},
  {Hasegawa}, {Johansson}, {Kwok}, {Larsson}, {Lecacheux}, {Liljestr{\"o}m},
  {Liseau}, {Mattila}, {Mitchell}, {Nordh}, {Olberg}, {Olofsson}, {Pagani},
  {Plume}, {Ristorcelli}, {Rydbeck}, {Sandqvist}, {von Sch{\'e}ele}, {Serra},
  {Tothill}, {Volk}, and {Wilson}]{Olofsson03}
{Olofsson},~A.~O.~H. et~al.  \emph{\aap} \textbf{2003}, \emph{402},
  L47--L54\relax
\mciteBstWouldAddEndPuncttrue
\mciteSetBstMidEndSepPunct{\mcitedefaultmidpunct}
{\mcitedefaultendpunct}{\mcitedefaultseppunct}\relax
\EndOfBibitem
\bibitem[{Franklin} et~al.(2008){Franklin}, {Snell}, {Kaufman}, {Melnick},
  {Neufeld}, {Hollenbach}, and {Bergin}]{Franklin08}
{Franklin},~J.; {Snell},~R.~L.; {Kaufman},~M.~J.; {Melnick},~G.~J.;
  {Neufeld},~D.~A.; {Hollenbach},~D.~J.; {Bergin},~E.~A. \emph{\apj}
  \textbf{2008}, \emph{674}, 1015--1031\relax
\mciteBstWouldAddEndPuncttrue
\mciteSetBstMidEndSepPunct{\mcitedefaultmidpunct}
{\mcitedefaultendpunct}{\mcitedefaultseppunct}\relax
\EndOfBibitem
\bibitem[{Lefloch} et~al.(2010){Lefloch}, {Cabrit}, {Codella}, {Melnick},
  {Cernicharo}, {Caux}, {Benedettini}, {Boogert}, {Caselli}, {Ceccarelli},
  {Gueth}, {Hily-Blant}, {Lorenzani}, {Neufeld}, {Nisini}, {Pacheco}, {Pagani},
  {Pardo}, {Parise}, {Salez}, {Schuster}, {Viti}, {Bacmann}, {Baudry}, {Bell},
  {Bergin}, {Blake}, {Bottinelli}, {Castets}, {Comito}, {Coutens}, {Crimier},
  {Dominik}, {Demyk}, {Encrenaz}, {Falgarone}, {Fuente}, {Gerin}, {Goldsmith},
  {Helmich}, {Hennebelle}, {Henning}, {Herbst}, {Jacq}, {Kahane}, {Kama},
  {Klotz}, {Langer}, {Lis}, {Lord}, {Maret}, {Pearson}, {Phillips}, {Saraceno},
  {Schilke}, {Tielens}, {van der Tak}, {van der Wiel}, {Vastel}, {Wakelam},
  {Walters}, {Wyrowski}, {Yorke}, {Bachiller}, {Borys}, {de Lange}, {Delorme},
  {Kramer}, {Larsson}, {Lai}, {Maiwald}, {Martin-Pintado}, {Mehdi},
  {Ossenkopf}, {Siegel}, {Stutzki}, and {Wunsch}]{Lefloch10}
{Lefloch},~B. et~al.  \emph{\aap} \textbf{2010}, \emph{518}, L113\relax
\mciteBstWouldAddEndPuncttrue
\mciteSetBstMidEndSepPunct{\mcitedefaultmidpunct}
{\mcitedefaultendpunct}{\mcitedefaultseppunct}\relax
\EndOfBibitem
\bibitem[{Goicoechea} et~al.(2012){Goicoechea}, {Cernicharo}, {Karska},
  {Herczeg}, {Polehampton}, {Wampfler}, {Kristensen}, {van Dishoeck},
  {Etxaluze}, {Bern{\'e}}, and {Visser}]{Goicoechea12}
{Goicoechea},~J.~R.; {Cernicharo},~J.; {Karska},~A.; {Herczeg},~G.~J.;
  {Polehampton},~E.~T.; {Wampfler},~S.~F.; {Kristensen},~L.~E.; {van
  Dishoeck},~E.~F.; {Etxaluze},~M.; {Bern{\'e}},~O.; {Visser},~R. \emph{\aap}
  \textbf{2012}, \emph{548}, A77\relax
\mciteBstWouldAddEndPuncttrue
\mciteSetBstMidEndSepPunct{\mcitedefaultmidpunct}
{\mcitedefaultendpunct}{\mcitedefaultseppunct}\relax
\EndOfBibitem
\bibitem[{Tafalla} et~al.(2013){Tafalla}, {Liseau}, {Nisini}, {Bachiller},
  {Santiago-Garc{\'{\i}}a}, {van Dishoeck}, {Kristensen}, {Herczeg}, and
  {Y{\i}ld{\i}z}]{Tafalla13}
{Tafalla},~M.; {Liseau},~R.; {Nisini},~B.; {Bachiller},~R.;
  {Santiago-Garc{\'{\i}}a},~J.; {van Dishoeck},~E.~F.; {Kristensen},~L.~E.;
  {Herczeg},~G.~J.; {Y{\i}ld{\i}z},~U.~A. \emph{\aap} \textbf{2013},
  \emph{551}, A116\relax
\mciteBstWouldAddEndPuncttrue
\mciteSetBstMidEndSepPunct{\mcitedefaultmidpunct}
{\mcitedefaultendpunct}{\mcitedefaultseppunct}\relax
\EndOfBibitem
\bibitem[{Nisini} et~al.(2013){Nisini}, {Santangelo}, {Antoniucci},
  {Benedettini}, {Codella}, {Giannini}, {Lorenzani}, {Liseau}, {Tafalla},
  {Bjerkeli}, {Cabrit}, {Caselli}, {Kristensen}, {Neufeld}, {Melnick}, and {van
  Dishoeck}]{Nisini13}
{Nisini},~B. et~al.  \emph{\aap} \textbf{2013}, \emph{549}, A16\relax
\mciteBstWouldAddEndPuncttrue
\mciteSetBstMidEndSepPunct{\mcitedefaultmidpunct}
{\mcitedefaultendpunct}{\mcitedefaultseppunct}\relax
\EndOfBibitem
\bibitem[{Santangelo} et~al.(2013){Santangelo}, {Nisini}, {Antoniucci},
  {Codella}, {Cabrit}, {Giannini}, {Herczeg}, {Liseau}, {Tafalla}, and {van
  Dishoeck}]{Santangelo13}
{Santangelo},~G.; {Nisini},~B.; {Antoniucci},~S.; {Codella},~C.; {Cabrit},~S.;
  {Giannini},~T.; {Herczeg},~G.; {Liseau},~R.; {Tafalla},~M.; {van
  Dishoeck},~E.~F. \emph{\aap} \textbf{2013}, \emph{557}, A22\relax
\mciteBstWouldAddEndPuncttrue
\mciteSetBstMidEndSepPunct{\mcitedefaultmidpunct}
{\mcitedefaultendpunct}{\mcitedefaultseppunct}\relax
\EndOfBibitem
\bibitem[{Draine}(2003)]{Draine03}
{Draine},~B.~T. \emph{\araa} \textbf{2003}, \emph{41}, 241--289\relax
\mciteBstWouldAddEndPuncttrue
\mciteSetBstMidEndSepPunct{\mcitedefaultmidpunct}
{\mcitedefaultendpunct}{\mcitedefaultseppunct}\relax
\EndOfBibitem
\bibitem[{Neufeld} and {Dalgarno}(1989){Neufeld}, and {Dalgarno}]{Neufeld89b}
{Neufeld},~D.~A.; {Dalgarno},~A. \emph{\apj} \textbf{1989}, \emph{340},
  869--893\relax
\mciteBstWouldAddEndPuncttrue
\mciteSetBstMidEndSepPunct{\mcitedefaultmidpunct}
{\mcitedefaultendpunct}{\mcitedefaultseppunct}\relax
\EndOfBibitem
\bibitem[{Flower} and {Pineau Des For{\^e}ts}(2010){Flower}, and {Pineau Des
  For{\^e}ts}]{Flower10}
{Flower},~D.~R.; {Pineau Des For{\^e}ts},~G. \emph{\mnras} \textbf{2010},
  \emph{406}, 1745--1758\relax
\mciteBstWouldAddEndPuncttrue
\mciteSetBstMidEndSepPunct{\mcitedefaultmidpunct}
{\mcitedefaultendpunct}{\mcitedefaultseppunct}\relax
\EndOfBibitem
\bibitem[{Wampfler} et~al.(2011){Wampfler}, {Bruderer}, {Kristensen},
  {Chavarr{\'{\i}}a}, {Bergin}, {Benz}, {van Dishoeck}, {Herczeg}, {van der
  Tak}, {Goicoechea}, {Doty}, and {Herpin}]{Wampfler11}
{Wampfler},~S.~F.; {Bruderer},~S.; {Kristensen},~L.~E.; {Chavarr{\'{\i}}a},~L.;
  {Bergin},~E.~A.; {Benz},~A.~O.; {van Dishoeck},~E.~F.; {Herczeg},~G.~J.; {van
  der Tak},~F.~F.~S.; {Goicoechea},~J.~R.; {Doty},~S.~D.; {Herpin},~F.
  \emph{\aap} \textbf{2011}, \emph{531}, L16\relax
\mciteBstWouldAddEndPuncttrue
\mciteSetBstMidEndSepPunct{\mcitedefaultmidpunct}
{\mcitedefaultendpunct}{\mcitedefaultseppunct}\relax
\EndOfBibitem
\bibitem[{van Kempen} et~al.(2010){van Kempen}, {Kristensen}, {Herczeg},
  {Visser}, {van Dishoeck}, {Wampfler}, {Bruderer}, {Benz}, {Doty}, {Brinch},
  {Hogerheijde}, {J{\o}rgensen}, {Tafalla}, {Neufeld}, {Bachiller}, {Baudry},
  {Benedettini}, {Bergin}, {Bjerkeli}, {Blake}, {Bontemps}, {Braine},
  {Caselli}, {Cernicharo}, {Codella}, {Daniel}, {di Giorgio}, {Dominik},
  {Encrenaz}, {Fich}, {Fuente}, {Giannini}, {Goicoechea}, {de Graauw},
  {Helmich}, {Herpin}, {Jacq}, {Johnstone}, {Kaufman}, {Larsson}, {Lis},
  {Liseau}, {Marseille}, {McCoey}, {Melnick}, {Nisini}, {Olberg}, {Parise},
  {Pearson}, {Plume}, {Risacher}, {Santiago-Garc{\'{\i}}a}, {Saraceno},
  {Shipman}, {van der Tak}, {Wyrowski}, {Y{\i}ld{\i}z}, {Ciechanowicz},
  {Dubbeldam}, {Glenz}, {Huisman}, {Lin}, {Morris}, {Murphy}, and
  {Trappe}]{vanKempen10}
{van Kempen},~T.~A. et~al.  \emph{\aap} \textbf{2010}, \emph{518}, L121\relax
\mciteBstWouldAddEndPuncttrue
\mciteSetBstMidEndSepPunct{\mcitedefaultmidpunct}
{\mcitedefaultendpunct}{\mcitedefaultseppunct}\relax
\EndOfBibitem
\bibitem[{Wampfler} et~al.(2013){Wampfler}, {Bruderer}, {Karska}, {Herczeg},
  {van Dishoeck}, {Kristensen}, {Goicoechea}, {Benz}, {Doty}, {McCoey},
  {Baudry}, {Giannini}, and {Larsson}]{Wampfler13}
{Wampfler},~S.~F.; {Bruderer},~S.; {Karska},~A.; {Herczeg},~G.~J.; {van
  Dishoeck},~E.~F.; {Kristensen},~L.~E.; {Goicoechea},~J.~R.; {Benz},~A.~O.;
  {Doty},~S.~D.; {McCoey},~C.; {Baudry},~A.; {Giannini},~T.; {Larsson},~B.
  \emph{\aap} \textbf{2013}, \emph{552}, A56\relax
\mciteBstWouldAddEndPuncttrue
\mciteSetBstMidEndSepPunct{\mcitedefaultmidpunct}
{\mcitedefaultendpunct}{\mcitedefaultseppunct}\relax
\EndOfBibitem
\bibitem[{Karska} et~al.(2013){Karska}, {Herczeg}, {van Dishoeck}, {Wampfler},
  {Kristensen}, {Goicoechea}, {Visser}, {Nisini}, {San Jos{\'e}-Garc{\'{\i}}a},
  {Bruderer}, {{\'S}niady}, {Doty}, {Fedele}, {Y{\i}ld{\i}z}, {Benz}, {Bergin},
  {Caselli}, {Herpin}, {Hogerheijde}, {Johnstone}, {J{\o}rgensen}, {Liseau},
  {Tafalla}, {van der Tak}, and {Wyrowski}]{Karska13}
{Karska},~A. et~al.  \emph{\aap} \textbf{2013}, \emph{552}, A141\relax
\mciteBstWouldAddEndPuncttrue
\mciteSetBstMidEndSepPunct{\mcitedefaultmidpunct}
{\mcitedefaultendpunct}{\mcitedefaultseppunct}\relax
\EndOfBibitem
\bibitem[{Green} et~al.(2013){Green}, {Evans}, {J{\o}rgensen}, {Herczeg},
  {Kristensen}, {Lee}, {Dionatos}, {Yildiz}, {Salyk}, {Meeus}, {Bouwman},
  {Visser}, {Bergin}, {van Dishoeck}, {Rascati}, {Karska}, {van Kempen},
  {Dunham}, {Lindberg}, {Fedele}, and {DIGIT Team}]{Green13}
{Green},~J.~D. et~al.  \emph{\apj} \textbf{2013}, \emph{770}, 123\relax
\mciteBstWouldAddEndPuncttrue
\mciteSetBstMidEndSepPunct{\mcitedefaultmidpunct}
{\mcitedefaultendpunct}{\mcitedefaultseppunct}\relax
\EndOfBibitem
\bibitem[{Neufeld} and {Dalgarno}(1989){Neufeld}, and {Dalgarno}]{Neufeld89a}
{Neufeld},~D.~A.; {Dalgarno},~A. \emph{\apj} \textbf{1989}, \emph{344},
  251--264\relax
\mciteBstWouldAddEndPuncttrue
\mciteSetBstMidEndSepPunct{\mcitedefaultmidpunct}
{\mcitedefaultendpunct}{\mcitedefaultseppunct}\relax
\EndOfBibitem
\bibitem[{Snell} et~al.(2005){Snell}, {Hollenbach}, {Howe}, {Neufeld},
  {Kaufman}, {Melnick}, {Bergin}, and {Wang}]{Snell05}
{Snell},~R.~L.; {Hollenbach},~D.; {Howe},~J.~E.; {Neufeld},~D.~A.;
  {Kaufman},~M.~J.; {Melnick},~G.~J.; {Bergin},~E.~A.; {Wang},~Z. \emph{\apj}
  \textbf{2005}, \emph{620}, 758--773\relax
\mciteBstWouldAddEndPuncttrue
\mciteSetBstMidEndSepPunct{\mcitedefaultmidpunct}
{\mcitedefaultendpunct}{\mcitedefaultseppunct}\relax
\EndOfBibitem
\bibitem[{Nisini} et~al.(2002){Nisini}, {Giannini}, and {Lorenzetti}]{Nisini02}
{Nisini},~B.; {Giannini},~T.; {Lorenzetti},~D. \emph{\apj} \textbf{2002},
  \emph{574}, 246--257\relax
\mciteBstWouldAddEndPuncttrue
\mciteSetBstMidEndSepPunct{\mcitedefaultmidpunct}
{\mcitedefaultendpunct}{\mcitedefaultseppunct}\relax
\EndOfBibitem
\bibitem[{Bjerkeli} et~al.(2009){Bjerkeli}, {Liseau}, {Olberg}, {Falgarone},
  {Frisk}, {Hjalmarson}, {Klotz}, {Larsson}, {Olofsson}, {Olofsson},
  {Ristorcelli}, and {Sandqvist}]{Bjerkeli09}
{Bjerkeli},~P.; {Liseau},~R.; {Olberg},~M.; {Falgarone},~E.; {Frisk},~U.;
  {Hjalmarson},~{\AA}.; {Klotz},~A.; {Larsson},~B.; {Olofsson},~A.~O.~H.;
  {Olofsson},~G.; {Ristorcelli},~I.; {Sandqvist},~A. \emph{\aap} \textbf{2009},
  \emph{507}, 1455--1466\relax
\mciteBstWouldAddEndPuncttrue
\mciteSetBstMidEndSepPunct{\mcitedefaultmidpunct}
{\mcitedefaultendpunct}{\mcitedefaultseppunct}\relax
\EndOfBibitem
\bibitem[{Liseau} et~al.(1996){Liseau}, {Ceccarelli}, {Larsson}, {Nisini},
  {White}, {Ade}, {Armand}, {Burgdorf}, {Caux}, {Cerulli}, {Church}, {Clegg},
  {Digorgio}, {Furniss}, {Giannini}, {Glencross}, {Gry}, {King}, {Lim},
  {Lorenzetti}, {Molinari}, {Naylor}, {Orfei}, {Saraceno}, {Sidher}, {Smith},
  {Spinoglio}, {Swinyard}, {Texier}, {Tommasi}, {Trams}, and {Unger}]{Liseau96}
{Liseau},~R. et~al.  \emph{\aap} \textbf{1996}, \emph{315}, L181--L184\relax
\mciteBstWouldAddEndPuncttrue
\mciteSetBstMidEndSepPunct{\mcitedefaultmidpunct}
{\mcitedefaultendpunct}{\mcitedefaultseppunct}\relax
\EndOfBibitem
\bibitem[{Bjerkeli} et~al.(2011){Bjerkeli}, {Liseau}, {Nisini}, {Tafalla},
  {Benedettini}, {Bergman}, {Dionatos}, {Giannini}, {Herczeg}, {Justtanont},
  {Larsson}, {McOey}, {Olberg}, and {Olofsson}]{Bjerkeli11}
{Bjerkeli},~P.; {Liseau},~R.; {Nisini},~B.; {Tafalla},~M.; {Benedettini},~M.;
  {Bergman},~P.; {Dionatos},~O.; {Giannini},~T.; {Herczeg},~G.;
  {Justtanont},~K.; {Larsson},~B.; {McOey},~C.; {Olberg},~M.;
  {Olofsson},~A.~O.~H. \emph{\aap} \textbf{2011}, \emph{533}, A80\relax
\mciteBstWouldAddEndPuncttrue
\mciteSetBstMidEndSepPunct{\mcitedefaultmidpunct}
{\mcitedefaultendpunct}{\mcitedefaultseppunct}\relax
\EndOfBibitem
\bibitem[{Vasta} et~al.(2012){Vasta}, {Codella}, {Lorenzani}, {Santangelo},
  {Nisini}, {Giannini}, {Tafalla}, {Liseau}, {van Dishoeck}, and
  {Kristensen}]{Vasta12}
{Vasta},~M.; {Codella},~C.; {Lorenzani},~A.; {Santangelo},~G.; {Nisini},~B.;
  {Giannini},~T.; {Tafalla},~M.; {Liseau},~R.; {van Dishoeck},~E.~F.;
  {Kristensen},~L. \emph{\aap} \textbf{2012}, \emph{537}, A98\relax
\mciteBstWouldAddEndPuncttrue
\mciteSetBstMidEndSepPunct{\mcitedefaultmidpunct}
{\mcitedefaultendpunct}{\mcitedefaultseppunct}\relax
\EndOfBibitem
\bibitem[{Santangelo} et~al.(2012){Santangelo}, {Nisini}, {Giannini},
  {Antoniucci}, {Vasta}, {Codella}, {Lorenzani}, {Tafalla}, {Liseau}, {van
  Dishoeck}, and {Kristensen}]{Santangelo12}
{Santangelo},~G.; {Nisini},~B.; {Giannini},~T.; {Antoniucci},~S.; {Vasta},~M.;
  {Codella},~C.; {Lorenzani},~A.; {Tafalla},~M.; {Liseau},~R.; {van
  Dishoeck},~E.~F.; {Kristensen},~L.~E. \emph{\aap} \textbf{2012}, \emph{538},
  A45\relax
\mciteBstWouldAddEndPuncttrue
\mciteSetBstMidEndSepPunct{\mcitedefaultmidpunct}
{\mcitedefaultendpunct}{\mcitedefaultseppunct}\relax
\EndOfBibitem
\bibitem[{Bjerkeli} et~al.(2012){Bjerkeli}, {Liseau}, {Larsson}, {Rydbeck},
  {Nisini}, {Tafalla}, {Antoniucci}, {Benedettini}, {Bergman}, {Cabrit},
  {Giannini}, {Melnick}, {Neufeld}, {Santangelo}, and {van
  Dishoeck}]{Bjerkeli12}
{Bjerkeli},~P.; {Liseau},~R.; {Larsson},~B.; {Rydbeck},~G.; {Nisini},~B.;
  {Tafalla},~M.; {Antoniucci},~S.; {Benedettini},~M.; {Bergman},~P.;
  {Cabrit},~S.; {Giannini},~T.; {Melnick},~G.; {Neufeld},~D.; {Santangelo},~G.;
  {van Dishoeck},~E.~F. \emph{\aap} \textbf{2012}, \emph{546}, A29\relax
\mciteBstWouldAddEndPuncttrue
\mciteSetBstMidEndSepPunct{\mcitedefaultmidpunct}
{\mcitedefaultendpunct}{\mcitedefaultseppunct}\relax
\EndOfBibitem
\bibitem[{Fich} et~al.(2010){Fich}, {Johnstone}, {van Kempen}, {McCoey},
  {Fuente}, {Caselli}, {Kristensen}, {Plume}, {Cernicharo}, {Herczeg}, {van
  Dishoeck}, {Wampfler}, {Gaufre}, {Gill}, {Javadi}, {Justen}, {Laauwen},
  {Luinge}, {Ossenkopf}, {Pearson}, {Bachiller}, {Baudry}, {Benedettini},
  {Bergin}, {Benz}, {Bjerkeli}, {Blake}, {Bontemps}, {Braine}, {Bruderer},
  {Codella}, {Daniel}, {di Giorgio}, {Dominik}, {Doty}, {Encrenaz}, {Giannini},
  {Goicoechea}, {de Graauw}, {Helmich}, {Herpin}, {Hogerheijde}, {Jacq},
  {J{\o}rgensen}, {Larsson}, {Lis}, {Liseau}, {Marseille}, {Melnick}, {Nisini},
  {Olberg}, {Parise}, {Risacher}, {Santiago}, {Saraceno}, {Shipman}, {Tafalla},
  {van der Tak}, {Visser}, {Wyrowski}, and {Y{\i}ld{\i}z}]{Fich10}
{Fich},~M. et~al.  \emph{\aap} \textbf{2010}, \emph{518}, L86\relax
\mciteBstWouldAddEndPuncttrue
\mciteSetBstMidEndSepPunct{\mcitedefaultmidpunct}
{\mcitedefaultendpunct}{\mcitedefaultseppunct}\relax
\EndOfBibitem
\bibitem[{Emprechtinger} et~al.(2010){Emprechtinger}, {Lis}, {Bell},
  {Phillips}, {Schilke}, {Comito}, {Rolffs}, {van der Tak}, {Ceccarelli},
  {Aarts}, {Bacmann}, {Baudry}, {Benedettini}, {Bergin}, {Blake}, {Boogert},
  {Bottinelli}, {Cabrit}, {Caselli}, {Castets}, {Caux}, {Cernicharo},
  {Codella}, {Coutens}, {Crimier}, {Demyk}, {Dominik}, {Encrenaz}, {Falgarone},
  {Fuente}, {Gerin}, {Goldsmith}, {Helmich}, {Hennebelle}, {Henning}, {Herbst},
  {Hily-Blant}, {Jacq}, {Kahane}, {Kama}, {Klotz}, {Kooi}, {Langer}, {Lefloch},
  {Loose}, {Lord}, {Lorenzani}, {Maret}, {Melnick}, {Neufeld}, {Nisini},
  {Ossenkopf}, {Pacheco}, {Pagani}, {Parise}, {Pearson}, {Risacher}, {Salez},
  {Saraceno}, {Schuster}, {Stutzki}, {Tielens}, {van der Wiel}, {Vastel},
  {Viti}, {Wakelam}, {Walters}, {Wyrowski}, and {Yorke}]{Emprechtinger10}
{Emprechtinger},~M. et~al.  \emph{\aap} \textbf{2010}, \emph{521}, L28\relax
\mciteBstWouldAddEndPuncttrue
\mciteSetBstMidEndSepPunct{\mcitedefaultmidpunct}
{\mcitedefaultendpunct}{\mcitedefaultseppunct}\relax
\EndOfBibitem
\bibitem[{van der Tak} et~al.(2010){van der Tak}, {Marseille}, {Herpin},
  {Wyrowski}, {Baudry}, {Bontemps}, {Braine}, {Doty}, {Frieswijk}, {Melnick},
  {Shipman}, {van Dishoeck}, {Benz}, {Caselli}, {Hogerheijde}, {Johnstone},
  {Liseau}, {Bachiller}, {Benedettini}, {Bergin}, {Bjerkeli}, {Blake},
  {Bruderer}, {Cernicharo}, {Codella}, {Daniel}, {di Giorgio}, {Dominik},
  {Encrenaz}, {Fich}, {Fuente}, {Giannini}, {Goicoechea}, {de Graauw},
  {Helmich}, {Herczeg}, {J{\o}rgensen}, {Kristensen}, {Larsson}, {Lis},
  {McCoey}, {Neufeld}, {Nisini}, {Olberg}, {Parise}, {Pearson}, {Plume},
  {Risacher}, {Santiago}, {Saraceno}, {Tafalla}, {van Kempen}, {Visser},
  {Wampfler}, {Y{\i}ld{\i}z}, {Ravera}, {Roelfsema}, {Siebertz}, and
  {Teyssier}]{vanderTak10}
{van der Tak},~F.~F.~S. et~al.  \emph{\aap} \textbf{2010}, \emph{518},
  L107\relax
\mciteBstWouldAddEndPuncttrue
\mciteSetBstMidEndSepPunct{\mcitedefaultmidpunct}
{\mcitedefaultendpunct}{\mcitedefaultseppunct}\relax
\EndOfBibitem
\bibitem[{Jim{\'e}nez-Serra} et~al.(2008){Jim{\'e}nez-Serra}, {Caselli},
  {Mart{\'{\i}}n-Pintado}, and {Hartquist}]{Jimenez08}
{Jim{\'e}nez-Serra},~I.; {Caselli},~P.; {Mart{\'{\i}}n-Pintado},~J.;
  {Hartquist},~T.~W. \emph{\aap} \textbf{2008}, \emph{482}, 549--559\relax
\mciteBstWouldAddEndPuncttrue
\mciteSetBstMidEndSepPunct{\mcitedefaultmidpunct}
{\mcitedefaultendpunct}{\mcitedefaultseppunct}\relax
\EndOfBibitem
\bibitem[{Jim{\'e}nez-Serra} et~al.(2005){Jim{\'e}nez-Serra},
  {Mart{\'{\i}}n-Pintado}, {Rodr{\'{\i}}guez-Franco}, and
  {Mart{\'{\i}}n}]{Jimenez05}
{Jim{\'e}nez-Serra},~I.; {Mart{\'{\i}}n-Pintado},~J.;
  {Rodr{\'{\i}}guez-Franco},~A.; {Mart{\'{\i}}n},~S. \emph{\apjl}
  \textbf{2005}, \emph{627}, L121--L124\relax
\mciteBstWouldAddEndPuncttrue
\mciteSetBstMidEndSepPunct{\mcitedefaultmidpunct}
{\mcitedefaultendpunct}{\mcitedefaultseppunct}\relax
\EndOfBibitem
\bibitem[{Codella} et~al.(2010){Codella}, {Lefloch}, {Ceccarelli},
  {Cernicharo}, {Caux}, {Lorenzani}, {Viti}, {Hily-Blant}, {Parise}, {Maret},
  {Nisini}, {Caselli}, {Cabrit}, {Pagani}, {Benedettini}, {Boogert}, {Gueth},
  {Melnick}, {Neufeld}, {Pacheco}, {Salez}, {Schuster}, {Bacmann}, {Baudry},
  {Bell}, {Bergin}, {Blake}, {Bottinelli}, {Castets}, {Comito}, {Coutens},
  {Crimier}, {Dominik}, {Demyk}, {Encrenaz}, {Falgarone}, {Fuente}, {Gerin},
  {Goldsmith}, {Helmich}, {Hennebelle}, {Henning}, {Herbst}, {Jacq}, {Kahane},
  {Kama}, {Klotz}, {Langer}, {Lis}, {Lord}, {Pearson}, {Phillips}, {Saraceno},
  {Schilke}, {Tielens}, {van der Tak}, {van der Wiel}, {Vastel}, {Wakelam},
  {Walters}, {Wyrowski}, {Yorke}, {Borys}, {Delorme}, {Kramer}, {Larsson},
  {Mehdi}, {Ossenkopf}, and {Stutzki}]{Codella10}
{Codella},~C. et~al.  \emph{\aap} \textbf{2010}, \emph{518}, L112\relax
\mciteBstWouldAddEndPuncttrue
\mciteSetBstMidEndSepPunct{\mcitedefaultmidpunct}
{\mcitedefaultendpunct}{\mcitedefaultseppunct}\relax
\EndOfBibitem
\bibitem[{Viti} et~al.(2011){Viti}, {Jimenez-Serra}, {Yates}, {Codella},
  {Vasta}, {Caselli}, {Lefloch}, and {Ceccarelli}]{Viti11}
{Viti},~S.; {Jimenez-Serra},~I.; {Yates},~J.~A.; {Codella},~C.; {Vasta},~M.;
  {Caselli},~P.; {Lefloch},~B.; {Ceccarelli},~C. \emph{\apjl} \textbf{2011},
  \emph{740}, L3\relax
\mciteBstWouldAddEndPuncttrue
\mciteSetBstMidEndSepPunct{\mcitedefaultmidpunct}
{\mcitedefaultendpunct}{\mcitedefaultseppunct}\relax
\EndOfBibitem
\bibitem[{Keene} et~al.(1997){Keene}, {Phillips}, and {van Dishoeck}]{Keene97}
{Keene},~J.; {Phillips},~T.~G.; {van Dishoeck},~E.~F. In \emph{{Pre-shock and
  post-shock abundance ratios of atomic carbon to CO in IC 443 G.}};
  {W.~B.~Latter, S.~J.~E.~Radford, P.~R.~Jewell, J.~G.~Mangum, \& J.~Bally},,
  Ed.; IAU Symposium; Kluwer: Dordrecht, 1997; Vol. 170; pp 382--384\relax
\mciteBstWouldAddEndPuncttrue
\mciteSetBstMidEndSepPunct{\mcitedefaultmidpunct}
{\mcitedefaultendpunct}{\mcitedefaultseppunct}\relax
\EndOfBibitem
\bibitem[{Indriolo} et~al.(2010){Indriolo}, {Blake}, {Goto}, {Usuda}, {Oka},
  {Geballe}, {Fields}, and {McCall}]{Indriolo10}
{Indriolo},~N.; {Blake},~G.~A.; {Goto},~M.; {Usuda},~T.; {Oka},~T.;
  {Geballe},~T.~R.; {Fields},~B.~D.; {McCall},~B.~J. \emph{\apj} \textbf{2010},
  \emph{724}, 1357--1365\relax
\mciteBstWouldAddEndPuncttrue
\mciteSetBstMidEndSepPunct{\mcitedefaultmidpunct}
{\mcitedefaultendpunct}{\mcitedefaultseppunct}\relax
\EndOfBibitem
\bibitem[{Reach} and {Rho}(1998){Reach}, and {Rho}]{Reach98}
{Reach},~W.~T.; {Rho},~J. \emph{\apjl} \textbf{1998}, \emph{507},
  L93--L97\relax
\mciteBstWouldAddEndPuncttrue
\mciteSetBstMidEndSepPunct{\mcitedefaultmidpunct}
{\mcitedefaultendpunct}{\mcitedefaultseppunct}\relax
\EndOfBibitem
\bibitem[{Lockett} et~al.(1999){Lockett}, {Gauthier}, and {Elitzur}]{Lockett99}
{Lockett},~P.; {Gauthier},~E.; {Elitzur},~M. \emph{\apj} \textbf{1999},
  \emph{511}, 235--241\relax
\mciteBstWouldAddEndPuncttrue
\mciteSetBstMidEndSepPunct{\mcitedefaultmidpunct}
{\mcitedefaultendpunct}{\mcitedefaultseppunct}\relax
\EndOfBibitem
\bibitem[{Gonz{\'a}lez-Alfonso} et~al.(2004){Gonz{\'a}lez-Alfonso}, {Smith},
  {Fischer}, and {Cernicharo}]{Gonzalez04}
{Gonz{\'a}lez-Alfonso},~E.; {Smith},~H.~A.; {Fischer},~J.; {Cernicharo},~J.
  \emph{\apj} \textbf{2004}, \emph{613}, 247--261\relax
\mciteBstWouldAddEndPuncttrue
\mciteSetBstMidEndSepPunct{\mcitedefaultmidpunct}
{\mcitedefaultendpunct}{\mcitedefaultseppunct}\relax
\EndOfBibitem
\bibitem[{Gonz{\'a}lez-Alfonso} et~al.(2008){Gonz{\'a}lez-Alfonso}, {Smith},
  {Ashby}, {Fischer}, {Spinoglio}, and {Grundy}]{Gonzalez08}
{Gonz{\'a}lez-Alfonso},~E.; {Smith},~H.~A.; {Ashby},~M.~L.~N.; {Fischer},~J.;
  {Spinoglio},~L.; {Grundy},~T.~W. \emph{\apj} \textbf{2008}, \emph{675},
  303--315\relax
\mciteBstWouldAddEndPuncttrue
\mciteSetBstMidEndSepPunct{\mcitedefaultmidpunct}
{\mcitedefaultendpunct}{\mcitedefaultseppunct}\relax
\EndOfBibitem
\bibitem[{van der Werf} et~al.(2010){van der Werf}, {Isaak}, {Meijerink},
  {Spaans}, {Rykala}, {Fulton}, {Loenen}, {Walter}, {Wei{\ss}}, {Armus},
  {Fischer}, {Israel}, {Harris}, {Veilleux}, {Henkel}, {Savini}, {Lord},
  {Smith}, {Gonz{\'a}lez-Alfonso}, {Naylor}, {Aalto}, {Charmandaris}, {Dasyra},
  {Evans}, {Gao}, {Greve}, {G{\"u}sten}, {Kramer}, {Mart{\'{\i}}n-Pintado},
  {Mazzarella}, {Papadopoulos}, {Sanders}, {Spinoglio}, {Stacey}, {Vlahakis},
  {Wiedner}, and {Xilouris}]{vanderWerf10}
{van der Werf},~P.~P. et~al.  \emph{\aap} \textbf{2010}, \emph{518}, L42\relax
\mciteBstWouldAddEndPuncttrue
\mciteSetBstMidEndSepPunct{\mcitedefaultmidpunct}
{\mcitedefaultendpunct}{\mcitedefaultseppunct}\relax
\EndOfBibitem
\bibitem[{Wei{\ss}} et~al.(2010){Wei{\ss}}, {Requena-Torres}, {G{\"u}sten},
  {Garc{\'{\i}}a-Burillo}, {Harris}, {Israel}, {Klein}, {Kramer}, {Lord},
  {Martin-Pintado}, {R{\"o}llig}, {Stutzki}, {Szczerba}, {van der Werf},
  {Philipp-May}, {Yorke}, {Akyilmaz}, {Gal}, {Higgins}, {Marston}, {Roberts},
  {Schl{\"o}der}, {Schultz}, {Teyssier}, {Whyborn}, and {Wunsch}]{Weiss10}
{Wei{\ss}},~A. et~al.  \emph{\aap} \textbf{2010}, \emph{521}, L1\relax
\mciteBstWouldAddEndPuncttrue
\mciteSetBstMidEndSepPunct{\mcitedefaultmidpunct}
{\mcitedefaultendpunct}{\mcitedefaultseppunct}\relax
\EndOfBibitem
\bibitem[{Rangwala} et~al.(2011){Rangwala}, {Maloney}, {Glenn}, {Wilson},
  {Rykala}, {Isaak}, {Baes}, {Bendo}, {Boselli}, {Bradford}, {Clements},
  {Cooray}, {Fulton}, {Imhof}, {Kamenetzky}, {Madden}, {Mentuch}, {Sacchi},
  {Sauvage}, {Schirm}, {Smith}, {Spinoglio}, and {Wolfire}]{Rangwala11}
{Rangwala},~N. et~al.  \emph{\apj} \textbf{2011}, \emph{743}, 94\relax
\mciteBstWouldAddEndPuncttrue
\mciteSetBstMidEndSepPunct{\mcitedefaultmidpunct}
{\mcitedefaultendpunct}{\mcitedefaultseppunct}\relax
\EndOfBibitem
\bibitem[{Spinoglio} et~al.(2012){Spinoglio}, {Pereira-Santaella}, {Busquet},
  {Schirm}, {Wilson}, {Glenn}, {Kamenetzky}, {Rangwala}, {Maloney}, {Parkin},
  {Bendo}, {Madden}, {Wolfire}, {Boselli}, {Cooray}, and {Page}]{Spinoglio12}
{Spinoglio},~L. et~al.  \emph{\apj} \textbf{2012}, \emph{758}, 108\relax
\mciteBstWouldAddEndPuncttrue
\mciteSetBstMidEndSepPunct{\mcitedefaultmidpunct}
{\mcitedefaultendpunct}{\mcitedefaultseppunct}\relax
\EndOfBibitem
\bibitem[{Kamenetzky} et~al.(2012){Kamenetzky}, {Glenn}, {Rangwala}, {Maloney},
  {Bradford}, {Wilson}, {Bendo}, {Baes}, {Boselli}, {Cooray}, {Isaak},
  {Lebouteiller}, {Madden}, {Panuzzo}, {Schirm}, {Spinoglio}, and
  {Wu}]{Kamenetzky12}
{Kamenetzky},~J. et~al.  \emph{\apj} \textbf{2012}, \emph{753}, 70\relax
\mciteBstWouldAddEndPuncttrue
\mciteSetBstMidEndSepPunct{\mcitedefaultmidpunct}
{\mcitedefaultendpunct}{\mcitedefaultseppunct}\relax
\EndOfBibitem
\bibitem[{Meijerink} et~al.(2013){Meijerink}, {Kristensen}, {Wei{\ss}}, {van
  der Werf}, {Walter}, {Spaans}, {Loenen}, {Fischer}, {Israel}, {Isaak},
  {Papadopoulos}, {Aalto}, {Armus}, {Charmandaris}, {Dasyra}, {Diaz-Santos},
  {Evans}, {Gao}, {Gonz{\'a}lez-Alfonso}, {G{\"u}sten}, {Henkel}, {Kramer},
  {Lord}, {Mart{\'{\i}}n-Pintado}, {Naylor}, {Sanders}, {Smith}, {Spinoglio},
  {Stacey}, {Veilleux}, and {Wiedner}]{Meijerink13}
{Meijerink},~R. et~al.  \emph{\apjl} \textbf{2013}, \emph{762}, L16\relax
\mciteBstWouldAddEndPuncttrue
\mciteSetBstMidEndSepPunct{\mcitedefaultmidpunct}
{\mcitedefaultendpunct}{\mcitedefaultseppunct}\relax
\EndOfBibitem
\bibitem[{Yang} et~al.(2013){Yang}, {Gao}, {Omont}, {Liu}, {Isaak}, {Downes},
  {van der Werf}, and {Lu}]{Yang13}
{Yang},~C.; {Gao},~Y.; {Omont},~A.; {Liu},~D.; {Isaak},~K.~G.; {Downes},~D.;
  {van der Werf},~P.~P.; {Lu},~N. \emph{\apjl} \textbf{2013}, \emph{771},
  L24\relax
\mciteBstWouldAddEndPuncttrue
\mciteSetBstMidEndSepPunct{\mcitedefaultmidpunct}
{\mcitedefaultendpunct}{\mcitedefaultseppunct}\relax
\EndOfBibitem
\bibitem[{Riechers} et~al.(2013){Riechers}, {Bradford}, {Clements}, {Dowell},
  {P{\'e}rez-Fournon}, {Ivison}, {Bridge}, {Conley}, {Fu}, {Vieira}, {Wardlow},
  {Calanog}, {Cooray}, {Hurley}, {Neri}, {Kamenetzky}, {Aguirre}, {Altieri},
  {Arumugam}, {Benford}, {B{\'e}thermin}, {Bock}, {Burgarella},
  {Cabrera-Lavers}, {Chapman}, {Cox}, {Dunlop}, {Earle}, {Farrah}, {Ferrero},
  {Franceschini}, {Gavazzi}, {Glenn}, {Solares}, {Gurwell}, {Halpern},
  {Hatziminaoglou}, {Hyde}, {Ibar}, {Kov{\'a}cs}, {Krips}, {Lupu}, {Maloney},
  {Martinez-Navajas}, {Matsuhara}, {Murphy}, {Naylor}, {Nguyen}, {Oliver},
  {Omont}, {Page}, {Petitpas}, {Rangwala}, {Roseboom}, {Scott}, {Smith},
  {Staguhn}, {Streblyanska}, {Thomson}, {Valtchanov}, {Viero}, {Wang},
  {Zemcov}, and {Zmuidzinas}]{Riechers13}
{Riechers},~D.~A. et~al.  \emph{\nat} \textbf{2013}, \emph{496}, 329--333\relax
\mciteBstWouldAddEndPuncttrue
\mciteSetBstMidEndSepPunct{\mcitedefaultmidpunct}
{\mcitedefaultendpunct}{\mcitedefaultseppunct}\relax
\EndOfBibitem
\bibitem[{Omont} et~al.(2011){Omont}, {Neri}, {Cox}, {Lupu}, {Gu{\'e}lin}, {van
  der Werf}, {Wei{\ss}}, {Ivison}, {Negrello}, {Leeuw}, {Lehnert}, {Smail},
  {Verma}, {Baker}, {Beelen}, {Aguirre}, {Baes}, {Bertoldi}, {Clements},
  {Cooray}, {Coppin}, {Dannerbauer}, {de Zotti}, {Dye}, {Fiolet}, {Frayer},
  {Gavazzi}, {Hughes}, {Jarvis}, {Krips}, {Micha{\l}owski}, {Murphy},
  {Riechers}, {Serjeant}, {Swinbank}, {Temi}, {Vaccari}, {Vieira}, {Auld},
  {Buttiglione}, {Cava}, {Dariush}, {Dunne}, {Eales}, {Fritz}, {Gomez}, {Ibar},
  {Maddox}, {Pascale}, {Pohlen}, {Rigby}, {Smith}, {Bock}, {Bradford}, {Glenn},
  {Scott}, and {Zmuidzinas}]{Omont11}
{Omont},~A. et~al.  \emph{\aap} \textbf{2011}, \emph{530}, L3\relax
\mciteBstWouldAddEndPuncttrue
\mciteSetBstMidEndSepPunct{\mcitedefaultmidpunct}
{\mcitedefaultendpunct}{\mcitedefaultseppunct}\relax
\EndOfBibitem
\bibitem[{van der Werf} et~al.(2011){van der Werf}, {Berciano Alba}, {Spaans},
  {Loenen}, {Meijerink}, {Riechers}, {Cox}, {Wei{\ss}}, and
  {Walter}]{vanderWerf11}
{van der Werf},~P.~P.; {Berciano Alba},~A.; {Spaans},~M.; {Loenen},~A.~F.;
  {Meijerink},~R.; {Riechers},~D.~A.; {Cox},~P.; {Wei{\ss}},~A.; {Walter},~F.
  \emph{\apjl} \textbf{2011}, \emph{741}, L38\relax
\mciteBstWouldAddEndPuncttrue
\mciteSetBstMidEndSepPunct{\mcitedefaultmidpunct}
{\mcitedefaultendpunct}{\mcitedefaultseppunct}\relax
\EndOfBibitem
\bibitem[{Combes} et~al.(2012){Combes}, {Rex}, {Rawle}, {Egami}, {Boone},
  {Smail}, {Richard}, {Ivison}, {Gurwell}, {Casey}, {Omont}, {Berciano Alba},
  {Dessauges-Zavadsky}, {Edge}, {Fazio}, {Kneib}, {Okabe}, {Pell{\'o}},
  {P{\'e}rez-Gonz{\'a}lez}, {Schaerer}, {Smith}, {Swinbank}, and {van der
  Werf}]{Combes12}
{Combes},~F. et~al.  \emph{\aap} \textbf{2012}, \emph{538}, L4\relax
\mciteBstWouldAddEndPuncttrue
\mciteSetBstMidEndSepPunct{\mcitedefaultmidpunct}
{\mcitedefaultendpunct}{\mcitedefaultseppunct}\relax
\EndOfBibitem
\bibitem[{Omont} et~al.(2013){Omont}, {Yang}, {Cox}, {Neri}, {Beelen},
  {Bussmann}, {Gavazzi}, {van der Werf}, {Riechers}, {Downes}, {Krips}, {Dye},
  {Ivison}, {Vieira}, {Wei{\ss}}, {Aguirre}, {Baes}, {Baker}, {Bertoldi},
  {Cooray}, {Dannerbauer}, {De Zotti}, {Eales}, {Fu}, {Gao}, {Gu{\'e}lin},
  {Harris}, {Jarvis}, {Lehnert}, {Leeuw}, {Lupu}, {Menten}, {Micha{\l}owski},
  {Negrello}, {Serjeant}, {Temi}, {Auld}, {Dariush}, {Dunne}, {Fritz},
  {Hopwood}, {Hoyos}, {Ibar}, {Maddox}, {Smith}, {Valiante}, {Bock},
  {Bradford}, {Glenn}, and {Scott}]{Omont13}
{Omont},~A. et~al.  \emph{\aap} \textbf{2013}, \emph{551}, A115\relax
\mciteBstWouldAddEndPuncttrue
\mciteSetBstMidEndSepPunct{\mcitedefaultmidpunct}
{\mcitedefaultendpunct}{\mcitedefaultseppunct}\relax
\EndOfBibitem
\bibitem[{Habing}(1996)]{Habing96}
{Habing},~H.~J. \emph{\aapr} \textbf{1996}, \emph{7}, 97--207\relax
\mciteBstWouldAddEndPuncttrue
\mciteSetBstMidEndSepPunct{\mcitedefaultmidpunct}
{\mcitedefaultendpunct}{\mcitedefaultseppunct}\relax
\EndOfBibitem
\bibitem[{Goldreich} and {Scoville}(1976){Goldreich}, and
  {Scoville}]{Goldreich76}
{Goldreich},~P.; {Scoville},~N. \emph{\apj} \textbf{1976}, \emph{205},
  144--154\relax
\mciteBstWouldAddEndPuncttrue
\mciteSetBstMidEndSepPunct{\mcitedefaultmidpunct}
{\mcitedefaultendpunct}{\mcitedefaultseppunct}\relax
\EndOfBibitem
\bibitem[{Glassgold}(1996)]{Glassgold96}
{Glassgold},~A.~E. \emph{\araa} \textbf{1996}, \emph{34}, 241--278\relax
\mciteBstWouldAddEndPuncttrue
\mciteSetBstMidEndSepPunct{\mcitedefaultmidpunct}
{\mcitedefaultendpunct}{\mcitedefaultseppunct}\relax
\EndOfBibitem
\bibitem[{Justtanont} et~al.(1994){Justtanont}, {Skinner}, and
  {Tielens}]{Justtanont94}
{Justtanont},~K.; {Skinner},~C.~J.; {Tielens},~A.~G.~G.~M. \emph{\apj}
  \textbf{1994}, \emph{435}, 852--863\relax
\mciteBstWouldAddEndPuncttrue
\mciteSetBstMidEndSepPunct{\mcitedefaultmidpunct}
{\mcitedefaultendpunct}{\mcitedefaultseppunct}\relax
\EndOfBibitem
\bibitem[{Willacy} and {Millar}(1997){Willacy}, and {Millar}]{Willacy97}
{Willacy},~K.; {Millar},~T.~J. \emph{\aap} \textbf{1997}, \emph{324},
  237--248\relax
\mciteBstWouldAddEndPuncttrue
\mciteSetBstMidEndSepPunct{\mcitedefaultmidpunct}
{\mcitedefaultendpunct}{\mcitedefaultseppunct}\relax
\EndOfBibitem
\bibitem[{Amiri} et~al.(2011){Amiri}, {Vlemmings}, and {van
  Langevelde}]{Amiri11}
{Amiri},~N.; {Vlemmings},~W.; {van Langevelde},~H.~J. \emph{\aap}
  \textbf{2011}, \emph{532}, A149\relax
\mciteBstWouldAddEndPuncttrue
\mciteSetBstMidEndSepPunct{\mcitedefaultmidpunct}
{\mcitedefaultendpunct}{\mcitedefaultseppunct}\relax
\EndOfBibitem
\bibitem[{Barlow} et~al.(1996){Barlow}, {Nguyen-Q-Rieu}, {Truong-Bach},
  {Cernicharo}, {Gonzalez-Alfonso}, {Liu}, {Cox}, {Sylvester}, {Clegg},
  {Griffin}, {Swinyard}, {Unger}, {Baluteau}, {Caux}, {Cohen}, {Cohen},
  {Emery}, {Fischer}, {Furniss}, {Glencross}, {Greenhouse}, {Gry}, {Joubert},
  {Lim}, {Lorenzetti}, {Nisini}, {Omont}, {Orfei}, {Pequignot}, {Saraceno},
  {Serra}, {Skinner}, {Smith}, {Walker}, {Armand}, {Burgdorf}, {Ewart}, {di
  Giorgio}, {Molinari}, {Price}, {Sidher}, {Texier}, and {Trams}]{Barlow96}
{Barlow},~M.~J. et~al.  \emph{\aap} \textbf{1996}, \emph{315}, L241--L244\relax
\mciteBstWouldAddEndPuncttrue
\mciteSetBstMidEndSepPunct{\mcitedefaultmidpunct}
{\mcitedefaultendpunct}{\mcitedefaultseppunct}\relax
\EndOfBibitem
\bibitem[{Neufeld} et~al.(1996){Neufeld}, {Chen}, {Melnick}, {de Graauw},
  {Feuchtgruber}, {Haser}, {Lutz}, and {Harwit}]{Neufeld96}
{Neufeld},~D.~A.; {Chen},~W.; {Melnick},~G.~J.; {de Graauw},~T.;
  {Feuchtgruber},~H.; {Haser},~L.; {Lutz},~D.; {Harwit},~M. \emph{\aap}
  \textbf{1996}, \emph{315}, L237--L240\relax
\mciteBstWouldAddEndPuncttrue
\mciteSetBstMidEndSepPunct{\mcitedefaultmidpunct}
{\mcitedefaultendpunct}{\mcitedefaultseppunct}\relax
\EndOfBibitem
\bibitem[{Neufeld} et~al.(1999){Neufeld}, {Feuchtgruber}, {Harwit}, and
  {Melnick}]{Neufeld99}
{Neufeld},~D.~A.; {Feuchtgruber},~H.; {Harwit},~M.; {Melnick},~G.~J.
  \emph{\apjl} \textbf{1999}, \emph{517}, L147--L150\relax
\mciteBstWouldAddEndPuncttrue
\mciteSetBstMidEndSepPunct{\mcitedefaultmidpunct}
{\mcitedefaultendpunct}{\mcitedefaultseppunct}\relax
\EndOfBibitem
\bibitem[{Justtanont} et~al.(2005){Justtanont}, {Bergman}, {Larsson},
  {Olofsson}, {Sch{\"o}ier}, {Frisk}, {Hasegawa}, {Hjalmarson}, {Kwok},
  {Olberg}, {Sandqvist}, {Volk}, and {Elitzur}]{Justtanont05}
{Justtanont},~K.; {Bergman},~P.; {Larsson},~B.; {Olofsson},~H.;
  {Sch{\"o}ier},~F.~L.; {Frisk},~U.; {Hasegawa},~T.; {Hjalmarson},~{\AA}.;
  {Kwok},~S.; {Olberg},~M.; {Sandqvist},~A.; {Volk},~K.; {Elitzur},~M.
  \emph{\aap} \textbf{2005}, \emph{439}, 627--633\relax
\mciteBstWouldAddEndPuncttrue
\mciteSetBstMidEndSepPunct{\mcitedefaultmidpunct}
{\mcitedefaultendpunct}{\mcitedefaultseppunct}\relax
\EndOfBibitem
\bibitem[{Maercker} et~al.(2008){Maercker}, {Sch{\"o}ier}, {Olofsson},
  {Bergman}, and {Ramstedt}]{Maercker08}
{Maercker},~M.; {Sch{\"o}ier},~F.~L.; {Olofsson},~H.; {Bergman},~P.;
  {Ramstedt},~S. \emph{\aap} \textbf{2008}, \emph{479}, 779--791\relax
\mciteBstWouldAddEndPuncttrue
\mciteSetBstMidEndSepPunct{\mcitedefaultmidpunct}
{\mcitedefaultendpunct}{\mcitedefaultseppunct}\relax
\EndOfBibitem
\bibitem[{Maercker} et~al.(2009){Maercker}, {Sch{\"o}ier}, {Olofsson},
  {Bergman}, {Frisk}, {.~Hjalmarson}, {Justtanont}, {Kwok}, {Larsson},
  {Olberg}, and {Sandqvist}]{Maercker09}
{Maercker},~M.; {Sch{\"o}ier},~F.~L.; {Olofsson},~H.; {Bergman},~P.;
  {Frisk},~U.; {.~Hjalmarson},~{\AA}.; {Justtanont},~K.; {Kwok},~S.;
  {Larsson},~B.; {Olberg},~M.; {Sandqvist},~A. \emph{\aap} \textbf{2009},
  \emph{494}, 243--252\relax
\mciteBstWouldAddEndPuncttrue
\mciteSetBstMidEndSepPunct{\mcitedefaultmidpunct}
{\mcitedefaultendpunct}{\mcitedefaultseppunct}\relax
\EndOfBibitem
\bibitem[{Decin} et~al.(2010){Decin}, {Justtanont}, {De Beck}, {Lombaert}, {de
  Koter}, {Waters}, {Marston}, {Teyssier}, {Sch{\"o}ier}, {Bujarrabal},
  {Alcolea}, {Cernicharo}, {Dominik}, {Melnick}, {Menten}, {Neufeld},
  {Olofsson}, {Planesas}, {Schmidt}, {Szczerba}, {de Graauw}, {Helmich},
  {Roelfsema}, {Dieleman}, {Morris}, {Gallego}, {D{\'{\i}}ez-Gonz{\'a}lez}, and
  {Caux}]{Decin10b}
{Decin},~L. et~al.  \emph{\aap} \textbf{2010}, \emph{521}, L4\relax
\mciteBstWouldAddEndPuncttrue
\mciteSetBstMidEndSepPunct{\mcitedefaultmidpunct}
{\mcitedefaultendpunct}{\mcitedefaultseppunct}\relax
\EndOfBibitem
\bibitem[{Willacy}(2004)]{Willacy04}
{Willacy},~K. \emph{\apjl} \textbf{2004}, \emph{600}, L87--L90\relax
\mciteBstWouldAddEndPuncttrue
\mciteSetBstMidEndSepPunct{\mcitedefaultmidpunct}
{\mcitedefaultendpunct}{\mcitedefaultseppunct}\relax
\EndOfBibitem
\bibitem[{Justtanont} et~al.(2010){Justtanont}, {Decin}, {Sch{\"o}ier},
  {Maercker}, {Olofsson}, {Bujarrabal}, {Marston}, {Teyssier}, {Alcolea},
  {Cernicharo}, {Dominik}, {de Koter}, {Melnick}, {Menten}, {Neufeld},
  {Planesas}, {Schmidt}, {Szczerba}, {Waters}, {de Graauw}, {Whyborn}, {Finn},
  {Helmich}, {Siebertz}, {Schm{\"u}lling}, {Ossenkopf}, and
  {Lai}]{Justtanont10}
{Justtanont},~K. et~al.  \emph{\aap} \textbf{2010}, \emph{521}, L6\relax
\mciteBstWouldAddEndPuncttrue
\mciteSetBstMidEndSepPunct{\mcitedefaultmidpunct}
{\mcitedefaultendpunct}{\mcitedefaultseppunct}\relax
\EndOfBibitem
\bibitem[{Melnick} et~al.(2001){Melnick}, {Neufeld}, {Ford}, {Hollenbach}, and
  {Ashby}]{Melnick01}
{Melnick},~G.~J.; {Neufeld},~D.~A.; {Ford},~K.~E.~S.; {Hollenbach},~D.~J.;
  {Ashby},~M.~L.~N. \emph{\nat} \textbf{2001}, \emph{412}, 160--163\relax
\mciteBstWouldAddEndPuncttrue
\mciteSetBstMidEndSepPunct{\mcitedefaultmidpunct}
{\mcitedefaultendpunct}{\mcitedefaultseppunct}\relax
\EndOfBibitem
\bibitem[{Hasegawa} et~al.(2006){Hasegawa}, {Kwok}, {Koning}, {Volk},
  {Justtanont}, {Olofsson}, {Sch{\"o}ier}, {Sandqvist}, {Hjalmarson}, {Olberg},
  {Winnberg}, {Nyman}, and {Frisk}]{Hasegawa06}
{Hasegawa},~T.~I.; {Kwok},~S.; {Koning},~N.; {Volk},~K.; {Justtanont},~K.;
  {Olofsson},~H.; {Sch{\"o}ier},~F.~L.; {Sandqvist},~A.; {Hjalmarson},~{\AA}.;
  {Olberg},~M.; {Winnberg},~A.; {Nyman},~L.-{\AA}.; {Frisk},~U. \emph{\apj}
  \textbf{2006}, \emph{637}, 791--797\relax
\mciteBstWouldAddEndPuncttrue
\mciteSetBstMidEndSepPunct{\mcitedefaultmidpunct}
{\mcitedefaultendpunct}{\mcitedefaultseppunct}\relax
\EndOfBibitem
\bibitem[{Saavik Ford} and {Neufeld}(2001){Saavik Ford}, and {Neufeld}]{Ford01}
{Saavik Ford},~K.~E.; {Neufeld},~D.~A. \emph{\apjl} \textbf{2001}, \emph{557},
  L113--L116\relax
\mciteBstWouldAddEndPuncttrue
\mciteSetBstMidEndSepPunct{\mcitedefaultmidpunct}
{\mcitedefaultendpunct}{\mcitedefaultseppunct}\relax
\EndOfBibitem
\bibitem[{Decin} et~al.(2010){Decin}, {Ag{\'u}ndez}, {Barlow}, {Daniel},
  {Cernicharo}, {Lombaert}, {De Beck}, {Royer}, {Vandenbussche}, {Wesson},
  {Polehampton}, {Blommaert}, {De Meester}, {Exter}, {Feuchtgruber}, {Gear},
  {Gomez}, {Groenewegen}, {Gu{\'e}lin}, {Hargrave}, {Huygen}, {Imhof},
  {Ivison}, {Jean}, {Kahane}, {Kerschbaum}, {Leeks}, {Lim}, {Matsuura},
  {Olofsson}, {Posch}, {Regibo}, {Savini}, {Sibthorpe}, {Swinyard}, {Yates},
  and {Waelkens}]{Decin10}
{Decin},~L. et~al.  \emph{\nat} \textbf{2010}, \emph{467}, 64--67\relax
\mciteBstWouldAddEndPuncttrue
\mciteSetBstMidEndSepPunct{\mcitedefaultmidpunct}
{\mcitedefaultendpunct}{\mcitedefaultseppunct}\relax
\EndOfBibitem
\bibitem[{Neufeld} et~al.(2011){Neufeld}, {Gonz{\'a}lez-Alfonso}, {Melnick},
  {Szczerba}, {Schmidt}, {Decin}, {de Koter}, {Sch{\"o}ier}, and
  {Cernicharo}]{Neufeld11a}
{Neufeld},~D.~A.; {Gonz{\'a}lez-Alfonso},~E.; {Melnick},~G.~J.; {Szczerba},~R.;
  {Schmidt},~M.; {Decin},~L.; {de Koter},~A.; {Sch{\"o}ier},~F.;
  {Cernicharo},~J. \emph{\apjl} \textbf{2011}, \emph{727}, L28\relax
\mciteBstWouldAddEndPuncttrue
\mciteSetBstMidEndSepPunct{\mcitedefaultmidpunct}
{\mcitedefaultendpunct}{\mcitedefaultseppunct}\relax
\EndOfBibitem
\bibitem[{Cherchneff}(2011)]{Cherchneff11}
{Cherchneff},~I. \emph{\aap} \textbf{2011}, \emph{526}, L11\relax
\mciteBstWouldAddEndPuncttrue
\mciteSetBstMidEndSepPunct{\mcitedefaultmidpunct}
{\mcitedefaultendpunct}{\mcitedefaultseppunct}\relax
\EndOfBibitem
\bibitem[{Neufeld} et~al.(2011){Neufeld}, {Gonz{\'a}lez-Alfonso}, {Melnick},
  {Szczerba}, {Schmidt}, {Decin}, {Alcolea}, {de Koter}, {Sch{\"o}ier},
  {Bujarrabal}, {Cernicharo}, {Dominik}, {Justtanont}, {Marston}, {Menten},
  {Olofsson}, {Planesas}, {Teyssier}, and {Waters}]{Neufeld11b}
{Neufeld},~D.~A. et~al.  \emph{\apjl} \textbf{2011}, \emph{727}, L29\relax
\mciteBstWouldAddEndPuncttrue
\mciteSetBstMidEndSepPunct{\mcitedefaultmidpunct}
{\mcitedefaultendpunct}{\mcitedefaultseppunct}\relax
\EndOfBibitem
\bibitem[{Woitke} et~al.(2009){Woitke}, {Thi}, {Kamp}, and
  {Hogerheijde}]{Woitke09h2o}
{Woitke},~P.; {Thi},~W.-F.; {Kamp},~I.; {Hogerheijde},~M.~R. \emph{\aap}
  \textbf{2009}, \emph{501}, L5--L8\relax
\mciteBstWouldAddEndPuncttrue
\mciteSetBstMidEndSepPunct{\mcitedefaultmidpunct}
{\mcitedefaultendpunct}{\mcitedefaultseppunct}\relax
\EndOfBibitem
\bibitem[{Bethell} and {Bergin}(2009){Bethell}, and {Bergin}]{Bethell09}
{Bethell},~T.; {Bergin},~E. \emph{Science} \textbf{2009}, \emph{326},
  1675\relax
\mciteBstWouldAddEndPuncttrue
\mciteSetBstMidEndSepPunct{\mcitedefaultmidpunct}
{\mcitedefaultendpunct}{\mcitedefaultseppunct}\relax
\EndOfBibitem
\bibitem[{Thi} and {Bik}(2005){Thi}, and {Bik}]{Thi05}
{Thi},~W.-F.; {Bik},~A. \emph{\aap} \textbf{2005}, \emph{438}, 557--570\relax
\mciteBstWouldAddEndPuncttrue
\mciteSetBstMidEndSepPunct{\mcitedefaultmidpunct}
{\mcitedefaultendpunct}{\mcitedefaultseppunct}\relax
\EndOfBibitem
\bibitem[{Woitke} et~al.(2009){Woitke}, {Kamp}, and {Thi}]{Woitke09}
{Woitke},~P.; {Kamp},~I.; {Thi},~W.-F. \emph{\aap} \textbf{2009}, \emph{501},
  383--406\relax
\mciteBstWouldAddEndPuncttrue
\mciteSetBstMidEndSepPunct{\mcitedefaultmidpunct}
{\mcitedefaultendpunct}{\mcitedefaultseppunct}\relax
\EndOfBibitem
\bibitem[{Glassgold} et~al.(2009){Glassgold}, {Meijerink}, and
  {Najita}]{Glassgold09}
{Glassgold},~A.~E.; {Meijerink},~R.; {Najita},~J.~R. \emph{\apj} \textbf{2009},
  \emph{701}, 142--153\relax
\mciteBstWouldAddEndPuncttrue
\mciteSetBstMidEndSepPunct{\mcitedefaultmidpunct}
{\mcitedefaultendpunct}{\mcitedefaultseppunct}\relax
\EndOfBibitem
\bibitem[{Salyk} et~al.(2011){Salyk}, {Pontoppidan}, {Blake}, {Najita}, and
  {Carr}]{Salyk11}
{Salyk},~C.; {Pontoppidan},~K.~M.; {Blake},~G.~A.; {Najita},~J.~R.;
  {Carr},~J.~S. \emph{\apj} \textbf{2011}, \emph{731}, 130\relax
\mciteBstWouldAddEndPuncttrue
\mciteSetBstMidEndSepPunct{\mcitedefaultmidpunct}
{\mcitedefaultendpunct}{\mcitedefaultseppunct}\relax
\EndOfBibitem
\bibitem[{Pontoppidan} et~al.(2010){Pontoppidan}, {Salyk}, {Blake}, and
  {K{\"a}ufl}]{Pontoppidan10visir}
{Pontoppidan},~K.~M.; {Salyk},~C.; {Blake},~G.~A.; {K{\"a}ufl},~H.~U.
  \emph{\apjl} \textbf{2010}, \emph{722}, L173--L177\relax
\mciteBstWouldAddEndPuncttrue
\mciteSetBstMidEndSepPunct{\mcitedefaultmidpunct}
{\mcitedefaultendpunct}{\mcitedefaultseppunct}\relax
\EndOfBibitem
\bibitem[{Mandell} et~al.(2012){Mandell}, {Bast}, {van Dishoeck}, {Blake},
  {Salyk}, {Mumma}, and {Villanueva}]{Mandell12}
{Mandell},~A.~M.; {Bast},~J.; {van Dishoeck},~E.~F.; {Blake},~G.~A.;
  {Salyk},~C.; {Mumma},~M.~J.; {Villanueva},~G. \emph{\apj} \textbf{2012},
  \emph{747}, 92\relax
\mciteBstWouldAddEndPuncttrue
\mciteSetBstMidEndSepPunct{\mcitedefaultmidpunct}
{\mcitedefaultendpunct}{\mcitedefaultseppunct}\relax
\EndOfBibitem
\bibitem[{Fedele} et~al.(2011){Fedele}, {Pascucci}, {Brittain}, {Kamp},
  {Woitke}, {Williams}, {Dent}, and {Thi}]{Fedele11}
{Fedele},~D.; {Pascucci},~I.; {Brittain},~S.; {Kamp},~I.; {Woitke},~P.;
  {Williams},~J.~P.; {Dent},~W.~R.~F.; {Thi},~W.-F. \emph{\apj} \textbf{2011},
  \emph{732}, 106\relax
\mciteBstWouldAddEndPuncttrue
\mciteSetBstMidEndSepPunct{\mcitedefaultmidpunct}
{\mcitedefaultendpunct}{\mcitedefaultseppunct}\relax
\EndOfBibitem
\bibitem[{Zhang} et~al.(2013){Zhang}, {Pontoppidan}, {Salyk}, and
  {Blake}]{Zhang13}
{Zhang},~K.; {Pontoppidan},~K.~M.; {Salyk},~C.; {Blake},~G.~A. \emph{\apj}
  \textbf{2013}, \emph{766}, 82\relax
\mciteBstWouldAddEndPuncttrue
\mciteSetBstMidEndSepPunct{\mcitedefaultmidpunct}
{\mcitedefaultendpunct}{\mcitedefaultseppunct}\relax
\EndOfBibitem
\bibitem[{Rivi\`ere-Marichalar} et~al.(2012){Rivi\`ere-Marichalar},
  {M{\'e}nard}, {Thi}, {Kamp}, {Montesinos}, {Meeus}, {Woitke}, {Howard},
  {Sandell}, {Podio}, {Dent}, {Mendigut{\'{\i}}a}, {Pinte}, {White}, and
  {Barrado}]{Riviere12}
{Rivi\`ere-Marichalar},~P.; {M{\'e}nard},~F.; {Thi},~W.~F.; {Kamp},~I.;
  {Montesinos},~B.; {Meeus},~G.; {Woitke},~P.; {Howard},~C.; {Sandell},~G.;
  {Podio},~L.; {Dent},~W.~R.~F.; {Mendigut{\'{\i}}a},~I.; {Pinte},~C.;
  {White},~G.~J.; {Barrado},~D. \emph{\aap} \textbf{2012}, \emph{538}, L3\relax
\mciteBstWouldAddEndPuncttrue
\mciteSetBstMidEndSepPunct{\mcitedefaultmidpunct}
{\mcitedefaultendpunct}{\mcitedefaultseppunct}\relax
\EndOfBibitem
\bibitem[{Fedele} et~al.(2012){Fedele}, {Bruderer}, {van Dishoeck}, {Herczeg},
  {Evans}, {Bouwman}, {Henning}, and {Green}]{Fedele12}
{Fedele},~D.; {Bruderer},~S.; {van Dishoeck},~E.~F.; {Herczeg},~G.~J.;
  {Evans},~N.~J.; {Bouwman},~J.; {Henning},~T.; {Green},~J. \emph{\aap}
  \textbf{2012}, \emph{544}, L9\relax
\mciteBstWouldAddEndPuncttrue
\mciteSetBstMidEndSepPunct{\mcitedefaultmidpunct}
{\mcitedefaultendpunct}{\mcitedefaultseppunct}\relax
\EndOfBibitem
\bibitem[{Meeus} et~al.(2012){Meeus}, {Montesinos}, {Mendigut{\'{\i}}a},
  {Kamp}, {Thi}, {Eiroa}, {Grady}, {Mathews}, {Sandell}, {Martin-Za{\"i}di},
  {Brittain}, {Dent}, {Howard}, {M{\'e}nard}, {Pinte}, {Roberge},
  {Vandenbussche}, and {Williams}]{Meeus12}
{Meeus},~G. et~al.  \emph{\aap} \textbf{2012}, \emph{544}, A78\relax
\mciteBstWouldAddEndPuncttrue
\mciteSetBstMidEndSepPunct{\mcitedefaultmidpunct}
{\mcitedefaultendpunct}{\mcitedefaultseppunct}\relax
\EndOfBibitem
\bibitem[{Fedele} et~al.(2013){Fedele}, {Bruderer}, {van Dishoeck}, {Carr},
  {Herczeg}, {Salyk}, {Evans}, {Bouwman}, {Meeus}, {Henning}, {Green},
  {Najita}, and {Guedel}]{Fedele13}
{Fedele},~D.; {Bruderer},~S.; {van Dishoeck},~E.~F.; {Carr},~J.;
  {Herczeg},~G.~J.; {Salyk},~C.; {Evans},~N.~J.,~II; {Bouwman},~J.;
  {Meeus},~G.; {Henning},~T.; {Green},~J.; {Najita},~J.~R.; {Guedel},~M.
  \emph{\aap} \textbf{2013}, in press\relax
\mciteBstWouldAddEndPuncttrue
\mciteSetBstMidEndSepPunct{\mcitedefaultmidpunct}
{\mcitedefaultendpunct}{\mcitedefaultseppunct}\relax
\EndOfBibitem
\bibitem[{Whipple}(1951)]{Whipple51}
{Whipple},~F.~L. \emph{\apj} \textbf{1951}, \emph{113}, 464\relax
\mciteBstWouldAddEndPuncttrue
\mciteSetBstMidEndSepPunct{\mcitedefaultmidpunct}
{\mcitedefaultendpunct}{\mcitedefaultseppunct}\relax
\EndOfBibitem
\bibitem[{Haser}(1957)]{Haser57}
{Haser},~L. \emph{{\it Bulletin de la Societe Royale des Sciences de Li\`ege}}
  \textbf{1957}, \emph{43}, 740\relax
\mciteBstWouldAddEndPuncttrue
\mciteSetBstMidEndSepPunct{\mcitedefaultmidpunct}
{\mcitedefaultendpunct}{\mcitedefaultseppunct}\relax
\EndOfBibitem
\bibitem[{Festou}(1981)]{Festou81}
{Festou},~M.~C. \emph{\aap} \textbf{1981}, \emph{95}, 69--79\relax
\mciteBstWouldAddEndPuncttrue
\mciteSetBstMidEndSepPunct{\mcitedefaultmidpunct}
{\mcitedefaultendpunct}{\mcitedefaultseppunct}\relax
\EndOfBibitem
\bibitem[{Crovisier}(1989)]{Crovisier89}
{Crovisier},~J. \emph{\aap} \textbf{1989}, \emph{213}, 459--464\relax
\mciteBstWouldAddEndPuncttrue
\mciteSetBstMidEndSepPunct{\mcitedefaultmidpunct}
{\mcitedefaultendpunct}{\mcitedefaultseppunct}\relax
\EndOfBibitem
\bibitem[{Crovisier} et~al.(1997){Crovisier}, {Leech}, {Bockelee-Morvan},
  {Brooke}, {Hanner}, {Altieri}, {Keller}, and {Lellouch}]{Crovisier97}
{Crovisier},~J.; {Leech},~K.; {Bockelee-Morvan},~D.; {Brooke},~T.~Y.;
  {Hanner},~M.~S.; {Altieri},~B.; {Keller},~H.~U.; {Lellouch},~E.
  \emph{\science} \textbf{1997}, \emph{275}, 1904--1907\relax
\mciteBstWouldAddEndPuncttrue
\mciteSetBstMidEndSepPunct{\mcitedefaultmidpunct}
{\mcitedefaultendpunct}{\mcitedefaultseppunct}\relax
\EndOfBibitem
\bibitem[{Mumma} et~al.(1993){Mumma}, {Weissman}, and {Stern}]{Mumma93}
{Mumma},~M.~J.; {Weissman},~P.~R.; {Stern},~S.~A. In \emph{Protostars and
  Planets III}; {Levy},~E.~H., {Lunine},~J.~I., Eds.; Univ. of Arizona Press:
  Tucson, 1993; pp 1177--1252\relax
\mciteBstWouldAddEndPuncttrue
\mciteSetBstMidEndSepPunct{\mcitedefaultmidpunct}
{\mcitedefaultendpunct}{\mcitedefaultseppunct}\relax
\EndOfBibitem
\bibitem[{Bockel{\'e}e-Morvan} et~al.(2004){Bockel{\'e}e-Morvan}, {Crovisier},
  {Mumma}, and {Weaver}]{Bockelee04}
{Bockel{\'e}e-Morvan},~D.; {Crovisier},~J.; {Mumma},~M.~J.; {Weaver},~H.~A. In
  \emph{Comets II}; {Festou, M.~C., Keller, H.~U., \& Weaver, H.~A.},, Ed.;
  Univ. of Arizona Press: Tucson, 2004; p 391\relax
\mciteBstWouldAddEndPuncttrue
\mciteSetBstMidEndSepPunct{\mcitedefaultmidpunct}
{\mcitedefaultendpunct}{\mcitedefaultseppunct}\relax
\EndOfBibitem
\bibitem[{Brooke} et~al.(1996){Brooke}, {Sellgren}, and {Smith}]{Brooke96}
{Brooke},~T.~Y.; {Sellgren},~K.; {Smith},~R.~G. \emph{\apj} \textbf{1996},
  \emph{459}, 209\relax
\mciteBstWouldAddEndPuncttrue
\mciteSetBstMidEndSepPunct{\mcitedefaultmidpunct}
{\mcitedefaultendpunct}{\mcitedefaultseppunct}\relax
\EndOfBibitem
\bibitem[Magee-Sauer et~al.({2002})Magee-Sauer, Mumma, DiSanti, and
  Dello~Russo]{Magee02}
Magee-Sauer,~K.; Mumma,~M.; DiSanti,~M.; Dello~Russo,~N. \emph{{\jgr}}
  \textbf{{2002}}, \emph{{107}}\relax
\mciteBstWouldAddEndPuncttrue
\mciteSetBstMidEndSepPunct{\mcitedefaultmidpunct}
{\mcitedefaultendpunct}{\mcitedefaultseppunct}\relax
\EndOfBibitem
\bibitem[{Gibb} et~al.(2003){Gibb}, {Mumma}, {dello Russo}, {Disanti}, and
  {Magee-Sauer}]{Gibb03}
{Gibb},~E.~L.; {Mumma},~M.~J.; {dello Russo},~N.; {Disanti},~M.~A.;
  {Magee-Sauer},~K. \emph{\icarus} \textbf{2003}, \emph{165}, 391--406\relax
\mciteBstWouldAddEndPuncttrue
\mciteSetBstMidEndSepPunct{\mcitedefaultmidpunct}
{\mcitedefaultendpunct}{\mcitedefaultseppunct}\relax
\EndOfBibitem
\bibitem[{Bonev} et~al.(2004){Bonev}, {Mumma}, {Dello Russo}, {Gibb},
  {DiSanti}, and {Magee-Sauer}]{Bonev04}
{Bonev},~B.~P.; {Mumma},~M.~J.; {Dello Russo},~N.; {Gibb},~E.~L.;
  {DiSanti},~M.~A.; {Magee-Sauer},~K. \emph{\apj} \textbf{2004}, \emph{615},
  1048--1053\relax
\mciteBstWouldAddEndPuncttrue
\mciteSetBstMidEndSepPunct{\mcitedefaultmidpunct}
{\mcitedefaultendpunct}{\mcitedefaultseppunct}\relax
\EndOfBibitem
\bibitem[{Bonev} and {Mumma}(2006){Bonev}, and {Mumma}]{Bonev06}
{Bonev},~B.~P.; {Mumma},~M.~J. \emph{\apj} \textbf{2006}, \emph{653},
  788--791\relax
\mciteBstWouldAddEndPuncttrue
\mciteSetBstMidEndSepPunct{\mcitedefaultmidpunct}
{\mcitedefaultendpunct}{\mcitedefaultseppunct}\relax
\EndOfBibitem
\bibitem[{Andresen} et~al.(1983){Andresen}, {Ondrey}, and {Titze}]{Andresen83}
{Andresen},~P.; {Ondrey},~G.~S.; {Titze},~B. \emph{\prl} \textbf{1983},
  \emph{50}, 486--488\relax
\mciteBstWouldAddEndPuncttrue
\mciteSetBstMidEndSepPunct{\mcitedefaultmidpunct}
{\mcitedefaultendpunct}{\mcitedefaultseppunct}\relax
\EndOfBibitem
\bibitem[{Raymond}(1979)]{Raymond79}
{Raymond},~J.~C. \emph{\apjs} \textbf{1979}, \emph{39}, 1--27\relax
\mciteBstWouldAddEndPuncttrue
\mciteSetBstMidEndSepPunct{\mcitedefaultmidpunct}
{\mcitedefaultendpunct}{\mcitedefaultseppunct}\relax
\EndOfBibitem
\bibitem[{Tappe} et~al.(2008){Tappe}, {Lada}, {Black}, and {Muench}]{Tappe08}
{Tappe},~A.; {Lada},~C.~J.; {Black},~J.~H.; {Muench},~A.~A. \emph{\apjl}
  \textbf{2008}, \emph{680}, L117--L120\relax
\mciteBstWouldAddEndPuncttrue
\mciteSetBstMidEndSepPunct{\mcitedefaultmidpunct}
{\mcitedefaultendpunct}{\mcitedefaultseppunct}\relax
\EndOfBibitem
\bibitem[{Tappe} et~al.(2012){Tappe}, {Forbrich}, {Mart{\'{\i}}n}, {Yuan}, and
  {Lada}]{Tappe12}
{Tappe},~A.; {Forbrich},~J.; {Mart{\'{\i}}n},~S.; {Yuan},~Y.; {Lada},~C.~J.
  \emph{\apj} \textbf{2012}, \emph{751}, 9\relax
\mciteBstWouldAddEndPuncttrue
\mciteSetBstMidEndSepPunct{\mcitedefaultmidpunct}
{\mcitedefaultendpunct}{\mcitedefaultseppunct}\relax
\EndOfBibitem
\bibitem[Buntkowsky et~al.({2008})Buntkowsky, Limbach, Walaszek, Adamczyk, Xu,
  Breitzke, Schweitzer, Gutmann, Waechtler, Frydel, Elnmler, Amadeu, Tietze,
  and Chaudret]{Buntkowsky08}
Buntkowsky,~G.; Limbach,~H.~H.; Walaszek,~B.; Adamczyk,~A.; Xu,~Y.;
  Breitzke,~H.; Schweitzer,~A.; Gutmann,~T.; Waechtler,~M.; Frydel,~J.;
  Elnmler,~T.; Amadeu,~N.; Tietze,~D.; Chaudret,~B. \emph{{\zpc}}
  \textbf{{2008}}, \emph{{222}}, {1049--1063}\relax
\mciteBstWouldAddEndPuncttrue
\mciteSetBstMidEndSepPunct{\mcitedefaultmidpunct}
{\mcitedefaultendpunct}{\mcitedefaultseppunct}\relax
\EndOfBibitem
\bibitem[Oka({2004})]{Oka04}
Oka,~T. \emph{{\jms}} \textbf{{2004}}, \emph{{228}}, {635--639}\relax
\mciteBstWouldAddEndPuncttrue
\mciteSetBstMidEndSepPunct{\mcitedefaultmidpunct}
{\mcitedefaultendpunct}{\mcitedefaultseppunct}\relax
\EndOfBibitem
\bibitem[{Mumma} and {Charnley}(2011){Mumma}, and {Charnley}]{Mumma11}
{Mumma},~M.~J.; {Charnley},~S.~B. \emph{\araa} \textbf{2011}, \emph{49},
  471--524\relax
\mciteBstWouldAddEndPuncttrue
\mciteSetBstMidEndSepPunct{\mcitedefaultmidpunct}
{\mcitedefaultendpunct}{\mcitedefaultseppunct}\relax
\EndOfBibitem
\bibitem[Sliter et~al.({2011})Sliter, Gish, and Vilesov]{Sliter11}
Sliter,~R.; Gish,~M.; Vilesov,~A.~F. \emph{{J. Phys. Chem. A}} \textbf{{2011}},
  \emph{{115}}, {9682--9688}\relax
\mciteBstWouldAddEndPuncttrue
\mciteSetBstMidEndSepPunct{\mcitedefaultmidpunct}
{\mcitedefaultendpunct}{\mcitedefaultseppunct}\relax
\EndOfBibitem
\bibitem[{Hama} et~al.(2011){Hama}, {Watanabe}, {Kouchi}, and
  {Yokoyama}]{Hama11}
{Hama},~T.; {Watanabe},~N.; {Kouchi},~A.; {Yokoyama},~M. \emph{\apjl}
  \textbf{2011}, \emph{738}, L15\relax
\mciteBstWouldAddEndPuncttrue
\mciteSetBstMidEndSepPunct{\mcitedefaultmidpunct}
{\mcitedefaultendpunct}{\mcitedefaultseppunct}\relax
\EndOfBibitem
\end{mcitethebibliography}

\end{document}